\documentclass[superscriptaddress,11pt]{article}
\DeclareMathAlphabet\mathbfcal{OMS}{cmsy}{b}{n}
\usepackage{authblk}
\usepackage{array}
\usepackage{type1cm}
\usepackage{lettrine}
\usepackage{graphicx}
\usepackage{geometry}
\usepackage{psfrag}
\usepackage{amsmath}
\usepackage{amsfonts}
\usepackage{appendix}
\usepackage[margin=10pt,font=footnotesize,labelfont=sc,format=hang,labelformat=parens]%
{caption}
\usepackage{subcaption}
\usepackage{float}
\usepackage{accents}
\usepackage{amssymb}%
\providecommand{\U}[1]{\protect\rule{.1in}{.1in}}
\numberwithin{equation}{section}
\def\be{\begin{equation}}
\def\ee{\end{equation}}
\def\ba{\begin{eqnarray}}
\def\ea{\end{eqnarray}}
\def\ei{\end{itemize}}

\newcommand*{\Perm}[2]{{}^{#1}\!P_{#2}}%

\def\J{\vec{\hat J}}
\def\tr{\text{Tr}}
\def\O{{\bf O}}

\def\F{{\cal F}}

\def\bra{\langle}
\def\ket{\rangle}

\def\bi{\bar i}

\def\bh{\bar h}
\def\bI{\bar I}
\def\bv{\bar v}
\def\bx{\bar x}
\def\bt{\bar t}
\def\bl{\bar l}

\def\bn{\bar n}
\def\bx{\bar x}
\def\by{\bar y}
\def\bz{\bar z}

\def\di{i^{\prime}}
\def\dj{j^{\prime}}
\def\dk{k^{\prime}}

\def\dm{m^{\prime}}

\def\e{\epsilon}
\def\bre{\bar{\epsilon}}

\def\d{\delta}

\begin{document}

\title{Anomaly free quantum dynamics  for Euclidean LQG}

%\author[a]{Abhay Ashtekar}
\author{Madhavan Varadarajan}
%\affil[a]{IGC, Pennsylvania State University, University Park, PA 16801, USA}
\affil{Raman Research Institute\\Bangalore-560 080, India}
\maketitle

\begin{abstract}
%Canonical Loop Quantum Gravity  is an attempt at non-perturbative canonical quantization of a classical Hamiltonian 
%description of gravity in terms of 
%$su(2)$ Electric fields  and conjugate connections. In this classical description,
Classical gravitational evolution  admits an elegant and compact re-expression in terms of gauge covariant generalizations of Lie derivatives
%{\em spatial} diffeomorphisms generated by a phase space dependent 
with respect to a spatial phase space dependent $su(2)$ valued vector field called the Electric Shift \cite{aame}.
A quantum dynamics for Euclidean Loop Quantum Gravity which ascribes a central role to the Electric Shift operator is 
derived in \cite{p4}. {\em Here we show that this quantum dynamics is non-trivially anomaly free}. Specifically, we show that on a suitable space of off shell states (a) the (non-vanishing) commutator between a pair of Hamiltonian 
constraint operators   mirrors the Poisson bracket between their classical correspondents, (b) the group of finite spatial diffeomorphisms is faithfully represented and
%, modulo certain technical assummptions, 
(c) the action of the Hamiltonian constraint operator  is diffeomorphism covariant with respect to the action of 
spatial diffeomorphisms.
%and (c) the group of finite spatial diffeomorphisms is faithfully represented. 

\end{abstract}

\thispagestyle{empty}
\let\oldthefootnote\thefootnote\renewcommand{\thefootnote}{\fnsymbol{footnote}}
\footnotetext{Email: madhavan@rri.res.in}
\let\thefootnote\oldthefootnote

\section{Introduction \label{sec1}}

The constraints of classical General Relativity in its Hamiltonian formulation admit a dramatic simplification when expressed in terms of the variables discovered by Ashtekar \cite{aanew}.
These variables comprise of a self dual $su(2)$ connection $A_a^i$ and its conjugate electric field $E^a_i$ with $i\in su(2)$ and $a$ being a tangent space index on the spatial  slice $\Sigma$.
While the connection $A_a^i$ corresponds to the pull back to $\Sigma$ of the self dual part of the spacetime spin connection, the electric field bears the interpretation of a densitized spatial triad and hence
determines the spatial geometry. Recently \cite{aame} it was shown that not only do the gravitational constraints simplify, but that the evolution equations also admit a simple and geometrically intriguing form
in the Ashtekar variables.  Specifically the evolution equations can be expressed in terms of certain gauge covariant generalizations of Lie derivatives with respect to an $su(2)$ valued spatial vector field
constructed from $E^a_i$ and the lapse. This phase space dependent object is called the Electric Shift.
While the considerations of \cite{aanew,aame} apply to both  Lorentzian and  Riemannian General Relativity,
%The constraints of vacuum  General Relativity assume a simple form in these variables.
for Lorentzian spacetime metrics the Ashtekar variables are complex and subject to certain `reality' conditions which are difficult to impose directly at the quantum level. In contrast, 
for Riemannian spacetime metrics (herein referred to as `Euclidean') the Ashtekar variables are real and can be directly employed in quantum theory. 
In \cite{p4}, motivated by the structure of the classical theory elucidated in \cite{aame},  we derived an action of the quantum Hamiltonian constraint for Euclidean Loop Quantum Gravity
which accords a central role to the quantum operator correspondent of the Electric Shift.
\footnote{For a first stab at the problem, see \cite{laddha}.}
Here we show that the action so derived is consistent with a non-trivial anomaly free algebra of quantum constraints. 
Our interest in Euclidean LQG in \cite{p4} and in this work
stems from the fact that progress in 
%that not only is the Euclidean theory and excellent toy model for the Lorentzian theory, but also that progress in 
Lorentzian LQG is contingent on a thorough understanding of Euclidean LQG \cite{qsd,ttcomplexifier,aacomplexifier,mvcomplexifier}.

In what follows, we shall assume familiarity with the contents of Reference \cite{p4}, which are summarized in Section \ref{sec2.1} for the convenience of the reader. Our starting point is the `Mixed Action' of section 3 of Reference \cite{p4}  and its generalization in Appendix B of that work.
This  Mixed Action version of the regulated Hamiltonian constraint at regulator value $\epsilon$ acts on spin network states at their vertices and deforms the graph structure in a neighborhood of coordinate size $\epsilon$ of each such vertex.
For our purposes it is necessary to modify these deformations slightly on scales much smaller than $\epsilon$. As we shall see these minor modifications on the one hand still  yield valid regulated constraint operators
and on the other, play a key simplifying role in our demonstration of an anomaly free constraint algebra. The tools and techniques we use in this work have been developed over the last decade 
in a series of papers \cite{p1,p2}, and most importantly, \cite{p3}, in the context of  a toy model for Euclidean gravity which can be seen to arise as a novel weak coupling limit of Euclidean gravity \cite{leeg0}.
While familiarity with these works would be helpful, we shall endeavour to make our presentation self contained except for (as mentioned above)  an assumed familiarity with the contents of \cite{p4}.
For readers familiar with \cite{p3}, an important simplification in this work concerns the use of Riemann Normal Coordinates as regulating structures instead of  the complicated network of coordinate patches used in \cite{p3}. 
%We shall elaborate on this choice
%as we go along.

We now  provide a rough summary of our considerations and results and then go on to a detailed technical description of the layout of the paper.
The intent of the rough summary is to convey only a qualitative feel for  the way in which our calculations work and the reader should not worry if certain details are not clear.
Indeed, we urge the reader to peruse this summary a second time after going through the technical aspects of the calculation described in subsequent sections of this work.
The mixed action of \cite{p4} of the Hamiltonian constraint, at regulator value $\epsilon$,  on a spin network is non trivial only at vertices
of the spin network which have non-zero volume. This action at each such vertex is a sum of two types of actions, each of which deforms the spin network graph and its labels in a small `$\e$ size'vicinity of the vertex. 
 The first type closely resembles the action of a diffeomorphism on the spin network which displaces the vertex and the structure around it  by an amount $\e$. 
 We shall refer to the deformed spin nets created by this action as ``electric diffeomorphism''
 type `children'  of the `parent' spin net being acted upon. 
 \footnote{The reason for the nomenclature `electric diffeomorphisms' is that this deformation is a more singular version of a diffeomorphism and corresponds 
 to the finite transformation at parameter $\e$ generated by the quantum counterpart of the electric shift \cite{p4}.}
 In contrast the second type does not result in a displaced vertex structure;
rather, it generates spin network deformations which  resemble  the `QSD' type deformations \cite{qsd} in that an `extraordinary edge' is placed in turn between each pair of existing edges emanating from the parent vertex. 
Similar to the QSD case \cite{ttme}, this second set of terms is responsible for the propagation of quantum gravitational perturbations and we refer to these terms as `propagation' type children.
Both the electric diffeomorphism and the propagation type actions also generate a set of kinks in each child in an $\e$ vicinity of the parental vertex being acted upon.

The off shell states we construct are linear combinations of `basis' off shell states. Each such basis state is a distributional state labelled by a  complex function $f$ on the Cauchy slice  $\Sigma$ and  a 3d Riemannian metric $h$ on $\Sigma$. 
%The vertex smooth function is a function on the spatial slice. 
Such a basis  off shell state  can be thought of as an infinite non-normalizable sum over spin network `bra states', these bra states being of a specific type,  with the coefficients in this sum being determined by $f, h$. 
The coefficient of a spin network bra summand is a product of an $h$ dependent coefficient and an $f$ dependent one. The $f$ dependent coefficient is simply the product of the evaluations of $f$ on each vertex of the 
spin network.
%which carries non-zero volume. 
The $h$ dependent coefficient is a specific combination of the distances between the kinks on the spin network graph.
We are interested in the evaluation of the (dual) action of the Hamiltonian constraint on such an  off shell state in the `continuum' limit of $\e\rightarrow 0$.

The (dual) action of the Hamiltonian constraint on such an off shell state has a diffeomorphism part and a propagation part as described above. Each of these has the effect of altering the coefficients in the sum over spin network bras.
The diffeomorphism part in the $\e\rightarrow 0$ continuum limit, at each vertex (of nonzero volume) of a bra, alters both the $f$ dependent part as well as the $h$ dependent part of the coefficient because as $\e\rightarrow 0$ 
the displaced vertex as well as the kinks in the `children'  bras shrink to a single point corresponding to the `parental' vertex. The net result is that this action 
replaces, in the coefficient of a parent `bra',  the evaluation of $f$ at a vertex of this  parent by 
a sum of its derivatives along  edges emanating from that vertex, multiplied by the evaluation of the lapse at that vertex. 
The propagation part primarily affects the $h$ dependent part of the coefficient through the shrinking of kinks and the result is to reduce the coefficient of the parent bra 
to zero. To summarise: the dual action of the Hamiltonian constraint on an off shell state is the sum of an electric diffeomorphism  part and a propagation part. In the continuum limit action on an off shell state, the propagation contribution vanishes
and we are left only with the electric diffeomorphism contribution. As a result the commutator  between a pair of Hamiltonian constraint actions acts in a manner very similar to a (sum over infinitesmal) diffeomorphisms
%Thus, roughly speaking, the action of the Hamiltonian constraint on an off shell state is very similar to that of a sum of certain (infinitesmal) diffeomorphisms. This results in a commutator which is also a sum over(infinitesmal) diffeomorphisms
and matches exactly with a quantization of the Poisson bracket between two Hamiltonian constraints, this Poisson bracket itself being, classically, proportional to the diffeomorphism constraint.

As remarked above the electric diffeomorphism part of the action resembles a diffeomorphism; indeed, it is not exactly a diffeomorphism but a more singular structure 
which falls into the class of `extended' diffeomorphisms
defined in \cite{r-f}. A crucial ingredient in our calculation is an assumed  volume non-triviality  of a child vertex created by this action from a  parent vertex with non-zero volume. A volume operator 
which satisfies this assumption is that defined by Rovelli and Smolin \cite{rsvol} and we shall use their volume operator in this work.  
\footnote{A preliminary analysis of physical states in the kernel of the constraint
hints at an effective removal of `moduli' parameters \cite{g-r} by virtue of the singular nature of electric diffeomorphisms alluded to here. We shall comment further on this in our concluding section.}

The layout of the paper is as follows. In section \ref{sec2} we start with a  review of the form of the mixed action of \cite{p4}. Next, we specify this action in more detail than in \cite{p4} through (a) a detailed
specification of regulating coordinate patches, (b) a precise specification of the positions of the displaced vertex and kinks created by this action and, (c) a specification of  `upward conical' vertex structure at the displaced vertex. 
As we shall see, these specifications are crucial (and in some sense, even  sufficient) for the working out of the calculation. 
The choice in (a)  of Riemann Normal Coordinates which are tied to the metric label $h$  of the off shell state
being acted upon, leads to diffeomorphism
covariance of the Hamiltonian constraint action through a mechanism very similar to that encountered in \cite{p3}.   The specification in (b) leads to the desired `derivative' term from the electric diffeormorphism part of 
the constraint action as well as
%on the one hand, and 
the trivialization of the propagation term described above in the evaluation of the constraint action on the off shell state. The specification in (c) plays a key role
in obtaining the correct combination of terms which allow the definition of a second constraint action and its (anomaly free) commutator.

In section \ref{sec3}, we construct the space of off shell states. As mentioned above this is achieved  through a specification of the set of spin network `bra' summands  associated with an off shell state and a specification
of the coefficient of each such summand in terms of evaluations of a vertex smooth function $f$ and a specific combination of distances between kinks in the spin network bra graph as computed by a Riemmanian metric $h$.
If we call the set of summands as the Bra Set $B$, we denote such an off shell state by $\Psi_{B,f,h}$.  In section \ref{sec4}, on this space of off shell states,  we compute the (dual) action of a 
single Hamiltonian constraint as well as  the action of a product
of Hamiltonian constraints and, thereby, their commutator. In section \ref{sec5}, we construct a quantization of the Poisson bracket between a pair of Hamiltonian constraints whose action is a sum of infinitesmal diffeomorphisms
(as opposed to electric diffeomorphisms) so that the resulting operator  kills diffeomorphism invariant states. We evaluate the (dual) action of this operator on the space of off shell states.  We show that on this space of off shell
states, this action 
equals $i\hbar $ times the  action of the constraint commutator   
computed in section \ref{sec4}. This  establishes a non-trivial anomaly free representation of the commutator between a pair of Hamiltonian constraints. 
In section \ref{sec6}, we show that 
the action of spatial diffeomorphisms is anomaly free on the space of off shell states. We also demonstrate consistency of the Hamiltonian constraint action 
with diffeomorphism covariance if a certain condition of linear independence on the space of off shell states holds. In section \ref{secl} we prove this condition of linear independence.
%under certain technical assumptions whose plausibility we discuss.
In section \ref{sec7} we discuss  an assortment of technicalities some of which touch on the consistency of the detailed specifications (a)- (c) described in section \ref{sec2} with the interpretation of the resulting 
constraint action as a valid approximant to the Hamiltonian constraint. In section \ref{sec8} 
%we discuss issues concerned with physical (i.e. on shell) states.
we comment on  the issue of regulator (in)dependence of physical states, then turn to an account of  possible (anomaly free) generalizations of the specific anomaly free constraint action discussed hitherto 
together with the effect of these generalizations on the nature of physical states and  conclude with a discussion of open problems and future directions of research.

Before we start our exposition, we introduce a useful notation  to distinguish between the notions of `term of leading {\em non-trivial} order in $\e$' and `term of order $\e$:\\

\noindent {\em Notation for term of order $\e$}: A term of $f$ of order $\e$ will be denoted as $O(\e)$ by which we mean that there exist  non-zero postive numbers $M,\e_0$  such that for all $0<\e<\e_0$, 
$|f| \leq M\e$.  \\

\noindent {\em Notation for term of leading  non-trivial order in $\e$}: A term $f$ of leading {\em non-trivial} order in $\e$ will be denoted as $\O(\e)$ by which we mean there exist non-zero positive numbers
$M_1,M_2, \e_0$ such that for all $0<\e<\e_0$, we have that $ M_1\e \leq  |f| \leq  M_2\e$. Note that a term of $\O(\e)$ is also of $O(\e)$ but a term  of $O(\e)$ may not be of $\O (\e)$.
\\

In what follows, we choose units in which $\hbar=G=c=1$.

\section{\label{sec2} The action of the Hamiltonian constraint operator}

Section \ref{sec2.1} provides a review of essential material from Reference \cite{p4}. Its main purpose is to set notation. Details may be found in Reference \cite{p4}. Section \ref{sec2.1.1} is devoted to a review of classical aspects
of Euclidean gravity. Section \ref{sec2.1.2} reviews the action of the regulated Hamiltonian constraint operator  at regulation parameter $\e$ constructed in \cite{p4}. 
The operator acts non-trivially only at vertices of spin network states. The construction requires a choice of regulating coordinate patches at these  vertices of the spin network being acted upon.
In \cite{p4} we constructed  this action in detail for a special class of vertices called `GR' vertices without a detailed specification of these regulating coordinates.
In section \ref{sec2.2} we provide such a detailed specification. In section \ref{sec2.3} we modify the action derived in \cite{p4} at scales much smaller than $\e$. As we shall see in subsequent sections, these modifications ensure that
the constraint action is anomaly free. In section \ref{sec2.4} we detail the constraint action on non-GR vertices.

\subsection{\label{sec2.1} Review}
In section \ref{sec2.1.1} we review key aspects of the classical phase space description of Euclidean gravity in terms of  $SU(2)$ connections and electric fields.
In section \ref{sec2.1.2} we review  key aspects of the action of the quantum  Hamiltonian constraint constructed in \cite{p4}. 
\subsubsection{\label{sec2.1.1}Classical Theory}

The canonically conjugate phase space variables are an $SU(2)$ connection $A_a^i$ and a densitised electric field $E^a_i$ with $\{A_a^i(x), E^b_j(y)\}= \delta^i_j \delta^b_a \delta (x,y)$ where
 $i, \in su(2)$ and $a,b$ are tangent space indices on the 3d Cauchy slice $\Sigma$.   The electric field defines a 3- metric $q_{ab}$ through 
 \be
 E^a_iE^{ai}= q q^{ab}
\label{defqab}
 \ee 
  with $q$ being the 
 determinant of $q_{ab}$. The phase space functions:
\begin{align}
G[\Lambda]  &  =\int\mathrm{d}^{3}x~\Lambda^{i}{\cal D}_{a}E_{i}^{a}\\
D[\vec{N}]  &  =\int\mathrm{d}^{3}x~N^{a}\left(  E_{i}^{b}F_{ab}^{i}-A_{a}%
^{i}{\cal D}_{b}E_{i}^{b}\right) \label{defclassd}\\
H[N]  &  =\tfrac{1}{2}\int\mathrm{d}^{3}x~{N}q^{-1/3}\epsilon^{ijk}E_{i}^{a}E_{j}%
^{b}F_{ab}^{k}, \label{defclassh}
\end{align}
are the Gauss law, diffeomorphism, and Hamiltonian constraints of Euclidean gravity. 
The  Hamiltonian constraint (\ref{defclassh}) is chosen to be of density weight $\frac{4}{3}$ with the lapse $N$ carrying a density weight of $-\frac{1}{3}$.
%Denoting the $su(2)$ Lie algebra commutator  by $[\;,\;]$, 
The Poisson brackets between the constraints are: 
\begin{align}
\{G[\Lambda],G[\Lambda^{\prime}]\} & = G[\Lambda, \Lambda^{\prime}] \;\;\;\;\;\; \{G[\Lambda],H[N]\}=0\\
\{D[\vec{N}],G[\Lambda]\}  &  =G[{\cal L} _{\vec{N}}\Lambda]\\
\{D[\vec{N}],D[\vec{M}]\}  &  =D[\pounds _{\vec{N}}\vec{M}] \label{classdd}\\
\{D[\vec{N}],H[N]\}  &  =H[\pounds _{\vec{N}}N]\label{classdh}\\
\{H[N],H[M]\}  &  =D[\vec{L}]+G[A\cdot\vec{L}],\qquad
L^{a}:=q^{-2/3}E_{i}^{a}E_{i}^{b}\left(  M\partial_{b}N-N\partial_{b}M\right) \label{classhh}
\end{align}
where in the first line $G[\Lambda, \Lambda^{\prime}]$ denotes the Gauss Law constraint smeared with the $su(2)$ Lie algebra commutator between $\Lambda, \Lambda^{\prime}$.
Note that the last Poisson bracket (between the Hamiltonian constraints) exhibits
structure functions just as in Lorentzian gravity albeit with a different overall sign.

The Electric Shift $N^a_i$ at point $p$ is defined by 
\be
N^a_i(p) = N(p)E^a_i(p) q^{-1/3}(p)
\label{defes}
\ee
As can be checked $N^a_i(p)$ transforms like a Lie algebra valued vector field.
%We are concerned exclusively with the Hamiltonian constraint in this paper. 
Segregating an `Electric Shift'
part in brackets, we write (\ref{defclassh}) as:
\be
H(N) = \frac{1}{2}\int \epsilon^{ijk} {(\frac{N E^a_i}{q^{\frac{1}{3}}})} F_{abk}E^b_j.
\label{hamclass}
\ee
On the constraint surface, the equations of motion generated by $H(N)$ can be rewritten in a geometrically intriguing and compact form in terms of generalizations of  Lie derivatives
with respect to the Electric Shift \cite{aame}. Corresponding to this, in the quantum theory \cite{p4}, the action of the  Hamiltonian constraint operator on a spin network state generates deformations of the underlying spin network graph,
these deformations being associated with the quantum Electric Shift operator. We review this action in the next section.

\subsubsection{\label{sec2.1.2} Hamiltonian Constraint Operator action on GR vertices}

%We work with the standard LQG kinematic Hilbert space spanned by $SU(2)$ gauge invariant spin networks.  

The Electric Shift operator at point $p$ is constructed through 
a quantization of a regulated version of (\ref{defes}) followed by the removal of the regulator. The regulation employs a regulating coordinate patch 
$\{x\}$ around the point $p$.
%The regulation procedure is to integrate the expression (\ref{defes}) over a  coordinate sphere of radius $\tau$, divide the result by the coordinate 
%volume of the sphere and then take the regulator $\tau$ to zero.
The regulated quantum correspondent ${\hat q}_{\tau}^{-\frac{1}{3}}$ of  $ q^{-\frac{1}{3}}$
at regulator value $\tau$ is ordered right most. Standard regularizations of this operator using either a Thiemann like trick \cite{qsd} or 
a Tychonoff trick (see for e.g. \cite{eugenio}) result in an action which is non-trivial only when $p$ coincides with a vertex $v$  of the gauge invariant spin network function
$S(A)$ 
being acted upon. This action changes the intertwiner at $v$.  
The final result is
\be
{\hat{N}}^{a}_{j}(v) S(A) =
\frac{3i}{4\pi}N(x(v))
\sum_{I=1}^N
{\hat { e}}^a_I  {\hat X}_{j\;I} S_{\lambda}(A)
\label{qshift}
\ee
Here $N(x(v))$ is the evaluation of the density weight $-\frac{1}{3}$ lapse $N$ at $v$ in the regulating coordinate patch. ${\hat { e}}^a_I$ denotes the unit coordinate (outward pointing) $I$th edge tangent of $S$ at $v$,
${\hat X}_{j\;I}$ is the left invariant $SU(2)$ vector field action at $v$ on the group element associated with the (outward pointing) $I$th edge of $S$. 
$S_{\lambda}(A)$ denotes a spin network identical to $S(A)$ except for the change of intertwiner at $v$ rendered by the 
operator $\lim_{\tau\rightarrow 0} \tau^{-2} {\hat q}_{\tau}^{-\frac{1}{3}}(v)$ with 
$\hat{q}_{\tau}^{-1/3}$ defined  either through Thiemann like identities \cite{qsd,ttbook} involving the Volume operator  or through 
the Tikhonov regularization.
\footnote{\label{fntycho} The regularization \cite{eugenio} defines the action of the  inverse of a positive operator ${\hat O}$  on a state $\psi$ as  $\lim_{\e \rightarrow 0^+}  ({\hat O} + \epsilon)^{-2} {\hat O} \psi$
where $({\hat O} + \epsilon)^{-1}$ is defined through spectral analysis. The result is an operator which annhilates the zero eigenvalue states of ${\hat O}$.}

\begin{figure}
\centering
\begin{subfigure}[h]{0.3\textwidth}
    \includegraphics[width=\textwidth]{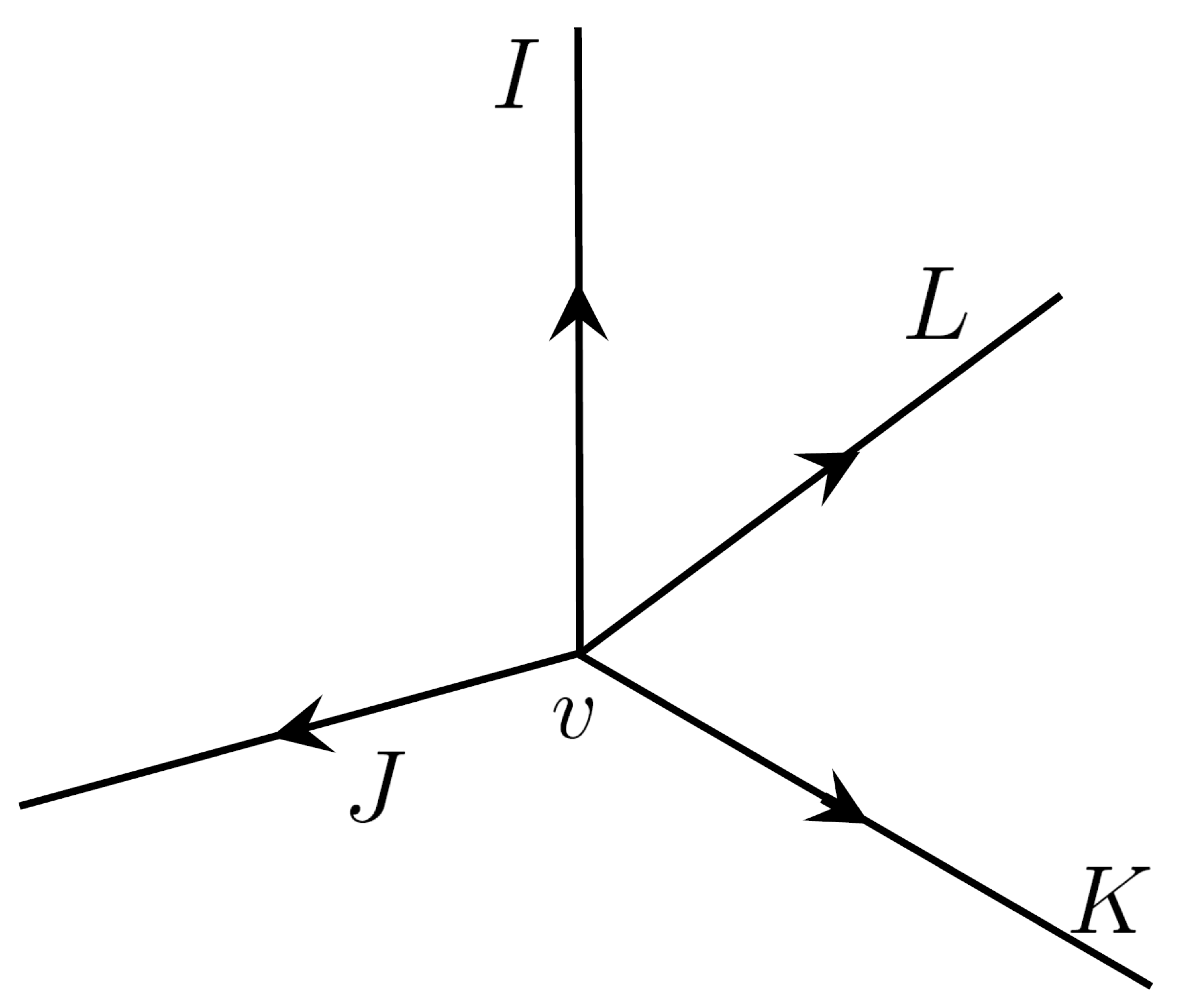}
    \caption{}
 \label{undef}
  \end{subfigure} \quad
  \begin{subfigure}[h]{0.3\textwidth}
    \includegraphics[width=\textwidth]{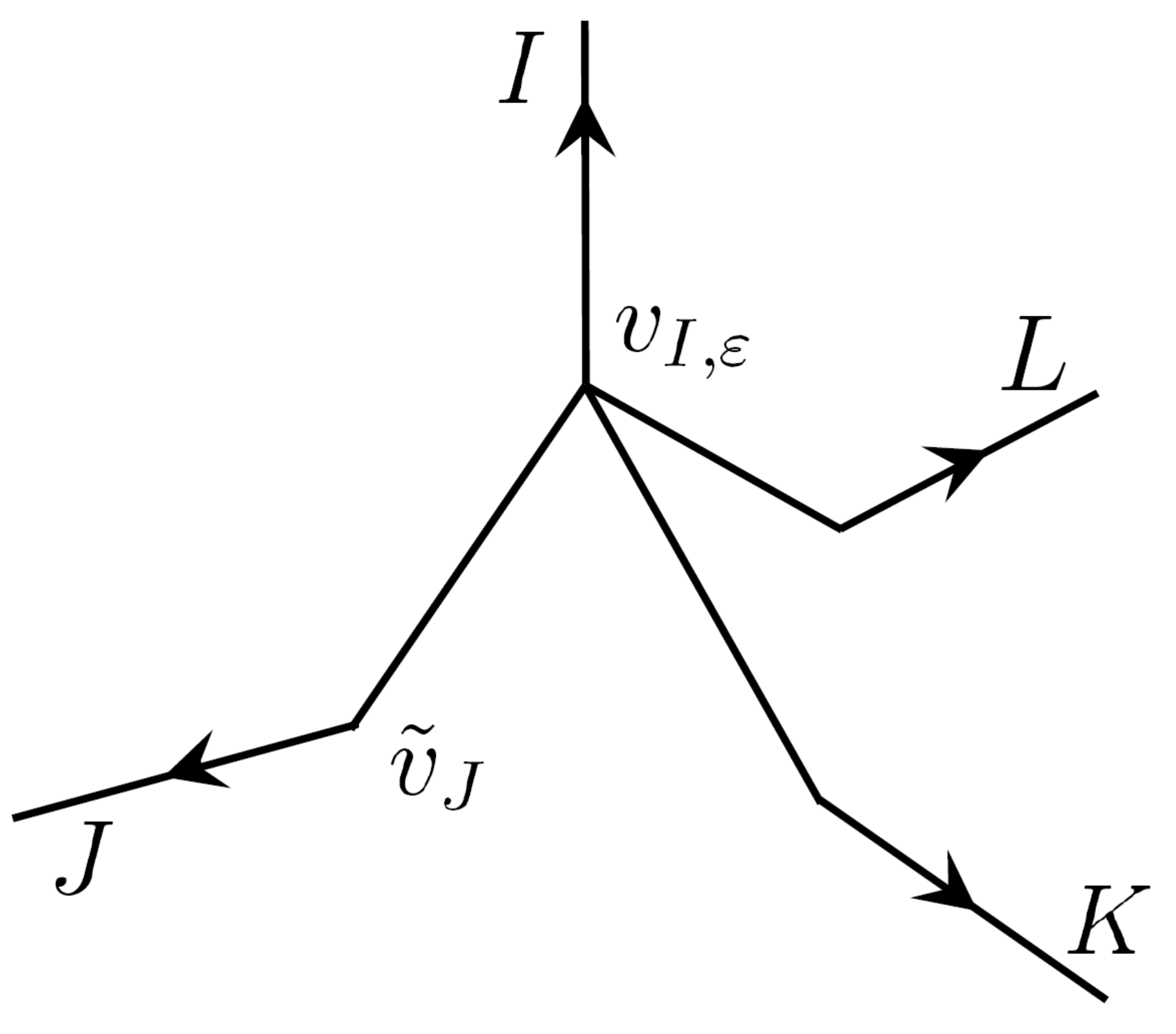}
    \caption{}
   \label{condef}
  \end{subfigure}
\begin{subfigure}[h]{0.10\textwidth}
    \includegraphics[width=\textwidth]{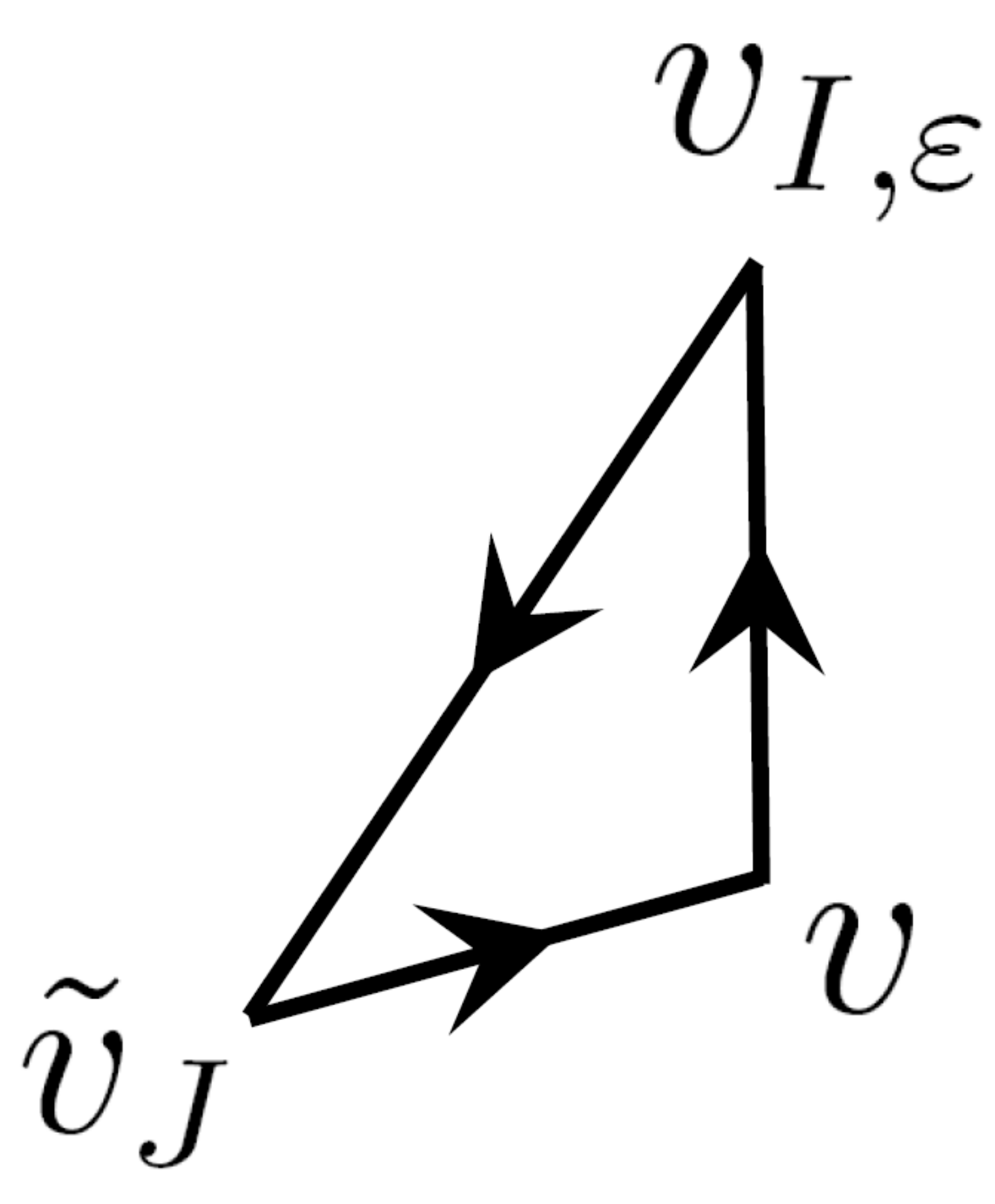}
    \caption{}
   \label{figloop}
  \end{subfigure} \quad
  \caption{ Fig \ref{undef} shows the undeformed vertex structure at $v$.  $\phi_{I,\e}$ is visualised to act on this vertex structure by deforming it  along its $I$th edge as shown  in Fig \ref{condef} wherein the displaced
vertex $v_{I,\e}$ and intersection point ${\tilde v}_J$ between the $J$th edge and its deformed image are labelled. Fig \ref{figloop} shows the loop 
$l_{IJ,\e}$ which starts from $v$,  runs along the $I$th edge to $v_{I,\e}$, moves along the $J$th displaced edge to ${\tilde v}_J$ and then back to $v$ along (and in the opposite
direction to) $e_J$.
}%
\label{fig1}%
\end{figure}

In the Hamiltonian constraint operator, the Electric Shift is ordered right most. In this work we shall use the {\em Tikhonov regularization} together with the {\em Rovelli-Smolin Volume Operator} \cite{rsvol}. 
The reason for these choices will become clear as we go along.
With these choices it follows from Footnote \ref{fntycho} that the Hamiltonian constraint operator has non-trivial action only at vertices of non-trivial volume.
In section 5 of \cite{p4} we derived an action of the Hamiltonian constraint operator ${\hat H}_{\e}(N)$ at regulator value $\e$ on spin network vertices which had a specific vertex structure which we call a 
Grot-Rovelli or `GR' vertex structure. A GR vertex structure is one for which any triple of edge tangent vectors is linearly independent. The action at a GR vertex $v$ is:
\ba
{\hat H}_{\epsilon}(N) S(A) &:= &\frac{3}{8\pi}N(x(v)) \big(\;\sum_{I=1}^N\frac{ j_I(j_I+1)(S_{\lambda,I, \e} - S_{\lambda})}{\e}
\nonumber\\ 
&-& \sum_{I=1}^N\frac{  (\sum_{J\neq I}\sum_{K\neq I,J}S_{\lambda,(1)I,J,K,\e}) +(\sum_{J\neq I} S_{\lambda,(2)I,J,\e}) - j_I(j_I+1)(N-1)S_{\lambda}}{\e}\;\big).\;\;\;\;\;\;\;\;\;\;
\label{mixed}
\ea
Here $S_{\lambda,I, \e}$  is obtained by deforming the entire vertex structure of $S_{\lambda}$ in an $\e$ vicinity of  $v$ along the $I$th edge in a specific way so that the vertex $v$
is itself displaced by an amount $\e$ to the point $v_{I,\e}$. 
This  deformation, denoted by $\phi_{I,\e}$  abruptly `pulls' the remaining edges along the $I$th one and results in a kink at each of the $J\neq I$ edges.
Thus if the vertex $v$ is $N$ valent $N-1$ kinks $\{{\tilde v}_J\}$ are created as depicted in Figures \ref{undef},\ref{condef}.
%turns out to lie in of the class of `extended' diffeomorphisms   

\begin{figure}[H]
\centering
  \begin{subfigure}[H]{0.3\textwidth}
    \centering
    \includegraphics[width=\textwidth]{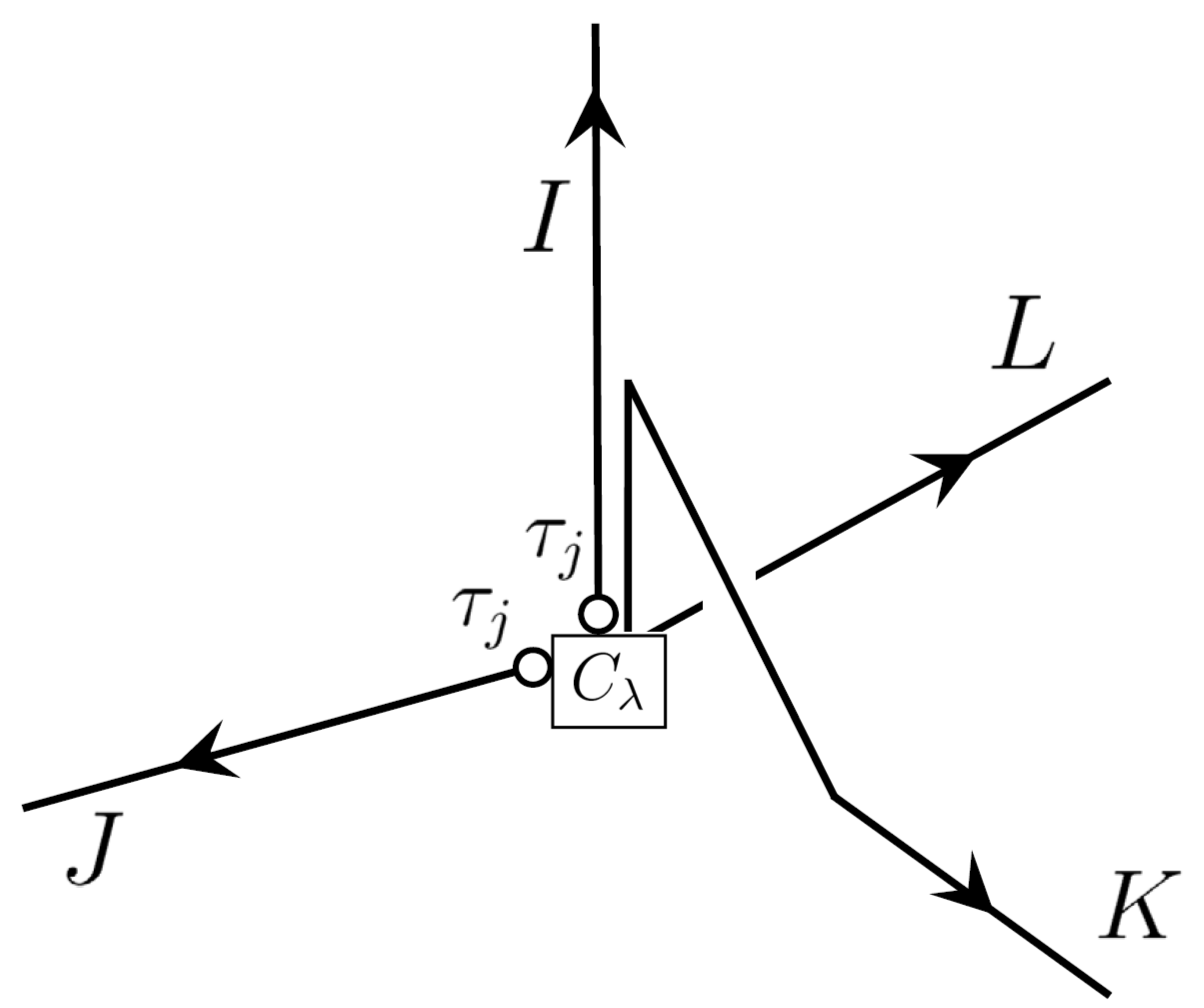}
    \caption{}
 \label{figmixed1}
  \end{subfigure} \quad
  \begin{subfigure}[H]{0.3\textwidth}
  \centering 
    \includegraphics[width=\textwidth]{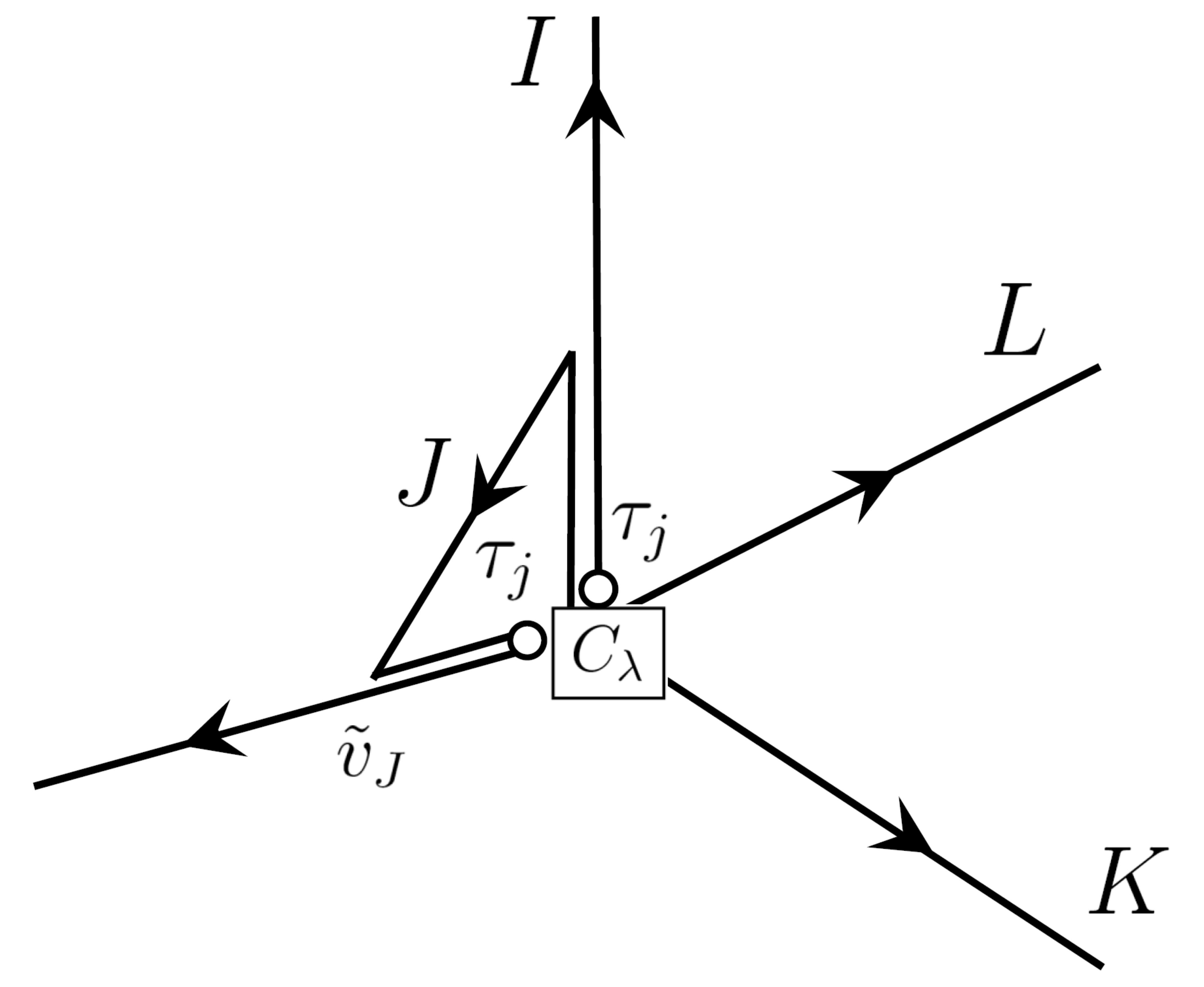}
    \caption{}
   \label{figmixed2}
  \end{subfigure} 
%\begin{subfigure}[h]{0.3\textwidth}
%    \includegraphics[width=\textwidth]{F1-3.pdf}
%    \caption{}
%   \label{grc}
%  \end{subfigure}
  \caption{ 
Figures \ref{figmixed1}, \ref{figmixed2}  depict the deformed vertex structure of $S_{\lambda,(1)I,J,K,\e},S_{\lambda,(2)I,J,\e} $ respectively. The box represents the intertwiner $C_{\lambda}$ located at the 
position of the undisplaced vertex $v$ of $S$ (see Figure \ref{undef}). Insertions of $\tau_j$ are represented by circles. The touching of an object with the interwtiner box indicates
an index contraction between an index of the object and that of the intertwiner so that whereas both the $\tau_j$'s have index contractions with $C_{\lambda}$ in Figure \ref{figmixed1}, in Figure \ref{figmixed2} only the 
one on the $I$th edge has such an index contraction.
}
 \label{figmixed}%
\end{figure}

In contrast the graph underlying the  states $S_{\lambda,(1)I,J,K,\e}, S_{\lambda,(2)I,J,\e}$  is  obtained by considering the action of $\phi_{I,\e}$ on one edge at a time rather than on the entire 
vertex structure. The deformed graph underlying each of these states then consists of the original graph together with this one deformed edge (the $K$th one in  $S_{\lambda,(1)I,J,K,\e}$ and the $J$th one in
$S_{\lambda,(2)I,J,\e}$. As a result, instead of point $v_{I,\e}$ being a new $N$ valent vertex, it is now a kink and instead of the  $N-1$ kinks 
only a single kink  is created (at ${\tilde v}_J$ in $S_{\lambda,(2)I,J,\e}$ and at  ${\tilde v}_K$ in $S_{\lambda,(1)I,J,K,\e}$). 
Here by a kink we mean either a bivalent vertex with linearly independent edge tangents or a trivalent vertex located in the interior of one semianalytic edge from which a second semianalytice edge emanates
with edge tangent linearly 
independent with respect to the edge tangent of the first edge at this vertex.
%in which two of the edge tangents are collinear and the third edge tangent is linearly 
%independent with respect to either of the two collinear ones. 
The deformed states $S_{\lambda,(1)I,J,K,\e}, S_{\lambda,(2)I,J,\e}$ are denoted in Figures \ref{figmixed1}, \ref{figmixed2}. 
%\nopagebreak

These states in general are not spin networks themselves but can be expanded in a spin network basis and the `one edge at a time' deformed graph is the (not necessarily coarsest) graph underlying 
each of these spin networks. In each of these spin networks the single deformed edge carries the same spin as its undeformed counterpart and the only two edges  which have altered spin labels
are the ones which connect $v$ to each of the two kinks.
%More in detail in $S_{\lambda,(1)I,J,K,\e}$ the $K$th edge is deformed from its position in $S$ so as to connect the kink at ${\tilde v}_K$ with that at $v_{I,\e}$.
Besides the generation of two extra kinks in each of these spin networks, it is useful for future purposes to understand how the point $v$ presents itself in these spin networks.
More in detail, the point $v$ is a non-degenerate GR vertex in $S$. Non-degeneracy implies that its valence is greater than 3. The question of interest is the nature of $v$ in the
spin networks generated by the `propagation' part of the constraint action on $S$ at $v$. Does $v$ remain a GR vertex of the same valence  or does its valence decrease? Is it possible 
for $v$ to present itself as a kink?  We turn to an account of the various possible presentations of $v$ in the spin network decompositions of $S_{\lambda,(1)I,J,K,\e}, S_{\lambda,(2)I,J,\e}$
below.

Let the valence of $v$ in $S$ be $N$ so that $N>3$ for non-degeneracy.
First consider the spin networks in the decomposition of $S_{\lambda,(1)I,J,K,\e}$. In each of these spin networks  the kink ${\tilde v}_K$ is a bivalent kink   (so that the part of $e_K$ between $v$ and ${\tilde v}_K$ in $S$ is absent).
However, the nature of the kink at $v_{I,\e}$  and the presentation of $v$ depends on details of the decomposition as follows. Let the $I$th and $K$th edge labels be  $j_I,j_K$  in $S$.
Then the spin  label $j$ of the edge between $v$ and $v_{I,\e}$  in  a spin network in the decomposition must arise as one of the representation labels in the 
product of the $j_I$ and $j_K$ representations.  If it so happens that  $S$ is such that $j_I=j_K$, then depending on the intertwiner at $v$ in $S$, it is possible that 
$j=0$ in one of the spin networks in the decomposition, in which case $v_{I,\e}$   presents itself in this spin network as a bivalent kink and $v$ reduces its valence by 2.
If $N=4$, $v$ presents itself as a bivalent kink  and for $N>4$ as an $N-2$ valent GR vertex. For the case that $j\neq 0$, $v_{I,\e}$ is a trivalent kink
%\footnote{By a  trivalent kink $k$ we mean  the point of intersection of an edge $e$ with an edge $e^{\prime}$ %such that $k$ is an interior point of $e$ and an endpoint of $e^{\prime}$; see section \ref{sec3.1} for a precise %definition} 
and $v$ is an $N-1$ valent GR vertex.

Next, consider $S_{\lambda,(2)I,J,\e}$.
It follows from \cite{p4}, that for  every spin network in the decomposition of $S_{\lambda,(2)I,J,\e}$,  the spin label of the edge between $v, {\tilde v}_J$ 
%and $v,v_{I,\e}$ are both 
is equal to
1 so that the kink at ${\tilde v}_J$ is  necessarily trivalent. On the other hand the spin label $j$ of the edge between $v,v_{I,\e}$ depends on the spin labels $j_I$ and $j_J$ of the $I$th and $J$th edges in $S$, and arises as one of the 
representation labels in the product of the $j_I$ and $j_J$ representations. If a spin network with $j=0$  manifests in the decomposition, then $v_{I,\e}$ is a bivalent kink and $v$ an $N-1$ valent GR vertex else $v_{I,\e}$ 
presents as a trivalent kink and $v$ as an $N$ valent GR vertex.
%and $v$ is a GR vertex with the same valence it has in $S$. 
This concludes our discussion of  possible  kink and vertex structures of spin networks
generated by the constraint action (\ref{mixed}) on a GR vertex.

Using the nomenclature developed in section \ref{sec1}, we refer to  the states $S_{\lambda,I, \e}$ as `diffeomorphism type children' and to the states $S_{\lambda,(1)I,J,K,\e}, S_{\lambda,(2)I,J,\e}$
as `propagation type children'. Equation (\ref{mixed}) may be seen to arise through a regularization of the components of $F_{ab}^i$ in the constraint  in terms of small loop holonomies (depicted in Figure \ref{figloop})
with the constraint action (\ref{mixed}) 
emerging as the  leading order term in an expansion in small loop areas, these areas being measured by the regulating coordinates.
By independently changing the size of these small loops for  the propagation terms  and the diffeomorphism terms  one can further modify the action (\ref{mixed})
along the lines of  Appendix C of \cite{p4}. More in detail, an application of the methods of section 5 of \cite{p4} to the final expression derived in Appendix C of \cite{p4}
results in the modified `mixed' action:
\ba
{\hat H}_{\epsilon}(N) S(A) &:= &\frac{3}{8\pi}N(x(v)) \big(\;\sum_{I=1}^N\frac{ j_I(j_I+1)(S_{\lambda,I,a_I, \e} - S_{\lambda})}{a_I\e}
\nonumber\\ 
&-& \sum_{I=1}^N\frac{  (\sum_{J\neq I}\sum_{K\neq I,J}S_{\lambda,(1)I,J,K,b_I, \e}) +(\sum_{J\neq I} S_{\lambda,(2)I,J,b_I, \e}) - j_I(j_I+1)(N-1)S_{\lambda}}{b_I\e}\;\big).\;\;\;\;\;\;\;\;\;\;
\label{abaction0}
\ea
%\be 
%{\hat H}_{\epsilon}(N) S(A) := \frac{3}{8\pi}N(x(v) \left(\sum_{I=1}^N\frac{ j_I(j_I+1)(S_{\lambda,I,a_I, \e} - S_{\lambda})}{a_I\e} + \sum_{I=1}^N\frac{(\sum_{J\neq I} S_{\lambda,I,J,b_I,\e})  - j_I(j_I+1)S_{\lambda}}{b_I\e}\;\right)
%\label{abaction0}
%\ee
Here the small loops underlying the diffeomorphism type child $S_{\lambda,I,a_I, \e}$ have (coordinate) areas which are $a_I$ times that of the corresponding small loops  underlying $S_{\lambda,I, \e}$ in ({\ref{mixed}) and those
underlying the propagation type children $S_{\lambda,(1) I,J,b_I,\e},   S_{\lambda,(2)I,J,b_I, \e}     $
have (coordinate) areas which are $b_I$ times that of the corresponding small loops  underlying  in $S_{\lambda,(1)I,J,K,\e}, S_{\lambda,(2)I,J,\e}$ in ({\ref{mixed}). 
The construction of these small loops of  prescribed  area is described in section \ref{sec2.3.1}.
%We shall describe the derivation of this expression in more detail in section \ref{sec7}.
{\em In this work we shall choose $a_I= j_I(j_I+1) $, $b_I=  \frac{4}{3}a_I(N-1)$ so that the action of ${\hat H}_\e (N)$ at a GR vertex $v$ of the spin network state $S(A)$ which will be used in this work is}:
\ba
{\hat H}_{\epsilon}(N) S(A) &:= &\frac{3}{8\pi}N(x(v)) \big(\;\sum_{I=1}^N\frac{ (S_{\lambda,I, \e} - {\frac{1}{4}}S_{\lambda})}{\e}
\nonumber\\ 
&-& \sum_{I=1}^N\frac{  (\sum_{J\neq I}\sum_{K\neq I,J}{\tilde S}_{\lambda,(1)I,J,K, \e}) +(\sum_{J\neq I} {\tilde S}_{\lambda,(2)I,J, \e})}{\e}\;\big).\;\;\;\;\;\;\;\;\;\;
\label{abaction}
\ea
where we have set 
\ba
S_{\lambda,I,\e} &:=& S_{\lambda,I,a_I=j_I(j_I+1), \e},
\label{defsdiffa}
\\
{\tilde S}_{\lambda,(1)I,J,K, \e} &:=& \frac{3}{4(N-1)j_I(j_I+1)}S_{\lambda,(1)I,J,K,b_I=\frac{4}{3}(N-1)a_I, \e}\;,
\label{sprop1b}\\
{\tilde S}_{\lambda,(2)I,J, \e}&:= & \frac{3}{4(N-1)j_I(j_I+1)}S_{\lambda,(2)I,J,b_I= \frac{4}{3}(N-1)a_I, \e}\;,
\label{sprop2b}
\ea

%\be 
%{\hat H}_{\epsilon}(N) S(A) := \frac{3}{8\pi}N(x(v) \left(\sum_{I=1}^N\frac{ j_I(j_I+1)(S_{\lambda,I,\e} - {\frac{1}{4}}S_{\lambda})}{\e} + {\frac{3}{4}}\sum_{I=1}^N\frac{(\sum_{J\neq I} S_{\lambda,I,J,\frac{4}{3},\e})}{\e}\;\right)
%\label{abaction}
%\ee
%where $S_{\lambda,I,\e}\equiv S_{\lambda,I,a_I=1, \e}$.

Beyond the factors multiplying  the propagation and diffeomorphism type children above, certain features of these deformed children will play a central role in our considerations. 
These features 
%of the deformed state which play a key role in this work 
are: (a) the location of the displaced vertex $v_{I,\e}$ in  $S_{\lambda,I, \e}$, (b) the edge tangent structure at this displaced vertex in $S_{\lambda,I, \e}$, 
and (c) the number and placement of the kinks in  
$S_{\lambda,I, \e}$,$S_{\lambda,(1)I,J,K,\e}, S_{\lambda,(2)I,J,\e}$.  While the feature (a) in this work is almost identical to that in \cite{p4}, we shall modify/further specify the features (b) and (c) relative to 
\cite{p4}.  The modifications will alter the structure of the deformed states on scales much smaller then $\e$ and hence (as discussed in section \ref{sec7}), the modified action will continue to be seen as 
arising from the quantization  of a regulated approximant constraint. We discuss these modifications/specifications in section \ref{sec2.3}. 
%he considerations of this section and of section \ref{sec2.3} apply to GR vertices of spin networks

\subsection{\label{sec2.2} Choice of Regulating Coordinates}

Since the Hamiltonian constraint action on a spin net state  is only non-trivial at vertices of non-degenerate volume, the lapse function is evaluated only at such vertices (\ref{abaction}). 
Since we work with a higher than unit density Hamiltonian constraint, the lapse is non-trivially density weighted
and its evaluation at such a spin net vertex requires a coordinate patch around that vertex. From \cite{p4} this coordinate patch serves as a regulating coordinate patch to define the action of the constraint operator at that vertex
of the spin net state.
As remarked in section \ref{sec1}, we are interested in the {\em dual} action of the constraint operator on a basis  off shell states,  each such basis  state $\Psi_{f,h}$  being  labelled by
a function $f$ and a Riemmannian metric $h$. Since any such state resides in the algebraic dual space
\footnote{ \label{fnalgdual}Elements of the algebraic dual are complex linear mappings on the finite span of spin network states. Each such element (denoted by $\Psi$) can be thought of as a (in general, formal) sum over spin network `bra' states.
The action of the complex linear map $\Psi$ on a spin net state $|s\ket$  is written as the `amplitude' $\Psi (|s\ket)$}
to the finite span of spin network states, this dual action is specified through the 
evaluation of the amplitude $\Psi_{f,h}({\hat H}_\e(N) |s\ket)$ for every spin net $|s\ket$.

Following \cite{p3}, we {\em tailor the choice of the regulating coordinates to the metric label $h$  of the basis state  $\Psi_{f,h}$}.  As detailed in \cite{p3} and shown in section \ref{sec6} of this work, 
such a choice ensures a transparent and simple
implementation of spatial diffeomorphism covariance. Our specific choice of regulating coordinates at a vertex $v$ is that of Riemann Normal Coordinates (RNCs) centered at $v$ and defined by $h$.
This is in contrast to the complicated network of coordinate patches used in \cite{p3} and simplifies the exposition considerably with respect to that of \cite{p3}. 

Two key properties of RNCs are as follows. First, the coordinate edge tangent vector ${\hat e}^a_I$ along the $I$th edge at the vertex $v$, when evaluated with respect to the RNCs at $v$ defined by the metric $h$,
is exactly the edge tangent vector with unit norm with respect to $h$ at $v$ i.e. 
\be
h_{ab}{\hat e}^a_I{\hat e}^b_I|_v=1. 
\label{unittngnt}
\ee
This follows directly from the definition of RNCs.
% Second, consider a 
Next let a point $p_{\e}$ be located at a  geodesic distance $\e$ away from the point $p$ for small enough $\e$ that $p_\e$ is in a convex normal neighborhood of $p$. Note that from the definition of the RNCs at $p$ this
distance is the same as the coordinate distance of $p_{\e}$ from $p$. Let the RNCs centered at $p$ be $\{x\}$ and at $p_{\e}$ be $\{x\}_\e$. Let the Jacobian of the former with respect to the latter be $J$ so that 
\be
J^{i}\;_{j} = \frac{\partial x^i}{\partial x^j_{\e}} 
\label{j}
\ee
Then at $p_\e$ we have that 
\be
\det{J}{|_{p_\e} }=1 + O(\e^2)
\label{detj}
\ee
where $ \det{J}$  is the determinant of $J$. For a proof, see Appendix \ref{seca1}.

\subsection{\label{sec2.3} Modified Action on GR vertices}

From \cite{p4}, the constraint action (\ref{abaction}) at a spin net vertex depends on the specification of the action of the `electric diffeomorphisms' defined by the quantum shift at that vertex, this action being non-trivial in a small vicinity
of the vertex. The specification of the action of these  electric diffeomorphisms  turns out to be  simple and straightforward for GR  vertices (recall that these are vertices at which any triple of edge tangents are linearly independent).
%Following \cite{p2,p3,p4}, we refer to these vertices as `Grot-Rovelli' or `GR' vertices. 
Similar to the QSD case \cite{qsd}, we would like the resultant deformation of the vertex structure to be such that the 
deformed edges and their undeformed counterparts have simple and intuitively reasonable intersection structure. Whereas this is  straightforward to achieve for GR vertex structures, for Non-GR (NGR) vertices,
it is quite an intricate affair to define a deformation and  the consequent routing of the deformed edges so that no undesired intersections occur.
In \cite{p4} we focussed on the GR case. Here we find it necessary to modify the deformations of GR vertices specified in \cite{p4} on scales much smaller than the regulating parameter scale $\e$ for reasons
described in section \ref{sec1}. Since the modifications are small scale they do not alter the validity of the derivation of the resulting action when viewed through the lense of approximants to leading order
in $\e$. More in detail, recall that the small loop holonomies discussed in section \ref{sec2.1.2} play a key role in the action of the constraint and each of them provide a leading order contribution proportional to 
their area, this area being of order $\e^2$. The  modifications described in this section can be thought of as very small scale deformations of these small loops. Since (as can be seen explicitly in Appendices  \ref{a2}, \ref{seca4}) these
deformations can be made arbitrarily small, they change the small loop area only by terms of higher order than $\e^2$. This is why they are called `small scale'
and this is why they do not change the validity of the derivation of the resulting constraint action. 
%For more details as to why this is so please see Appendix 

In section \ref{sec2.3.1} we describe the deformations of a GR vertex $v$  as constructed in \cite{p4}.
Our description will be slightly qualitative in the interest of pedagogy. A technically complete construction suitable for our purposes (with a slightly altered placement of $v_{I,\e}$ in accordance with  the latter part of 
section \ref{sec2.3.1})   may be found in Appendix \ref{seca2.0}.
In sections \ref{sec2.3.2} and \ref{sec2.3.3} we detail the small scale  modifications of these deformations. We relegate attendant technicalities to Appendices \ref{seca2.1}, \ref{seca2.2} and \ref{seca4}.

%In our discussion,  we shall provide a  detailed specification of only those aspects of the modified deformations which are necessary for our computations of anomaly free quantum constraint actions in section \ref{sec}.
%Beyond this, it is essential to provide a specification of all aspects of the modified deformations and to demonstrate that the modifications can be thought of as small scale
%and `higher order in $\e$'. We postpone this demonstration to section \ref{sec7} for reasons of pedagogy and, in 
%In  our exposition below,  we aim only for a  qualitative, pictorial  justification of why these modifications can be thought of as small scale.
%For pedagogical reasons we postpone a  technical justification to section \ref{sec7}.
%and `higher order in $\e$'. 
%For pedagogical reasons, we flesh out this  qualitaive account in full technical detail later  in section \ref{sec7}.

\subsubsection{\label{sec2.3.1} Deformation of a GR Vertex in Reference \cite{p4}}

From section \ref{sec2.2} the  regulating coordinates are RNCs centered at $v$ and associated with the metric $h$. 
From the discussion in section \ref{sec2.2} and as depicted in Figure \ref{condef}, the deformation associated with the $I$th edge tangent contribution to the Electric shift  
displaces the original vertex  $v$ along  the edge $e_I$  to $v_{I,\e}$. In \cite{p4} $v_{I,\e}$ is located  a coordinate distance $\e$ along the $I$th edge from  the vertex $v$ i.e.
the length of the part of the $I$th edge  between the points $v$ and $v_{I,\e}$ as measured by the regulating coordinates is $\e$.

The $J$th deformed edge meets its undeformed counterpart $e_J$  at the kink ${\tilde v}_J$ and joins this kink to $v_{I,\e}$.
In the tangent space $T_v$ at $v$ consider the  $N-1$ planes $P_{IJ}, J\neq I$ formed by the $I$th and the $J$th edge tangents. Since $v$ is GR, these planes are 
distinct. It follows that for small enough $\e$, we can confine the $J$th deformed edge to  within a small vicinity of the $IJ$ coordinate plane and in this way,
other than the intersection structure at ${\tilde v}_J, v_{I,\e}$ described above,  
avoid any further intersections between the undeformed and deformed edges as well between the deformed edges themselves. It also follows that the vertex $v_{I,\e}$ remains
GR. 

The small loops $l_{IJ,\e}$ alluded to in section \ref{sec2.2} and depicted in Figure \ref{figloop} run from $v$ to $v_{I,\e}$ along $e_I$, from $v_{I,\e}$ to ${\tilde v}_J$ along the $J$th deformed edge
and then back from  ${\tilde v}_J$ to $v$ along the undeformed $J$th edge.
As discussed in section \ref{sec2.2}, the areas of these loops are adjusted so as to obtain the expression (\ref{abaction}). From the discussion in \cite{p4}  these loops have areas of 
$\O(\e^2)$ so that ${\tilde v}_J$ are located at distances of $\O(\e)$ from $v$. The desired loop areas may then be obtained by adjusting the locations of the kinks ${\tilde v}_J$.
This concludes our brief (and slightly more precise)  description  of the deformation of the graph structure around a GR vertex as defined in \cite{p4}.

%while leaving the placement of $v_{I,\e}$ unchanged.

In what follows we shall change the placement of $v_{I,\e}$ by an amount $O(\e^2)$. This, by itself, does not change the small loop areas to 
$\O(\e^2)$. We shall also 
`smoothen' all but 3 of the kinks and, further, change the placement of these  remaining kinks. These modifications
in the kink structure and placement are implemented in such a way that  the loop areas are unchanged to $\O(\e^2)$.
As a consequence the resulting constraint action retains its validity as a leading order in $\e$ approximant.
Moreover the modifications can be seen to be on scales much smaller than $\e$. 
%As mentioned above, we shall only provide an intuitive and pictorial justification for these statements in our discussion below
%with a technical  account postponed to section \ref{sec7}.

\subsubsection{\label{sec2.3.2} Modified placement of $v_{I,\e}$  and  modified choice and placement of kinks}

We modify the placement of $v_{I,\e}$ as follows. Recall that in \cite{p4}  $v_{I,\e}$ is placed such that the coordinate length of the $I$th edge between $v$ and $v_{I,\e}$ is $\e$.
Since this edge is not necessarily a coordinate straight line   this edge length can differ from the  coordinate distance between the two points $v_{I,\e}$ and $v$. However, since
the edge is $C^{r}, r>>1$, this difference is of $O(\e^2)$. We modify the placement of $v_{I,\e}$ by a coordinate distance $O(\e^2)$ so that the  coordinate distance between 
the new position of $v_{I,\e}$ and $v$ is $\e$. Thus, $v_{I,\e}$ is now  placed on $e_I$ in such a way that the coordinate length of the coordinate straight line
connecting $v$ to $v_{I,\e}$ is $\e$. Since the coordinates are RNCs centered at $v$, this coordinate straight line distance is also the geodesic distance between $v$ and $v_{I,\e}$.

While the modifed placement of $v_{I,\e}$ is implemented for the `diffeomorphism' and the `propagation' type deformations,  we shall 
modify  the `diffeomorphism' deformations in a slightly different manner from the `propagation' deformations with regard to their ${\tilde v}_J$ kink structure.
Accordingly first consider the `diffemorphism'  deformation along the edge $e_I$  depicted in Fig \ref{condef}
with $N-1$  kinks $\{{\tilde v}_J,\; J\neq I\}$. As depicted in Fig \ref{Figkinksmooth},
it is possible to alter the deformed edge emanating from the kink ${\tilde v}_J$ in an arbitrary small vicinity of 
${\tilde v}_J$ so as to make it meet its undeformed counterpart `smoothly'. 
At a technical level,  by `smoothly' we mean in a manner which preserves the first $r$ derivatives.
In this regard, recall that the edges are $C^r$ semianalytic. This smoothening can then be achieved by suitable `polynomial' fits discussed in, for example, $\cite{ttbook}$. For a detailed
account see Appendix \ref{a2} of this paper.

\begin{figure}[H]
\centering
%  \begin{subfigure}[H]{0.3\textwidth}
    \centering
    \includegraphics[width=0.3\textwidth]{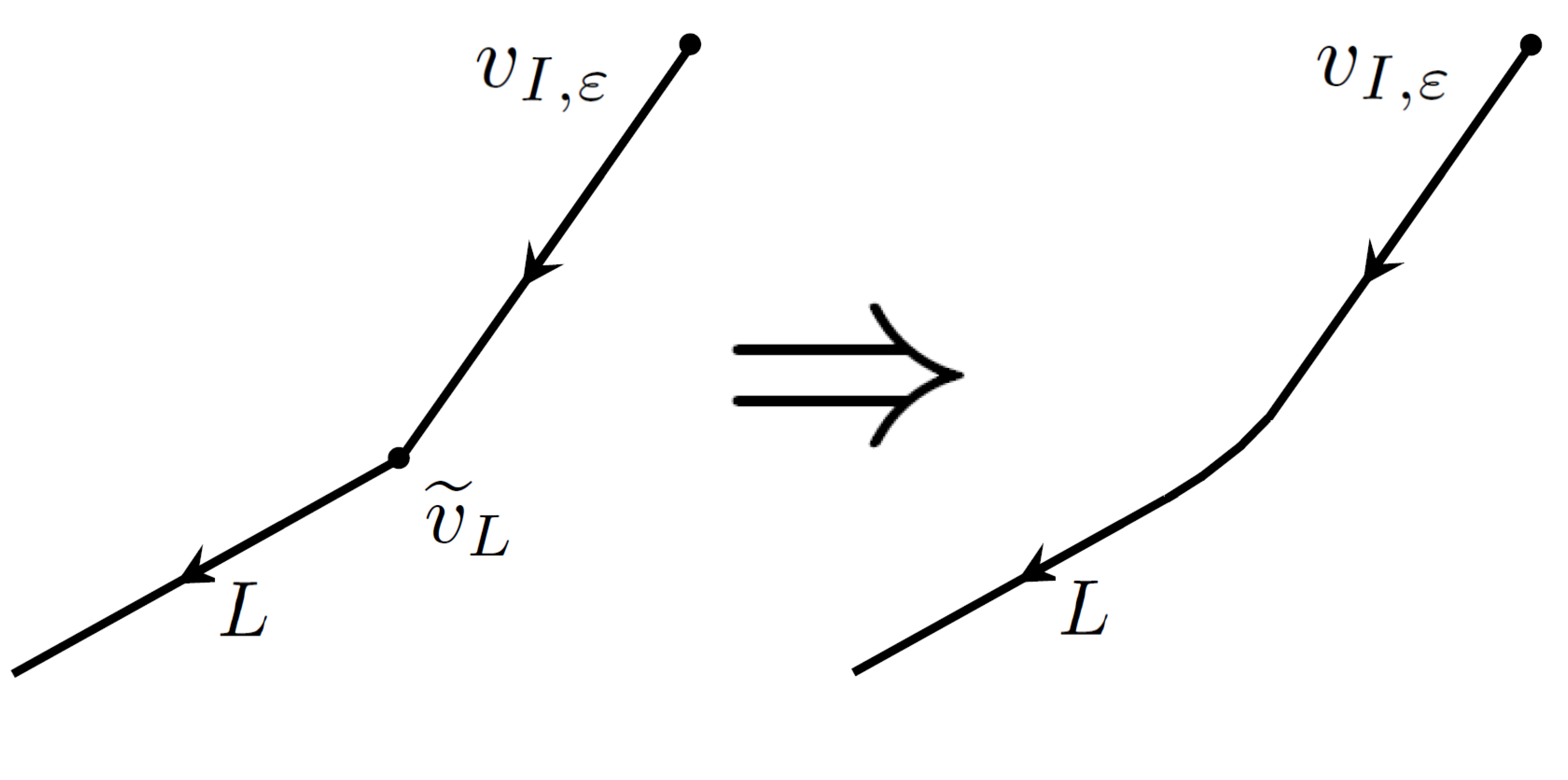}
    \caption{The kink ${\tilde v}_L$ between the pair of edges on the left is smoothened to give the single semianalytic edge on the right   }
 \label{Figkinksmooth}
%  \end{subfigure} \quad
\end{figure}
 %
%  \begin{subfigure}[H]{0.3\textwidth}
%  \centering 
%    \includegraphics[width=\textwidth]{Figkinkmove.pdf}
%    \caption{}
%   \label{Figkinkmove}
%  \end{subfigure} 
%\begin{subfigure}[h]{0.3\textwidth}
%    \includegraphics[width=\textwidth]{F1-3.pdf}
%    \caption{}
%   \label{grc}
%  \end{subfigure}
%  \caption{ 
%Figures \ref{figmixed1}, \ref{figmixed2}  depict the deformed vertex structure of $S_{\lambda,(1)I,J,K,\e},S_{\lambda,(2)I,J,\e} $ respectively. The box represents the intertwiner $C_{\lambda}$ located at the 
%position of the undisplaced vertex $v$ of $S$ (see Figure \ref{undef}). Insertions of $\tau_j$ are represented by circles. The touching of an object with the interwtiner box indicates
%an index contraction between an index of the object and that of the intertwiner so that whereas both the $\tau_j$'s have index contractions with $C_{\lambda}$ in Figure \ref{figmixed1}, in Figure \ref{figmixed2} only the 
%one on the $I$th edge has such an index contraction.
%}
% \label{figkink}%
%\end{figure}

In this manner we smoothen all but 3 of the kinks. Let the remaining kinks be  ${\tilde v}_{\hat J_i}, i=1,2,3$
As depicted in Figure \ref{Figkinkmove}, we can then move each of these kinks along $e_{J_i}$ as close to $v$ as we desire with an arbitrarily small area change of the loop $l_{IJ_i,\e}$.
In the modified loop, the deformed edge emanates from the new kink position and  runs `almost' along $e_{J_i}$ till it reaches the vicinity of the old position of the 
kink, and, subsequently moves towards $v_{I,\e}$ as before.
In detail this can be achieved in one of two ways: either by  polynomial $C^r$ joins or by the action of appropriate semianalytic diffeomorphisms of compact support (see Appendices \ref{a2} and \ref{seca4}).
%and the discussion in section \ref{sec7}).

\begin{figure}[H]
\centering
%  \begin{subfigure}[H]{0.3\textwidth}
    \centering
    \includegraphics[width=0.3\textwidth]{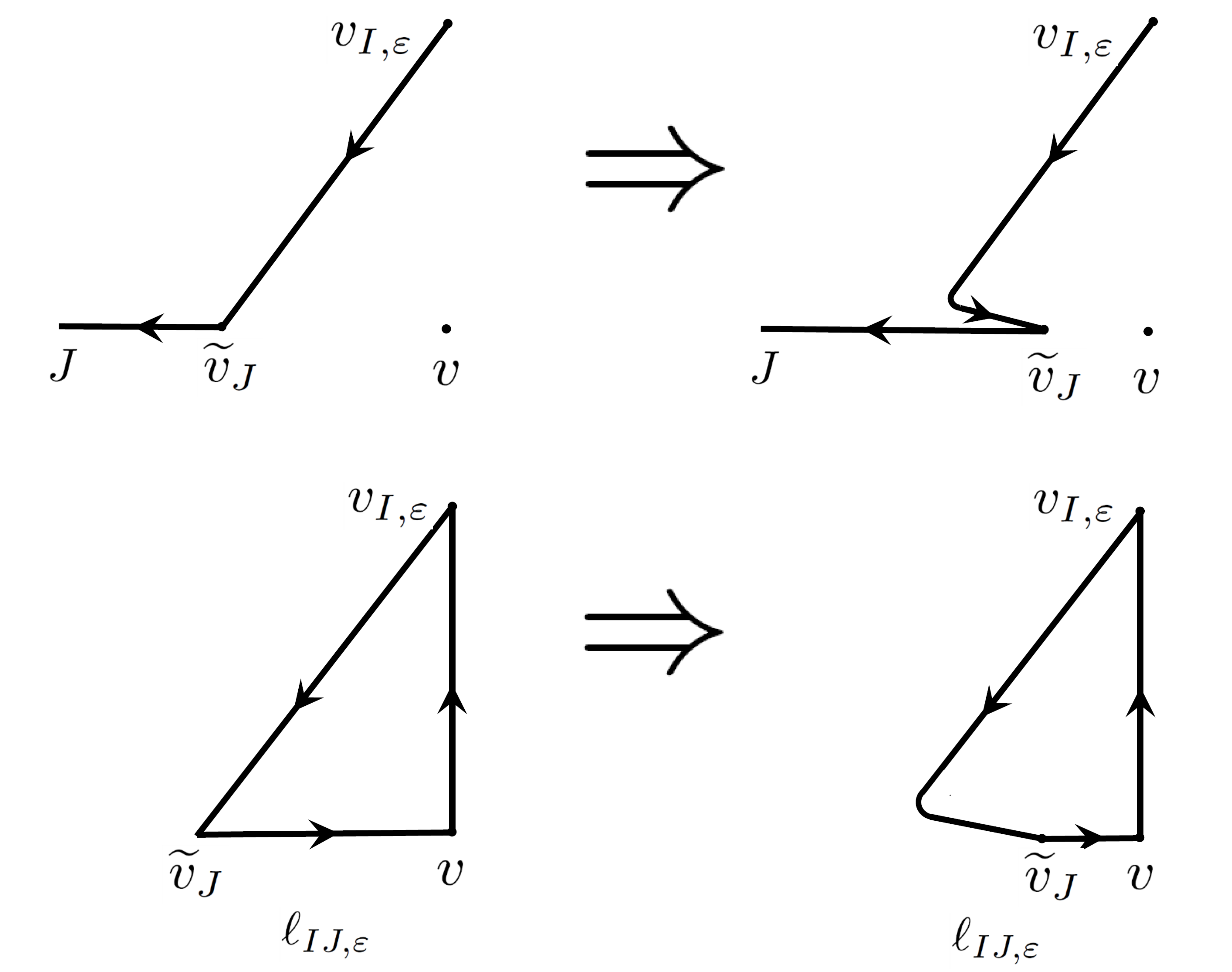}
    \caption{}
 \label{Figkinkmove}
%  \end{subfigure} \quad
\caption{ The first figure depicts the movement of the kink  ${\tilde v}_J$  towards $v$ through the procedure outlined in Appendices \ref{a2} and \ref{seca4}. The second figure shows the modification in the 
small loop $l_{IJ,\e}$ required to achieve this movement. The lower strand of the modified loop on the right curves in the vicinity of the original position of the kink  and connects to the displaced kink. 
The closer this strand is to the undeformed edge $e_J$, the smaller is the change in area of the modified loop relative to the original one on the left.}
%is the 
%Figures \ref{figmixed1}, \ref{figmixed2}  depict the deformed vertex structure of $S_{\lambda,(1)I,J,K,\e},S_{\lambda,(2)I,J,\e} $ respectively. The box represents the intertwiner $C_{\lambda}$ located at the 
%position of the undisplaced vertex $v$ of $S$ (see Figure \ref{undef}). Insertions of $\tau_j$ are represented by circles. The touching of an object with the interwtiner box indicates
%an index contraction between an index of the object and that of the intertwiner so that whereas both the $\tau_j$'s have index contractions with $C_{\lambda}$ in Figure \ref{figmixed1}, in Figure \ref{figmixed2} only the 
%one on the $I$th edge has such an index contraction.
%}
\end{figure}

We shall place these 3 kinks as follows.
%in  a  different manner for the `diffeomorphism' and  for the `propagation' deformations.
%For the diffeomorphism type deformation we place the kinks  as follows.
Let $\e$ be small enough that the 3 kinks are in a single Convex Normal Neighborhood (with respect to the metric $h$) of $v$.
We shall place the 3 kinks, such that two of them are at a distance of $\O(\e^q),q\geq 2$ from $v$ with the third at a distance of $\O(\e^{p}), p>>q$, in such a way that  the 3 interkink geodesic distances (with respect to $h$) are unequal,  with the ratio of the smallest
interkink distance to the largest being equal to $1/2$  to order $\e^q$ i.e.
\be
\frac{d_{min}}{d_{max}} = \frac{1}{2} \;+\;O(\e^q),\;\;\;\;\;
d_{min}:= \min_{i\neq j} d({{\tilde v}_{J_i} ,{\tilde v}_{J_j}}) \;\;\;\; d_{max}:= \max_{i, j} d({{\tilde v}_{J_i} ,{\tilde v}_{J_j}}) 
\label{d/d=1/2}
\ee
with 
\be
d({{\tilde v}_{J_i} ,{\tilde v}_{J_j}})\sim \O(\e^{q \geq 2}) \;\;\;, i,j \in \{1,2,3\}
\label{doe2}
\ee
where $d({p_1,p_2})$ denotes the geodesic distance between the points $p_1,p_2$ (see Appendix \ref{seca3} for a demonstration that such a placement exists).
Recall that the state obtained by an electric diffeomorphism  of $S_{\lambda}$  was denoted by $S_{\lambda, I,\e}$ in (\ref{defsdiffa}),section \ref{sec2.1}.
We shall denote its modification (with only 3 kinks placed as above and with the modified placement of $v_{I,\e}$ described above) by 
\be
 S_{\lambda, I,{\vec{\hat J}}, \e}  
\label{3kinks}
\ee
where ${\vec {\hat J}}:= ({\hat J}_1,{\hat J}_2, {\hat J}_3)$ denotes the 3 chosen edges on which the kinks are placed. In addition, the notation 
will also be assumed to encode the placement of each of the kinks as follows. The kink closest to $v$ is denoted by  ${\tilde v}_{\hat J_3}$  Of the two remaining kinks, the kink closer
to ${v}$ is chosen to be ${\tilde v}_{\hat J_2}$ (this corresponds to $v_1$ in Appendix \ref{seca3})
The third (remaining) kink, corresponding to $v_2$ in Appendix \ref{seca3} is then at ${\tilde v}_{\hat J_1}$.

Next, consider the `propagation' deformations associated with the $I$th component of the electric shift as depicted in Fig \ref{figmixed}. Each such  deformation creates a pair of kinks, one at $v_{I,\e}$ and the other at $\tilde{v_J}$.
While the location of $v_{I,\e}$ has been described above and is the same as for the diffeomorphism deformation, we move the kink $\tilde{v_J}$ to a coordinate distance of  $\O(\e^q),q\geq 2$ from $v$, 
where we remind the reader that these coordinates are RNC's centered at $v$.
As for the `diffeomorphism'  type deformation this can be done with an arbitarily small change in the coordinate area of the loop $l_{IJ,\e}$.
Since the metric is $C^{r-1}, r>>1$, it is straightforward to conclude that 
the geodesic distance of this kink from $v$ is   of the same order as its coordinate distance from $v$. Since $v_{I,\e}$ is at a geodesic distance $\e$ from $v$, it then follows that for the propagation deformation with a kink at ${\tilde v}_J$ and for small enough $\e$
that
\be
d( {\tilde v}_{J}, v_{I,\e}) = \e +O(\e^{q\geq 2}) .
\label{dprop}
\ee
From here on the states (\ref{sprop1b}), (\ref{sprop2b}) will be assumed to have kink placements as detailed  in (\ref{dprop}).

\subsubsection{\label{sec2.3.3} Modified edge tangent structure at displaced vertex}

From section \ref{sec2.3.1} and Appendix \ref{seca2.0}, the `diffeomorphism' type deformation of a GR vertex $v$  along the edge $e_I$ preserves the GR property so that $v_{I,\e}$ is also GR.
We shall modify this deformation in an arbitrarily small vicinity of $v_{I,\e}$ in a manner which preserves the  GR property as well as  the location of $v_{I,\e}$,  and endows the edge tangent configuration with  
%the deformed edges $e^{\prime}_J$ meet at $v_{I,\e}$ such that their edge tangent form 
a stiff `upward' conical structure around the edge $e_I$.
By a `stiff upward cone'  we mean that the component of the $J$th edge tangent at $v_{I,\e}$  orthogonal to ${\hat e}^a_I$ is  much smaller than the component along $e_I$. The small orthogonal contribution to each of the $N-1$ edge 
tangents is required to be such that the vertex $v_{I,\e}$ remains GR. 
More precisely, we require that for small enough $\e$:
\be
{\hat e^{\prime}}^a_{J}|_{v_{I,\e}} = {\hat e}^a_I|_{v_{I,\e}} + O(\e^2) v^a_J .
\label{upward}
\ee
where $O(\e^2)v^a_J$ is non-vanishing for sufficiently small $\e,\; \e>0$,  $v^a_J$ is  of metric norm $\O(1)$ and the set of vectors $\{{\hat e}^a_I|_{v_{I,\e}},  v^a_J, J\neq I\}$  is such that no triple  is linearly dependent.
We depict the desired modification of the edge tangent structure  in Fig \ref{Figupward}.

\begin{figure}[H]
\centering
%  \begin{subfigure}[H]{0.3\textwidth}
    \centering
    \includegraphics[width=0.3\textwidth]{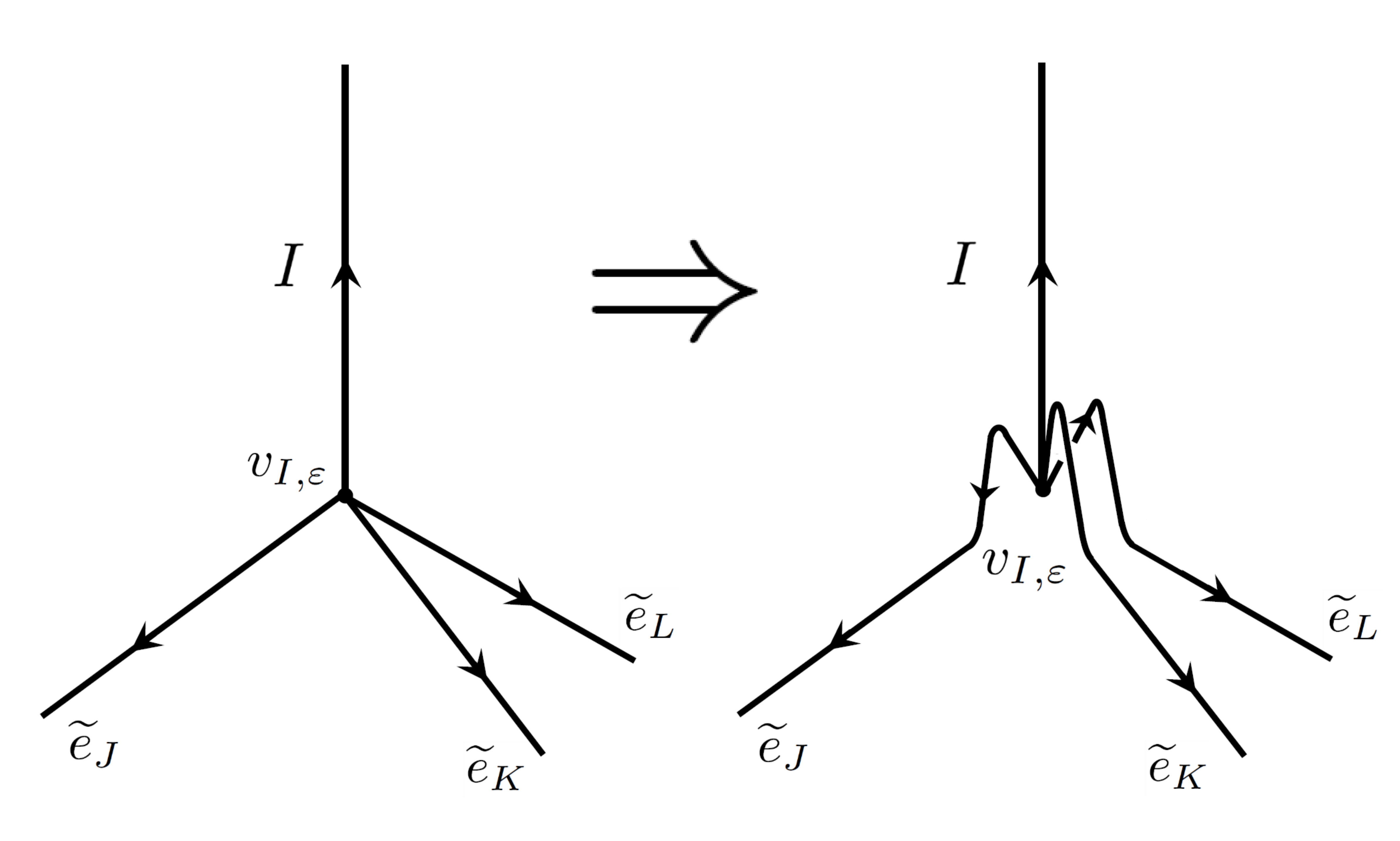}
    \caption{The edge tangent configuration at the vertex $v_{I,\e}$ on the left is modified to the upward pointing configuration shown on the right.}
 \label{Figupward}
%  \end{subfigure} \quad
\end{figure}

As shown in Appendices \ref{a2}, \ref{seca4} this modification  can be implemented  both by using exclusively polynomial $C^r$ joins as well as through the use of such joins in conjunction 
with semianalytic diffeomorphisms of compact support.

\subsubsection{\label{sec2.3.4} Modified constraint action on GR vertices}

The `upward conical' modification  of edge tangent structure described in the previous section is applied to the vicinity of the vertex $v_{I,\e}$ of the 3-kink modified state $S_{\lambda, I,{\vec {\hat J}}\e}$ (\ref{3kinks}).
To summarise: The modification of the `diffeomorphism child' $S_{I,\e}$ is three fold: the first modification  is  the  $O(\e^2)$ adjustment of section \ref{sec2.3.2}  to the placement of $v_{I,\e}$, the second  is to the kink structure as
described in section \ref{sec2.3.2}  and the third
modification is the upward conical modification described above. The modification of each of the `propagation' children is two fold ${\tilde S}_{\lambda,(1)I,J,K, \e}, {\tilde S}_{\lambda,(2)I,J, \e}$:  
first, we have the  modified placement of the kink $v_{I,\e}$,  identical to that for the diffeomorphism child and second,  the kinks (at ${\tilde v}_K$ , respectively ${\tilde v}_J$) are placed as described in section \ref{sec2.3.2}.
We shall abuse notation  slightly and continue to denote the resulting `vertex placement, upward conical, 3-kink' modification of $S_{\lambda, I,\e}$ by  $S_{\lambda,I, {\vec {\hat J}}, \e}$ and also 
continue to denote the `kink placement' modifications of the propagation children by 
${\tilde S}_{\lambda,(1)I,J,K, \e},{\tilde S}_{\lambda,(2)I,J, \e}$.

The modified version of the constraint action is then:
\ba
{\hat H}_{\epsilon}(N) S(A) &:= &\frac{3}{8\pi}N(x(v)) \big(\;\sum_{I=1}^N\frac{ (S_{\lambda,I, {\vec {\hat J}}\e} - {\frac{1}{4}}S_{\lambda})}{\e}
\nonumber\\ 
&-& \sum_{I=1}^N\frac{  (\sum_{J\neq I}\sum_{K\neq I,J}{\tilde S}_{\lambda,(1)I,J,K, \e}) +(\sum_{J\neq I} {\tilde S}_{\lambda,(2)I,J, \e})}{\e}\;\big).\;\;\;\;\;\;\;\;\;\;
\label{abactionm1}
\ea
In the above expression, the diffeomorphism type deformation depends on an arbitrary choice of kink triples specified through ${\vec {\hat J}}$. Since there is no natural choice for  ${\vec {\hat J}}$
we sum over all such choices. 
%Denote the number of permutations of choices of $m$ objects out of a total of $M$ objects by $\Perm{M}{m}$. Then 
For fixed $I$ there are $\Perm{N-1}{3}$ choices of ${\vec {\hat J}}$ for a GR vertex of valence $N$. Each such choice yields a legitimate constraint action (\ref{abactionm1}).
Hence we sum over such actions and divide by the factor $\Perm{N-1}{3}$ to obtain our final expression for the constraint action on a GR vertex $v$:
\ba
{\hat H}_{\epsilon}(N) S(A) &:= &\frac{3}{8\pi}N(x(v)) \big(\;\sum_{I=1}^N \frac{  ([\frac{1}{\Perm{N-1}{3}}\sum_{\J}S_{\lambda,I, {\vec {\hat J}}\e}] - {\frac{1}{4}}S_{\lambda})}{\e}
\nonumber\\ 
&-& \sum_{I=1}^N\frac{  (\sum_{J\neq I}\sum_{K\neq I,J}{\tilde S}_{\lambda,(1)I,J,K, \e}) +(\sum_{J\neq I} {\tilde S}_{\lambda,(2)I,J, \e})}{\e}\;\big).\;\;\;\;\;\;\;\;\;\;
\label{abactionm}
\ea

\subsection{\label{sec2.4} Form of assumed action on non-GR vertices}

We assume the following form for the constraint action on an $N$ valent non GR vertex $v$:
\be
{\hat H}_{\epsilon}(N) S(A) = \frac{N(x(v))}{\e} \big(\;(\sum_{I=1}^N A_IS_{\lambda,I, \e})  + B S_{\lambda}+ (\sum_{I=1}^N\sum_{J\neq I}\sum_{\alpha_{IJ}} P_{\alpha_{IJ}} S_{\alpha_{IJ},\e})\;\;\big)
\label{ngr}
\ee
Here $S_{\lambda,I, \e}$ is a diffeomorphism type deformation of $S_{\lambda}$ which displaces the vertex $v$ along $e_I$ to $v_{I,\e}$ by a coordinate distance of $\O(\e)$. 
We shall assume that this deformation preserves the NGR nature of the vertex so that 
$v_{I,\e}$ is also an NGR vertex.  In addition we make a further assumption which we state using certain nomenclature defined in section \ref{sec3.1} and which we justify through a construction
described in Appendices \ref{sec4.0b} and \ref{seca2.2}. 
%The reader may return to a perusal of this assumption and its justification after reading these Appendices.
%The assumption  and justification is as follows.
We  assume that, similar to the GR case,  three of  the deformed edges of  $S_{\lambda,I, \e}$ meet their undeformed counterparts at kinks 
%so that in the terminology of section \ref{sec3.1},
 %the abstract decorated  graph structure underlying $S_{\lambda,I, \e}$ is obtained by the embellishment of the %decorated abstract 
%graph structure underlying $S$ through the introduction of bivalent kinks on some of its edge interiors.
Such a deformation may be obtained by the action on  $S_{\lambda}$ of  a semianalytic diffeomorphism of the type constructed in section \ref{sec4.0b}  to yield a state $S_{diff}$ followed by the introduction of the desired
kinks. Since the diffeomorphism is identity outside a small neighborhood $U$  of the vertex, we may place kinks $k$ at the chosen edges in $S_{diff}$ at the intersection of these edges with   the boundary of this  neighborhood 
through the following procedure. Choose a small neighborhood of $k$ such that no other edge than the edge of interest intersects this neighborhood and fix a semianalytic coordinate patch thereon .  Remove part of the edge connecting $k$ to 
a point $p$ in this neighborhood and join $k$ to $p$ by a a pair of coordinate straight  lines one from $k$ to a suitable point $q$ in this neighborhood and then from $q$ to $p$ so that these lines do not intersect the 
edge of interest except at $k,p$ and such that $k, q,p$ are kinks. Finally,  smoothen the kinks at $q,p$  using the techniques of Appendix \ref{seca2.2} thereby leaving a kink at $k$.
This concludes our account of the state $S_{\lambda,I, \e}$. 

$S_{\lambda }$ is, as defined for the GR case, a spin net identical to $S$ except for a different intertwiner at the NGR vertex  $v$. 
The subscript ${\alpha}_{IJ}$ denotes a `propagation' type deformation of the vertex structure at $v$ in an $\e$ size vicinity of $v$. 
By `propagation' type we mean that, similar to the GR case of propagation type deformations, the deformation structure is as follows. 
As in the GR case, the deformations which contribute to the constraint action are expected to arise from the nontriviality of the curvature components $F_{ab}^i{\hat e}^a_I {\hat e}^b_J$. Hence the $IJ$ label
ranges over pairs of edges with {\em linearly independent} tangents at $v$. 
%In  deformation index $\alpah_{IJ}$, the  edge label $J$ ranges over values such that 
Consider the points   $v_{I,\e}$ along the $I$th edge $e_I$ of $S$ and ${\tilde v}_J$ on the edge $e_J$ of $S$ with ${\tilde v}_J$ at a coordinate distance of $\O(\e^{q\geq 2})$ from $v$.
\footnote{Similar to the GR case, we assume that this kink positioning can be implemented without affecting the interpretation of the constraint action  at parameter $\e$ as that of a leading order approximant to the 
constraint.}
Join these 2 points with a semianalytic edge $e^{\prime}_J$ which does not intersect the graph $\gamma$ underlying $S$ except at these two points and whose edge tangents at ${\tilde v}_J,v_{I,\e}$ are linearly independent
with respect to the edge tangents along $e_J,e_I$ at  ${\tilde v}_J,v_{I,\e}$ respectively.
\footnote{This can be achieved, for example, through the procedure detailed in (ii) section  \ref{sec7}. }
Then the  graph
underlying each of the deformed states $S_{\alpha_{IJ},\e}$ is 
\be
\gamma_{IJ} = \gamma \cup e^{\prime}_J
\label{gammaij}
\ee
Using the notation $e(p,q)$ to denote the semianalytic edge containing $p$ to $q$, consider the edges $e(v,v_{I,\e}), e(v_{I,\e}, {\tilde v}_J),  e({\tilde v}_J, v) \subset \gamma_{IJ}$.
These edges combine to form the loop $l_{IJ,\e}$.  The colorings of  $\gamma_{IJ} -l_{IJ,\e} \subset \gamma$ in each $S_{\alpha_{IJ},\e}$ are the same as for their counterparts in $S$. While the coloring of 
$e(v_{I,\e}, {\tilde v}_J)\equiv e^{\prime}_J$ in 
$S_{\alpha_{IJ},\e}$ is the same as that of $e_J$ in $S$, the colorings of the remaning edges of 
$l_{IJ,\e} \subset \gamma_{IJ}$ are different from those in $S$ and can include the trivial $j=0$ coloring,
\footnote{\label{fnngrtogr}Since any edge with  $j=0$ coloring is absent, it is possible that the removal of such an edge from the set of edges at $v$ in $S$ can convert $v$ from an NGR vertex in $S$ to a GR vertex in  $S_{\alpha_{IJ},\e}$.}
in which case 
$\gamma_{IJ}$ is not the coarsest graph underling $S_{\alpha_{IJ},\e}$. The points $v_{I,\e}, {\tilde v_J}$ 
are kinks in $S_{\alpha_{IJ},\e}$ (for a precise definition of kinks see section \ref{sec3.1}) and depending on the edge colorings of $l_{IJ,\e}$ have uniquely defined intertwiners by virtue of their bivalent or trivalent nature. 
Finally, the intertwiner at $v$  in $S_{\alpha_{IJ},\e}$ differs from that in $S$.
Thus the  label ${\alpha_{IJ},\e}$  denotes a deformation which introduces the pair of kinks  $v_{I,\e}, {\tilde v}_J$ together with the  colored edge $e(v_{I,\e}, {\tilde v}_J)$ and 
encodes the changes in colorings on the edges $e(v,v_{I,\e}), e({\tilde v}_J, v)$ and the change  in the intertwiner at $v$ relative to $S$. If  $e(v,v_{I,\e})$ is colored with $j=0$, $v_{I,\e}$ is a bivalent kink. If 
$e(v_{I,\e}, {\tilde v}_J)$  is colored with $j=0$,  ${\tilde v}_J$ is a bivalent kink. Clearly the presence of each such bivalent kink reduces the valence of $v$ relative to its valence in $S$ by 1.

Since the vertex $v$ is nondegenerate in $S$ it has a minimum valence of 4. Since it is NGR  in $S$, it is straightforward to see that reduction of its valence by 1 or 2 can lead to its presentation in $S_{\alpha_{IJ},\e}$
as any of the following: a GR or NGR vertex, a $C^{m<r}$ bivalent or trivalent kink, a $C^r$ trivalent kink  
\footnote{See section \ref{sec3.1} for the definition of a $C^{m}$ kink.}
or  an interior point on a single semianalytic  edge. Note that if both $v_{I,\e}$ and ${\tilde v}_J$ are trivalent there is no reduction in valence and $v$ continues to be an NGR vertex with unchanged edge 
tangent structure relative to its presentation in  $S$.

It follows  from the positioning of $v_{I, \e}, {\tilde v}_J$ described above that the coordinate distance between ${\tilde v}_J$ and $v_{I,\e}$ is of $\O(\e)$. It also follows straightforwardly 
that the geodesic distances as measured by the metric  $h$ are:
\ba
d( v_{I,\e}, v) &=& \O(\e)  \label{ngr1}\\
d( {\tilde v}_{J}, v) &=& \O(\e^{q\geq 2}) \label{ngr2}\\
d( {\tilde v}_{J}, v_{I,\e}) &=& \O(\e)  \label{ngr3}
\ea
Note that, for sufficiently small $\e$,  as $\e$ is decreased, the deformed spin network $S_{\alpha_{IJ},\e}$ has the same
%the deformed spin networks $S_{\alpha_{IJ},\e}$ for sufficiently small $\e_0$,  have the same 
$\e$-independent vertex intertwiner at $v$ as well as the same  $\e$ independent abstract colored graph structure
(see section \ref{sec3.1} for a definition of this abstract structure). This  $\e$ independent  
abstract  structure  is encoded by the index $\alpha_{IJ}$. The extra label $\e$ in $\alpha_{IJ},\e$ indicates that the deformation is confined in an $\e$ vicinity of $v$ with the  placements of the kinks
at $v_{I,\e}, {\tilde v}_J$  as detailed above.
\footnote{More precisely $\alpha_{IJ}$ is a label for the embeddable abstract spin network whose embedding into $S_{\alpha_{IJ},\e}$ is specified by ${\alpha_{IJ},\e}$ (see section \ref{sec3.1} for the definition of 
an embeddable abstract spin network).}
%Thus $\alpha_{IJ}$ is an index which runs through a finite number of distinct deformations involving the creation of a pair of kinks as discussed above and $S_{\alpha_{IJ},\e}$ is the deformed spin net corresponding to the deformation
%$\alpha_{IJ}$.
%In section \ref{sec7}

$A_I, B, P_{\alpha_{IJ}}$ are (not necessarily non-vanishing) complex coefficients. 
%Note that constraint action on a GR vertex is of the type (\ref{ngr}).
While  it is not necessary for our demonstration of anomaly free constraint action to specify the deformations and coefficients in (\ref{ngr}) beyond our discussion above, we shall present one possible  concrete and complete such specification
and justify its validity in section \ref{sec7}.
\\

\noindent{\bf Note}: We note here that the deformed spin network states  (in the spin network decomposition of the deformed states) which arise in the  constraint action (\ref{abaction})  on GR vertices are also of the type $S_{\alpha_{IJ},\e}$ so that (\ref{abaction}) is of the 
type (\ref{ngr}). We reiterate this observation in the first part of section \ref{sec4.2}.

\subsection{\label{sec2.5} Constraint action on any spin net}
Equations (\ref{abaction}) and (\ref{ngr}) imply that the constraint action on any spin net $|s\ket$ is:
\ba
{\hat H}_{\epsilon}(N) S(A) &:= &\frac{3}{8\pi}\sum_{v\in V_{GR}(s)}N(x(v)) \big(\;\sum_{I_v=1}^{N_v}\frac{ ([\frac{1}{\Perm{N_v-1}{3}}\sum_{\J_v}S_{\lambda_v,I_v, {\vec {\hat J}}_v,\e}]  - {\frac{1}{4}}S_{\lambda_v})}{\e}
\nonumber\\ 
&-& \sum_{I_v=1}^{N_v}\frac{  (\sum_{J_v\neq I_v}\sum_{K_v\neq I_v,J_v}{\tilde S}_{\lambda_v,(1)I_v,J_v,K_v, \e}) +(\sum_{J_v\neq I_v} {\tilde S}_{\lambda_v,(2)I_v,J_v, \e})}{\e}\;\big)\;\;\;\;\;\;\;\;\;\;
\nonumber\\
&+& \sum_{v\in V_{NGR}(s)}\frac{N(x(v))}{\e} \big(\;(\sum_{I_v=1}^{N_v} A_{vI_v}S_{\lambda_v,I_v, \e})  + B_v S_{\lambda_v}+ (\sum_{I_v=1}^{N_v}\sum_{J_v\neq I_v}\sum_{\alpha_{vI_vJ_v}} P_{\alpha_{vI_vJ_v}} S_{\alpha_{vI_vJ_v},\e})\big)\;\;\;\;\;\;\;\;\;\;\;\;\;\;
\label{action}
\ea
Here $V_{GR}(s)$ is the set of volume non-degenerate GR vertices of $s$ and $V_{NGR}(s)$ is the set of volume non-degenerate NGR vertices of $s$. 
The $v$ subscripts are used in obvious notation to denote  dependence on the vertex $v$.

It is useful to denote the contribution from the GR vertices in the first two lines as ${\hat H}_{\epsilon}(N) S(A)|_{V_{GR}(s)}$ and that from the NGR vertices as 
${\hat H}_{\epsilon}(N) S(A)|_{V_{NGR}(s)}$ so that 
\be
{\hat H}_{\epsilon}(N) S(A) = {\hat H}_{\epsilon}(N) S(A)|_{V_{GR}(s)}+ {\hat H}_{\epsilon}(N) S(A)|_{V_{NGR}(s)} \;.
\label{action1}
\ee
For future purposes it is important to emphasise certain features of the  deformed states generated by the constraint action (\ref{action}). These features follow directly from the discussion and analysis in sections \ref{sec2.1.2}, 
\ref{sec2.3} and \ref{sec2.4} :\\
\noindent (i) In  the first line of (\ref{action}) the  GR vertex $v$ in $S$ is replaced by one  at $v_{I,\e}$  in $S_{\lambda_v,I_v,\J_v,\e}$. \\
\noindent (ii) In the second line of (\ref{action}), 
in the state ${\tilde S}_{\lambda_v,(2)I_v,J_v, \e}$  can be decomposed into spin networks. In each such spin network, there is a trivalent kink on the $J_v$th edge  and either a  trivalent or a bivalent kink  at  $v_{I_v,\e}$ in which case the   $v$ remains GR and has either the same  valence as in $S$
or this valence reduces by 1.
The state  ${\tilde S}_{\lambda_v,(1)I_v,J_v, K_v,\e}$ can also  be decomposed into spin networks. In each of these spin networks  there are a pair of kinks, a bivalent one along the $K_v$th edge emanating from $v$ and
a bivalent or trivalent one at $v_{I,\e}$. The point $v$ is either a bivalent kink or a GR vertex of valence 3 or more.
\\
\noindent (iii) In the 3rd line of (\ref{action}), the  NGR vertex $v$ in $S$ is replaced by an NGR vertex   at $v_{I,\e}$  in $S_{\lambda_v,I_v, \e}$. 
\\
\noindent (iv) In the 3rd line of (\ref{action}), the deformations $\alpha_{vI_vJ_v}$ create a pair of kinks one along the $J_v$th edge and another at $v_{I,\e}$. 
Depending on the properties of $\alpha_{vI_vJ_v}$, either or both of these kinks could be bivalent or trivalent. Consequently the valence of $v$ in $S_{\alpha_{vI_vJ_v},\e}$ could reduce
by  0 or 1 or 2 relative to its valence in $S$. Depending on the original vertex structure at $v$ in $S$, this could result in $v$ presenting itsef in $S_{\alpha_{vI_vJ_v},\e}$ as a GR vertex of valence 3 or 
more,  an NGR vertex of valence 3 or more, a trivalent kink or a bivalent kink or an interior point of a single semianalytic edge.
%If  both kinks are trivalent, $v$ presents itself as an NGR vertex with unchanged valence and edge tangent structure relative to
%its presentation in $S$.

\section{\label{sec3} Off Shell Basis States $\Psi_{B,f,h}$}

The off shell states are finite linear combinations of basis states. Each basis state is labelled by a $C^{r-1}$ 3d Riemannian metric $h$ and a $C^r$ function $f$ on the semianalytic $C^r$ Cauchy slice $\Sigma$.
Permissible metric labels are  elements of the set   ${\cal H}_{h_0}$ of distinct diffeomorphic images of the  metric $h_0$ where $h_0$ is a metric with no conformal isometries.
An off shell basis state is an element of the algebraic dual to the finite span of spin net states and can be thought of as a (in general) non-normalizable formal sum over spin network bra states (see Footnote \ref{fnalgdual}) .
The set of these  bra state summands $B$  is referred to as the Bra Set and the Bra Set serves as an additional label for basis off shell states.  Thus $\Psi_{B,f,h}$ denotes an off shell basis state labelled by 
a  $C^r$ function $f$, a metric $h$ with $h\in {\cal H}_{h_0}$ and a Bra Set $B$. The state is then specified by the coefficients of each bra in the Bra Set, each such coefficient being determined by $f,h$ and the bra summand
being multiplied by that coefficient. The coefficient of a bra $\bra s| \in B$ is then simply the evaluation of the algebraic dual off shell basis state on the ket counterpart of $\bra s|$ i.e. the coefficient  of $\bra s|$ is
$\Psi_{B,f,h}(|s\ket)$.
While the $f$ dependence of the coefficient is just the product of the evaluations of $f$ on every
%non-degenerate 
GR vertex of $\bra s|$, the $h$ dependence is more complicated and depends on the structure of the set of kinks
in the  graph underlying the spin net label $s$. 

The layout of this section is as follows. Subsection \ref{sec3.1} introduces nomenclature and definitions which  play a key role in the articulation of the subsequent sections.
In subsection \ref{sec3.2} we specify the properties of permissible Bra Set labels. In subsection \ref{sec3.3}
%we provide a precise definition of a kink in a graph which is suited to our requirements
%and 
we isolate a particular $h$ dependent structural feature of the set of kinks on a graph which is of use in specifying the $h$ dependence of the coefficent. 
In subsection \ref{sec3.4} we specify $\Psi_{B,f,h}$ through a specification of the coefficients $\Psi_{B,f,h}(|s\ket)$ for every $ \bra s| \in B$. In what follows we shall often refer to 
  $\Psi_{B,f,h}(|s\ket)$ as the amplitude of the off shell state on $|s\ket$.

\subsection{\label{sec3.1} Preliminaries}

In order to specify the off shell state it is necessary to define the following:
\begin{enumerate}
%\begin{itemize}
%\noindent (ii) {\em GR Intertwiner Equivalence Class $[s]_{\I}$}:  Let $s$ be a spin net label consisting of the (coarsest) graph $\gamma (s)$ underlying $s$, the colorings of its edges and the collection of its
%vertex intertwiners. Then $[s]_{\I}$ is defined to be the set of all spin nets with the same $\gamma (s)$, the same colorings as $s$ and the same intertwiners as $s$ at all non-GR vertices of $\gamma(s)$.
%Thus $s_1 \neq s$ is equivalent to $s$ iff it differs in its intertwiners at one or more  GR vertices of $\gamma (s)$ and the equivalence class of all such $s_1$ together with $s$ constitute $[s]_{\I}$.
%\\

\item{\em Bivalent Kink}: A Bivalent $C^{m}$ kink (with $0\leq m <r$)  is an intersection $k$ between two unoriented semianalytic edges
\footnote{\label{fnsa}A semianalytic edge is defined \cite{lost} to be a 1d semianalytic manifold with 2 point boundary.}
$e_i, i=1,2$ such that $k$ is a common endpoint for these edges and  which has the following property. 
Orient one of the edges to be incoming and 
the other outgoing at $k$.
%with the following properties.
Fix  a  semianalytic $C^r$ coordinate patch around 
$k$. Then for the case $m>0$\\
\noindent (a) there exists a parameterization of each edge consistent with its orientation in which the first $m$ derivatives at $k$ satisfy $\frac{d^m}{dt_1} e_1^{\mu}= \frac{d^m}{dt_2^m} e_2^{\mu}$  
 and in which at $k$  the $(m+1)$th derivatives are unequal i.e. $\frac{d^{m+1}}{dt_1^{m+1}} e_1^{\mu}\neq \frac{d^{m+1}}{dt_2^{m+1}} e_2^{\mu}$,  and,\\
\noindent (b) there is no choice of parameterizations
 in which the first $m+1$ derivatives are equal at $k$.   For $m=0$ a bivalent $C^0$ kink is one at which the edge tangents at $k$ are linearly independent.  
 It is straightforward to see that these definitions are independent of the choice of coordinates and the choice of which edge is taken to be incoming.
Consequently the notion of a $C^{m<r}$ bivalent kink is diffeomorphism  invariant with respect to $C^r$ semianalytic diffeomorphisms.

\item {\em Trivalent Kink}: A trivalent kink $k$ is the intersection of an unoriented  semianalytic edge $e$ located in its interior, with a distinct unoriented semianalytic edge $e^{\prime}$ with an end point at  $k$.
%with the following property. 
Trivalent kinks can be of type $C^m$, $0\leq m\leq r$. These kink types are defined as follows:

\noindent 2.1 Choose an orientation for $e,e^{\prime}$.
%and choose an outgoing orientation for $e^{\prime}$ at $k$.
If 
the edge tangents of $e$ and $e^{\prime}$ are linearly independent at  $k$ the kink is defined to be a $C^0$ kink. Clearly this notion is invariant under the action of $C^r$ semianalytic diffeomorphisms.

\noindent 2.2 The point $k$ is defined to be a trivalent $C^m$  kink for $0<m<r$ if 
there exists an orientation assignment for $e$ with $e^{\prime}$ assigned an outgoing orientation such that the incoming part of $e$ at $k$ together with $e^{\prime}$ (which has been chosen to be  outgoing at $k$)
form a $C^m$ bivalent kink. Clearly the notion of such a $C^m$ kink is invariant under the action of $C^r$ semianalytic diffeomorphisms.
%and parameterizations of $e,e^{\prime}$ 
%consistent with their chosen orientations such that either \\
%(i) the incoming part of $e$ together with 
%$e^{\prime}$ forms  a bivalent $C^{m}$ kink with $0<m<r$  (so that the edge tangents to $e^{\prime}$ and the outgoing part of $e$  point in opposite directions) or,\\
%(ii) the outgoing part of $e$ together with 
%$e^{\prime}$ forms  a bivalent $C^{m}$ kink with $0<m<r$  (so that the edge tangents to $e^{\prime}$ and the incoming part of $e$  point in opposite directions).

%Finally, if $e,e^{\prime}$ are such that there exists an orientation of $e^{\prime}$  together with  parameterizations of $e,e^{\prime}$ consistent with their chosen orientations in which the first $r$ derivatives along the  outgoing part 
%of $e$ and those of $e^{\prime}$ agree or the first $r$ derivatives 
%along the  ingoing part of $e$ and those of $e^{\prime}$ agree at $k$, then $k$ is defined to be a $C^r$ trivalent kink.

\noindent 2.3 The point $k$ is defined to be a trivalent $C^r$ kink  if\\
(i) $e^{\prime}$ is assigned an outgoing orientation, and\\
(ii) there exists an orientation assignment and parameterization $t$ consistent with this assignment for $e$ together a paramaterization $t^{\prime}$ of  $e^{\prime}$ consistent with its  assigned outgoing orientation at $k$
such that 
the first $r$ derivatives, $\frac{d^m}{dt^{\prime m}} e^{\prime\mu}, m=1,..,r$   
along the  outgoing part 
of $e^{\prime}$ agree  with their counterparts, 
$\frac{d^m}{dt^m} e^{\mu}, m=1,..,r$
along the  incoming  part  (as well as, by virtue of $e$ being a single edge,  the outgoing part) of $e$ at $k$.
A $C^r$ kink $k$ can also be seen as the intersection of 3 edges $e_1,e_2,e_3$ in which two of the edges (say $e_2,e_3$) merge  into the remaining edge $e_1$ in a $C^r$ manner
(i.e. with $e_2,e_3$ incoming and $e_1$ outgoing there exist parameterizations $t_1,t_2,t_3$ of the three edges consistent with their orientations such that 
%the first $r$ derivatives agree with each other so  
$\frac{d^m}{dt_i^m} e_i^{\mu},i=1,2,3$ agree  with each other for each $m \in \{1,..,r\}$).\\
It can be checked that this definition is  coordinate independent so that the notion of a $C^r$ trivalent kink is  invariant under the action of $C^r$ semianalytic diffeomorphisms.

\item{\em Vertex}: An $N$ valent vertex $v$ for $N>3$ is the intersection of $N$ distinct semianalytic edges each with a single endpoint at $v$ (thus a loop through $v$ would contribute 2 distinct edges).
A 3 valent vertex is the intersection of 3 distinct semianalytic edges which is not a trivalent kink.
Note that the trivalent vertex can be NGR (i.e. have linearly dependent edge tangents) without being a kink.

\item{\label{easn}\em Embeddable Abstract Spin Network}:  We develop the notion of an embeddable abstract spin network as an abstract  labelled structure with the labels specifying the nature of  realizations (i.e. embeddings) of this
structure as an embedded spin network. 

Our starting point is the abstract spin network defined by Baez \cite{baez}. We review that definition here, for details see section 2 of Reference \cite{baez}.
An abstract (oriented) graph is specified by its edges, vertices and connectivity i.e. which edges meet at which vertices.
More in detail an  abstract graph 
$\gamma$ comprises of 
a finite set $E$ of edges, a finite set $V$ of vertices, and functions
$s: E\rightarrow V$, $t: E\rightarrow V$
where  the vertex $s(e)$ is called the source of the edge $e$, and the vertex $t(e)$ the target of
the $e$. An edge is outgoing at its source and incoming at its target. An abstract  spin network consists of an abstract graph $\gamma$, a labelling of each  its edges $e_I$ by  spin quantum numbers $j_I$ 
(i.e. by representations of $SU(2)$) and
an assignation of invariant tensors called intertwiners ${\cal I}_v$ , one for every vertex $v$, which map the product of incoming edge representations to that of outgoing edge representations. Note that there
is a slight degeneracy in  the set of labels as follows.
%\\
%\noindent{\em Caveat:}
Consider an abstract spin network state specified by $\gamma, \{j_I,\}, \{{\cal I}_v\}$. Change $\gamma$ to $\gamma^{\prime}$  
by flipping  some or all of the orientations of the edges of $\gamma$. Retain the same spin labels for the edges in $\gamma^{\prime}$ as for their counterparts  in $\gamma$. Then from standard $SU(2)$ representation theory 
their exists a unique (at most upto rescaling by a constant $c_v$ such that $\prod_vc_v=1$) choice of vertex intertwiners $\{{\cal I}_v^{\prime}\}$ such that the same spin network state is 
specified by the labels $\gamma^{\prime}, \{j_I,\}, \{{\cal I}^{\prime}_v\}$.
In what follows, we shall append the qualifier `abstract' to any element of the abstract graph whenever there is a possibility of confusion between abstract and embedded structures. In particular for abstract vertices
as defined above, we do not distinguish between vertices and kinks, whereas we do distinguish between embedded images of such vertices as kinks and vertices as defined above.

We now embellish the definition of an abstract spin network by augmenting the intertwiner specification  with additional finitely specified information (called a `decoration') at each of its vertices.
To this end, given an abstract graph we define its decorated counter part 
%at each of its vertices 
as follows.

First consider the case of bivalent vertices. We shall always work with a representative of the abstract spin network with orientation such that at any bivalent kink there is one incoming and one outgoing edge (it is straightforward to see that
this is always possible by suitable orientation flips and intertwiner choices) with the intertwiner simply being the identity matrix.
%We choose the standard Clebsch-Gordon normalization to specify this intertwiner, thereby fixing the intertwiner completely.
%We then  drop the intertwiner label for any such vertex  as this  label is redundant and completely determined by the edge orientation and choice of normalization.
We tailor our definitions below in such a way that the only abstract vertices which are permitted to be bivalent are those which are embeddable as $C^{m<r}$ kinks.
Accordingly, for a bivalent vertex
the decoration consists of a single integer valued between $0$ and $r-1$, the idea being that  permissible embeddings would embed the two edges at this vertex 
(more, precisely, their unoriented counterparts) in such a way that the decorated vertex is embedded  
as a bivalent $C^{m<r}$ kink.

Next consider trivalent vertices. 
%In these cases there is a {unique} intertwiner (upto rescaling by a constant) consistent with the specified  orientation of the edges at these vertices in the abstract graph
%and their spin labels. 
For each trivalent vertex, we specify an intertwiner (note that for this 3 valent case  the intertwiner choice is limited to rescaling of the standard Clebsch-Gordon coefficient appropriate to the   edge colorings and orientations at such
a vertex). 
The decoration of a trivalent vertex  consists of the following labels. Let the 3 {\em unoriented} edges intersecting at the abstract vertex  be $e_{I_1},e_{I_2},e_{I_3}$. 
The first label is an integer $-2 \leq m_0 \leq r$.
If $m_0= -2$, the vertex is to be embedded as an NGR vertex, if $m_0=-1$ the vertex is to be embedded as a GR vertex. If $ m_0\neq -1,-2$,  the abstract vertex
is to be embedded as a $C^{m_0}$ kink. 
To complete the specification of this embedding,  we append an   additional  label  comprising of one of the edge labels $e_{I_i}$.  
If $m_0<r$   the abstract vertex is to be embedded as a trivalent $C^{m_0<r}$ 
kink such that 2.1 (for $m_0=0$) or 2.2 (for $r>m_0>0$) above is satisfied with $e_{I_i}= e^{\prime}$ and  $\cup_{j\neq i}e_{I_j}=e$  embedded as a single $C^r$ edge (see 2.1, 2.2 above).
If $m_0=r$  the vertex is to be embedded as a $C^r$ trivalent kink, with 2.3 satisfied with $e_{I_i}= e_1$  and the two remaining edges  
merging in a $C^r$ manner with this
${I_i}$th edge (see 2.3 above).

If the abstract vertex has valence greater than 3, we specify an intertwiner. In contrast to the trivalent case,  there is a finite (and greater than one) dimensional vector space of such intertwiners.
The decoration label  consists of a single integer $n$,  $n\in \{0,1\}$. If $n=0$ the vertex is to be embedded as a GR vertex, else as an NGR vertex.
%In this case an explicit intertwiner specification is necessary.

A decorated abstract spin network consists of a decorated graph, a labelling of its edges by  spin quantum numbers (i.e. by representations of $SU(2)$) and
an assignation of  intertwiners, one for every vertex of valence greater than 2,  which maps the product of incoming edge representations to that of outgoing edge representations (modulo the degeneracy in this labelling through 
orientation and intertwiner change as discussed above). 

We refer to the (usual) spin networks    defined on graphs in $\Sigma$ as embedded spin networks. 
Consider a representative of an embedded spin network defined on
the coarsest graph underlying it (in which case the only remaining  choice of representative is its  edge orientation and the associated intertwiners, this choice being the exact counterpart of the one discussed
above for abstract spin networks). Such a  representative, through its label set,  naturally defines an associated decorated abstract spin network.
A decorated abstract spin network ${\bf S}$ is defined to be an {\em embeddable abstract spin network} iff there exists an embedded spin network $S$ with 
associated decorated abstract spin network ${\bf S}$. In such a case, $S$ will be called an {\em embedding} of ${\bf S}$ and ${\bf S}$ will be said to embed into $S$.
Thus an {embeddable abstract spin network} is a decorated abstract spin network which admits at least one  embedding into $\Sigma$ as an embedded spin network.
Clearly, while an embeddable abstract spin network will have infinitely many distinct embeddings,  two distinct (i.e. non-isomorphic
\footnote{An isomorphism ${\iota}$ between two embeddable abstract spin networks  ${\bf S_1}, {\bf S_2}$ is a bijection from the  vertex set $V_1$ of ${\bf S_1}$ to the vertex set $V_2$ of ${\bf S_2}$ 
in such a way that (a) $v_2= {\iota (v_1})$ has the same decorations as $v_1$, (b) for every incoming (outgoing) colored edge at $v_1$ there is a corresponding identically colored and oriented edge at $v_2$ and vice versa,
and (c) $v_2$ and $v_1$ have identical intertwiners.}
and inequivalent in the sense of the orientation-intertwiner label degeneracy described above)
embeddable abstract spin networks cannot embed into the same embedded spin network.

\item{\em Asymmetric Colored Graph}:
%{\em Asymmetric decorated abstract spin network}: 
Consider the {\em unoriented colored graph} $\alpha_{c}({\bf S})$ of a decorated abstract  spin network ${\bf S}$ obtained by dropping the orientation,  intertwiner and decoration information from ${\bf S}$.
Thus $\alpha_{c} ({\bf S}) $ is specified by (a) the vertex set $V(S)$, (b) edge set $E$, (c) a map $b$ from $E$ to $V\times V$ such that $b(e\in E)= \{x,y\} =\{s(e), t(e)\}$ with $x,y$ called the endpoints of $e$ and (d) a labelling of 
elements $e$ of $E$ by spins $j(e)$.  Let the number of edges with end points $\{x,y\}$ be $\mu (x,y)$.

A {\em colored graph automorphism} of $\alpha_c$  is a bijection of $\alpha_c$ which is defined through  a permutation $\sigma$  on $V ({\bf S})$  such that for each edge $e$ with end points $\{x,y\}$ 
there exists an edge $e_{\sigma}$ with end points $b(e_{\sigma})= \{ \sigma (x), \sigma (y)\}$ with $j(e_{\sigma})= j (e)$ and $\mu (\sigma(x), \sigma (y)) = \mu (x,y)$.
An {\em asymmetric colored graph} is a colored graph which has no automorphisms apart from the identity.
%
%decorated abstract spin network is one  for which its colored graph has no automorphisms apart from the identity. The colored graph of an { asymmetric} decorated abstract spin network will be
%called an {\em asymmetric colored graph}. Since automorphisms map bivalent vertices to bivalent ones, it is straightforward to see that if a
%colored graph is asymmetric, the graph obtained by placing a bivalent vertex in the interior of any of its edges is also asymmetric.

\item{\label{vib}\em Asymmetric Spin Network Basis}: The Volume operator acts at vertices of spin networks as a self adjoint operator on vertex intertwiners \cite{rsvol,aajurekvol}.
For fixed vertex valence and edge colorings and edge orientations, the intertwiner space is finite dimensional and the self adjoint operator preserves each such finite dimensional space and thereby restricts to 
a self adjoint operator on this finite dimensional vector space of intertwiners.
The eigen states of the operator action in this finite dimensional space  constitute an orthonormal basis of intertwiners for non-degenerate eigen values and for degenerate eigen values some orthonormal basis
spanning the eigen space can be chosen and fixed once and for all. By an orthonormal basis we mean the following. Consider an $N$ valent vertex at which there are $M$ outgoing edges with spins $j_1,..,j_M$ 
and 
$N-M$ incoming edges with spins $j_{M+1},..j_N$ so that an intertwiner $C$ at this vertex has the index structure $C^{A_1..A_M}{}_{A_{M+1}...A_N}$ with $A_I$ being a spin $j_I$ representation space label. 
Then an orthornormal basis $\{C_{(\mu)}\}$ of intertwiners for this 
configuration of oriented colored edges  satisfies:
\be
\sum_{A_I, I=1,..N} C^{A_1..A_M}_{(\mu)}{}_{A_{M+1}...A_N}(C^{A_1..A_M}_{(\nu)}{}_{A_{M+1}...A_N})^*
= (\prod_{I=1}^{N} \sqrt{2j_I+1}) \;\; \delta_{\mu,\nu}
\label{intnorm}
\ee
We call this basis of intertwiners as a  {\em volume intertwiner basis}. It is straightforward to check that the normalization condition (\ref{intnorm}) for $\mu=\nu$ ensures 
that the embeddings of abstract spin networks with intertwiners so normalised are of unit norm with respect to the  Ashtekar Lewandowski  inner product on the kinematic Hilbert space
of LQG.

Our aim is to construct a well defined basis for embedded spin network states. For future purposes we require the basis set to map into itself under the action of diffeomorphisms and electric diffeomorphisms.
Even just for diffeomorphisms this does not seem possible for arbitrary states because of the possible existence of graph symmetries \cite{alm2t} which can map a spin network on 
a graph with such symmetries to a distinct spin network on the same graph. However for sufficiently asymmetric spin networks it turns out that we can construct the desired basis as follows.

Consider a colored unoriented graph $\gamma_c$ underlying some  embeddable abstract spin network.  Restrict attention to $\gamma_c$ with the following properties:\\
(a) $\gamma_c$ is an asymmetric colored graph.\\
(b) If $v$ is a bivalent vertex of $\gamma_c$ then $v$ is connected to itself by a single edge i.e. the only bivalent vertices in $\gamma_c$ are the single vertices in loops
with each loop corresponding to a single connected component of $\gamma_c$.\\
%The only bivalent vertices in $\gamma_c$ are those connected by a single edge 
(c) $\gamma_c$ has no loops which start and end at  vertices with valence greater than 2    i.e. 
there is no edge with identical endpoints such that this endpoint has valence greater than 2.\\
%(c) $\gamma_c$ has no bivalent vertices.\\
%Pick a representative from each isomorphism class of $\gamma_c$ and abuse notation to call it $\gamma_c$. 
%By virtue of property (a) and (b), even if we endow the edges of $\gamma_c$ with orientations, the resulting oriented graph is also asymmetric (i.e. there are no orientation preserving automorphisms
%of such a gra

%In order to accomodate the action of electric diffeomorphisms, we are interested in defining a basis for the space of embeddings of all possible abstract decorated  spin networks living on the (by definition) undecorated graph $\gamma_c$
%as well as on any graph $\gamma{\prime}_c$ obtained from $\gamma_c$ by the introduction of bivalent vertices in the `interior' of its edges (we explain what we mean by this below).
%We proceed as follows. 
Note from (b)  that $\gamma_c$ consists of the disjoint union of one or more connected components which are not single loops with a set of zero or  one or more single loops
\footnote{ Reference \cite{lost} defines a semianalytic edge as one with distinct endpoints; this necessitates  a slightly involved definition of embedding of a loop. We refrain from 
commenting on this further as we will be interested in loops with multiple kinks,  for which there is no tension between  abstract and embedded notions.} 
i.e.
$\gamma_c= \gamma_{0c} \cup L$ where $\gamma_{0c}$ has no bivalent vertices and $L$ is the disjoint union of a set of loops or $L$ is empty. If $L$ is not empty, the asymmetry of $\gamma_c$ implies the edge in each distinct loop must be of a distinct
color. In order to accomodate the action of electric diffeomorphisms, we are interested in defining a basis for the space of embeddings of all possible abstract decorated  spin networks living on the (by definition) undecorated graph $\gamma_{0c}$
as well as on any graph $\gamma{\prime}_{0c}$ obtained from $\gamma_{0c}$ by the introduction of bivalent vertices in the `interior' of its edges (we explain what we mean by this below).
Since electric diffeomorphisms act non-trivially only on non-degenerate vertices, we shall treat the the `loop' sector seperately.
Accordingly we focus first on spin network labels associated with $\gamma_{0c}$ and subsequently  discuss the spin net labels associated with the loop sector.

We choose an orientation of the edges of $\gamma_{0c}$  and {\em fix this choice once and for all}.
As discussed in \ref{easn}, intertwiners at trivalent vertices are then fixed by $SU(2)$ representation theory upto a constant rescaling. The normalization (\ref{intnorm}) fixes this constant upto phase
and we fix this phase once and for all by demanding that the intertwiners be real (see for e.g \cite{makkinen}). With this choice fixed, we omit explicit  intertwiner specificaton  for 3 valent vertices, it being
understood implicitly that we use the above fixed choice.
At every $N>3$ valent vertex  $v$ in the vertex set of $\gamma_{0c}$  we fix an orthonormal volume intertwiner basis $\{C_{v(\mu_v)}\}$ with appropriate index structure and choose the intertwiner at $v$
 to be one of the volume intertwiner basis intertwiners. 

Next, introduce  bivalent vertices in the `interior' of each edge i.e.  in an edge $e$ with color $j(e)$ and  end points $v,\bv$ introduce bivalent vertices $k_1,k_2,..,k_n, n\geq 1$ such that  $k_j$ 
is connected to $k_{j-1}, k_{j+1}$ by an edge with color $j(e)$ with $k_0:=v, k_{n+1}:= \bv$. Let the orientation of the edges be consistent with that of $e$ and fix the intertwiners in accordance with \ref{easn} above.
It is straightforward to check that the `identity' intertwiner so fixed satisfies (\ref{intnorm}).
Decorate the vertices of the resulting spin nets and for simplicity restrict the decorations of the bivalent vertices to correspond to their embedding as $C^0$ bivalent kinks.  
Note that the above introduction of bivalent vertices changes the colored graph $\gamma_{0c}$. 
It is straightforward to see that the new colored graph $\gamma^{\prime}_{0c}$ so obtained is also asymmetric and hence satisfies (a), (c) above.
\footnote{\label{fnbiv}If not there exists an automorphism $\sigma$ which sends the sequence of vertices $v,k_1,..k_n,\bv\neq v$ connected to each other by single edges to another such sequence. By virtue of (b),(c) above this sequence must  start 
at $\sigma(v) \in \gamma_{0c}$ and end at $\sigma (\bv)\neq \sigma (v) \in \gamma_{0c}$ with
identical connectivity so that this sequence must arise through the introduction of bivalent kinks on some edge $\sigma (e)$.  
Restricting the automorphism to the vertex set of $\gamma_{0c}$ (so that no bivalent kinks are involved) yields an automorphism of $\gamma_{0c}$ which must be the identity so that 
$\sigma (v)= v, \sigma (\bv)= \bv, \sigma (e)=e$, which in turn implies that the sequence $ \{ \sigma (k_i) i=0,..,n\}$ is associated with the edge $e$.
This implies that 
$\sigma (k_1)= k_1, \sigma (k_n)= k_n$, which implies $\sigma (k_2)=k_2,\sigma (k_{n-1}) =k_{n-1}$ and so on, so that the automorphism $\sigma$ is necessarily the identity, and the new graph is necessarily asymmetric.}
As implied by Footnote \ref{fnbiv} and discussed further below every $N\geq 3$ valent vertex of $\gamma^{\prime}_{0c}$ can be identified uniquely with a vertex in $\gamma_{0c}$ and as we shall see below, this identification
allows us to choose the same orthornormal volume intertwiner basis at the  $N>3$ valent vertices so identified. 
%We fix an orthornormal volume intertwiner basis at every  $N>3$ valent vertex. this basis being the same that chosen for
Varying over all possible basis intertwiners consistent with this choice, and all possible decorations, we obtain a set of abstract decorated spin network states living on the colored graph $\gamma_{0c}$ and colored graphs  obtained from it 
through the introduction of bivalent $C^0$ kinks in the manner detailed above.  
Let the subset 
 of embeddable elements of this set be ${\bf E}_{\gamma_{0c}}$.
\footnote{It seems straightforward to argue that all members of the set are embeddable but this is not necessary for our purposes.}
Since the embellishment of an abstract spin network with decorations does not, by definition, change its underlying colored graph,
the graphs underlying the set ${\bf E}_{\gamma_{0c}}$ of abstract decorated spin networks also  satisfy (a),(c).

We  briefly digress here to explain the role of property (c). Let $\gamma_{0c}$ be endowed with an orientation as above. 
This orientation naturally induces an orientation on the  graphs $\gamma^{\prime}_{0c}$ in the manner described above. Since the only automorphism of $\gamma_{0c}, \gamma^{\prime}_{0c}$ is the identity
and since, by virtue of (c), every edge has distinct vertices, the identity automorphism is orientation preserving. In contrast if property (c) did not hold, the identity 
automorphism of any loop  could still be consistent with orientation reversal of its embedding and this could change the embedded spin network function while leaving its underlying embedded
colored graph untouched. 
%In contrast for any single loop labelled with color $j$, the only gauge invariant embedded spin network corresponds to the trace of its holonomy in the spin $j$ representation
%and  from the theory of $SU(2)$ representations, the evaluation of this  trace is {\em independent} of the orientation of the loop. 
This ends the digression and we return to the consideration of 
the set ${\bf E}_{\gamma_{0c}}$ of decorated  abstract spin networks on $\gamma_{0c}, \gamma^{\prime}_{0c}$ and turn to a discussion of  their embeddings.

%Accordingly, 
%let the subset 
% of embeddable elements of this set be ${\bf E}_{\gamma_{0c}}$.
%\footnote{It seems straightforward to argue that all members of the set are embeddable but this is not necessary for our purposes.}
%These elements are non-isomorphic as a result of the choice of fixed representative $\gamma_{0c}$ above.
Accordingly, let the set of   embedded spin networks obtained as all possible embeddings of elements of ${\bf E}_{\gamma_{0c}}$ be  ${E}_{\gamma_{0c}}$.
Consider ${\bf s} \in {\bf E}_{\gamma_{0c}}$ which happens to have colored graph $\gamma_{0c}$.
Let ${\bf s}$
embed to $s \in {E}_{\gamma_{0c}}$ via the embedding map $F: {\bf s} \rightarrow s$ and call  the colored graph underlying $s$ as $\gamma_{0c} (s)$.
It follows that  $s$ is the unique embedding of ${\bf s}$ on $\gamma_{0c} (s)$ i.e. there exists no $ {\bar s} \in {E}_{\gamma_{0c}}, {\bar s}\neq s$ such that $\gamma_{0c}({\bar s}) = \gamma_{0c} (s)$.
%\footnote{Note that since the elements of ${\bf E}_{asym}$  are non-isomorphic, distinct elements embed to distinct images.}
To see this, proceed as follows. 
%First note that for any embedding map $G$,  the embedding $G(\gamma_{0c})$ also satisfies (a),(b). This follows from the fact that if $a$ is a nontrivial  automorphism of $G(\gamma_{0c})$, ${\bf a}= G^{-1}a G$ 
%is a nontrivial automorphism of $\gamma_{0c}$ and from the fact that $G^{-1}l$ is a loop in $\gamma_{0c}$ if $l$ is a loop in $(\gamma_{0c})$. Next, 
Let there exist ${\bar s}= {\bar F} ({\bf s})$ for some embedding ${\bar F}\neq F$ with $\gamma_{0c}({\bar s}) = \gamma_{0c}(s)$ and  ${\bar s}\neq s$. 
The absence of nontrivial automorphisms of $\gamma_{0c}$ implies that ${\bar F}^{-1}F \gamma_{0c} = \gamma_{0c}$ so that 
if $V$ is the vertex set  and $E$ the set of  edges of $\gamma_{0c}$, we have that
%the absence of non-trivial automorphisms of $\gamma_{0c} (s)$ implies that
$F(v)={\bar F}(v) \forall v\in V, F(e)= {\bar F}(e) \forall e \in E$. By virtue of 
our digression above $F(e),{\bar F}(e)$ have the same orientations. Since $F,{\bar F}$ yield the same embedded oriented colored graph  and map  the intertwiner on every vertex  $v$ of ${\bf s}$ to $F(v)={\bar F}(v)$, 
it must be the case that $s={\bar s}$. Similar comments apply to embeddable abstract spin nets with colored graphs of type $\gamma^{\prime}_{0c}$ by virtue of their asymmetry. 
Moreover, since any $\gamma^{\prime}_{0c}$ is obtained by the introduction of bivalent 
vertices in the `interior' of edges of $\gamma_{0c}$, there is a natural 1-1 mapping of those vertices of $\gamma^{\prime}_{0c}$ which are of valence greater than 2 to the vertex set of $\gamma_{0c}$ 
so that we may label these vertices of $\gamma^{\prime}_{0c}$ with elements of the vertex set $V$ of $\gamma_{0c}$. As indicated in Footnote \ref{fnbiv},  with this labeling of such vertices
there is  a unique association of edge connected sequences of bivalent vertices between end points $v,\bv$ in $\gamma^{\prime}_{0c}$ with an edge with these end points in $\gamma_{0c}$.
It follows that at any such vertex $v$, one can uniquely identify incoming and outgoing edges on the colored  graph $\gamma^{\prime}_{0c}$ with the chosen fixed orientation above 
with their counterparts in $\gamma_{0c}$ with the chosen fixed orientation. If the embeddable abstract spin networks ${\bf s}, {\bf s}^{\prime}$ with colored graphs $\gamma_{0c}, \gamma^{\prime}_{0c}$ both have  intertwiner $C_v$ at vertex $v$, 
we shall say that ${\bf s},{\bf s}^{\prime}$ have the {\em same} intertwiner at $v$ and it is in this sense that our choice of orthornormal volume intertwiner basis is the same for $v \in \gamma_{0c}, \gamma^{\prime}_{0c}$.
It follows that these identifications are also induced onto their embedded images. In particular if 
these spin networks are embedded through embeddings $F, F^{\prime}$, we say that  $F({\bf s}), F({\bf s}^{\prime})$ have the same
intertwiners at $F(v), F^{\prime}(v)$.

Let us now consider the loop component  $L$ of $\gamma_{c}$ (when non-empty) and color each loop $l_j$  with spin $j$ so that $l_1$  is colored with $j=1$, $l_2$ with $j=2$ and so on. 
Let us denote $L$ by $L_m$ when $L$ has $m$ such loops with $L_0$ being the empty set.
We restrict attention to the case in which the bivalent vertex $v$ is to be decorated with a $C^0$ kink decoration.
Next we introduce a pair of bivalent vertices  on each $l_j$, decorate one of them as a  $C^1$ kink and the second as a $C^2$ kink. We  call  $l_j$ embellished with these two kinks as $l^{(3)}_j$.
We call the collection $\{l^{(3)}_j\}$ so obtained as   $L^{(3)}_m$. Clearly $L^{(3)}_m$ has no decoration preserving colored graph automorphisms except the identity. We fix an orientation of each 
$l^{(3)}_j$ once and for all corresponding to a traversal from the $C^0$ kink $v$ to the $C^1$ kink and thence to the $C^2$ kink and finally back to $v$.
%Introduce a set of bivalent vertices to yield $l^{\prime}_j$,  and in obvious notation, $L^{\prime}$,  and let them all be decorated as $C^0$ kinks. In what follows we shall fix this choice of decorations once and for all
%and abuse notation to denote these decorated loops and the sets  thereof  by the same symbols as their undecorated counterparts i.e. by  $l_j, l^{\prime}_{j}, L, L^{\prime}$. Hereon these symbols refer to 
%loops decorated with $C^0$ bivalent kinks.
%, with $k_2^0$ signifying that all vertices
%are to be embedded as $C^0$ bivalent kinks. 
%The argumentation of Footnote \ref{fnbiv} can be repeated for the case that $v=\bv$ with the conclusion that (i) the only automorphism of $L$ is the identity (ii) the only automorphisms of $L^{\prime}$ are generated by
%independent cyclic permutations of the mutually connected bivalent vertices of each $l^{\prime}_j$. 
Next consider any embedding $F$ of $l^{(3)}_j$ into $F(l^{(3)}_j)$.  Note that $F(l^{(3)}_j)$ is a continuous oriented loop  with orientation acquired from  the fixed orientation chosen for $l^{(3)}_j$. 
With $F(l^{(3)}_j)$ associate
the spin network function  $\tr h_{F(l^{(3)}_j)}(F(v),F(v))$ where $h_{F(l{(3)}_j)}$ is the holonomy, in the spin $j$ representation,  around the loop $F(l^{(3)}_j)$ from its starting point $F(v)$ to its end point which is also $F(v)$ with $\tr$
denoting the trace in the spin $j$ representation. %Note that this trace is independent of orientation.
Since $l^{(3)}_j$ has no nontrivial decoration preserving colored graph automorphisms, similar to $\gamma_{0c}$ it admits a unique embedding to any of its embedded images. Finally, with the embedding $F(L^{(3)}_m)$ associate the 
spin network function $\prod_{j=1}^m\tr h_{F(l^{(3)}_j)}(F(v),F(v))$.

Let 
${\bf E}_{L^{(3)}_m}, m>0$ be the set containing  $L^{(3)}_m$  and let the set of   embedded spin networks obtained as above through holonomy traces from this single element of  ${\bf E}_{L_m}$ be $E_{L_m}$.
%and call the set consisting of 
%(iii) if we decorate each vertex of $l^{\prime}_j$ the automorphism group of 
%arbitrarily the comments of (ii) apply only if these decorations
%for fixed $l^{\prime}_j$ are identical.
%Next note that any fixed orientation of $l_j$  induces a unique orientation of $l^{\prime}_j$ with unique intertwiners in accordance with \ref{easn} above.  Our digression above then implies that any embedding of any decoration of  
%$L, L^{\prime}$ onto a fixed embedded graph $L^{\Sigma}, L^{\prime, \Sigma}$ yields a unique embedded spin network.  Call the resulting space of embedded spin networks as ${\bf E}_L$.
%We  briefly digress here to explain the role of property (b). Let $\gamma_c$ be endowed with an orientation as above. 
%This orientation naturally induces an orientation on the  graphs $\gamma^{\prime}_c$ in the manner described above. Since the only automorphism of $\gamma_c, \gamma^{\prime}_c$ is the identity
%and since, by virtue of (b), every edge has distinct vertices, the identity automorphism is orientation preserving. In contrast if property (b) did not hold, the identity 
%automorphism of any loop  could still be consistent with orientation reversal of the loop. We have imposed (b) to rule out this possibility. This ends the digression.
%We return to the consideration of the set of ${\bf E}_{\gamma_{m,c}}$ abstract decorated spin networks of the previous paragraph.%and by virtue of the uniqueness of embeddings, this statement only depends on the emdedded images of $\gamma_c, \gamma^{\prime}_c$.
Define  ${\bf E}_{\gamma_{m,c}}:= {\bf E}_{\gamma_{0c}}\otimes {\bf E}_{L^{(3)}_m}$ so that ${\bf E}_{\gamma_c}$
is the set of ordered pairs $({\bf s} \in {\bf E}_{\gamma_{0c}}, {\bf l} \in {\bf E}_{L^{(3)}_m})$ 
with  ${\bf l}=L^{(3)}_m$. Such an ordered pair naturally defines an  abstract spin network on the union of some decoration $d$ of $\gamma_{0c}$ (resp. $\gamma^{\prime}_{0c}$) with $L^{(3)}_m$.
Let us call the $\gamma_{0c}, \gamma^{\prime}_{0c}$ so decorated as $\gamma_{0c,d}, \gamma^{\prime}_{0c,d}$
\footnote{Note that the bivalent vertices on $\gamma^{\prime}_{0c}$ are restricted to be decorated so as to embed to $C^0$ kinks. Hence the only freedom left is the decoration of the vertices
of $\gamma^{\prime}_{0c}$ which are also vertices of $\gamma_{0c}$ and the notation indicates that these vertices are decorated exactly as in $\gamma_{0c}$.} 
and define the decorated colored graphs
$\gamma_{m,c,d}= \gamma_{0c,d}\cup L^{(3)}_m$, $\gamma^{\prime}_{m,c,d}= \gamma^{\prime}_{0c,d}\cup L^{(3)}_m$,
%An element of  ${\bf E}_{\gamma_{m,c}}$ lives either on the colored graph  $\gamma_{m,c}$ or a colored  graph of type $\gamma^{\prime}_{m,c}$.
%or ${\bf l}$  of type $L^{\prime}$. 
Given such an element of ${\bf E}_{\gamma_{m,c}}$  and an embedding $F$ of $\gamma_{m,c,d}$ (resp. $\gamma^{\prime}_{m,c,d}$), we define the embedding of this element  $({\bf s}, {\bf l})$ by $F$ as 
the embedded spin network function obtained as a product of the embedded spin network function $s = F ({\bf s})$   with the embedded spin network function defined by $F({\bf l})$.
The set of all distinct  embedded spin network functions obtained for all possible embeddings  of elements of ${\bf E}_{\gamma_{m,c}}$ is denoted as $E_{\gamma_{m,c}}$
where we have set ${\bf E}_{\gamma_{0,c}}= {\bf E}_{\gamma_{0c}}$.
%Call this set 
%obtained as the product of the spin network functions corresponding to each element of ${\bf E}_{\gamma_{0c}}$ with each element of ${\bf E}_L$.
%These define a set of embedded spin networks ${\bf E}_{\gamma_{c}}$ with each element having a `loop' component (corresponing to an element of ${\bf E}_L$)  and a `non-loop component'%
%corresponding to an element of ${\bf E}_{\gamma_{0c}}$.
Finally, note that embedded spin networks whose underlying embedded colored graphs do not coincide in $\Sigma$,  are orthogonal.

From the discussion above it is straightforward to see that the following statements hold:\\

\noindent (i) The elements of ${ E}_{\gamma_{m,c}}$  provide a well defined basis of states on spin networks 
living on all possible embeddings of (a) decorations $\gamma_{0c,d}$ of  the  undecorated colored graph 
$\gamma_{0c}$ together with $L^{(3)}_m$ 
as well as of (b) any  decorated graph $\gamma{\prime}_{0c,d}$ obtained from $\gamma_{0c,d}$ by the introduction of bivalent `$C^0$ kink'  vertices in the `interior' of its  edges together with $L^{(3)}_m$.
\\

\noindent (ii) Since the image of any embedding by a diffeomorphism is also an embedding, the set ${E}_{\gamma_{m,c}}$ is stable under the action of diffeomorphisms. Moreover vertices,  oriented 
edges and interwtiners are naturally identifed between any of these  images. In particular we may identify, in obvious notation 
$F(v), F(e), F({\cal I}_v) \in F({\bf s})$  with $F_{\phi}(v), F_{\phi}(e), F_{\phi}({\cal I}_v) \in F_{\phi}({\bf s})$ for $v,e \in \gamma_{0c}$
where the embedding $F_{\phi}$ is defined as $\phi\circ F$  for some diffeomorphism $\phi$.\\

\noindent (iii) 
%{\em Stability of ${ E}_{\gamma_{m,c}}$}: 
Let $S\in { E}_{\gamma_{m,c}}$ have a  non-degenerate vertex $v$ (so that $v$ lies in the non-loop component of ${S}$). Let $S_{\lambda_v, I_v,\e}$ be an  electric diffeomorphism deformation of $S_{\lambda_v}$ at $v$. Then 
$S_{\lambda_v}$ and $S_{\lambda_v, I_v, \e}$ have the 
same intertwiner at $v, v_{I_v,\e}$. Moreoever, since the intertwiners are volume eigenstates, it follows that modulo the same factor, $S_{\lambda_v}, S_{\lambda_v, I_v,\e}$ are elements of ${ E}_{\gamma_{m,c}}$. 
Accordingly, the abstract correspondent of $S_{\lambda_v, I_v,\e}$ (upto an overall factor) is in ${\bf E}_{\gamma_{m,c}}$ so that any electric diffeomorphism child of $S$, in the sense of \ref{childdef} below is in ${ E}_{\gamma_{m,c}}$.

Finally, any possible parent (in the sense of $\ref{pbleparentdef}$ below) of  $S_{\lambda_v, I_v,\e}$ must be an element of ${ E}_{\gamma_{m,c}}$. To see this, consider  a state ${\bar S}$ which when acted upon by the constraint
yields an electric diffeomorphism child $C$ with abstract colored graph isomorphic to that underlying  $S_{\lambda_v, I_v,\e}$. Since these graphs are asymmetric this isomorphism is unique and we may identify these graphs
using this isomorphism. Note that the embedded colored graph underlying ${\bar S}$ is obtained   by the removal of a triplet of kinks from the non-loop part of the colored graph underlying $C$ so that 
the abstract colored graph obtained from removing the corresponding set of bivalent vertices from the abstract colored  graph which embeds to the colored graph  underlying $C,S_{\lambda_v, I_v,\e}$
must embed to the embedded colored graph underlying ${\bar S}$ which implies that ${\bar S}\in   { E}_{\gamma_{m,c}}$.

Thus, given a state  in $S$ all its children by electric diffeomorphism in the sense of \ref{childdef}, and all its possible parents by electric diffeomorphism in the sense of \ref{pbleparentdef}, are in ${ E}_{\gamma_{m,c}}$.
We refer to this property as the  {\em stability} of the set ${ E}_{\gamma_{m,c}}$ under the action of electric diffeomorphisms.
\\
%
%The  set of embedded spin networks obtained as all possible embeddings  of these embeddable abstract spin networks constitute the desired basis.
%Endow each edge with an orientation which we fix once and for all such that  every bivalent vertex has an incoming and an outgoing edge.
%Intertwiners at trivalent vertices are fixed as discussed above.
%Consider any abstract decorated spin network state living on this oriented colored graph with  (a) intertwiners at bivalent and trivalent vertices  fixed as discussed in \ref{easn} above and (b) the intertwiner  at every $N>3$ valent vertex
%chosen to be one of the basis intertwiners.  The resulting decorated abstract spin network state is called an volume intertwiner basis state. 
%The space of such states provides a basis for all abstract decorated spin networks living on $\gamma_c$ and hence on their embeddings.

By considering all distinct (i.e. non-isomorphic) choices of $\gamma_{0c}$, all choices of decorations $d$ thereof and  all values  of $m\geq 0$, we obtain a basis on the space of all embeddings of embeddable abstract spin networks  living on all
corresponding choices of $\gamma_{m,c,d},\gamma^{\prime}_{m,c,d}$ 
%subject to (a),(b),(c) 
%as well as those
%living on  all $\gamma^{\prime}_{m,c,d}$ with $\gamma^{\prime}_{0c}$ denoting the embellishment of $\gamma_{0c}$ by the addition of bivalent vertices in edge interiors as in (ii) above. 
This basis is stable under the action of diffeormorphisms and electric diffeomorphisms. We call this basis as the {\em Asymmetric Spin Network Basis}.
%We call the space of such  embedded spin networks as {\em asymmetric (embedded) spin networks} and the basis we have defined on this space as the {\em Asymmetric Spin Network Basis}.
An embeddable abstract spin network which embeds into any element of this basis will be called an {\em  abstract asymmetric  basis spin network}

\item{\label{childdef}\em Electric diffeomorphism Child of an Asymmetric Spin Network Basis element}: Let $S$ be an asymmetric spin network basis element and let its abstract asymmetric counterpart be ${\bf S}$. 
%We define `children' in terms of the constraint action on a `parent'. To this end, recall that 
The action of the electric diffeomorphism part of the constraint (\ref{action}) at sufficiently small $\e$ on $S$ yields, besides $S_{\lambda}$  
states $S_{\lambda_v, I_v,{\hat J_v},\e}$ which are related to $S$ through the action of electric diffeomorphisms at nondegenerate vertices of $S$. 
Since each of these children have a decorated  abstract colored graph structure which is obtained by the embellishment of some of the edges of $S$ with bivalent  $C^0$ kinks, the discussion in \ref{vib} above
implies that each of these children (up to overall factors) are also asymmetric spin network basis elements.
Clearly for  all sufficiently small $\e$ the  states $S_{\lambda_v, I_v,{\vec J}_v,\e}$ for fixed $I_v, {\vec J}_v$ 
(upto overall factors) are embeddings of the same  abstract asymmetric spin network ${\bf C}( I_v, {\vec J}_v)$. We shall refer to any embedding of any ${\bf C} ( I_v, {\vec J}_v)  $
as a {\em child} of $S$ (note that ${\bf C}( I_v, {\vec J}_v)$ for distinct $(I_v, {\vec J}_v)$ need not be distinct). 
 %a number of spin networks each of whose graphs are deformations
%of the graph underlying $S$.  Let $C_\e$ be a spin network state which has a nontrivial inner product with any of these deformed spin network states (which do not include $S_{\lambda}$). 

%We start with a tentative definition of a child and after discussing its (de)merits provide an improved working definition. 

\item{\label{pbleparentdef}\em Possible parent by electric  diffeomorphism of an Asymmetric Spin Network Basis element}: Let $S$ be an asymmetric spin network basis element with abstract asymmetric counterpart
 ${\bf S}$. $P$ is called a {\em possible parent} of $S$ if all   embeddings of  ${\bf S}$ are children of $P$ in the sense of  \ref{childdef} above.
Since the abstract colored graph structure of $P$ is obtained by removal of a triple of  bivalent $C^0$ kinks from the non loop component of ${\bf S}$, the discussion in \ref{vib} above shows that 
$P$ is also an asymmetric spin network basis element.

\item{\label{un}\em Useful Notation}:
For the purposes of the next section  and section \ref{secl} it is useful to define the following notation which builds on that introduced in \ref{vib} above.
Accordingly, within the set ${\bf E}_{\gamma_{m,c}}$ defined above consider those elements which correspond to a fix decoration $d$ of the vertices of $\gamma_{0c}$.
%(recall that if $L$ is empty $\gamma_c= \gamma_{0c}$).
Call this subset  of ${\bf E}_{\gamma_{m,c}}$ as ${\bf E}_{\gamma_{m,c,d}}$. Let the set of embeddings of elements of this set be ${ E}_{\gamma_{m,c,d}}$.

%Next, recall that 
%\be
%\gamma_c = \gamma_{0c} \cup L
%\label{gul}
%\ee
%with $L$ being the loop component of $\gamma_c$. 
%Let us choose for $L$
%Let $L_m$ denote  
%the disjoint union of $m$ loops labelled $l_i, i=,1,..m$, with $l_j$ colored  by spin $j$. Set $L=L_m$ in (\ref{gul}) and denote
%$\gamma_c$ with this choice of $L$ as $\gamma_{(m)c}$ so that $\gamma_{(m) c}= \gamma_{0c} \cup L_m$. Accordingly, denote 
%${\bf E}_{\gamma_c}$  with $\gamma_c := \gamma_{(m)c}$ by ${\bf E}_{\gamma_{(m)c}}$  and the set of embeddings of elements of 
%${\bf E}_{\gamma_{(m)c}}$  by ${ E}_{\gamma_{(m)c}}$. Finally within  ${\bf E}_{\gamma_{(m)c}}$ restrict attention to elements with fixed decorations $d$ of $\gamma_{0c}$ and 
%call the resulting set ${\bf E}_{\gamma_{(m)c,d}}$  and the set of embeddings of elements of ${\bf E}_{\gamma_{(m)c,d}}$ as ${ E}_{\gamma_{(m)c,d}}$.

Define the union of the sets ${\bf E}_{\gamma_{m,c}}$ for all finite $m$  to be ${ {\mathbfcal E}}_{\gamma_{0c}}$ and the set of embeddings of elements of this union to be
${ {\cal E}}_{\gamma_{0c}}$ so that ${ {\mathbfcal E}}_{\gamma_{0c}} = \cup_{m=0}^{\infty} {\bf E}_{\gamma_{m,c}}$, ${\cal E}_{\gamma_{0c}} = \cup_{m=0}^{\infty} { E}_{\gamma_{m,c}}$, 
where we recall from \ref{vib} above that we have set $\gamma_{(0)c}\equiv \gamma_{0c}$.

Similarly define ${ {\mathbfcal E}}_{\gamma_{0c,d}}, { {\cal E}}_{\gamma_{0c,d}}$ as the fixed decoration (with decoration $d$ of $\gamma_{0c}$) elements  of ${{\mathbfcal E}}_{\gamma_{0c}}, { {\cal E}}_{\gamma_{0c}}$
so that ${{\mathbfcal E}}_{\gamma_{0c,d}} = \cup_{m=0}^{\infty} {\bf E}_{\gamma_{m,c,d}}$, ${\cal E}_{\gamma_{0c,d}} = \cup_{m=0}^{\infty} { E}_{\gamma_{m,c,d}}$.

%Fix some asymmetric colored graph $\gamma_{0c}$ of the type discussed above (so that $\gamma_{0c}$ has no connected components consisting of single loops) and 
%for $m>1$ define $\gamma_{m, c}= \gamma_{0c} \cup L_m$. Define the set of embedded spin networks ${\cal E}_{\gamma_{0c}}$ as
% ${\cal E}_{\gamma_{0c}} = \cup_{m=0}^{\infty} {\bf E}_{\gamma_{m,c}}$  (with $\gamma_{0,c}\equiv \gamma_{0c}$).
%Consider within this set those elements which correspond to a fix decoration $d$ of the vertices of $\gamma_{0c}$.
%Call this set ${\cal E}_{\gamma_{0c,d}}$.

\end{enumerate}

\subsection{\label{sec3.2} The Bra Set $B$}

Recall that the Bra Set label $B$ of an off shell state contains all the bra summands which when summed over with appropriate coefficients 
yield a formal sum which represents the off shell state $\Psi_{B,f,h}$. The dual action of the constraint ${\hat H}_{\e}(N)$ on this off shell state
is defined through its `amplitude', $\Psi_{B,f,h}({\hat H}_{\e}(N) S)$, for any spin net state $S$.  The $\e\rightarrow 0$ limit of these amplitudes
defines the continuum limit constraint action. Clearly only those spin net states generated by the constraint action ${\hat H}_{\e}(N) S$
which have non-trivial inner product with one or more elements of  $B$ contribute to the amplitude for $S$. The spin net states which do contribute in this way will 
be said to have overlap in $B$. In order to have adequate control on the calculations
of these amplitudes it suffices, that for any  $S$ with nondegenerate vertices,  either all the spin nets generated by the electric diffeomorphism part of the constraint action on nondegenerate vertices of $S$, together with $S$ have overlap in $B$ 
%on $S$  have overlap in $B$ 
or that none of them do (in which case the amplitudes will be seen to vanish in the continuum limit).
\footnote{In a previous version of this work, similar to \cite{p3} we attempted to impose a similar condition by also including propagation type children; however it turns out the 
weaker condition here suffices and can be imposed consistently.}
Further, it is reasonable to anticipate that this feature of overlap or lack thereof should persist
in an $\e$-independent way for sufficiently small $\e$ so that the continuum limit amplitudes exist. 
The properties (i)- (iii) below  of permissible Bra Sets are motivated by the  discussion above.
In their articulation,  we use the definitions of the previous section.
%in the specification of properties of permissible Bra Sets below. 
In particular we distinguish between vertices and kinks.\\

\noindent We require that  the  Bra Set label $B$ of an off shell basis state  have   the following properties:\\

\noindent (i) Each element $S$ of $B$ is an {\em asymmetric  spin network basis element}   whose vertices are all GR and which has at least one such vertex.
%We also require that every connected component of the colored graph underlying $S$ has a vertex of valence greater than 3.
%The intertwiners  at these  GR vertices  of $S$  are  chosen to be basis intertwiners so that each bra is an element of   the  Asymmetric Spin Network basis (see \ref{vib} of section \ref{sec3.1}).
If $S\in B$ then we can  replace  the  volume intertwiner basis element at each vertex of valence greater than 3   by a distinct such basis element to get a state $S^{\prime}$. 
(Recall that the volume intertwiner  basis is associated with abstract asymmetric basis spin networks which embed to their (embedded) asymmetric  spin network basis counterparts).
We require that any  $S^{\prime}$ obtained by such a replacement is also in $B$.
\\

\noindent (ii)  If $S\in B$ and ${\bf S}$ is the embeddable abstract spin net which embeds as $S$, then $B$ contains {\em all} embeddings of ${\bf S}$.
%\footnote{Note that this is consistent with (i) by virtue of the definition of the Asymmetric Spin Network basis in terms of embeddings of abstract structures in section \ref{sec3.1}
%}
\\

\noindent (iii)  Let  $C$ be  an electric diffeomorphism type  child of $S \in B$  as defined in \ref{childdef} (Clearly, all vertices of   $C$ are GR and $C$ has at least one such vertex).
%and each of its connected components have at least one vertex of
%valence greater than 3). 
%
%Replace the intertwiners at these vertices by some choice of basis intertwiners to get a state $C^{\prime}$.
Then $C \in B$.  If $P$ is a possible parent of $S \in B$ with parentage via electric diffeomorphism  as defined in \ref{pbleparentdef} then $P \in B$ (Clearly, all vertices of   $P$ are GR and $P$ has at least one such vertex).
%and 
%each of its connected components have at least one vertex of
%valence greater than 3).  .
%GR, replace the intertwiners at such vertices of $P$  by some choice of basis intertwiners to get a state $P^{\prime}$. Then $P^{\prime} \in B$.  Note that the choice of states $C^{\prime}, P^{\prime}$ in the above articulation 
%is irrelevant by virtue of (i).
\\

The asymmetry requirement  
in (i) in conjunction with the role of abstract structures in (ii) together with closure of $B$ with respect to appropriate parent-child relations in (iii) ensures that  $B$ is stable under the action of diffeomorphisms and 
electric diffeomorphisms as discussed in section \ref{sec3.1}.
The role of property (i) is also  to ensure sufficent overlap in $B$ given the fact that there is a finite dimensional space of intertwiners. 
Property (ii) ensures $\e$-independent overlap as well as $h$-independent overlap (see the 
discussion in section \ref{sec4.3}); the former is necessary for the continuum limit constraint action to exist and the latter is an important ingredient of our proof of diffeomorphism covariance of the constraint action 
in section \ref{sec6}. Property (iii) ensures that $B$ is closed under suitable  `parent-child' relations which in turn ensures the kind of overlap required in the discussion in the first paragraph of this section.
%While these properties suffice for our demonstration of anomaly freedom in  sections \ref{sec4} -\ref{sec6},  weaker properties may suffice as well. We comment on one possible weaking 
%in section \ref{sec7}. 
The reason we insist here on `at least one GR vertex' in (i) is for the ensuing simplicity of expression for the off-shell amplitudes in section \ref{sec3.4}.
%
%The insistence on at least one $N>3$ valent vertex for each connected component is to provide for the possibility of non-degenerate vertices (recall that gauge invariant vertices of valence less than 4 are degenerate)
%as well as for technical reasons discussed in section \ref{secl}.

%The existence of `at least one GR vertex' in (iii) is already implied by (i) by virtue of conditions (a)-(c) in \ref{vib} of section \ref{sec3.1}, those conditions deriving from the technical discussion therein. 
%Notwithstanding this we 
%retain the slightly redundant articulation above of this property of `at least one GR vertex' as it plays a direct role in the  simplicity of expression for the off-shell amplitudes in section \ref{sec3.4}. 
%it plays a direct role in 
%Note also that (i) permits degenerate GR vertices; this also does not seem to be a crucial feature and we believe that further restriction to non-degenerate GR vertices would not alter our results.

Our demonstration of  anomaly freedom of  hamiltonian constraint commutators in section \ref{sec4} and \ref{sec5}  uses only properties (i)-(iii)
without any further specification of Bra Sets. 
The considerations of section \ref{sec6}
which demonstrate diffeomorphism covariance require a certain property of linear independence of off -shell states  to hold. This property  is proved in section \ref{secl}. The  proof of linear independence for  off shell states with a single Bra Set label
but different metric and `vertex smooth' function labels requires Bra Sets to satisfy an additional property (iv)(a). The proof for the general case of different Bra Set labels requires a further property (iv)(b).
It seems plausible that the properties (iv)(a,b) are implied by property (iii) above. However a putative proof is beyond the scope of this work. Instead in section \ref{secl} we explicitly construct 
a rich family of Bra Sets which satisfy properties (i)-(iv).

More in detail:
It is straightforward to see that properties  (i)-(iii) hold if we choose as our Bra Set, the set 
${E}_{\gamma_{m,c,d}}$ (see \ref{un} of section \ref{sec3.1} above) with $\gamma_{m,c,d}$ chosen so that all vertices of states in this set are GR and there is at least one such vertex.
Properties (i),(ii) hold by virtue of this choice and due to the realization of elements of ${E}_{\gamma_{m,c,d}}$ as embeddings of abstract structures and (iii) holds by virtue of the procedure of 
adding bivalent kinks to the interior of edges of $\gamma_{0c}$ as outlined in \ref{vib} of the previous section.
Clearly identical argumentation applies to the set ${\cal E}_{\gamma_{0c,d}}$  with ${\gamma_{0c,d}}$ chosen so that all vertices of states in this set are GR and there is at least one such vertex. 
Thus ${\cal E}_{\gamma_{0c,d}}$ for any such ${\gamma_{0c,d}}$ 
%satisfies (i)-(iii). 
also satisfies (i)-(iii)
As we shall see in section \ref{secl7}, a family of  Bra Sets for which we are able to conclusively  demonstrate  satisfaction of properties (i)-(iii) {\em and} (iv), 
is provided by  the family of sets ${\cal E}_{\gamma_{0c,d}}$ (one for each distinct isomorphism class of $\gamma_{0,c}$ with appropriate $d$).\\
%so as to ensure (i)) as Bra Sets.

%We shall refer to ${\bf E}_{\gamma_{c,d}}$ in its role as a Bra Set label by $B_{\gamma_{c,d}}$  or simply by $B$.
%In particular in sections \ref{sec4} to \ref{sec6} our considerations involve only a single such label with
%the properties (i)-(iii) and we use the less detailed notation $B$. 
%However, in our considerations of section \ref{secl}, we shall need to distinguish
%between different Bra Set labels and in that section we shall employ the more detailed notation $B_{\gamma_{c,d}}$ to denote the Bra Set ${\bf E}_{\gamma_{c,d}}$.
{\noindent}NOTE: In the next and subsequent sections, we shall refer to a certain `distance' function between two points also by $d$. However from the context it will be amply clear 
whether the symbol $d$ refers to this distance or to the decorations of graph defined above and there will be no scope for confusion.

\subsection{\label{sec3.3} Kink Sets}

Let $S$ be a spin net. 
The set of bivalent and trivalent kinks in the graph underlying $S$ is called the {Kink Set} ${\bf K}$ of $S$.
We are interested in segregating elements of  ${\bf K}$ into subsets of `closest kinks' with respect to the metric $h$.

To this end, we first define the function $d(a_1,a_2)$ between any two distinct points $a_1,a_2 \in \Sigma$ as follows:
If  there exists a unique geodesic (with respect to the metric $h$) with length $l, l<1$ which joins $a_1$ to $a_2$ then we define  $d=l$ else we set $d=1$.
We shall refer to $d$ as a `distance' function. 
Next, compute the distances $d(k,k^{\prime})$ between all pairs of elements $k\neq k^{\prime}$ of ${\bf K}$ and follow the following algorithm to segregate elements of ${\bf K}$ into subsets.

Look for a set of 3 kinks such that their interkink distances are smaller than those between the remaining kinks and also smaller than the distance between any remaining kink 
and any member of this triplet. If such a set exists call it $K^{(1)}_3$.
If no such set exists, look for a pair of closest kinks so that the distance between them is smaller than the distances  between the remaining kinks and also smaller than the distance between any remaining kink 
and any member of this pair. If such a set exists, call it $K^{(1)}_2$ and stop the procedure
so that there are only two subsets, namely $K^{(1)}_2$ and its complement $K^{\prime}$.
If no such set exists  there is only one relevant subset ${\bf K}$.

%If there is a unique least such distance, call the pair of closest kinks $k^{(1)}_{1}, k^{(1)}_2$. We look for a 3rd kink such that its distance from  each of $k^{(1)_1, k^{(1)}_2$ 
%is smaller than its distance to any of the remaining kinks and also smaller than the distances between any of the remaining kinks.
%If there is no such kink we call the 2 point set of  closest kinks as $K^{(1)}_2$ and stop the procedure so that there are only two subsets, namely $K^{(1)}_2$ and its complement $K^{\prime}$
%If there is such a kink, call it $k^{(1)}_3$ and call the set of elements $k^{(1)}_i, i=1,2,3$ 
%as $K^{(1)_3$. 

If $K^{(1)}_3$ exists, remove $K^{(1)}_3$ from ${\bf K}$ and repeat the above procedure for the resulting set of kinks. 
Clearly, the  procedure then comes to a halt after some number $n \geq 1$ of iterations with the final `closest' kink set being  either a 2 kink set or a 3 kink set. 
Denote by $K^{(i)}_3$ the 3- kink set obtained at the $i$th iteration and by $K^{(n)}_2$  the 2-kink set (if it exists) obtained at  the (necessarily) $n$th iteration.  It follows that 
we have a segregation of ${\bf K}$ into either of \\

\noindent Case 0: ${\bf K}$ (here the procedure ends necessarily  at the first step $n=1$ with no `closest' 2 kink or 3 kink sets) or \\

\noindent Case 1: $K^{(i)}_3, i=1, \ldots n-1$, $K^{(n)}_2$, ${\bf K} - (\cup_{i=1}^{n-1}  K^{(i)}_3 \cup K^{(n)}_2)$ or \\

\noindent Case 2: $K^{(i)}_3, i=1,\ldots n$,  ${\bf K} - (\cup_{i=1}^{n}  K^{(i)}_3) $.\\

Above, for Case 1, it is understood that if $n=1$ the procedure yields a segregation of ${\bf K}$ into a single 2 point kink set and its complement.

For future purposes it is useful to define the minimum and maximum interkink distances in any 3 kink set $K^{(i)}_3$ as
\be
d^{(i)}_{\max} = \max_{k,k^{\prime} \in K^{(i)}_3} d(k,k^{\prime}), \;\;\;\;
d^{(i)}_{\min} = \min_{k\neq k^{\prime} \in K^{(i)}_3} d(k,k^{\prime})
\label{defminmax}
\ee
Here,  if  all 3 interkink distances are equal we define the maximum and minimum distance both  to be equal to this interkink distance.

\subsection{\label{sec3.4} The coefficients}

Consider an off shell state $\Psi_{B,f,h}$. The coefficient of the bra $\bra s|$ (also referred to as the off shell amplitude for $s$) in the formal sum over bras associated with $\Psi_{B,f,h}$ is
$\Psi_{B,f,h}(|s\ket)$.  Consider the set of spin network states $B_{\perp}$ such that every element of $B_{\perp}$ is orthogonal to every element of $B$.
Then since $\Psi_{B,f,h}$  is a sum over states in $B$, the amplitude of this off shell state on any element of $B_{\perp}$ vanishes.
%but trivial on spin networks which are not asymmetric,  
and, since $\Psi_{B,f,h}$ is a linear map on the finite span of spin nets,  we may specify $\Psi_{B,f,h}$ through its amplitudes on elements of $B$ (each of which are asymmetric spin network basis states, see \ref{vib}, section \ref{sec3.1}
and (i), section \ref{sec3.2}).
%\footnote{Note  that spin nets which are not asymmetric (i.e. they do not live on asymmetric colored graphs) are orthogonal to those which are asymmetric. For the relevant definitions see \ref{vib}, section \ref{sec3.1}}
%The basis we choose is one for which vertex intertwiners are elements of the  Volu Basis. Abusing terminology, we shall refer to this basis of states as the volume intertwiner basis.
In section \ref{sec3.4.1} we specify the amplitude for any element of $B$. In section \ref{sec3.4.2} we derive  the amplitude for any  spin network state based on its
specification in section \ref{sec3.4.1}. In section \ref{sec3.4.3} we discuss a  useful property of the amplitude of section \ref{sec3.4.2}.

\subsubsection{\label{sec3.4.1} Amplitude for states in $B$ and perpendicular to $B$}

The amplitude $\Psi_{B,f,h}(|s\ket)$ for any  asymmetric spin network (ASN) basis state $|s\ket$ is:
\ba 
\Psi_{B,f,h}(|s\ket) &=& 0 \;\;\;\;\;{\rm if} \;\; \bra s| \in B_{\perp}  \label{psiamp0}\\
&=&  (\prod_{v\in V(s)} f(v)) g_{s,h} \;\;\;\;\;{\rm if} \;\; \bra s| \in B 
\label{psiamp}
\ea
where we recall  again that any state in $B$ is necessarily an asymmetric spin network basis state.
Here $V(s)$ is the set of vertices of $s\in B$, each such vertex being GR by virtue of property (i), section \ref{sec3.2}.
$f$ is a $C^r$ function on $\Sigma$ so that the off shell amplitude depends on the product of evaluations of $f$ at each vertex of $s$. $g_{s,h}$
is a  (positive) function  $g ({\bf K}, h) $ of the Kink Set ${\bf K}$ of $s$ segregated with respect to the metric $h$ 
%endowed with the $h$ dependent segregation structure defined in section \ref{sec3.2}.
which  depends on the network of interkink distances defined in terms of $h$ between elements of ${\bf K}$ endowed with the $h$ dependent segregation structure defined in 
section \ref{sec3.2}.

Let the Kink Set segregation procedure of section \ref{sec3.2} terminate in $n$ steps, $n\geq 1$ yielding the (exhaustive) segregation structures defined in Cases 0 to 2 of that section. 
For Case 0, ${\bf K}$  does not admit any non-trivial segregation and we set 
\be
 g({\bf K}, h)=1
\label{defgcase0}
\ee
For Case 1, we define $g$ to be:
\ba
 g({\bf K}, h) &:=&   \left(\prod_{i=1}^{n-1}   \left(\frac{d^{(i)}_{\min} }{d^{(i)}_{\max}}\right )^2\right) \left(d( k^{(n)}_1, k^{(n)}_2)\right)^2 \;\;{\rm if}\; n>1 \label{defgcase1a}\\
    &:=&  \big( d( k^{(1)}_1, k^{(1)}_2 ) \big)^2 \;\;{\rm if}\; n=1 \label{defgcase1b}
\ea   
Finally, for  Case 2, we define $ g({\bf K}, h)$ to be:
\be
 g({\bf K}, h)  := \prod_{i=1}^{n}  \left(\frac{d^{(i)}_{\min} }{d^{(i)}_{\max}}\right)^2
\label{defgcase2}
\ee

Thus, we set $g_{s,h} = g({\bf K}, h)$ with $g({\bf K}, h)$ defined through equations (\ref{defgcase0})-(\ref{defgcase2}).

\subsubsection{\label{sec3.4.2} Amplitude for arbitrary spin network state}
Consider any expansion of a spin network state $|s\ket$ in terms of other spin network states. Clearly, the only spin networks which contribute non-trivially to this expansion are those
for which all spin net labels agree with those of $s$ except, possibly, the vertex intertwiner labels. Hence all these spin networks live on the same {\em embedded colored graph} as $s$ (by embedded colored graph we mean
the graph as a subset of $\Sigma$ with specified edge colorings).
This immediately implies that 
any  spin network state $|s\ket$  may either  be expanded in terms of a {\em finite} number of  asymmetric spin network basis states (see \ref{vib}, section \ref{sec3.1}) or is orthogonal to every such basis state. In the former case 
we have that:
%spin networks $\{|s_i\ket\}$ whose  vertex intertwiners are elements of the volume intertwiner basis (see 7., section \ref{sec3.1}).
%We shall refer to each such $|s_i\ket$ as a  basis intertwiner state.
\be
|s\ket = \sum_{i=1}^p c^s_i |s_i\ket .
\label{vexps}
\ee
where $|s_i\ket$ are asymmetric spin network basis states.  
%We shall say that $|s\ket$ has overlap in $B$ if  the bra correspondent of at least one of the asymmetric spin network basis states (with non-trivial coefficient) in its expansion is an element of $B$.
%\footnote{This is just an equivalent rephrasing of the notion of overlap introduced in section \ref{sec3.2}.} 
%Depending on our convenience we may on occassion omit the `ket' notation and say that $s$ has overlap in $B$ or even that $\bra s|$ has overlap in $B$ (by which we mean that the ket correspondent of 
%this bra has overlap in $B$). 
Recall from section \ref{sec3.2} that a spin net $s$ will be said to have overlap in $B$ if $s$ has non-vanishing inner product with at least one element of $B$, where for convenience we have 
ommitted to represent the spin nets involved  in bra-ket notation.  Let $s$ has overlap in $B$. Hence it shares all its labels except perhaps intertwiners with some element of $B$. The discussion above
in cojunction with 
 property (i), section \ref{sec3.2}, implies that (a) this element of $B$ is an element of the asymmetric spin network basis and (b) all states living on the same embedded colored graph as $s$ with all possible choices of 
volume basis intertwiner labels at vertices are in $B$.  It follows that if $s$ has overlap in $B$, it admits an expansion (\ref{vexps}) in which every $s_i$ is  in $B$.
%From the remarks above together with 
%Thus, if $s$ has no overlaeither $s$ does not admit an expansion (\ref{vexps})
%property (i), section \ref{sec3.2}, it follows that if  $s$ has overlap in $B$ then it admits an expansion (\ref{vexps})  {\em every}
%asymmetric spin net basis state in its expansion (\ref{vexps}) is in $B$. 
Conversely, if $s$ does not have overlap in $B$, then either $s$ admits an expansion (\ref{vexps}) in which no $s_i$ with $c^s_i\neq 0$ is in $B$ or 
$s$ is  orthogonal to every asymmetric spin net basis element  
It then follows from (\ref{psiamp}) that:
\ba
\Psi_{B,f,h}(|s\ket) &=& 0 \;\;\;\;\;{\rm if} \;\; |s\ket \;\;{\rm has}\;{\rm no}\;{\rm overlap}\; {\rm in} \;\; B  \label{psiamps0}\\
&=& (\prod_{v\in V(s)} f(v)) g_{s,h} \gamma_s   \;\;\;\;\;{\rm if} \;\; |s\ket \;\;{\rm has}\;{\rm overlap}\; {\rm in} \;\; B  \label{psiamps}\\
\ea
where, referring to the expansion (\ref{vexps}),  
\be
\gamma_s  :=   \sum_{i=1}^p { c^s_i}\;\;\;\;  s_i \in B, \; i=1,..,p
\ee

%and $\bar c^s_i$ denotes the complex conjugate of $c^s_i$.

\subsubsection{\label{sec3.4.3} Useful properties of $\gamma_s$}

Let $v$ be a vertex of the spin network $s$. Consider a neighborhood $R_v$ of  $v$ which is small enough that it contains no vertex of $s$ other than $v$.
Denote the Rovelli-Smolin (RS) Volume operator associated with  this region by ${\hat V}_{R_v}$. 
From 7., section \ref{sec3.1}, independent of the choice of $R_v$, ${\hat V}_{R_v}$ acts as a self adjoint  operator on the intertwiner at $v$. 
Denote this `RS vertex volume operator' by ${\hat V}_v$. Clearly, every asymmetric spin network  basis element  with a vertex at $v$ is an eigen state of ${\hat V_v}$.

Next, consider the  action of the inverse determinant metric operator $\lim_{\tau\rightarrow 0} \tau^{-2} {\hat q}_{\tau}^{-\frac{1}{3}}(v)$ on $|s\ket$. As discussed in 
Reference \cite{p4} and reiterated in section \ref{sec2.1.2}, this operator also acts as a linear operator on the space of intertwiners at $v$ in $s$. Let us call this
linear operator as ${\hat {\lambda}}_v$. Recall that in  this work the operator $\lim_{\tau\rightarrow 0} \tau^{-2} {\hat q}_{\tau}^{-\frac{1}{3}}(v)$
is defined via a Tychonoff regulation. 
It then follows from the construction of this operator in \cite{p4} that upto a factor ${\hat \lambda}_v$ is defined via Footnote \ref{fntycho} with ${\hat O}:= ({\hat V}_v)^{\frac{2}{3}}$
so that eigenstates of ${\hat V}_v$ with eigen values $\nu_v\neq 0$  are eigen states of ${\hat \lambda}_v$  with upto a numerical factor) eigen values $(\nu_v)^{-\frac{2}{3}}$
and so that the zero eigen value eigen states  of ${\hat V}_v$  are annihilated by ${\hat \lambda}_v$. We shall refer to ${\hat \lambda}_v$ as the RS inverse volume operator.

Note that for a spin net $s$ with vertex $v$, just as the action of ${\hat V}_{R_v}$ on $s$ yields an intertwiner change through the action of ${\hat V}_v$, the 
 discussion above indicates that the action of $\lim_{\tau\rightarrow 0} \tau^{-2} {\hat q}_{\tau}^{-\frac{1}{3}}(v)$ on $s$ also yields an intertwiner change through the action of 
 ${\hat \lambda}_v$.

Next, let $|s\ket$ be a spin network state  with basis expansion (\ref{vexps}). Let  $s$ have a GR vertex at $v$. Consider an electric diffeomorphism type deformation,  without first acting with ${\hat \lambda}_v$, 
of $|s\ket$ along its  $I_v$th
edge at $v$ which yields the state $|s_{I,\e}\ket$.  The vertex $v$ and its vertex structure is displaced to $v_{I,\e}$  as detailed in sections \ref{sec2.1.2}, \ref{sec2.3.1} - \ref{sec2.3.3}. Recall  that:\\
\noindent (i) $s_{I,\e}$ differs from $s$ only in an $\e$ size neighborhood of $v$.\\
\noindent (ii) In this neighborhood  of $v$, the vertex $v$ in $s$ is replaced by the vertex $v_{I,\e}$ in $s_{I,\e}$. The displaced vertex $v_{I,\e}$ in  $s_{I,\e}$ has the same intertwiner as that of $v$ in 
$s$ and the edges emanating from $v_{I,\e}$ in $s_{I,\e}$ are the deformed counterparts of those from $v$ in $s$ with the same edge colorings.\\
\noindent (iii) Bivalent kinks are created in this $\e$-neighborhood of $v$.\\
%with, from gauge invariance, necessarily trivial (i.e. equal to identity) intertwiners.\\
\noindent (iv) While the edge tangent structure at $v_{I,\e}$  in $s_{I,\e}$ is, in general, not diffeomorphic to that at $v$ in $s$, the RS Volume operator is insensitive to this difference so that as
operators on intertwiners, the  RS vertex volume operator  at $v_{I,\e}$ in $|s_{I,\e}\ket$ and that at $v$ in $|s\ket$ are {\em identical}. Similar conclusions hold for the RS vertex inverse volume operator as well
for vertex operators composed of products of powers of the RS vertex volume and inverse volume operators.

It follows from (i)-(iii) above together with the detailed construction of the ASN basis (see (iii) of \ref{vib}, section \ref{sec3.1}) that $|s_{I,\e}\ket$  has the basis expansion:
\be 
|s_{I,\e} \ket = \sum_{i=1}^p c^s_i |(s_i)_{I,\e}\ket
\label{vexpds}
\ee
where $|(s_i)_{I,\e}\ket$ is the electric diffeomorphism type deformation (i.e. without first acting with ${\hat \lambda}$)  of $|s_i\ket$ and its  coefficient  $c^s_i$ in the expansion (\ref{vexpds}) is identical to the coefficient of 
$|s_i\ket$ in the expansion (\ref{vexps}). An immediate consequence of this is that:
%Next, note that if $s$ has overlap in $B$, it follows from (iii), section \ref{sec3.2} that $s_{I,\e}$ has overlap in $B$ and that for such $s$ we have the following property satisfied by $\gamma$:
%
%\be
%\gamma_{s_{I,\e}}= \gamma_s 
%\label{gamma=}
%\ee

Next let $v$ be a non-degenerate vertex of $s$.  From the discussion above it follows that the action of $\lim_{\tau\rightarrow 0} \tau^{-2} {\hat q}_{\tau}^{-\frac{1}{3}}(v)$ 
on $|s\ket$ results in a change of intertwiner at $v$ through the action of ${\hat \lambda}_v$. As in the rest of this paper, we denote the resulting state by 
$|s_{\lambda_v}\ket$. In this notation, $|(s_{I,\e})_{\lambda_{v_{I,\e}}}\ket $ is obtained from the action of $\lim_{\tau\rightarrow 0} \tau^{-2} {\hat q}_{\tau}^{-\frac{1}{3}}(v_{I,\e})$ 
on $|s_{I,\e} \ket$ with the intertwiner change in  $|(s_{I,\e})_{\lambda_{v_{I,\e}}}\ket $ effected through the action of ${\hat \lambda}_{v_{I,\e}}$ on the intertwiner at $v_{I,\e}$ in 
$|s_{I,\e} \ket$. From (i)-(iv) above the new intertwiner in $|s_{\lambda_v}\ket$ at $v$  and in $|(s_{I,\e})_{\lambda_{v_{I,\e}}}\ket $ at $v_{I,\e}$ are identical.
It follows that the state obtained through  an electric diffeomorphism type deformation at $v$ of $|s_{\lambda_v}\ket$ is the same as $|(s_{I,\e})_{\lambda_{v_{I,\e}}}\ket $. This relationship is 
expressed in our notation as:
\be
|s_{\lambda_v, I,\e}\ket = |(s_{I,\e})_{\lambda_{v_{I,\e}}}\ket .
\label{gmmadiffcom}
\ee

Similar argumentation  may be applied to each $s_i$ in the expansion (\ref{vexps}). Since $s_i$ is an 
eigen vector of ${\hat \lambda}_v$ with some eigen value $\lambda_{i,v}$, it then follows that $s_{I,\e}$
is an eigen vector of ${\hat \lambda}_{v_{I,\e}}$ with same eigen value.
\footnote{Here, as above, we have suppressed the 
`$v$' subscripts on the edge index $I$, as well as the additional specification of edge triples ${\vec J}$ to avoid notational clutter.} Accordingly we have:
\ba 
{\hat \lambda}_v | s_i \ket &=& \lambda_{i,v} | s_i \ket \label{may1}\\
|((s_i)_{I,\e})_{\lambda_{v_{I,\e}}}\ket &:=& {\hat \lambda}_{v_{I,\e}}|(s_i)_{I,\e} \ket
= \lambda_{i,v}|(s_i)_{I,\e} \ket \label{may2}
\ea
From (\ref{gmmadiffcom}) and (\ref{may2}) we have that:
\be
|(s_i)_{\lambda_v, I,\e}\ket = \lambda_{i,v}|(s_i)_{I,\e} \ket \label{may2.1}
\ee
From (\ref{vexps})  and (\ref{may1}) we have that:
\be
|s_{\lambda_v}\ket = \sum_{i} c_i^{s} \lambda_{i,v} |s_i\ket
\label{may3.0}
\ee
Replacing $s$ by $s_{\lambda_v}$ in (\ref{vexpds}) yields:
\ba
|s_{\lambda_v, I,\e}\ket &=& \sum_{i} c_i^{s} \lambda_{i,v}|(s_i)_{I,\e} \ket \label{may3.1}\\
& = &\sum_{i} c_i^{s}|(s_i)_{\lambda_v, I,\e}\ket \label{may3.2}
\ea
where we have used (\ref{may2.1}) in the last line above.
Note that  for $v$ non-degenerate in $s_i$, $(s_i)_{\lambda_v, I,\e}$ is, upto a factor of $\lambda_{i,v}$  an electric diffeomorphism  child of $s_i$. 
From (iii), section \ref{sec3.2}, it follows that if $s_i \in B$  and $v$ is non-degenerate in $s_i$, then 
$(s_i)_{I,\e} $ is in $B$. Hence $s_{\lambda_v, I,\e}$ has overlap in $B$ if $s$ has overlap in $B$
and $v$ is non-degenerate in $s$. From (\ref{may3.1}), (\ref{may3.0}), (\ref{gmmadiffcom}) it follows that for $s$ with overlap in $B$:
%Note that if $v$ is a non-degenerate vertex of $s$ and if $s$ has overlap in $B$ then (i), section \ref{sec3.2} implies that  $s_{\lambda_v}$ has overlap in $B$ 
%and (iii), section \ref{sec3.2} implies that $s_{\lambda_v, I,\e}$ also has overlap in $B$.
%It then follows  from (\ref{gmmadiffcom}) and (\ref{gamma=}) that for such $s$,
\be
\gamma_{    (s_{I,\e})_{ \lambda_{v_{I,\e}}   }}= 
\gamma_{   s_{\lambda_v, I,\e} }=
\gamma_{s_{\lambda_v} } =  \sum_{i} c_i^{s} \lambda_{i,v}  \;.
\label{gammal=}
\ee
Repeating these arguments for an  application of  the inverse determinant metric operator twice at $v$ in $|s\ket$ (instead of once as above) and denoting the result by 
$|s_{\lambda_v,\lambda_v}\ket$, we have in obvious notation that:
\be
\gamma_{ (s_{I,\e})_{\lambda_{v_{I,\e}}, \lambda_{v_{I,\e}}}}= \gamma_{s_{\lambda_v,\lambda_v} }\;.
\label{gammagamma}
\ee

Finally, consider the action of the operator  $\lim_{\tau\rightarrow 0} \tau^{-2} {\hat q}_{\tau}^{-\frac{1}{3}}({\bar v})$, with ${\bar v}\neq v$,  on $|s_{\lambda_v}\ket$ with ${\bar v},v$ nondegenerate.
Since this operator changes  the intertwiner exclusively at ${\bar v}$ it commutes with the action of an electric diffeomorphism at $v$ which is supported 
away from ${\bar v}$. It follows that:
\be
|(s_{\lambda_v, I,\e})_{\lambda_{\bar v}}\ket = |((s_{\lambda_v})_{\lambda_{\bar v}})_{I_v,\e}\ket, \;\;\;{\bar v}\neq v
\label{gammavvbar1}
\ee
where the notation signifies that the state on  the left hand side state is obtained by an electric diffeomorphism type deformation of $|s_{\lambda_v}\ket$ along its $I_v$th edge followed by the action 
of $\lim_{\tau\rightarrow 0} \tau^{-2} {\hat q}_{\tau}^{-\frac{1}{3}}({\bar v})$ and that  the state on the right hand side is obtained by the action of 
$\lim_{\tau\rightarrow 0} \tau^{-2} {\hat q}_{\tau}^{-\frac{1}{3}}({\bar v})$ on $|s_{\lambda_v}\ket$ to yield 
$|(s_{\lambda_v})_{\bar v}\ket$ 
followed by  an electric diffeomorphism type deformation   along the $I_v$th edge at $v$ of $|(s_{\lambda_v})_{\bar v}\ket$.
If $s$ has overlap in $B$ then similar arguments as above imply that the states in (\ref{gammavvbar1}) have overlap in $B$. Next, an application of ${\hat \lambda}_{\bv}$ on (\ref{may3.0}), (\ref{may3.1}) yields:
\ba
{\hat \lambda}_{\bv}|s_{\lambda_v}\ket  &= &\sum_{i} c_i^{s} \lambda_{i,\bv}\lambda_{i,v} |s_i\ket  \label{may4.0}\\
{\hat \lambda}_{\bv}|s_{\lambda_v, I,\e}\ket &=& \sum_{i} c_i^{s} \lambda_{i,\bv}\lambda_{i,v}|(s_i)_{I,\e} \ket \label{may4.1}
\ea

It then follows from (\ref{gammavvbar1}), (\ref{may4.0}), (\ref{may4.1}) that:
\be
\gamma_{(s_{\lambda_v, I,\e})_{\lambda_{\bar v}}}= \gamma_{((s_{\lambda_v})_{  \lambda_{\bar v}  })_{I_v,\e}} = \gamma_{(s_{\lambda_v})_{   \lambda_{\bar v}   }}= \sum_{i} c_i^{s} \lambda_{i,\bv}\lambda_{i,v}, \;\;\;{\bar v}\neq v .
\label{gammavvbar2}
\ee
%where in the last equality we have used (\ref{gamma=}).
%\\

%\noindent{\bf Note}: Clearly, if $v$ is a degenerate vertex of $s$ the inverse metric vertex operator at $v$, ${\hat \lambda}_v$, annhilates $s$ so that, in our notation, $s_{\lambda_v}=0$ which in turn implies that 
% $\gamma_{s_{\lambda_v} } =0$ for such $v$.
%In what follows, rather than explicitly mention the non-degenerate or degenerate
%nature of the vertices encountered, we simply define $\gamma_{s_{\lambda_v} } =0$ if $v$ is a degenerate vertex of $s$.

%d/d=1/2 dprop

Finally, given $s$ with expansion (\ref{vexps}) and  its image $s_{\phi}$ by a diffeomorphism $\phi$, we have that:
\be
|s_{\phi}\ket = {\hat U} (\phi) |s\ket = \sum_{i} c^s_i |(s_i)_{\phi}\ket
\label{may5.0}
\ee
where we have defined $|(s_i)_{\phi}\ket ={\hat U} (\phi) |s_i\ket $. From (ii), \ref{vib} of section \ref{sec3.1}
we have that $\{(s_i)_{\phi}\}$ are also asymmetric spin net basis elements and each $(s_i)_{\phi}$ is
an embedding of the same embeddable abstract spin network which embeds to $s_i$. It then follows from
(ii), section \ref{sec3.2}  that $s_{\phi}$ has overlap in $B$ if $s$ has overlap in $B$ (and conversely
using the fact that $\phi^{-1}$ is a diffeomorphism). From (\ref{vexps}), (\ref{may5.0}) we have that for such
$s$:
\be
\gamma_s = \gamma_{s_{\phi}} = \sum_i c^s_i
\label{may5.1}
\ee

\section{\label{sec4} Continuum Limit Action on  $\Psi_{B,f,h}$}
As described in section \ref{sec2},  the action of the  constraint operator at parameter $\e$ on a state $S$  deforms the state to give `diffeomorphism' type children and `propagation' type children.
The  deformations endow these children with  extra kinks relative to their parent $S$. As $\e\rightarrow 0$, these kinks contract towards each other and induce a certain contraction behavior onto the 
off shell amplitudes for these children through the dependence of these amplitudes on the interkink distance function $g$ defined in the previous section.
Clearly, an evaluation of the 
(dual) continuum action of the constraint on an off shell basis state then depends on the contraction behavior of $g$. We describe this contraction behavior in section \ref{sec4.0}.
We use this contraction behavior to evaluate the continuum limit action of a single Hamiltonian constraint in section \ref{sec4.1} and of a pair of constraints and their commutator in section \ref{sec4.2}.

\subsection{\label{sec4.0}Contraction behavior of the interkink distance function}

First consider a `diffeomorphism' type deformation of a state $S\in B$ along its $I$th edge at its (necessarily GR) nondegnerate vertex $v$. 
From property (iii), section \ref{sec3.2}, $S_{\lambda, I,\J,\e} \in B$.
Relative to $S$, the deformed spin net $S_{\lambda, I,\J,\e}$ has 3 extra kinks.
Clearly, for small enough $\e$ these kinks  form 
the 3-kink set  $K^{(1)}_3$ defined in section \ref{sec3.3}. From section \ref{sec2.3}, specifically, equation (\ref{d/d=1/2}) it follows that for small enough $\e$, and {\em independent} of the choice of kink triple $\J$, 
\be
g_{S(\lambda, I,\J, \e)} = \frac{1}{4}g_S  + O(\e^{q\geq 2})
\label{congdiff}
\ee
where we have suppressed the $h$ subscript of $g$ to avoid notational clutter.

Next consider the case of a propagation type deformation of a state $S$, not necessarily in $B$, at its nondegenerate (but from Footnote \ref{fnngrtogr}, not necessarily GR) vertex $v$ which yields a child $C_\e = S_{\alpha_{IJ},\e} \in B$. 
From the Note at the end of section \ref{sec2.4}, recall that propagation type deformations in both the GR and NGR case can be denoted in this manner.
From (ii), (iv) at the end of  section \ref{sec2.5} it follows that independent of the particular type of propagation deformation, this child is endowed with 2 extra kinks, one at $v_{I,\e}$ and one at ${\tilde v}_J$ relative to its parent $S$.
Additionally,  depending on the particular type of propagation deformation, 
the vertex $v$ of $S$ can be either a GR vertex in $C_\e$ (since $C_\e \in B$) or a point in the interior of an edge in $C_\e$ or a (bivalent or trivalent) kink in $C_\e$.   
In the  case of a bivalent or trivalent kink, it follows that for small enough $\e$,  the kinks at $v_{I,\e}, {\tilde v}_J, v$ form the 3 kink set $K^{(1)}_3$ defined in section \ref{sec3.3}. From the discussion around (\ref{dprop})
and from (\ref{ngr2}) we have that for this set:
\be 
d_{\min} = d (v,{\tilde v}_J)= O(\e^{q\geq 2})\;\;\;\; d_{max}= \max (d(v,v_{I,\e}), d({\tilde v}_J, v_{I,\e})) = \O(\e)
\label{d3kink}
\ee
from which it follows that 
\be
g_{C_\e} = O(\e^{2})
\label{4.3}
\ee

In the case that $v$ is a GR vertex or an interior point of an edge in $C_\e$, for small enough $\e$, clearly, the kinks at $v_{I,\e}, {\tilde v}_J$ are the closest pair of kinks in the kink set of $C_\e$.
Two possibilities manifest at any $\e$. Either there is no 3rd kink $k$ in $C_\e$ which can form a 3-kink set $K^{(1)}_3$ 
%in which case 
%we have a segregation of the kink set of $C_\e$  single 2 kink set
or there is such a kink $k$.  In the former case we have a segregation of the kink set of $C_\e$ into a single 2 kink set and its complement.
%so 
In the latter case, clearly
$k$ is already present in $S$ and we have the 3 kink set $K^{(1)}_3= \{v_{I,\e}, {\tilde v}_J, k\}$.
While no kinks in $C_\e$ other than  $v_{I,\e}, {\tilde v}_J$  move as $\e$ decreases, it is possible that 
the segregation structure of the Kink Set of $C_\e$  could change from one in which the set of nearest kinks is a 2 kink set to one in which the set of
nearest kinks is a 3-kink set and vice versa. Moreover in the case that 3-kink sets are encountered at different values of $\e$,
it is possible that the identity of the nearest `spectator' kink  $k$  in these 3 kink sets also  changes. 
%It is straightforward to see that this could happen only  
%if there are 2 or more kinks appropriately positioned equidistant from $v$ in $S$. Indeed, in the case of  equidistant positioning from $v$ of 2 or more kinks in $S$ (and only in that case), it 
%could happen that as $\e$ decreases the segregation structure of the Kink Set of $C_\e$  could change from one in which the set of nearest kinks is a 2 kink set to one in which the se% of
%nearest kinks is a 3-kink set and vice versa.
Despite these diverse possibilities, we show next, that  the interkink distance function vanishes as $O(\e^2)$.

Note that from 
%If follows from 
(\ref{ngr1})-(\ref{ngr3}) and the Note in section \ref{sec2.4},  it follows for all small enough $\e$ that the distance from any fixed non-contracting `spectator' kink ${\bar k}$ (so that  $\bar k \in C_\e$ {\em and} ${\bar k}\in  S$),  
to the contracting pair satisfies:
\be
\max (d({\bar k},v_{I,\e}), d({\bar k},{\tilde v}_J) > {\bar D_{\bar k}}
\ee
for some $\e$ independent positive constant $0< {\bar D_{\bar k}} \leq 1$. 
To see this, note that the kinks $v_{I,e}, {\tilde v}_J$ contract to the parental vertex $v\in S$. It follows that we may set ${\bar D_{\bar k}} := \frac{1}{2}d({\bar k},v)$.
Since there are a finite number of kinks in $S$, we may define 
\be
\min_{{\bar k}\in S} {\bar D_{\bar k}} = {\bar D}
\label{dspectator}
\ee
so that ${\bar D}>0$ is also an $\e$ independent constant bounded above by 1.
It also follows that for small enough 
$\e$, equations (\ref{dprop}), (\ref{ngr1}), (\ref{ngr3})  imply that:  
%that in this 2 kink set the interkink distance satisfies:
\be 
\left( d(v_{I,\e}, {\tilde v}_J)\right)^2 \leq D_1 \e^2 
\label{d2kink}
\ee
for some $\e$ independent $D_1>0$.
Next, for  any (small enough) $\e$, define the set ${\cal S}_\e$ as:
\be
{\cal S}_\e = \{ \left( d(v_{I,\e}, {\tilde v}_J)\right)^2,\;\; 
d_{\bar k}, {\bar k} \in S\}
\ee
where 
\be
d_{\bar k}:= g({\bf K}_{\bar k}, h)\left(\frac{ d(v_{I,\e}, {\tilde v}_J)}{\max (d({\bar k},v_{I,\e}), d({\bar k},{\tilde v}_J))}\right)^2 .
\ee
Here ${\bf K}_{\bar k} $ is the kink set obtained by removing the kinks $v_{I,\e}, {\tilde v}_J, {\bar k}$
from the kink set of $C_\e$ and 
$g({\bf K}_{\bar k}, h)$ is the interkink distance function defined for the kink set ${\bf K}_{\bar k} $ segregated with respect to $h$ and defined in  section \ref{sec3.4} through equations 
(\ref{defgcase0})-(\ref{defgcase2}). Note that the absence  of $v_{I,\e}, {\tilde v}_J$ in  ${\bf K}_{\bar k}$ implies that $g({\bf K}_{\bar k}, h)$ is some strictly positive $\e$ independent 
number bounded by 1. Define:
\be 
\max_{{\bar k}\in S}  g({\bf K}_{\bar k}, h) = {\bar D}_1
\label{leftover}
\ee
so that ${\bar D}_1$ is also a strictly positive $\e$ independent number bounded by 1: 
\be 
0< {\bar D_1} \leq 1 .
\label{leftover1}
\ee
Next, note that from the discussion of the segregation behavior of the kink set above, it follows that for any small enough  $\e$, taking into account the possible diversity of 
kink segregation behaviors discussed above, we have that:
\be
g_{C_\e} \in {\cal S}_\e \;.
\ee
It follows from  (\ref{dspectator}), (\ref{d2kink}) and (\ref{leftover1})  that  there exists an $\e$ independent constant $D:= \frac{D_1}{{\bar D}^2}$ such that 
\be
\max_{x_\e\in {\cal S}_\e} x_\e  \leq D\e^2 \;.
\ee
from which we obtain the bound:
\be 
g_{C_\e} \leq D\e^2 \;.
\label{4.4}
\ee
%\be
%d_{\min} = d({\tilde v}_J, v_{I,\e}) \leq D_2\e  \;\;\; d_{max}= \max (d(k,v_{I,\e}), d(k,{\tilde v}_J) \geq D_3
%\label{k-3}
%\ee
%where $D_2,D_3>0$ are  independent of $\e$.%
%Hence in both these cases (\ref{k-2}), (\ref{k-3})
%we have that 
%\be 
%g_{C_\e} \leq  D\e^2 
%\ee
%for some $\e$ independent $D$.
Equations (\ref{4.3}) and (\ref{4.4}) imply that  for any propagation child we have the contraction behavior:
\be
g_{C_\e} \leq E \e^2 
%\label{congprop1}
\ee
for some $E>0$ which is independent of $\e$ so that 
\be
\lim_{\e\rightarrow 0 } \frac{g_{C_\e}}{\e} =0.
\label{congprop}
\ee

\subsection{\label{sec4.1} Single Hamiltonian Constraint}

The (dual) action of a single Hamiltonian constraint at parameter $\e$ on $\Psi_{B,f,h}$ is defined through the evaluations, for all spin networks $S$, of the amplitude:
\be 
\Psi_{B,f,h} ({\hat H}_\e(N) S)
\label{hepsi}
\ee
The continuum limit action of a single Hamiltonian constraint on $\Psi_{B,f,h}$, is  defined to be the $\e\rightarrow 0$ limit of the evaluations, for all $S$, of the amplitude (\ref{hepsi}): 
\be 
\lim_{\e\rightarrow 0} \Psi_{B,f,h} ({\hat H}_\e(N) S), 
\label{hpsi}
\ee
Before embarking on the calculation proper of this continuum limit action, we describe the underlying mechanism through which the calculation leads to a well defined
continuum limit.  First note that the continuum limit contribution of the action on GR vertices to  (\ref{action1}) vanishes if $S$ has any NGR vertices because the Bra Set $B$ only 
contains states with exclusively GR vertices.  Similarly the continuum limit contribution of the `diffeomorphism' part of the constraint action on NGR vertices also vanishes
for the same reason.  Next, since the contraction behavior of the kink distance function in the case of propagation type deformations leads an overall factor of $O(\e^2)$ (\ref{congprop}), this 
overpowers the factor of $\e$ in the denominator of (\ref{action}) and renders the propagation part of the constraint  action on any vertex (GR or NGR)  trivial in the continuum 
limit. Thus the only non-trivial amplitudes are those for states $S$ with exclusively GR vertices and the contribution to the continuum limit action on these states
comes only from the `diffeomorphism' part of the action i.e only from the first line of   (\ref{action}).  The contribution from $S_{\lambda_v,I_v,\e}$ leads to the evaluation of $f$
at the displaced vertex $v_{I,\e}$ together with a factor of $1/4$ coming from the contraction behavior of the interkink distance function (\ref{congdiff}). 
The contribution from $S_{\lambda_v}$ leads to an evaluation of $f$ at the original vertex $v$ with the factor $1/4$. The difference of these two is then clearly of $O(\e)$ and
yields, in the continuum limit,  a derivative of $f$ along the $I_v$th edge tangent at $v$.

We now proceed with the calcuation and flesh out the rough picture presented above in full technical detail.
It is convenient to employ a notation for the propagation type states generated by the constraint action on nondegenerate GR vertices which is similar to that 
employed to denote propagation type states generated by the constraint action on nondegenerate NGR vertices.
Accordingly, we decompose the propagation type children in equation (\ref{action}),
${\tilde S}_{\lambda,(1)I,J,K, \e}, {\tilde S}_{\lambda,(2)I,J, \e}$,   into their constituent spin networks. Each such spin network has one of the kink and vertex 
structures detailed in (ii) of section \ref{sec2.5}. We enumerate these  spin networks by a `deformation' index $\beta_{I,J}, I\neq J$ similar to the deformation index 
$\alpha_{IJ}$ used in section \ref{sec2.4}  for the spin networks created by propagation type deformations of an NGR vertex $v$. In particular it is immediate to see that the deformations indexed by $\beta_{IJ}$ satisfy exactly the same
general properties outlined in the second paragraph of section \ref{sec2.4}, the only difference from the NGR case being that the possible presentations of the  vertex $v$ in these deformed spin networks are 
as detailed in (ii), section \ref{sec2.5} rather than in the 3rd paragraph of section \ref{sec2.4}.

Thus the  label $\beta_{IJ}$  denotes a deformation of $S$ in the vicinity of its nondegenerate GR vertex $v$ which introduces the pair of kinks  $v_{I,\e}, {\tilde v}_J$ together with the  edge $e(v_{I,\e}, {\tilde v}_J)$ 
\footnote{Here $e(p,q)$ refers to an edge $e$ with endpoints $p,q$} 
colored with
spin $j_J$ and which 
encodes the changes in colorings on the edges $e(v,v_{I,\e}), e({\tilde v}_J, v)$ and the change  in the intertwiner at $v$ relative to $S$. Depending on $\beta_{IJ}$ the  colorings of either or both of $e(v,v_{I,\e}), e({\tilde v}_J, v)$ 
can be the trivial $j=0$ coloring.
In this notation  the  constitutent spin networks in the decomposition of ${\tilde S}_{\lambda,(1)I,J,K, \e}$ are denoted by $S_{\beta_{IK}},\e$ for appopriate $\beta_{IK}$ and those in the decomposition of 
${\tilde S}_{\lambda,(2)I,J, \e}$  by $S_{\beta_{IJ},\e}$ for appropriate $\beta_{IJ}$.
As for $\alpha_{IJ},\e$  the deformation label $\beta_{IJ}$ encodes the abstract deformation structure (namely that of the colored graph structure  and intertwiner change) whereas the label $\e$ indicates
that this deformation is restricted to an $\e$ size vicinity of $v$.

Accordingly, augmenting $\beta_{IJ}$ with an index $v$ denoting the nondegenerate GR vertex of $S$ at which the constraint acts,   we re-express the second line of (\ref{action}) as:
\be
\frac{3}{4} \sum_{I_v=1}^{N_v}\frac{  (\sum_{J_v\neq I_v}\sum_{K_v\neq I_v,J_v}{\tilde S}_{\lambda_v,(1)I_v,J_v,K_v, \e}) +(\sum_{J_v\neq I_v} {\tilde S}_{\lambda_v,(2)I_v,J_v, \e})}{\e}
= (\sum_{I_v=1}^{N_v}\sum_{J_v\neq I_v}\sum_{\beta_{vI_vJ_v}} P_{\beta_{vI_vJ_v}} S_{\beta_{vI_vJ_v},\e})
\ee
for appropriately defined deformations $\beta_{vI_vJ_v}$ and complex coefficients $P_{\beta_{vI_vJ_v}}$.
In this notation, the constraint action (\ref{action}) can be written as:
\ba
{\hat H}_{\epsilon}(N) S(A) &:= &\frac{3}{8\pi}\sum_{v\in V_{GR}(s)}N(x(v)) 
\big(\;\sum_{I_v=1}^{N_v}\frac{ ([\frac{1}{\Perm{N_v-1}{3}}\sum_{\J_v}S_{\lambda_v,I_v, {\vec {\hat J}}_v,\e}]  - {\frac{1}{4}}S_{\lambda_v})}{\e}
\nonumber\\ 
%\;\sum_{I_v=1}^{N_v}\frac{ (S_{\lambda_v,I_v,\J_v \e} - {\frac{1}{4}}S_{\lambda_v})}{\e}
%\nonumber\\ 
%&-& 
&-&(\sum_{I_v=1}^{N_v}\sum_{J_v\neq I_v}\sum_{\beta_{vI_vJ_v}} P_{\beta_{vI_vJ_v}} \frac{S_{\beta_{vI_vJ_v},\e}}{\e})\;\big)\;\;\;\;\;\;\;\;\;\;
\nonumber\\
&+& \sum_{v\in V_{NGR}(s)}\frac{N(x(v))}{\e} \big(\;(\sum_{I_v=1}^{N_v} A_{vI_v}S_{\lambda_v,I_v, \e})  + B_v S_{\lambda_v}+ (\sum_{I_v=1}^{N_v}\sum_{J_v\neq I_v}\sum_{\alpha_{vI_vJ_v}} P_{\alpha_{vI_vJ_v}} S_{\alpha_{vI_vJ_v},\e})\big).\;\;\;\;\;\;\;\;\;\;\;\;\;\;
\label{actionf}
\ea
Here, as in section \ref{sec2.5}, $V_{GR}$ and $V_{NGR}$ refer, respectively,  to the set of {\em nondegenerate} GR and NGR vertices of $S$. 
In contrast, the set of all vertices of $S$,  degenerate or nondegenerate, GR or NGR, is denoted hereon by $V(S)$. In this regard, note that for any $S$ with overlap in $B$,
$V(S)$ has exclusively GR vertices some or all of which could be degenerate/nodegenerate.

Next, we show that the if $S$ has one or more NGR vertices, its continuum limit amplitude (\ref{hpsi}) vanishes.
First consider the contribution to this amplitude from the GR vertices of $S$. The relevant constraint action in the first two lines of (\ref{actionf}) above.
Since in each of the states in these two lines, there is no action on NGR vertices, these vertices and their edge  tangent structure remain unchanged.
Since the Bra Set $B$  contains states {\em all} of whose vertices are GR, it follows that for small enough $\e$, the amplitude (\ref{hepsi}) vanishes for $S$ for which $V(S)$ has at least one NGR vertex,
and, hence so does its continuum limit amplitude (\ref{hpsi}).
Next, consider the contribution from NGR vertices of $S$. The same argument as above implies that this contribution would vanish if $S$ has 2 or more NGR vertices.
Let us then consider the case where $S$ has a single NGR vertex at $v$. If  $v$ is degenerate, this contribution vanishes.  If  $v$ is nondegenerate,
it follows from the above discussion together with (\ref{actionf}) that:
\be
\Psi_{B,f,h} ({\hat H}_\e(N)  S) = 
\Psi_{B,f,h} (\frac{N(x(v))}{\e} \big(\;(\sum_{I_v=1}^{N_v} A_{vI_v}S_{\lambda_v,I_v, \e})  + B_v S_{\lambda_v}+ (\sum_{I_v=1}^{N_v}\sum_{J_v\neq I_v}\sum_{\alpha_{vI_vJ_v}} P_{\alpha_{vI_vJ_v}} S_{\alpha_{vI_vJ_v},\e})\big).
\ee
Note that $v_{I_v,\e}$ is NGR in $S_{\lambda_v,I_v, \e}$,   and that $v$ is NGR in $S_{\lambda_v}$. Hence the amplitudes for these states vanish and we have that 
\be
\Psi_{B,f,h} ({\hat H}_\e (N)S) = \frac{N(x(v))}{\e} 
 (\sum_{I_v=1}^{N_v}\sum_{J_v\neq I_v}\sum_{\alpha_{vI_vJ_v}} P_{\alpha_{vI_vJ_v}} \Psi_{B,f,h}(S_{\alpha_{vI_vJ_v},\e})).
\label{4.0}
 \ee
Clearly, the only nontrivial contributions arise from those $S_{\alpha_{vI_vJ_v},\e}$ which have overlap in $B$. From (\ref{psiamps}),  we have for such $S_{\alpha_{vI_vJ_v},\e}$ that:
\be
\Psi_{B,f,h}(  S_{\alpha_{vI_vJ_v},\e}) = (\prod_{\bv\in V( S_{\alpha_{vI_vJ_v},\e}   )} f(\bv)) g_{S_{\alpha_{vI_vJ_v},\e},h} \gamma_{S_{\alpha_{vI_vJ_v},\e}}.
\label{4.1}
\ee
Note that  $S_{\alpha_{vI_vJ_v},\e}$ for fixed $\alpha_{vI_vJ_v}$ and  for all small enough $\e$ has the same abstract graph structure and   the same  $\e$ independent vertex intertwiners.
%can be obtained as embeddings of the same abstract embeddable spin network.
It follows that $\gamma_{S_{\alpha_{vI_vJ_v},\e}}$
is independent of $\e$.
Next, note that from (\ref{congprop}) we have that 
\be 
g_{S_{\alpha_{vI_vJ_v},\e},h} = O(\e^2).
\label{4.2}
\ee
Using the fact that $f$ is a $C^r$  function (so that its evaluations are bounded independent of $\e$) together with  (\ref{4.2})  in (\ref{4.1}) implies that  for small enough $\e$:
\be
\Psi_{B,f,h} (S_{\alpha_{vI_vJ_v},\e}) =  \frac{O(\e^2)}{\e} = O(\e)
\label{4.13}
\ee
which implies that the $\e\rightarrow 0$ limit of (\ref{4.13}), and, hence of  (\ref{4.0}) vanishes.

Next, let  $S$  be a spin network which has exclusively GR vertices so that $V(S)$ comprises only of (not necessarily nondegenerate) GR vertices.
We have that:
\ba
{\hat H}_{\epsilon}(N) S(A) := \frac{3}{8\pi}\sum_{v\in V_{GR}(S)}N(x(v)) \big(\;\sum_{I_v=1}^{N_v}\frac{ ([\frac{1}{\Perm{N_v-1}{3}}\sum_{\J_v}     S_{\lambda_v,I_v,\J_v \e}] - {\frac{1}{4}}S_{\lambda_v})}{\e}
\nonumber\\ 
%&-& 
-(\sum_{I_v=1}^{N_v}\sum_{J_v\neq I_v}\sum_{\beta_{vI_vJ_v}} P_{\beta_{vI_vJ_v}} \frac{S_{\beta_{vI_vJ_v},\e}}{\e})\;\big)
\label{4.14}
\ea
where we remind the reader that $V_{GR}$ is the set of all {\em nondegenerate} GR vertices in $V(S)$.
%in the first line we have continued to append the subscript $GR$ to $V$ to emphasize that nontrivial contributions  arise from those parental vertices which are .
Note that if $S$ does not have overlap in $B$, none of the states in the first line of (\ref{4.14}) have overlap in $B$. To see this recall from section \ref{sec3.4.2} that if $S$ 
has no overlap in $B$ and has an expansion (\ref{vexps}), none of the asymmetric spin network basis states in this expansion are in $B$. 
Thus, if $\{S_i, i=1,..,n\}$ are the set of these states (with non-zero expansion coefficients),
we have that $S_i \notin B, \;i=1,..,n$. If
 $S_i \notin B$, none of its children  are in $B$
 else $S_i$ would be a possible parent, by electric diffeomorphism, of a child in $B$ and hence, itself in $B$ by 
(iii), section \ref{sec3.2}.  However, equation (\ref{may3.2}) shows that every electric diffeomorphism 
child of $S$ is a linear combination of children of $\{S_i, i=1,..,n\}$. This shows that no electric diffeomorphism
child of $S$ in the first line of (\ref{4.14}) can have overlap in $B$. Note also that (\ref{may3.0}) shows that
$S_{\lambda_v}$ also cannot have overlap in $B$ if $S$ has no overlap in $B$.
%Note also that any  child, by electric diffeomorphism of any asymmetric spin network basis state element is also (upto a factor)
%also an asymmetric basis state element. This implies that any child $S_{\lambda_v,I_v,\J_v \e}$ has an expansion of the type (\ref{vexps}) in terms of children 
%of $S_i$. 
%\footnote{To see this, recall that the embedded colored graph underlying $S_{\lambda_v,I_v,\J_v \e}$ is obtained by an electric diffeomorphism type deformation (as detailed in section \ref{sec2.3}) of that underlying $S$
%with the latter shared by $S_i$  and the former by the child $(S_i)_{\lambda_v,I_v,\J_v \e}$ of $S_i$ for those $S_i$ for which $v$ is nondegenerate.}
%Also none of its children with no GR vertex can be in $B$  by (iii), section \ref{sec3.2}.
%Also if $S_i\notin B$, then (i), section \ref{sec3.2} implies that no intertwiner basis state in the expansion
%of $S_{\lambda_v}$ can be in $B$. 
It follows that none of the  states in the first line of  (\ref{4.14}) can have any overlap  with states in $B$, if $S$ has no overlap in $B$ but admits the expansion (\ref{vexps}).

Next, let $S$ not admit such an expansion.
Then it must be the case that the embedded colored  graph underlying $S$ is distinct from that underlying 
any asymmetric spin net basis element so that $S$ necessarily cannot have overlap in $B$ by virtue of the fact that every element of $B$ is such an asymmetric spin net basis element (see (i), section \ref{sec3.2}).
%(which agrees with the fact that $S$ does not have overlap in $B$ by (i), section \ref{sec3.2}). 
Since $S_{\lambda_v}$ lives on the same colored decorated graph as $S$, it also does not admit such an expansion and hence also does not have overlap in $B$.
Similarly if $S_{\lambda_v,I_v,\J_v \e}$ also does not admit an expansion of type (\ref{vexps}), its decorated colored graph is also distinct from that underlying any asymmetric spin network basis element and hence 
$S_{\lambda_v,I_v,\J_v \e}$ does not have overlap in $B$.
Next, recall again that in any expansion of a spin network in terms of other spin networks, all spin networks contributing non-trivially to the expansion necessarily live on the same 
embedded colored graph.
Suppose  now that  $S_{\lambda_v,I_v,\J_v \e}$ does admit an expansion (\ref{vexps}) so that its embedded colored graph is a permissible embedding of some abstract decorated graph $\gamma^{\prime}_{m,c,d}$ which necessarily has (at least) a triple of
appropriately placed bivalent kinks 
in its non-loop 
component. Clearly $S$ must live on an embedded  colored graph  obtained by removing a triple of kinks from the non-loop component of that underlying  $S_{\lambda_v,I_v,\J_v \e}$, which implies that 
the embedded colored graph underlying $S$ is a permissible embedding of the abstract decorated graph obtained by removing an appropriate triple of kinks from  edges in the non-loop component of $\gamma^{\prime}_{m,c,d}$.
Since such a graph is of type $\gamma_{m,c,d}$ or $\gamma^{\prime}_{m,c,d}$, $S$  necessarily admits an expansion  in  terms of  embedded spin networks each of which is an  embedding of an element of 
${\bf E}_{\gamma_{m,c}}$. Thus $S$
must admit 
an expansion of type (\ref{vexps}) which is a contradiction so that $S_{\lambda_v,I_v,\J_v \e}$ cannot have overlap in $B$.
Thus, if  $S$ does not have overlap in $B$, none of the  states in the first line of  (\ref{4.14}) can have any overlap   in $B$, which proves the assertion.
As a result, the contribution from the first line of (\ref{4.14}) 
vanishes. 

Next, consider the `propagation' contributions in  the second line of (\ref{4.14}). 
These contributions  can be seen vanish in the continuum limit by   
arguments identical to those which lead to (\ref{4.2}), (\ref{4.13})
{\em whether or not $S$ has overlap
in $B$}. More in detail  for any  nondegenerate vertex $v$ in $S$, we have, once again, that:
\be 
g_{S_{\alpha_{vI_vJ_v},\e},h} = O(\e^2),  \;\;\; \Psi_{B,f,h} (S_{\alpha_{vI_vJ_v},\e}) =  \frac{O(\e^2)}{\e} = O(\e) .
\ee
so   that the continuum limit of the propagation part of the constraint vanishes. 
%From arguments identical to those leading to (\ref{4.2}), these contributions  vanish in the continuum limit 

As a result, the amplitude (\ref{hepsi}) for  $S$  with no overlap in $B$ vanishes in its continuum limit (\ref{hpsi})
%for small enough $\e$, as does its continuum limit (\ref{hpsi}) 
and the only possibly non-trivial contributions to this continuum limit  amplitude for $S$ which has  overlap
in $B$ arise solely from the first line of (\ref{4.14}).
Therefore, in (\ref{4.14}) we restrict attention to  $S$  which 
%has   exclusively GR vertices and  
has overlap in $B$ and consider contributions only from its first line.  From arguments similar to those above
it then follows that all the states in the first
line of 
%see (iii), section \ref{sec3.2}) in
(\ref{4.14}) have overlap in $B$ so that the contributions of interest evaluate to:
\be
\Psi_{B,f,h}( {\hat H}_{\epsilon}(N) S) =
   \frac{3}{8\pi}\sum_{v\in V_{GR}(s)}N(x(v)) \sum_{I_v=1}^{N_v}\frac{ ([\frac{1}{\Perm{N_v-1}{3}}\sum_{\J_v}\Psi_{B,f,h}(S_{\lambda_v,I_v, \e})] - {\frac{1}{4}}\Psi_{B,f,h}(S_{\lambda_v}))}{\e}  .
\label{4.15}
   \ee
Omitting the subscript `$h$' on the interkink distance function $g$, we have that:
\ba 
\Psi_{B,f,h}(S_{\lambda_v,I_v, \e}) &=&  (\prod_{v^{\prime}\neq v_{I,\e}} f(v^{\prime}) f (v_{I,\e}) g_{S_{\lambda_v,I_v,\J_v, \e}} \gamma_{S_{\lambda_v,I_v,\J_v, \e}} \nonumber\\
&=& 
(\prod_{v^{\prime}\neq v} f(v^{\prime})f (v_{I_v,\e}) (\frac{1}{4}g_{S}) \gamma_{S_{\lambda_v}} + O(\e^2)
\label{4.16}
\ea
where we have used (\ref{psiamps}) in the first line and 
(\ref{congdiff}) and (\ref{gammal=}) in the second line of (\ref{4.16}). Here, in the first line $v^{\prime}$ ranges over all vertices of $S_{\lambda_v,I_v,\J_v, \e}$ other than $v_{I,\e}$. Equivalently, 
in the second line $v^{\prime}$ ranges over all elements of $S$ other than the parental vertex  of interest, $v$.

Next, note that in the RNC's at $v$ we have that 
%\footnote{ The RNC's are $C^{r-1}$ so that the edge $e_{I_v}$ is also $C^{r-1}$ in these coordinates. Tay 
\be
x^{\mu}(v_{I_v,\e}) = x^\mu (v) + \e {\hat e}_{I_v}^{\mu} + O(\e^2)
\label{4.17}
\ee
where ${\hat e}_{I_v}^{\mu}$ is the unit (with respect to the metric $h$) tangent to the $I_v$th edge at $v$.
To see this, note that $I_v$th edge is a $C^r, r>>1$ curve and hence a $C^{r-1}$ curve in terms of the $C^{r-1}$ RNCs at $v$.
In a $C^{r-1}$ parameterization by some parameter $t$ such that $v$ lies at parameter $t=0$ and $v_{I,\e}$ at parameter $t_{\e}$, Taylor expansion yields 
\be
x^{\mu}(v_{I_v,\e})- x^\mu (v) = \frac{dx^{\mu}}{dt}(t=0) t_\e+ O(t_\e^2) .
\label{t0}
\ee
Since $v_{I_v,\e}$ lies at an RNC coordinate distance $\e$ from $v$ and since the  metric at $v$ in RNCs is $h_{\mu \nu}=\delta_{\mu \nu}$ we have that 
\ba
\e^2 &=& \sum_{\mu=1}^3(x^{\mu}(v_{I_v,\e})- x^\mu (v))^2 \label{t1}\\
&=& (t_\e)^2(\delta_{\mu \nu}(\frac{dx^{\mu}}{dt}(t=0)\frac{dx^{\nu}}{dt}(t=0))  + O(t_\e))\label{t2}\\
\Rightarrow \e &= & {\bf O}(t_\e) \Rightarrow O(t_\e) = O(\e) \label{t3} \\
\Rightarrow  \e + O(\e^2) &= & t_\e (h_{\mu\nu}\frac{dx^{\mu}}{dt}(t=0)\frac{dx^{\nu}}{dt}(t=0))^{\frac{1}{2}}
\label{t4}
\ea
where we substituted (\ref{t3})  in (the square root of) (\ref{t2}) to obtain (\ref{t4}). Using (\ref{t4}) in (\ref{t0}) together with the fact that 
$\frac{dx^{\mu}}{dt}(t=0)$ is the $I_v$th edge tangent  in the parameterization $t$ yields the desired result (\ref{4.17}).
Note that by virtue of the defining property of RNCs the unit edge tangent with respect to the metric $h$ agrees with the unit coordinate edge tangent (see also (\ref{unittngnt})).
%Here, the $O(\e^2)$ term captures the deviation of the $I_v$th edge from a coordinate straight line and  we have used the fact that the RNC's are $C^{r-1}, r>>1$.
%Note that by virtue of the defining property of RNC's ${\hat e}_{I_v}^{a}$ is the unit (with respect to the metric $h$)  $I_v$th edge tangent vector at $v$.
Using the fact that $f$ is $C^r$, (\ref{4.17})  implies that:
\be
f(v_{I_v,\e})= f(v) + \e {\hat e}_{I_v}^{\mu} \partial_{\mu} f(v) +  O(\e^2) .
\label{4.17f}
\ee
Using this in conjunction with (\ref{4.16}) and (\ref{gammal=}) we have that:
\ba
&&{ [\frac{1}{\Perm{N_v-1}{3}}\sum_{\J_v}\Psi_{B,f,h}(S_{\lambda_v,I_v, \e})] - {\frac{1}{4}}\Psi_{B,f,h}(S_{\lambda_v})} \nonumber\\
&=&
\big(\;\;(\prod_{v^{\prime}\neq v} f(v^{\prime})) \; \frac{1}{4}g_{S} \gamma_{S_{\lambda_v}} (f(v) + \e {\hat e}_{I_v}^{\mu} \partial f(v)\; + \;\; O(\e^2))\;\;\big)
- \big((\prod_{v^{\prime}\neq v} f(v^{\prime}))\frac{1}{4}g_{S} \gamma_{S_{\lambda_v}} f(v)\big) \nonumber\\
&=& 
 \big(\;\frac{1}{4}g_{S} \gamma_{S_{\lambda_v}}
(\prod_{v^{\prime}\neq v} f(v^{\prime})) {\hat e}_{I_v}^{\mu} \partial_{\mu} f(v)\;\big)\;\e  \; \; + \;\;O(\e^2)
\ea
from which it follows that:
\be
\lim_{\e\rightarrow 0}\Psi_{B,f,h}( {\hat H}_{\epsilon}(N) S) =
  \frac{3}{8\pi}\sum_{v\in V_{GR}(s)}N(x(v) g_{S}
 \gamma_{S_{\lambda_v}}(\prod_{v^{\prime}\neq v} f(v^{\prime}))(\sum_{I_v=1}^{N_v}   \frac{1}{4}{\hat e}_{I_v}^{\mu} \partial_{\mu} f(v))
\label{derivative}
 \ee
where we remind the reader that the vertex set $V(S)$ of $S$ has only GR vertices, that 
$V_{GR}(S)$ is the set of nondegenerate  vertices of $S$ and that in (\ref{derivative}), $v^{\prime}$ ranges over all elements of $V(S)$ other than $v$.
%$S$ is such that $V(S)= V_{GR}(S)$.
To summarise, the continuum limit of the single action of the Hamiltonian constraint on the off shell basis state $\Psi_{B,f,h}$ is given by the following exhaustive set of
amplitudes:
\ba 
\Psi_{B,f,h}( {\hat H}(N) S) &:=& \lim_{\e\rightarrow 0}\Psi_{B,f,h}( {\hat H}_{\epsilon}(N) S)\nonumber \\
&=& 0\;\; {\rm if}\;\;S\;{\rm has}\;{\rm no}\;{\rm overlap}\; {\rm in}\; B \label{first0}\\
&=& 
\frac{3}{8\pi}\sum_{v\in V_{GR}(s)}N(x(v) g_{S}
 \gamma_{S_{\lambda_v}}(\prod_{v^{\prime} \in V(S), v^{\prime}\neq v} f(v^{\prime}))(\sum_{I_v=1}^{N_v}   \frac{1}{4}{\hat e}_{I_v}^{a} \partial_{a} f(v))\;\; \nonumber\\
&&      \;\;\;\;\;\;\;\;\;\;\;  \;\;\;\;\;\;\;\;\;\;\ \;\;\;\;\;\;\;\;\;\;\;\;\;\;\;\;\;\;\;\;\;\;  \;\;\;\;\;\;\;\;\;\;\ \;\;\;\;\;\;\;\;\;\;\;
{\rm if}\;\;S\;{\rm has}\;{\rm overlap}\; {\rm in} \;B                \label{first1}
\ea
where 
%we have retained the subscript $GR$ in (\ref{first1}) to emphasize that all vertices of any state in $B$ are GR and 
we have replaced the coordinate index $\mu$ by an abstract index to 
emphasize the coordinate independence of the derivative of $f$ along the unit (with respect to $h$) $I_v$th edge tangent at $v$.

\subsection{\label{sec4.2} Constraint Product and Commutator}
The (dual) action of a product of  Hamiltonian constraints at parameters ${\bar \e}, \e$ on $\Psi_{B,f,h}$ is defined through the evaluations, for all spin networks $S$, of the amplitude:
\be 
\Psi_{B,f,h} ({\hat H}_{\bar \e}(M) {\hat H}_\e (N)S) .
\label{2hepsi}
\ee
The continuum limit action of the product of Hamiltonian constraints on $\Psi_{B,f,h}$, is  defined to be the ${\bar \e}, \e\rightarrow 0$ limit of the evaluations, for all $S$, of the amplitude (\ref{2hepsi}): 
\be 
\lim_{\e\rightarrow 0} (\lim_{{\bar\e}\rightarrow 0}\Psi_{B,f,h} ({\hat H}_{\bar \e}(M) {\hat H}_\e (N)S)).
\label{2hpsi}
\ee
The commutator between a pair of  Hamiltonian constraints can then be inferred directly from (\ref{2hpsi}) to be 
\be
\lim_{\e\rightarrow 0} (\lim_{{\bar\e}\rightarrow 0}\Psi_{B,f,h} ({\hat H}_{\bar \e}(M) {\hat H}_\e (N)S)) 
-\lim_{\e\rightarrow 0} (\lim_{{\bar\e}\rightarrow 0}\Psi_{B,f,h} ({\hat H}_{\bar \e}(N) {\hat H}_\e (M)S) ).
\label{commhpsi}
\ee

%The continuum limit of a product of two constraint actions is defined to the evaluations, for all spin networks $S$, of the continuum limit of the following amplitude:
%\be
In section \ref{sec4.2.1} we show that the continuum limit of the constraint  product amplitude (\ref{2hpsi}) vanishes for $S$ which has one or more NGR vertices.
In section \ref{sec4.2.2} we evaluate the continuum limit of the constraint product amplitude for $S$ which has exclusively GR vertices.
In section \ref{sec4.2.3} we infer  the commutator (\ref{commhpsi}) from the evaluation of (\ref{2hpsi}) in sections \ref{sec4.2.1} and \ref{sec4.2.2}.

\subsubsection{\label{sec4.2.1} Vanishing Product Amplitude for the case that $S$ has NGR vertices}

We have that:
\be
{\hat H}_{\bar \e}(M) {\hat H}_\e (N)S  = {\hat H}_{\bre}(M)({\hat H}_\e (N)S|_{V_{GR}(S)}) + {\hat H}_{\bre}(M)({\hat H}_\e (N)S|_{V_{NGR}(S)}) .
\label{24.1}
\ee
The first contribution above, from nondegenerate GR vertices in $S$ results in spin networks which we denote by $\{S_{\chi,\e}\}$ for some appropriate enumeration index $\chi$.
%Every $S_{\chi,\e}$  has  NGR vertices where ${\hat H}_\e (N)$ has not acted.
From (\ref{first0}) the $\bre\rightarrow 0$ limit of the amplitude $\Psi_{B,f,h}({\hat H}_{\bre}(M)(S_{\chi,\e}))$ vanishes 
because every $S_{\chi,\e}$  has  NGR vertices where ${\hat H}_\e (N)$ has not acted.
The same argument shows that the contribution of the second term in (\ref{24.1}) to the continuum limit in $\bre$ also vanishes if there are 2 or more NGR vertices in $S$.

Consider the remaining case, in which $S$ has a single $NGR$ vertex at $v_0$.
Let $v_0$ be nondegenerate (else there is no contribution to the second term of (\ref{24.1}) so that both terms on the right hand side of (\ref{24.1}) vanish).
From section \ref{sec2.4}, the diffeomorphism part of the action of ${\hat H}_{\e}(N)$ at $v_0$ results in spin nets which continue to have a single
NGR vertex. It follows that the $\bre$ continuum limit vanishes on these spin networks. Hence we only need to consider the propagation type children generated by ${\hat H}_{\e}(N)$ at $v_0$ for which the propagation deformation
results in $v_0$ changing from an NGR vertex in $S$ to a GR vertex or a point in the interior of an edge
or to a (trivalent or bivalent) kink in the child. Amongst these children, from 
(\ref{first0}), we only need to consider children which happen to have overlap in $B$. Let us denote such a child by $C_\e$. 
The continuum limit action of ${\hat H}_{\bar \e}(M)$ on such a state is given by (\ref{first1}):
\be
\Psi_{B,f,h}( {\hat H}(M) C_\e)=
\frac{3}{8\pi}\sum_{v\in V_{GR}(C_\e)}M(x(v))g_{C_\e}
 \gamma_{{C_\e}_{\lambda_v}}(\prod_{v^{\prime}\in V(C_\e),v^{\prime}\neq v} f(v^{\prime}))(\sum_{I_v=1}^{N_v}   \frac{1}{4}{\hat e}_{I_v}^{\mu} \partial_{\mu} f(v)).
\label{24.2}
\ee
Note that none of the vertices of $C_\e$ change their positions as $\e$ decreases, only the kinks contract.
Also note that $\gamma_{{C_\e}_{\lambda_v}}$ only depends on the colored graph structure and vertex intertwiners of $C_\e$ and these are independent of $\e$.
Thus  $\gamma_{{C_\e}_{\lambda_v}}$  is independent of $\e$. The only  $\e$ dependent contribution to (\ref{24.2}) is then   
$g_{C_\e}$. From (\ref{congprop}) this contribution is of $O(\e^2)$ for small enough $\e$.
Since the propagation children generated from an NGR vertex come with a factor of $\e^{-1}$ (see (\ref{ngr}), we have that:
\be
\Psi_{B,f,h}( {\hat H}(M) {\frac{C_\e}{\e}}) = O(\e)
\ee 
which implies that the contribution of the second term of (\ref{24.1}) to the  continuum limit product  amplitude in (\ref{2hpsi}) vanishes because it involves the double limit in which 
the $\e\rightarrow 0$ limit   is taken after the  $\bre\rightarrow 0$ limit.

To summarise: the continuum limit product  amplitude vanishes for spin networks $S$ with at least one NGR vertex.
Accordingly, in  the next section we focus on the case in which $S$ has exclusively GR vertices.

\subsubsection{\label{sec4.2.2} Product Amplitude for the case of exclusively GR vertices }
Consider the continuum limit product amplitude (\ref{2hpsi})  
%${\hat H}_{\bar \e}(M) {\hat H}_\e (N)S$ 
in the case that $S$ has exclusively GR vertices.
First consider the propagation type states generated by the action 
of ${\hat H}_\e (N)$ on such $S$.  It is immediate to see that the  arguments of section \ref{sec4.2.1} for the contribution of such propagation children continue to hold if  the vertex $v_0$
in those arguments is GR in $S$ instead of NGR in $S$. Hence if the propagation part of  action of ${\hat H}_\e (N)$ on $S$ results children with overlap in  $B$, 
the contribution of such propagation children to the product amplitude vanishes in the $\e\rightarrow 0$ continuum limit. If such propagation children do not have overlap in $B$, (\ref{first0}) implies that the action of ${\hat H}(M)$
on such children vanishes.  Hence the only remaining contributions  to the product amplitude arise from the action of the diffeomorphism part of ${\hat H}_\e(N)$ on $S$ i.e from the states in the first line of (\ref{4.14}).
First, consider the case in which $S$ has no overlap with $B$. Recall from the arguments immediately after (\ref{4.14}) that if $S$ has exclusively GR vertices and has no overlap with $B$, none of the states in the first line of (\ref{4.14}) have overlap in $S$. Equation (\ref{first0}) then implies that the contribution of the electric diffeomorphism part of ${\hat H}_\e(N)$ to the product amplitude vanishes for such $S$.
It follows that the continuum limit  product amplitude vanishes for $S$ with no overlap in $B$ and for $S$ with overlap in $B$ only the children generated by the diffeomorphism part of ${\hat H}_\e(N)$  i.e. the states 
in the first line of (\ref{4.14})contribute to this
amplitude.  Recall, from the discussion after (\ref{4.14}) that for $S$ with overlap in $B$ all the states in the 
first line of (\ref{4.14}) have overlap in $B$. 
%Hence none of the states which result from 
%the action ${\hat H}_\e(N)$ on $S$ can have overlap with $B$. From (i),(ii) at the end of section \ref{sec2.5} it follows  that every such state also has either exclusively GR vertices ( or no vertices at all, on%
%the action of ${\hat H}_{\bar \e}(M)$ on any such state also results in states with no overlap in $B$ (or annihilates such a state).
%It then follows that the continuum limit product amplitude vanishes  for all  $S$ such that  $S$  has no overlap with $B$  and $S$ has exclusively GR vertices.
%On the other hand, if $S$ has overlap with $B$ (so that it necessarily has exclusively GR vertices), similar argumentation (see the lines preceding (\ref{4.15})) implies that 
%every state (with at least one GR vertex) which is  obtained by the action of the product ${\hat H}_\e (N)S$ has overlap with $B$. Consider the propagation type states generated by the action 
%of ${\hat H}_\e (N)$ on such $S$.  It is immediate to see that the  arguments of section \ref{sec4.2.1} for the contribution of such propagation children continue to hold if  the vertex $v_0$
%in those arguments is GR in $S$ instead of NGR in $S$. Hence the contribution of such children to the continuum limit product amplitude (\ref{2hpsi}) vanishes.
%Hence the only possible non-trivial contribution to the continuum limit product limit arises  from  the diffeomorphism part of ${\hat H}_\e(N)S$. 
From these arguments together with (\ref{action}) it follows that for $S$ with overlap in $B$:
\ba
&\lim_{\e\rightarrow 0} (\lim_{{\bar\e}\rightarrow 0}\Psi_{B,f,h} ({\hat H}_{\bar \e}(M) {\hat H}_\e (N)S)) \;\;\;\;\;\;\;\;\;\;\;\;\;\;\;\;\;\;\;\;\;\;\;\;\;\;\;\;\;\;\;\;\;\;\;\;\;\;\;\;\;\;\;\;
\nonumber \\
&=\lim_{\e\rightarrow 0}
\big(\lim_{\bre\rightarrow 0}\Psi_{B,f,h}\left({\hat H}_{\bre}(M) \frac{3}{8\pi}\sum_{v\in V_{GR}(S)}N(x(v)) \;\sum_{I_v=1}^{N_v}\frac{ ([\frac{1}{\Perm{N_v-1}{3}}\sum_{\J_v}     S_{\lambda_v,I_v,\J_v \e}] - {\frac{1}{4}}S_{\lambda_v})}{\e}
\right)\big) \nonumber
\\
&{}\\
&= \frac{3}{8\pi}\sum_{v\in V_{GR}(S)}N(x(v)) \lim_{\e\rightarrow 0}
%\nonumber \\
%& 
\;\left(
\frac{ [\sum_{I_v=1}^{N_v}\frac{1}{\Perm{N_v-1}{3}}\sum_{\J_v}    \lim_{{\bar\e}\rightarrow 0}\Psi_{B,f,h} ({\hat H}_{\bar \e}(M)S_{\lambda_v,I_v,\J_v \e})] - N_v{\frac{1}{4}}\lim_{{\bar\e}\rightarrow 0}\Psi_{B,f,h}({\hat H}_{\bar \e}(M)S_{\lambda_v})}{\e}
\right)\nonumber \\ 
%\;\;\;\;\;\;
&{}
\label{22.1}
\ea
where in the last line we have used the independence of $\Psi_{B,f,h}({\hat H}_{\bar \e}(M)S_{\lambda_v})$ from the label $I_v$ to replace the sum over $I_v$ by a factor of $N_v$.

From (\ref{first1}) we have that:
\ba
&\lim_{{\bar\e}\rightarrow 0}\Psi_{B,f,h} ({\hat H}_{\bar \e}(M)S_{\lambda_v,I_v,\J_v \e})=: \Psi_{B,f,h} ({\hat H}(M)S_{\lambda_v,I_v,\J_v \e}) \nonumber \\
&=\frac{3}{8\pi}\sum_{\bv\in V_{GR}(S_{\lambda_v,I_v,\J_v \e})}M(x_{\bv}(\bv)) g_{S_{\lambda_v,I_v,\J_v \e}}
 \gamma_{ (S_{\lambda_v,I_v,\J_v \e} )_{\lambda_{\bv}}}
 (\prod_{\bv^{\prime}\neq \bv} f(\bv^{\prime}))(\sum_{\bI_{\bv}=1}^{N_{\bv}}   \frac{1}{4}{\hat e}_{\bI_{\bv}}^{a} \partial_{a} f(\bv)). 
\label{22.2}
 \ea
Here $\bv,\bv^{\prime} $ refer to vertices of the state $S_{\lambda_v,I_v,\J_v \e}$. The vertex label $\bv$ ranges over the set of nondegenerate (necessarily GR) vertices $V_{GR}(S_{\lambda_v,I_v,\J_v \e})$
and can either be $v_{I_v,\e}$ or any  of the  undeformed nondegenerate vertices of $S$ (i.e nondegenerate vertices other than $v$).
The vertex label $\bv^{\prime}$ ranges over the entire vertex set of necessarily GR but not necessarily nondegenerate vertices, $V(S_{\lambda_v,I_v,\J_v \e})$. The index 
${\bar I}_{\bv}$ is an edge index for the edges emanating from $\bv$ in $S_{\lambda_v,I_v,\J_v \e}$. 
%A vertex $\bv$ of $S_{\lambda_v,I_v,\J_v \e}$
%can either be $v_{I_v,\e}$ or any  of the  undeformed vertices of $S$ (i.e vertices other than $v$).

In the argument of the lapse $M$ we have found it useful to add an explicit subscript $\bv$ to the Riemann Normal Coordinate notation adopted so far.
Thus $\{x_{\bv}\}$ refers to the Riemann Normal Coordinates centered at the point $\bv$. $M(x_{\bv}(\bv))$ refers to 
 the evaluation of the density weighted lapse $M$ at the point $\bv$ in the RNCs $\{x_{\bv}\}$ centered at $\bv$.
If $\bv = v_{I_v, \e}$  we shall find it useful to relate the evaluation of the lapse  $M$  at $v_{I_v,\e}$ in the RNCs centered at $v_{I_v,\e}$ to
its evaluation at  $v_{I_v,\e}$ in the RNCs centered at $v$. The evaluations in these two sets of coordinates are related by an appropriate factor
of the determinant of the Jacobian between these two sets of coordinates. From (\ref{detj}) the determinant of this Jacobian is unity to $O(\e^2)$.
It follows that the two evaluations, in the  notation used above, are related as follows:
\be
M(x_{v_{I_v,\e}}(v_{I_v,\e})) = M(x_{v}(v_{I_v,\e})) + O(\e^2).
\label{mvev}
\ee 

$\gamma_{ (S_{\lambda_v,I_v,\J_v \e} )_{\lambda_{\bv}}}$ is the $\gamma$ factor for the spin network $(S_{\lambda_v,I_v,\J_v \e} )_{\lambda_{\bv}}$ which is obtained from 
$S_{\lambda_v,I_v,\J_v \e}$ by the action of the inverse determinant operator at $\bv$ which changes the intertwiner at $\bv \in S_{\lambda_v,I_v,\J_v \e}$  
(see the paragraph after (\ref{qshift}) for a precise definition of this inverse determinant metric operator).
%Since $S_{\lambda_v,I_v,\J_v \e}$  is obtained from $S_{\lambda}$ by an electric diffeomorphism type deformation, 
%equation (\ref{gammal=}) implies that 
%\be
%\gamma_{ (S_{\lambda_v,I_v,\J_v \e} )_{\lambda_{\bv}}} = \gamma_{ (S_{\lambda_v} )_{\lambda_{\bv}}}
%\label{gammadiff}
%\ee
%Using (\ref{gammadiff}) and 
Separating out the vertex contributions to (\ref{22.2}) from $v_{I_v,\e}$ and the remaining vertices of $S_{\lambda_v,I_v,\J_v \e}$ we write the right hand side of (\ref{22.2}) as:
\ba
\frac{3}{8\pi}\sum_{\bv\in V_{GR}(S_{\lambda_v,I_v,\J_v \e})}M(x_{\bv}(\bv)) g_{S_{\lambda_v,I_v,\J_v \e}}
 \gamma_{ (S_{\lambda_v,I_v,\J_v \e} )_{\lambda_{\bv}}}
 (\prod_{\bv^{\prime}\neq \bv} f(\bv^{\prime}))(\sum_{\bI_{\bv}=1}^{N_{\bv}}   \frac{1}{4}{\hat e}_{\bI_{\bv}}^{a} \partial_{a} f(\bv))\nonumber\\
= \frac{3}{8\pi}\sum_{\bv\neq v_{I_v,\e}}    M(x_{\bv}(\bv)) g_{S_{\lambda_v,I_v,\J_v \e}}
 \gamma_{ (S_{\lambda_v,I_v,\J_v \e} )_{\lambda_{\bv}}}
 (\prod_{\bv^{\prime}\neq \bv, v_{I_v,\e}} f(\bv^{\prime}))f(v_{I_v,\e}) (\sum_{\bI_{\bv}=1}^{N_{\bv}}   \frac{1}{4}{\hat e}_{\bI_{\bv}}^{a} \partial_{a} f(\bv))
\nonumber \\
+  \frac{3}{8\pi}    M(x_{v_{I_v,\e}}(v_{I_v,\e})) g_{S_{\lambda_v,I_v,\J_v \e}}
 \gamma_{ (S_{\lambda_v,I_v,\J_v \e} )_{\lambda_{v_{I_v,\e}}}}
 (\prod_{v^{\prime}\neq v_{I_v,\e}} f(v^{\prime}))(\sum_{J_{v_{I_v,\e}}=1}^{N_{v}}   \frac{1}{4}{\hat e}_{J_{v_{I,\e}}}^{a} \partial_{a} f(v_{I,\e})).
\label{A1+A2}
 \ea
In the second line,  since $\bv \ne v_{I_v,\e}$, equation (\ref{gammavvbar2}) implies that 
\be
\gamma_{ (S_{\lambda_v,I_v,\J_v \e} )_{\lambda_{\bv}}} = \gamma_{ (S_{\lambda_v} )_{\lambda_{\bv}}}.
\label{gammadiffvvbar}
\ee
In the third line,  equations (\ref{gmmadiffcom}) and (\ref{gammagamma}) imply that:
\be
\gamma_{ (S_{\lambda_v,I_v,\J_v \e} )_{\lambda_{v_{I_v,\e}}}} = \gamma_{(S_{\lambda_v})_{\lambda_v}}.
\label{gammadiffvv}
\ee
Let us now simplify the second line of (\ref{A1+A2}) as follows:
\ba
&&\frac{3}{8\pi}\sum_{\bv\neq v_{I_v,\e}}    M(x_{\bv}(\bv)) g_{S_{\lambda_v,I_v,\J_v \e}}
 \gamma_{ (S_{\lambda_v,I_v,\J_v \e} )_{\lambda_{\bv}}}
 (\prod_{\bv^{\prime}\neq \bv, v_{I_v,\e}} f(\bv^{\prime}))f(v_{I_v,\e}) (\sum_{\bI_{\bv}=1}^{N_{\bv}}   \frac{1}{4}{\hat e}_{\bI_{\bv}}^{a} \partial_{a} f(\bv))
\nonumber \\
&=&\frac{3}{8\pi}\sum_{\bv\neq v_{I_v,\e}}    M(x_{\bv}(\bv)) g_{S_{\lambda_v,I_v,\J_v \e}}
 \gamma_{ (S_{\lambda_v} )_{\lambda_{\bv}}}
 (\prod_{\bv^{\prime}\neq \bv, v_{I_v,\e}} f(\bv^{\prime}))f(v_{I_v,\e}) (\sum_{\bI_{\bv}=1}^{N_{\bv}}   \frac{1}{4}{\hat e}_{\bI_{\bv}}^{a} \partial_{a} f(\bv))\nonumber\\
&=& 
\frac{3}{8\pi}\sum_{v^{\prime}\neq v}    M(x_{v^{\prime}}(v^{\prime})) g_{S_{\lambda_v,I_v,\J_v \e}}
 \gamma_{ (S_{\lambda_v} )_{\lambda_{v^{\prime}}}}
 (\prod_{v^{\prime\prime}\neq v^{\prime}} f(v^{\prime\prime})) (\sum_{I_{v^{\prime}}=1}^{N_{v^{\prime}}}  ( \frac{1}{4}{\hat e}_{I_{v^{\prime}}}^{a} \partial_{a} f(v^{\prime}))\nonumber \\
&+& \e \;
\frac{3}{8\pi}\sum_{v^{\prime}\neq v}    M(x_{v^{\prime}}(v^{\prime})) g_{S_{\lambda_v,I_v,\J_v \e}}
 \gamma_{ (S_{\lambda_v} )_{\lambda_{v^{\prime}}}}
 (\prod_{v^{\prime\prime}\neq v^{\prime}, v} f(v^{\prime\prime})) 
  \frac{1}{4}{\hat e}_{I_{v}}^{a} \partial_{a} f(v))
 (\sum_{I_{v^{\prime}}=1}^{N_{v^{\prime}}}   \frac{1}{4}{\hat e}_{I_{v^{\prime}}}^{a} \partial_{a} f(v^{\prime}))\nonumber \\
&+& O (\e^2) \nonumber \\
&=& 
\frac{3}{8\pi}\sum_{v^{\prime}\neq v}    M(x_{v^{\prime}}(v^{\prime})) \frac{1}{4}g_{S_{\lambda_v}}
 \gamma_{ (S_{\lambda_v} )_{\lambda_{v^{\prime}}}}
 (\prod_{v^{\prime\prime}\neq v^{\prime}} f(v^{\prime\prime})) (\sum_{I_{v^{\prime}}=1}^{N_{v^{\prime}}}   \frac{1}{4}{\hat e}_{I_{v^{\prime}}}^{a} \partial_{a} f(v^{\prime}))\nonumber \\
&+& \e \;
\frac{3}{8\pi}\sum_{v^{\prime}\neq v}    M(x_{v^{\prime}}(v^{\prime})) \frac{1}{4}g_{S_{\lambda_v}}
 \gamma_{ (S_{\lambda_v} )_{\lambda_{v^{\prime}}}}
 (\prod_{v^{\prime\prime}\neq v^{\prime}, v} f(v^{\prime\prime})) 
    (\frac{1}{4}{\hat e}_{I_{v}}^{a} \partial_{a} f(v))
 (\sum_{I_{v^{\prime}}=1}^{N_{v^{\prime}}}   \frac{1}{4}{\hat e}_{I_{v^{\prime}}}^{a} \partial_{a} f(v^{\prime}))\nonumber \\
&+& O (\e^2)
\label{A1}
\ea
where we have used (\ref{gammadiffvvbar}) to simplify the gamma factor of  the first line to obtain the second,  (\ref{4.17f}) to expand the second line and finally, (\ref{congdiff}) to expand the interkink distance function 
of the diffeomorphism child  $S_{\lambda_v,I_v,\J_v \e}$ in terms of that of its parent.  In the first two lines of the above set of equations, $\bv$ ranges over all nondegenerate vertices of 
$S_{\lambda_v,I_v,\J_v \e}$ 
other than $v_{I_v,\e}$. This range is  the same as that of all nondegenerate vertices of $S$  other than $v$. Accordingly, in the remaining part of (\ref{A1}), we have renamed $\bv$ as $v^{\prime}$
with $v^{\prime}$ ranging over all vertices in  $V_{GR}(S)$ other than $v$.
On the other hand, in the first two lines of the above equations
$\bv^{\prime}$ ranges over the entire vertex set of
$S_{\lambda_v,I_v,\J_v \e}$ modulo the restrictions under the product sign. 
Renaming $\bv^{\prime}$ as $v^{\prime\prime}$ in the remaining part of (\ref{A1}) it is easy to see that this range is the same as that of all vertices in $V(S)$ modulo the restrictions under the
product sign.
 
Next, we simplify the third line of (\ref{A1+A2}). In order to do we adopt the following  simplifying edge enumeration  for edges at $v_{I_v,\e}$
which assigns the same identifying number to the deformed counterpart of any edge at $v$ as that of its undeformed counterpart.  Accordingly we  shall assume an enumeration of
these  edges such that the $J{}_{v_{I,\e}}$th edge $e_{J{}_{v_{I,\e}}}$ at $v_{I,\e}$ is the deformed  counterpart
of the edge $e_{J_v= J{}_{v_{I,\e}}}$ at $v$ (here the deformation is that corresponding to an electric diffeomorphism along the $I_v$th edge of $v$).
We have that:
\ba
&&\frac{3}{8\pi}    M(x_{v_{I_v,\e}}(v_{I_v,\e})) g_{S_{\lambda_v,I_v,\J_v \e}}
 \gamma_{ (S_{\lambda_v,I_v,\J_v \e} )_{\lambda_{v_{I_v,\e}}}}
 (\prod_{v^{\prime}\neq v_{I_v,\e}} f(v^{\prime}))(\sum_{J_{v_{I_v,\e}}=1}^{N_{v}}   \frac{1}{4}{\hat e}_{J_{v_{I,\e}}}^{a} \partial_{a} f(v_{I,\e}))
 \nonumber \\
&=&\frac{3}{8\pi}    (M(x_{v}(v_{I_v,\e}))+ O(\e^2) ) g_{S_{\lambda_v,I_v,\J_v \e}}
 \gamma_{ (S_{\lambda_v} )_{\lambda_{v}}}
 (\prod_{v^{\prime}\neq v_{I_v,\e}} f(v^{\prime}))(\sum_{J_{v_{I_v,\e}}=1}^{N_{v}}   \frac{1}{4}{\hat e}_{J_{v_{I,\e}}}^{a} \partial_{a} f(v_{I,\e}))
\label{A2.1}\\
&=&\frac{3}{8\pi}    (M(x_{v}(v_{I_v,\e}))+ O(\e^2) ) g_{S_{\lambda_v,I_v,\J_v \e}}
 \gamma_{ (S_{\lambda_v} )_{\lambda_{v}}}
 (\prod_{v^{\prime}\neq v_{I_v,\e}} f(v^{\prime}))  \frac{1}{4}(N_v{\hat e}_{I_{v_{I,\e}}}^{a}  \partial_{a} f(v_{I,\e})) +O(\e^2))\;\;\;\;\;\;\;\;\;
\label{A2.2}\\
&=&\frac{3}{8\pi}    g_{S_{\lambda_v,I_v,\J_v \e}}
 \gamma_{ (S_{\lambda_v} )_{\lambda_{v}}}
 (\prod_{v^{\prime}\neq v_{I_v,\e}} f(v^{\prime}))  \frac{1}{4}N_v M(x_{v}(v_{I_v,\e}) ){\hat e}_{I_{v_{I,\e}}}^{a}  \partial_{a} f(v_{I,\e})) \;\; + \;\; \;\;\; O(\e^2)
\label{A2.3}\\
&=&\frac{3}{8\pi}   \frac{1}{4} g_{S_{\lambda_v}}
 \gamma_{ (S_{\lambda_v} )_{\lambda_{v}}}
 (\prod_{v^{\prime}\neq v_{I_v,\e}} f(v^{\prime}))  \frac{1}{4}N_v M(x_{v}(v_{I_v,\e}) ){\hat e}_{I_{v_{I,\e}}}^{a}  \partial_{a} f(v_{I,\e})) \;\; + \;\; \;\;\; O(\e^2).
\label{A2.4}
\ea
where $v^{\prime}$ ranges over the entire vertex set of $S_{\lambda_v,I_v,\J_v \e}$ other than $v_{I_v,\e}$, this range coinciding with the entire vertex set of $S$ other than $v$.

In (\ref{A2.1}) we have used  (\ref{gammadiffvv})  to simplify the $\gamma$ factor.  We have also used (\ref{mvev}) in (\ref{A2.1}) to transit from the $\{x_{v_{I_v,\e}}\}$ RNCs centered at $v_{I_v,\e}$ to the 
$\{x_v\}$ RNCs centered at $v$ in the evaluation of the lapse $M$.  Since the parameter $\e$ is measured with respect to the $\{x_v\}$ RNCs, this will enable us, subsequently, to expand  the lapse
via a Taylor series expansion  of the type (\ref{4.17f}).
%Recall that the parameter $\e$ is measured with respect to the $\{x_v\}$ RNCs and that 
%equations 
%(\ref{4.17}),  (\ref{4.17f}) and (\ref{upward})  encode the behavior of various quantities of interest as $\e$ decreases.  The transition from $\{x_{v_{I_v,\e}}\}$  to $\{x_v\}$ is a crucial step 
%in order to make use of these equations to simplify our expressions, it is crucial
%that we transit to the $\{x_v\}$ RNCs from the $\{x_{v_{I_v,\e}}\}$ ones.

In (\ref{A2.2}) we use the `upward conical stiffening' of the edge tangent structure at $v_{I_v,\e}$ as encoded in (\ref{upward})  to simplify the sum over edge tangents.
%Note that ${\hat e}_{J_{v_{I,\e}}=I_v}^{a}$ is the  unit (with respect to the metrac $h$) edge tangent along the $I_v$th 
In (\ref{A2.3}) we use the boundedness of the various pieces of (\ref{A2.2}) to separate out the $O(\e^2)$ piece. We shall continue to do this as we go along without explicit mention.
In (\ref{A2.4}) we use the contraction behaviour (\ref{congdiff}) of the interkink distance function $g$.  

Next, we expand the quantity $M(x_{v}(v_{I_v,\e}) ){\hat e}_{I_{v_{I,\e}}}^{a}  \partial_{a} f(v_{I,\e})$ (\ref{A2.4}) in a  Taylor series about $v$ using  (\ref{4.17})-(\ref{t4}).
Recall that the RNCs are $C^{r-1}$ coordinates, the metric $h$ is  $C^{r-1}$  tensor,  $f$ is a  $C^r$ function and $e_{I_v}$ is a semianalytic $C^r$ edge with, 
as in  (\ref{4.17})-(\ref{t4}), semianalytic parameterization $t$
such that $v_{I,\e}$ is located on the edge $e_{I_v}$ at parameter value $t=t_\e = \O (\e)$ and $v$ at $t=0$. From (\ref{upward}) and the numbering convention of edges discussed above, it follows that 
${\hat e}_{I_{v_{I,\e}}}^{a}$ in (\ref{A2.4}) is the unit (with respect to the metric $h$ at  $v_{I_v,\e}$) edge tangent to the edge $e_{I_v}$ at the point $v_{I_v,\e}$ located at parameter value $t_\e$.

It is convenient to use the following notation.  Denote the point on the edge $e_{I_v}$ at parameter $t$ by $e_{I_v}(t)$, 
and the edge tangent at the point $e_{I_v}(t)$ by ${{{\dot e}^a}}_{I_v}(t)$. Then  the unit (with respect to $h$) edge tangent at this point is:
\be
{\hat e}_{I_{v}^{a}} (t) =    \frac{{{{\dot e}^a}}_{I_v}(t)}{|   {\vec{{\dot e}}}_{I_v}(t)|}
\label{A2.41}
\ee
where $|   {\vec{{\dot e}}}_{I_v}(t)|$ is the metric norm of ${{{\dot e}^a}}_{I_v}(t)$ i.e. 
\be
|   {\vec{{\dot e}}}_{I_v}(t)|^2:= h_{ab}(e_{I_v}(t)){{{\dot e}^a}}_{I_v}(t){{{\dot e}^b}}_{I_v}(t).
\label{A2.42}
\ee
Finally denote  $M(x_{v}(e_{I_v}(t)) )$ by $M(t)$.  

It then follows that 
$M(x_{v}(p))({\hat e}_{I_{v}}^{a}(p)  \partial_{a} f(p))|_{p=e_{I_v}(t)} = (M(t) \frac{1}{|{\vec{{\dot e}}}_{I_v}(t)|}\frac{df}{dt})$ is a $C^{r-2}$ function of $t$ 
and admits a Taylor expansion about $t=0$:
\ba
&& M(x_{v}(v_{I_v,\e}) ){\hat e}_{I_{v_{I,\e}}}^{a}  \partial_{a} f(v_{I,\e}) = (M(t)\frac{1}{|{\vec{{\dot e}}}_{I_v}(t)|} \frac{df}{dt})|_{t=t\e} \\
&=& M(t=0)\frac{df}{dt}|_{t=0} + t_\e (\frac{d}{dt} (M(t)\frac{1}{|{\vec{{\dot e}}}_{I_v}(t)|} \frac{df}{dt}))|_{t=0} + O(t_\e^2) \\
&=& M(x_{v}(v) )({\hat e}_{I_{v}}^{a}  \partial_{a} f(v))
\;+\; \frac{\e}{|   {\vec{{\dot e}}}_{I_v}(t=0)|}(\frac{d}{dt} (M(t) \frac{1}{|{\vec{{\dot e}}}_{I_v}(t)|}\frac{df}{dt}))|_{t=0} + O(\e^2)
\label{A2.43}
\ea
where in the last line we have used $e_{I_v}(t=0)=v$ in the first term  and (\ref{A2.41}) together with (\ref{t4}).
Finally using the obvious notation:
\be
{\hat e}^a_{I_v} \partial_a F(p)|_{p=e_{I_v}(t)}\equiv  \frac{1}{|{\vec{{\dot e}}}_{I_v}(t)|}{{{\dot e}^a}}_{I_v}(t)\partial_aF(e_{I_v}(t))=
\frac{1}{|{\vec{{\dot e}}}_{I_v}(t)|}\frac{dF(e_{I_v}(t))}{dt} 
\label{A2.43a}
\ee
where $F= F(e_{I_v}(t))$ is any differentiable function  $F:e_{I_v}\rightarrow {\bf C}$, 
we can use the Taylor expansion (\ref{A2.43}) in (\ref{A2.4}) to obtain:
\ba
&&\frac{3}{8\pi}    M(x_{v_{I_v,\e}}(v_{I_v,\e})) g_{S_{\lambda_v,I_v,\J_v \e}}
 \gamma_{ (S_{\lambda_v,I_v,\J_v \e} )_{\lambda_{v_{I_v,\e}}}}
 (\prod_{v^{\prime}\neq v_{I_v,\e}} f(v^{\prime}))(\sum_{J_{v_{I_v,\e}}=1}^{N_{v}}   \frac{1}{4}{\hat e}_{J_{v_{I,\e}}}^{a} \partial_{a} f(v_{I,\e}))
 \nonumber \\
&=& \frac{3}{8\pi}   \frac{1}{4} g_{S_{\lambda_v}}
\gamma_{ (S_{\lambda_v} )_{\lambda_{v}}}
 (\prod_{v^{\prime}\neq v} f(v^{\prime})) \frac{1}{4}N_v 
\nonumber\\
&& \big(M(x_{v}(v)) {\hat e}_{I_v}^{a}  \partial_{a} f(v)) + \e      {\hat e}_{I_v}^{b}  \partial_{b}(M(x_{v}(v)) {\hat e}_{I_v}^{a}  \partial_{a} f(v)))\big)
 \;\; + \;\; \;\;\; O(\e^2).
\label{A2.5}
\ea
%Finally, in  (\ref{A2.5}) we use a Taylor series expansion of the type (\ref{4.17f}) 
%to expand the evaluation of the combination of lapse and derivative of $f$ terms in terms of $\e$.
Using  (\ref{A1}) and (\ref{A2.5}) in (\ref{A1+A2}) and recalling that (\ref{A1+A2}) is the right hand side of (\ref{22.2}), it follows that equation (\ref{22.2}) can be written as:
\ba
&&\lim_{{\bar\e}\rightarrow 0}\Psi_{B,f,h} ({\hat H}_{\bar \e}(M)S_{\lambda_v,I_v,\J_v \e})\nonumber \\
&=& 
\frac{3}{8\pi}\sum_{v^{\prime}\neq v}    M(x_{v^{\prime}}(v^{\prime})) \frac{1}{4}g_{S_{\lambda_v}}
 \gamma_{ (S_{\lambda_v} )_{\lambda_{v^{\prime}}}}
 (\prod_{v^{\prime\prime}\neq v^{\prime}} f(v^{\prime\prime})) (\sum_{I_{v^{\prime}}=1}^{N_{v^{\prime}}}   \frac{1}{4}{\hat e}_{I_{v^{\prime}}}^{a} \partial_{a} f(v^{\prime}))\nonumber \\
&+&
\frac{3}{8\pi}   \frac{1}{4} g_{S_{\lambda_v}}
 \gamma_{ (S_{\lambda_v} )_{\lambda_{v}}}
 (\prod_{v^{\prime}\neq v} f(v^{\prime})) \frac{1}{4}N_v M(x_{v}(v)) {\hat e}_{I_v}^{a}  \partial_{a} f(v)\nonumber \\
&+& \e\;\big( 
\;
\frac{3}{8\pi}\sum_{v^{\prime}\neq v}    M(x_{v^{\prime}}(v^{\prime})) \frac{1}{4}g_{S_{\lambda_v}}
 \gamma_{ (S_{\lambda_v} )_{\lambda_{v^{\prime}}}}
 (\prod_{v^{\prime\prime}\neq v^{\prime}, v} f(v^{\prime\prime})) 
 ( \frac{1}{4}{\hat e}_{I_{v}}^{a} \partial_{a} f(v))
 (\sum_{I_{v^{\prime}}=1}^{N_{v^{\prime}}}   \frac{1}{4}{\hat e}_{I_{v^{\prime}}}^{a} \partial_{a} f(v^{\prime}))\nonumber \\
&+&
\frac{3}{8\pi}   \frac{1}{4} g_{S_{\lambda_v}}
 \gamma_{ (S_{\lambda_v} )_{\lambda_{v}}}
 (\prod_{v^{\prime}\neq v} f(v^{\prime})) \frac{1}{4}N_v {\hat e}_{I_v}^{b}  \partial_{b}(M(x_{v}(v)) {\hat e}_{I_v}^{a}  \partial_{a} f(v))\big) \nonumber \\
&+& O(\e^2) 
\label{A1A2}
\ea
Note that the $0$th order and $1$st order contributions in $\e$ to (\ref{A1A2}) are independent of the $\J_v$  label of the first line of (\ref{A1A2}). Hence in the expression in square brackets in 
(\ref{22.1}) the combinatorial factor in the denominator cancels with the factor obtained by summing over  $\J_v$  and up to terms of $O(\e^2)$ we have that:
\be
 \frac{1}{\Perm{N_v-1}{3}}\sum_{\J_v}    \lim_{{\bar\e}\rightarrow 0}\Psi_{B,f,h} ({\hat H}_{\bar \e}(M)S_{\lambda_v,I_v,\J_v \e})
=
\lim_{{\bar\e}\rightarrow 0}\Psi_{B,f,h} ({\hat H}_{\bar \e}(M)S_{\lambda_v,I_v,\J_v \e})
\label{square}
\ee
with the right hand side of (\ref{square}) being given by that of (\ref{A1A2}). In (\ref{22.1}) there is also an additional sum over $I_v$ in the square brackets for this amplitude. 
Performing this sum, we obtain the zeroth order in $\e$ contribution to the square bracketed term to be 
\ba
&& [\sum_{I_v=1}^{N_v}\frac{1}{\Perm{N_v-1}{3}}\sum_{\J_v}    \lim_{{\bar\e}\rightarrow 0}\Psi_{B,f,h} ({\hat H}_{\bar \e}(M)S_{\lambda_v,I_v,\J_v \e})]|_{0}
\nonumber\\
&=&
N_v\frac{3}{8\pi}\sum_{v^{\prime}\neq v}    M(x_{v^{\prime}}(v^{\prime})) \frac{1}{4}g_{S_{\lambda_v}}
 \gamma_{ (S_{\lambda_v} )_{\lambda_{v^{\prime}}}}
 (\prod_{v^{\prime\prime}\neq v^{\prime}} f(v^{\prime\prime})) (\sum_{I_{v^{\prime}}=1}^{N_{v^{\prime}}}   \frac{1}{4}{\hat e}_{I_{v^{\prime}}}^{a} \partial_{a} f(v^{\prime}))\nonumber \\
&+&
N_v\frac{3}{8\pi}   \frac{1}{4} g_{S_{\lambda_v}}
 \gamma_{ (S_{\lambda_v} )_{\lambda_{v}}}
 (\prod_{v^{\prime}\neq v} f(v^{\prime})) M(x_{v}(v)) \sum_{I_v=1}^{N_v} \frac{1}{4}{\hat e}_{I_v}^{a}  \partial_{a} f(v).
\label{zeroth}
\ea
On the other hand, using (\ref{first1}) we have that:
\ba
&&\lim_{{\bar\e}\rightarrow 0}\Psi_{B,f,h}({\hat H}_{\bar \e}(M)S_{\lambda_v})\nonumber\\
&=&
\frac{3}{8\pi}\sum_{v^{\prime}}M(x_{v^{\prime}}(v^{\prime})) g_{(S_{\lambda_v})}
 \gamma_{ (S_{\lambda_v})_{  \lambda_{v^{\prime}}     }    }
 (\prod_{ v^{\prime\prime}\neq v^{\prime} } 
 f(v^{\prime\prime}))
 (\sum_{ I_{v^{\prime}}=1}^{ N_{v^{\prime}}} 
 \frac{1}{4}{\hat e}_{I_{ v^{\prime}} }^{a} \partial_{a} f(v^{\prime})).
\label{cmslambda}
 \ea
Separating out the contributions from $v$ and $v^{\prime}\neq v$ in (\ref{cmslambda}) it is immediate to see from (\ref{zeroth}) that:
\be
[\sum_{I_v=1}^{N_v}\frac{1}{\Perm{N_v-1}{3}}\sum_{\J_v}    \lim_{{\bar\e}\rightarrow 0}\Psi_{B,f,h} ({\hat H}_{\bar \e}(M)S_{\lambda_v,I_v,\J_v \e})]|_{0}
= N_v
\frac{1}{4}\lim_{{\bar\e}\rightarrow 0}\Psi_{B,f,h}({\hat H}_{\bar \e}(M)S_{\lambda_v}).
\label{zerothcancel}
\ee
This implies that in the numerator of the term inside the limit operation in (\ref{22.1}),  the zeroth order in $\e$ contribution to the square brackets cancels with the  term following it.
Since the denominator contains a factor of $\e$, $O(\e^2)$ contributions to the square bracket vanish in the $\e\rightarrow 0$ limit and the only nontrivial contributions come from a sum over $I_v$
of $\O(\e)$ contributions to the right hand side of (\ref{square}). These contributions can be read off from (\ref{A1A2}). Putting all this together, we obtain:
\ba
&&    \Psi_{B,f,h} ({\hat H}(M) {\hat H} (N)S) :=        \lim_{\e\rightarrow 0} (\lim_{{\bar\e}\rightarrow 0}\Psi_{B,f,h} ({\hat H}_{\bar \e}(M) {\hat H}_\e (N)S)) \nonumber\\
&=& 
\frac{3}{8\pi}\sum_{v\in V_{GR}(S)}N(x_v(v))\nonumber\\
&&\lim_{\e\rightarrow 0}
\left(
\frac{ [\sum_{I_v=1}^{N_v}\frac{1}{\Perm{N_v-1}{3}}\sum_{\J_v}    \lim_{{\bar\e}\rightarrow 0}\Psi_{B,f,h} ({\hat H}_{\bar \e}(M)S_{\lambda_v,I_v,\J_v \e})] - N_v{\frac{1}{4}}\lim_{{\bar\e}\rightarrow 0}\Psi_{B,f,h}({\hat H}_{\bar \e}(M)S_{\lambda_v})}{\e}
\right)
\nonumber\\
&=& (\frac{3}{8\pi})^2\sum_{v\in V_{GR}(S)}N(x_v(v))  \frac{1}{4}g_{S}
\nonumber\\
&& 
\Big(\;\;\sum_{v^{\prime}\neq v}    M(x_{v^{\prime}}(v^{\prime})) 
 \gamma_{ (S_{\lambda_v} )_{\lambda_{v^{\prime}}}}
 (\prod_{v^{\prime\prime}\neq v^{\prime}, v} f(v^{\prime\prime})) 
 (\sum_{I_v=1}^{N_v} \frac{1}{4}{\hat e}_{I_{v}}^{a} \partial_{a} f(v))
 (\sum_{I_{v^{\prime}}=1}^{N_{v^{\prime}}}   \frac{1}{4}{\hat e}_{I_{v^{\prime}}}^{a} \partial_{a} f(v^{\prime}))\nonumber \\
&+&
 \gamma_{ (S_{\lambda_v} )_{\lambda_{v}}}
 (\prod_{v^{\prime}\neq v} f(v^{\prime})) \frac{1}{4}N_v \sum_{I_v=1}^{N_v}{\hat e}_{I_v}^{b}  \partial_{b}(M(x_{v}(v)) {\hat e}_{I_v}^{a}  \partial_{a} f(v))\;\;\Big) .
 \label{hmhnfinal}
 \ea 
Here we have denoted $N(x(v))$ (see (\ref{22.1})) by $N(x_v(v))$ to bring the notation in line with that used in the analysis of (\ref{A1+A2}). Also 
we have replaced $g_{S_{\lambda_v}}$ by $g_S$ because $S_{\lambda_v}$ and $S$ only differ by their intertwiners at $v$ and the interkink distance function is insensitive
to this difference as it only depends on the kink structure of the state. For the  equations from (\ref{A2.4}) on, we have not explicitly mentioned the ranges of vertex labels
under the summation and product signs as these are straightforward to infer as we go along so that in the final  equation (\ref{hmhnfinal}) above these ranges are as follows.
The vertex label $v$ as indicated ranges over the set $V_{GR}(S)$ of all nondegenerate (and necessarily GR, since $S$ has overlap with $B$) vertices of $S$.
The vertex label $v^{\prime}$ also ranges over $V_{GR}(S)$ whereas the vertex label $v^{\prime\prime}$ ranges over the set $V(S)$ of all (necessarily GR but not necessarily nondegenerate)
vertices of $S$.

\subsubsection{\label{sec4.2.3} Constraint Commutator}

The continuum limit commutator amplitude  can be read off from the part of the continuum limit product amplitude which is antisymmetric under interchange of the lapse functions $M,N$
(see (\ref{2hpsi}), (\ref{commhpsi})).
It follows from the considerations of sections \ref{sec4.2.1} - \ref{sec4.2.3} that the continuum limit commutator amplitude is only non-trivial for spin networks states which have overlap in $B$,
in which case this amplitude commutator can be read off from the part of  (\ref{hmhnfinal}). 

Our final result is then

\ba
&&\Psi_{B,f,h} ([{\hat H}(M), {\hat H} (N)]S) \nonumber \\
&=& 0
%\nonumber\\
%&&
\;\;\;\;\;\;\;\;\;\;\;\;\;\;\;\;\;\;\;\;\;\;\;\;\;{\rm if} \; S \; {\rm has} \; {\rm no} \; {\rm overlap} \; {\rm in} \; B
\label{comm1}\\
&=&  (\frac{3}{8\pi})^2   \frac{1}{16}g_{S}    \sum_{v\in V_{GR}(S)}\gamma_{ (S_{\lambda_v} )_{\lambda_{v}}}N_v
 (\prod_{v^{\prime}\in V(S),v^{\prime}\neq v} f(v^{\prime})) 
\nonumber \\
&&
 \Big( \;\;N(x_v(v)){\hat e}_{I_v}^{b}  \partial_{b}M(x_{v}(v)) -
 M(x_v(v)){\hat e}_{I_v}^{b}  \partial_{b}N(x_{v}(v))\;\; \Big)
 {\hat e}_{I_v}^{a}  \partial_{a} f(v) \nonumber \\
%\nonumber \\
%&&\;\;\;\;\;\;\;\;\;\;\;\;
&& \;\;\;\;\;\;\;\;\;\;\;\;\;\;\;\;\;\;\;\;\;\;\;\;\;\;\;\;\;\;\;\;\;\;\;\;\;{\rm if} \; S \; {\rm has} \; {\rm overlap} \; {\rm in} \; B\;\;\;\;\;\;\;\;\;\;\;\;
 \label{commfinal}
 \ea
 
\subsection{\label{sec4.3} On the importance of embeddable abstract spin networks}

Consider the  off shell state $\Psi_{B,f,h}$.  Let us fix the Bra Set $B$.
Note  that the calculations and results of sections \ref{sec4.1}, \ref{sec4.2} go through for fixed $B$ and {\em any} choice of $f,h$. In particular, even if we change $h$,
the calculations hold despite the fact that we are changing the regulating RNCs. The reason for this is tied to the role of embeddable abstract spin networks in the definition of $B$ in section \ref{sec3}.
Specifically, the associated decorated abstract spin network (see section \ref{sec3.1}) for each deformed child generated by the action of the constraint on a parent is {\em independent} of the metric $h$ used to define the deformation.
This implies that if a child generated by the constraint action with respect to RNCs defined by a metric $h$ has overlap with $B$, the corresponding child generated with respect to RNCs defined with respect to a different metric $h^{\prime}$
will also have overlap with $B$, this overlap only depending on the (same)  underlying decorated abstract spin network associated with the $h$-child and the $h^{\prime}$-child.
It is important to emphasize that while we can, using the techniques of Appendix \ref{seca4}, relate constraint actions at different $\e$ with respect to the {\em same} metric $h$ by diffeomorphisms if we so wish,
we do not know if constraint actions defined with respect to different metrics are diffeomorphic. Since the metrics under consideration are all diffeomorphic to each other, it may be possible to define these actions
so that they are indeed diffeomorphic to each other. 
Instead, 
by going  beyond diffeomorphism classes to `decorated abstract spin network'-classes 
in the definition of Bra Sets, we bypass this issue. The consequent fact, that  the results of sections \ref{sec4.1}, \ref{sec4.2} go through for fixed $B$ regardless of $h$, is a crucial ingredient in our proof
of diffeomorphism covariance of the constraint action in section \ref{sec6}. A potential advantage of the use of abstract structures over a putative attempt to incorporate the desired property of the constraint
actions for different $h$ being diffeomorphic, is that while the latter is expected to leave an imprint of the diffemorphism class of metrics ${\cal H}_{h_0}$ on the off-shell space, the former  could possibly lead to an 
off shell space independent of the choice of  (diffeomorphism class of) $h_0$. In a similar fashion  an insistence on abstract structures may conceivably also lead to a  manifestly  $h_0$ independent
physical state space; this would be desireable as it would signal independence of physics from the choice of regulating structure.

\section{\label{sec5} Quantization of $\{H(M), H(N)\}$}

For notational convenience denote the Poisson bracket $\{H(M), H(N)\}$ by $O(M,N)$.
From (\ref{classhh}), we have that on the Gauss Law constraint surface:
\ba
\{H[M],H[N]\}:= O(M,N)   & =& \int\mathrm{d}^{3}x      F_{ab}^{k}E_{k}^{b} \omega_c q^{-2/3}E_{i}^{a}E_{i}^{c}, 
\label{pbhh}
\\
&&\omega_c= \left(  N\partial_{c}M-M\partial_{c}N\right) 
\label{om}
\ea
The expression on the right hand side is the diffeomorphism constraint smeared with a  metric dependent `shift' vector field:
\be
L^a = \omega_c q^{-2/3}E_{i}^{a}E_{i}^{c}.
\label{mshift}
\ee
We shall construct the operator ${\hat O}(M,N)$ corresponding to $\{H(M), H(N)\}$ by regulating the right hand side of (\ref{pbhh}) in three  steps. First, in section \ref{sec5.1} we construct a regulated  
operator  ${\hat L}^a_\e$ corresponding to the `metric shift'  $L^a$. Next, in section \ref{sec5.2} we use this regulated version to construct a regulated version ${\hat O}_\e(M,N)$ of ${\hat O}(M,N)$
as a $\frac{1}{\e}$  times the difference between a finite `$\e$ size'  diffeomorphism and  the identity. Importantly, in line with the classical expression, this diffeomorphism 
will be an ordinary  {\em semianalytic} diffeomorphism rather than an electric diffeomorphism.
Finally in section \ref{sec5.3}   we compute the continuum limit action of ${\hat O}_\e(M,N)$ on the basis state $\Psi_{B,f,h}$ and show that this continuum limit agrees with the 
constraint commutator continuum limit (\ref{commfinal}).

\subsection{\label{sec5.1} Construction of  ${\hat L}^a_\e$}

Denote the metric shift at point $p$ by $L^a(p)$.
In this section we construct the  corresponding regulated metric shift operator ${\hat L}^a_\e (p)$.  Our final result is 
as follows. 
For every nondegenerate vertex  $v$ of $S$, consider a small neighborhood $B_{2\e}(v)$ of coordinate size $\e$ as measured by the RNC's at $v$.
Let $\e$ be small enough that $B_{2\e}(v)$ contains no other vertex of $S$. Then ${\hat L}^a_\e (p)$ is non-trivial only if $p\in B_{2\e}(v)$ for some
non-degenerate vertex $v$ of $S$ with action
\be
{\hat L}^a_\e (p) S =  \frac{2}{ (\frac{4\pi}{3})^2}
\sum_{I_v=1}^{N_v} (j_I)(j_I+1) (\omega_c(v){\hat e}^c_I(v)) {\hat e}^a_{I_v,\e}(p) (S_{\lambda_v})_{\lambda_v}
\label{lhate}
\ee
where  ${\hat e}^a_{I_v,\e}(p)$ is a regulated version of the unit coordinate edge tangent ${\hat e}^a_{I_v}(v)$  supported only in 
$B_{2\e}(v)$ and defined through equations (\ref{a4.04}),  (\ref{a4.041}) with the identification $f^{\mu}\equiv {\hat e}^{\mu}_{I_v,v}$ in the coordinate system defined in section \ref{sec4.0b}. 
It follows from this definition that  as in the regularization of the electric shift  \cite{p4}, for  every small enough $\e$,  ${\hat e}^a_{I_v,\e}(p)$  is chosen such that on the intersection of 
the $I_v$th edge $e_{I_v}$ with $B_{2\e}(v)$,  ${\hat e}^a_{I_v,\e}(p)$ is parallel to the edge tangent to $e_{I_v}$ i.e.
\be
{\hat e}^a_{I_v,\e}(p) = \lambda (p) {\dot e}^a_{I_v} (p) \;{\rm for}\; p \in e_{I_v} \cap B_{2\e}(v), 
\label{propeie}
\ee
and that, as in \cite{p4},  $\lim_{\e \rightarrow 0}{\hat e}^a_{I_v,\e}(p)$ vanishes everywhere except at $v$ where it is equal to  ${\hat e}^a_{I_v}(v)$.
It is in this sense that ${\hat e}^a_{I_v,\e}(p)$ is a regularization of ${\hat e}^a_{I_v}(v)$.

The rest of this section is devoted to a derivation of (\ref{lhate}).
%the detailed behavior of  ${\hat e}^a_{I_v,\e}(p)$ has been described in \cite{p4}. For our purposes in section \ref{sec5.2}  a key property 

Fix RNC's at a point $q \in\Sigma$. Let $B_\e(q)$ denote an open  coordinate ball of radius $\e$ centered at $q$. Let ${\bar B}_{\e^n}(q)$ be the closure of 
the open ball of radius $\e^n, \;n>>1$ centered at $q$. Define $B^{\prime}_\e(q)$ as the exterior of ${\bar B}_{\e^n}(q)$ in $B_\e(q)$:
\be
B^{\prime}_\e(q)= B_\e(q)- {\bar B}_{\e^n}(q).
\label{defbp}
\ee
In what follows we shall often replace the argument consisting of the point  $q$  by its coordinate value  i.e. we write  $B^{\prime}_\e(x)$ 
and we implicitly mean $B^{\prime}_\e(q)$ for $q$ with coordinates $x^{\mu}, \mu=1,2,3$. We shall also often employ a  similar notation for  arguments of  fields so that, for e.g.
$E^{a}_i (x)$  will mean the evaluation of $E^a_i(q)$ at a point $q$ which has coordinates $x^{\mu}, \mu=1,2,3$ with density weights also evaluated with respect to $\{x\}$.

Next, recall from \cite{p4} and the discussion in section \ref{sec3.4.3}, we may regulate ${\hat q}^{-\frac{1}{3}}(p)$ 
to have non-trivial action only if $p$ coincides with a nondegenerate vertex $v$ of $S$ in which case we have for all small enough $\delta$:
\be
({\hat q}_{\delta}^{-\frac{1}{3}}(v)) S
:=  \delta^{2} S_{\lambda_v}.
\label{defqd}
\ee
Next, we define the   classical approximant $L^a_{\tau,\eta}(p)\;\eta <<\tau$:
\be
L^a_{\tau,\eta}(p) := \omega_c (p)\left(\frac{1}{{\frac{4}{3}}\pi\tau^3}\int_{B_{\tau}^{\prime}(p)}
 d^3x \;(\int_{B_{{\eta}^{\prime}(x)}}d^3\bx \frac{ E^c_i(\bx)}{\frac{4}{3}\pi\eta^3})E^a_i(x)\right) ({ q_{\tau}^{-\frac{1}{3}}}(p))({ q_{\eta}^{-\frac{1}{3}}}(p))
%(\frac{1}{ q^{\frac{1}{3}}}(v))_{\tau}
\label{defltlnc}
\ee
where $({ q_{\delta}^{-\frac{1}{3}}}(p))$ is the classical precursor of (\ref{defqd}) 
%$(\frac{1}{ q^{\frac{1}{3}}}(v))_{\e}$ 
defined  through an appropriate power of the volume of $B_\delta(v)$ measured with respect to the metric $q_{ab}$ as in Reference \cite{p4} and Footnote \ref{fntycho}.

It is straightforward to check that:
\be
\lim_{\tau\rightarrow 0}( \lim_{\eta \rightarrow 0}L^a_{\tau,\eta}(p)) = L^a (p)
\ee
so that $L^a_{\tau,\eta}(p)\;\eta <<\tau$ is indeed an approximant to the metric shift. Next, we construct a quantization of this approximant. We choose an operator ordering in which
quantum operators appear in the same order in which their  classical correspondents appear in (\ref{defltlnc}) so that ${\hat{q}_{\eta}^{-\frac{1}{3}}}(p)$ appears right most.
Then if $p$ is not a nondegenerate vertex of  $S$, we have, for all small enough $\eta$, that
\be
{\hat L}^a_{\tau,\eta}(p) S=0 .
\label{lpneqv}
\ee
If $p=v$ is a nondegenerate vertex of $S$, we have that:
\ba
{\hat L}^a_{\tau,\eta}(v) S
&:=& \omega_c(v)\left(\frac{1}{{\frac{4}{3}}\pi\tau^3}\int_{B_{\tau}^{\prime}(v)}
 d^3x \;(\int_{B_{{\eta}^{\prime}(x)}}d^3\bx \frac{ {\hat E}^c_i(\bx)}{\frac{4}{3}\pi\eta^3}){\hat E}^a_i(x)\right) ({ {\hat q}_{\tau}^{-\frac{1}{3}}}(v))({{\hat q}_{\eta}^{-\frac{1}{3}}}(v)) S
\label{l1}\\
&=&
(\frac{1}{{\frac{4}{3}}\pi})^2\frac{1}{\tau\eta}\omega_c(v)\int_{B_{\tau}^{\prime}(v)}d^3x \int_{B_{{\eta}^{\prime}(x)}}d^3\bx {\hat E}^c_i(\bx)\;\; {\hat E}^a_i(x)(S_{\lambda_v})_{\lambda_v}
\label{l2}\\
&=& 
(\frac{1}{{\frac{4}{3}}\pi})^2\frac{1}{\tau\eta }\omega_c(v)\int_{B_{\tau}^{\prime}(v)}d^3x \int_{B_{{\eta}^{\prime}(x)}}d^3\bx {\hat E}^c_i(\bx)\nonumber\\
&&i\sum_I\int_{e_{I}}dt_I{\dot e}^a_I \delta^3(x, e_I(t_I)) (h_{e_I}(1,t_I) \tau_ih_{e_I}(t_I, 0))^C{}_D \frac{\partial  (S_{\lambda_v})_{\lambda_v}}{\partial h_{e_I}(1,0)^C{}_D}
\label{l3}
\\
&=& 
(\frac{1}{{\frac{4}{3}}\pi})^2\frac{i}{\tau\eta}\omega_c(v)\sum_I\int_{\tau_{n,I}}^{\tau_I}dt_I{\dot e}^a_I \int_{B_{{\eta}^{\prime}(e_I(t_I))}}d^3\bx {\hat E}^c_i(\bx)
(h_{e_I}(1,t_I) \tau_ih_{e_I}(t_I, 0))^C{}_D \frac{\partial  (S_{\lambda_v})_{\lambda_v}}{\partial h_{e_I}(1,0)^C{}_D}.\;\;\;\;\;\;\;\;\;\;
\label{l4}
\ea
Above, we have used (\ref{defqd}) in (\ref{l2}).  The standard action of the triad operator in (\ref{l2}) yields (\ref{l3}).  In (\ref{l3}), similar to our considerations in \cite{p4},  $\tau_i$ is the matrix representative 
of the $i$th generator of 
$su(2)$ in the $j_I$ spin representation which colors the edge $e_I$ in $S$. For notational convenience we have suppressed the subscript $v$ in the edge index $I$ and edge parameter $t_I$. 
Integration over the Dirac delta function in (\ref{l3}) yields (\ref{l4}). For small enough $\tau$, this leads to an integration over the  part of the edge $e_I$ between the parameter values
$\tau_{n,I}$ and $\tau_I$ with $e(\tau_{n, I})$ being the point of intersection of $e_I$ with the boundary of ${\bar B}_{\tau^n}(v)$ and $e(\tau_I)$ the point of intersection of $e_I$ with the boundary 
of ${\bar B}_{{\tau} (v)}$.

In what follows we  retain only leading order 
contributions  in the small parameters $\eta,\tau$ to (\ref{l4}). We follow the standard practice employed in LQG derivations of operator actions,  of estimating these contributions by assuming that the
spin network wave function  $S\equiv S(A)$ is evaluated on smooth (i.e. $C^r$) connections rather than distributional elements of the quantum configuration space. We proceed as follows.

Next,  we evaluate the action of $\int_{B_{{\eta}^{\prime}(e_I(t_I))}}d^3\bx {\hat E}^c_i(\bx)$ in (\ref{l4}) on the holonomy combination to its right. 
Note that $S$ (and hence $(S_{\lambda_v})_{\lambda_v}$ )
is a monomial in each of its constituent edge
holonomies  and for small enough $\tau, \eta$,  ${B_{{\eta}^{\prime}(e_I(t_I))}}$ intersects $S$ only in a subset of its $I$th edge. Note also 
the argument $\bx$ of the triad operator is not coincident with $e_I({t_I})$ in this subset. It follows that:
\ba
&&\int_{B_{{\eta}^{\prime}(e_I(t_I))}}d^3\bx {\hat E}^c_i(\bx)
(h_{e_I}(1,t_I) \tau_ih_{e_I}(t_I, 0))^C{}_D \frac{\partial  (S_{\lambda_v})_{\lambda_v}}{\partial h_{e_I}(1,0)^C{}_D} 
\nonumber \\
&=& i \int_{B_{{\eta}^{\prime}(e_I(t_I))}}d^3\bx
\Big(\int_{t_I}^1 d\bt_I{\dot e}^c_I \delta^3(\bx, e_I(\bt_I)) (h_{e_I}(1,\bt_I)\tau_i h_{e_I}( \bt_I,t_I)    \tau_ih_{e_I}(t_I, 0))^C{}_D \;\;\;\;\;\;\;\;\; \nonumber\\
&+& \int_{0}^{t_I} d\bt_I{\dot e}^c_I \delta^3(\bx, e_I(\bt_I)) (h_{e_I}(1,t_I)\tau_i  h_{e_I}(t_I,\bt_I)\tau_i h_{e_I}(\bt_I, 0))^C{}_D \Big)\frac{\partial  (S_{\lambda_v})_{\lambda_v}}{\partial h_{e_I}(1,0)^C{}_D}.
\label{l5}
\ea
Next,  consider $e(t_I)$ which is a coordinate distance of at least $2\eta$ away from $e_I(\tau_{n,I})$ as well as $e_I({\tau_{I}})$.
The parameter range for which this is satisfied is almost all of  $\tau_{n,I} <t_I <\tau_{I}$ except for a parameter range of $O(\eta)$ near the endpoints $\tau_{n,I}, \tau_I$ (the fact that the parameter
range is of the same order of the distance between the end points of the parameter range is implied by the $C^r$ property of the edge $e_I$ through  (\ref{t4})). 
For such $t_I$,  the intersection of $e_I$ with ${B_{{\eta}^{\prime}(e_I(t_I))}}$ 
is the union of two disjoint segments of $e_I$, one
between the parameter values $t_I+\eta^+_{n,I}, t_I+ \eta_I^+$ and  one between the parameter values $t_I-  \eta_I^-, t_I-\eta^-_{n,I}$ 
where $\eta_{\pm}, \eta_{n, I}^{\pm}$ are parameter values such that:\\
\noindent (i) $e_I$ intersects the boundary of ${\bar B}_{\eta^n}(e_I(t_I))$ in the pair of points $e_I(t_I+\eta^+_{n,I}), e_I(t_I- \eta^+_{n,I})$.\\
\noindent (ii) $e_I$ intersects the boundary of ${\bar B}_{\eta}(e_I(t_I))$ in the pair of points $e_I(t_I+\eta^+_{I}), e_I(t_I- \eta^+_{I})$.\\
For such $e_I(t_I)$ we  continue on from (\ref{l5}). Integrating the delta function over $d^3\bx$ and using (i), (ii) above we get:
\ba
&&\int_{B_{{\eta}^{\prime}(e_I(t_I))}}d^3\bx {\hat E}^c_i(\bx)
(h_{e_I}(1,t_I) \tau_ih_{e_I}(t_I, 0))^C{}_D \frac{\partial  (S_{\lambda_v})_{\lambda_v}}{\partial h_{e_I}(1,0)^C{}_D} 
\nonumber \\
&=& i 
\Big(\int_{ t_I+\eta^+_{n,I}   }^{  t_I+ \eta_I^+   } d\bt_I{\dot e}^c_I (h_{e_I}(1,\bt_I)\tau_i h_{e_I}(\bt_I,t_I)    \tau_ih_{e_I}(t_I, 0))^C{}_D \;\;\;\;\;\;\;\;\; \nonumber\\
&+& \int_{  t_I-  \eta_I^-     }^{   t_I-\eta^-_{n,I}    } d\bt_I{\dot e}^c_I  (h_{e_I}(1,t_I)\tau_i  h_{e_I}(t_I,\bt_I)\tau_i h_{e_I}(\bt_I, 0))^C{}_D \Big)\frac{\partial  (S_{\lambda_v})_{\lambda_v}}{\partial h_{e_I}(1,0)^C{}_D}
\label{l6}\\
&=& i 
\Big(\int_{ t_I+\eta^+_{n,I}   }^{  t_I+ \eta_I^+   } d\bt_I{\dot e}^c_I (h_{e_I}(1,t_I)\tau_i   \tau_ih_{e_I}(t_I, 0))^C{}_D \;\;\;\;\;\;\;\;\; \nonumber\\
&+& \int_{  t_I-  \eta_I^-     }^{   t_I-\eta^-_{n,I}    } d\bt_I{\dot e}^c_I  (h_{e_I}(1,t_I)\tau_i  \tau_i h_{e_I}(t_I, 0))^C{}_D \Big)\frac{\partial  (S_{\lambda_v})_{\lambda_v}}{\partial h_{e_I}(1,0)^C{}_D}
\;\;\;+\;O(\eta^2)
%\nonumber \\
%&+& O(\eta^2)
\label{l7}\\
&=&-i(j_I)(j_I+1) \Big(\int_{ t_I+\eta^+_{n,I}   }^{  t_I+ \eta_I^+   } d\bt_I{\dot e}^c_I 
+ \int_{  t_I-  \eta_I^-     }^{   t_I-\eta^-_{n,I}    } d\bt_I{\dot e}^c_I \Big)
(h_{e_I}(1, 0)^C{}_D  \frac{\partial  (S_{\lambda_v})_{\lambda_v}}{\partial h_{e_I}(1,0)^C{}_D})
+ O(\eta^2)\; \label{l8}
\\
&=&-i(j_I)(j_I+1) \big(  (e_I^c({  t_I+ \eta_I^+   })-  e_I^c(t_I+\eta^+_{n,I}))
+ (e^c(  t_I-\eta^-_{n,I} ) - e^c ( t_I-  \eta_I^-  )) \big) \nonumber\\
&& h_{e_I}(1, 0)^C{}_D  \frac{\partial  (S_{\lambda_v})_{\lambda_v}}{\partial h_{e_I}(1,0)^C{}_D}\; + \;O(\eta^2)
\label{l9}
\\
&=& - i(j_I)(j_I+1) (2{\hat e}^c (t_I) \eta ) (S_{\lambda_v})_{\lambda_v} \;+O\;(\eta^2) .
\label{l10}
\ea
Here,  in the first integral of (\ref{l6})  we have used  $h_{e_I}(1,\bt_I)= h_{e_I}(1,t_I)(h_{e_I}(\bt_I, t_I))^{-1}$, the fact that parameter and distance orders are the same for $C^r$ edges (\ref{t4}) and 
the smooth connection expansion of holonomies  $h_{e_I}(\bt_I,t_I) = {\bf 1} +O(\eta )$, and similarly for the second integral. This results in (\ref{l7}) from (\ref{l6}).
To obtain (\ref{l8}) we use the standard $su(2)$ representation property  $\tau_i\tau_i = -j_I(j_I+1){\bf 1}$. 
The integrations in (\ref{l8}) yield the differences between the RN coordinates of the $I$th edge at the parameter end points for each of the two integrals therein so as to yield (\ref{l9}). It is straightforward to use expansions of the type
(\ref{4.17}) together with $n>>1$ and  (i)-(ii) above to show that the contribution of these coordinate differences in (\ref{l9}) yields, to $O(\eta^2)$,  the  $(2{\hat e}^c (t_I) \eta   ) $ factor in (\ref{l10}).
Finally from the fact  that $(S_{\lambda_v})_{\lambda_v}$ is a monomial in its edge holonomies, we replace 
$h_{e_I}(1, 0)^C{}_D  \frac{\partial  (S_{\lambda_v})_{\lambda_v}}{\partial h_{e_I}(1,0)^C{}_D}$ in (\ref{l9}) by  $(S_{\lambda_v})_{\lambda_v}$ in (\ref{l10}).

Next, let us estimate the contribution to (\ref{l4}) from those $e(t_I)$ which are within a distance of $2\eta$ from   $e_I(\tau_{n,I})$ or $e_I({\tau_{I}})$.
Integrating over the delta functions in  (\ref{l5}) yields:
\ba
&&\int_{B_{{\eta}^{\prime}(e_I(t_I))}}d^3\bx {\hat E}^c_i(\bx)
(h_{e_I}(1,t_I) \tau_ih_{e_I}(t_I, 0))^C{}_D \frac{\partial  (S_{\lambda_v})_{\lambda_v}}{\partial h_{e_I}(1,0)^C{}_D} 
\nonumber \\
&=& i \Big(\int_{B_{{\eta}^{\prime}(e_I(t_I))} \cap e_{1, t_I} }d\bt_I{\dot e}^c_I  (h_{e_I}(1,\bt_I) \tau_ih_{e_I}(\bt_I,t_I)    \tau_ih_{e_I}(t_I, 0))^C{}_D \;\;\;\;\;\;\;\;\; \nonumber\\
&+& i \int_{B_{{\eta}^{\prime}(e_I(t_I))} \cap e_{0, t_I} } d\bt_I{\dot e}^c_I (h_{e_I}(1,t_I)\tau_i  h_{e_I}(t_I,\bt_I)\tau_i h_{e_I}(\bt_I, 0))^C{}_D \Big)
\frac{\partial  (S_{\lambda_v})_{\lambda_v}}{\partial h_{e_I}(1,0)^C{}_D} 
\label{l10.1}
\ea
Using the boundedness of the functions of the connection encountered in (\ref{l10.1}) together with (\ref{4.17})  we have that:
\ba
&&|\int_{B_{{\eta}^{\prime}(e_I(t_I))} \cap e_{(1, t_I)} }d\bt_I{\dot e}^{\mu}_I  (h_{e_I}(1,\bt_I)\tau_i h_{e_I}(\bt_I,t_I)    \tau_ih_{e_I}(t_I, 0))^C{}_D 
\frac{\partial  (S_{\lambda_v})_{\lambda_v}}{\partial h_{e_I}(1,0)^C{}_D} |
\nonumber\\
&+&
|\int_{B_{{\eta}^{\prime}(e_I(t_I))} \cap e_{(0, t_I)} } d\bt_I{\dot e}^{\mu}_I (h_{e_I}(1,t_I)\tau_i  h_{e_I}(t_I,\bt_I)\tau_i h_{e_I}(\bt_I, 0))^C{}_D
\frac{\partial  (S_{\lambda_v})_{\lambda_v}}{\partial h_{e_I}(1,0)^C{}_D} |
\nonumber\\
&&\leq {\bar C} \eta
\label{l11}
\ea
for some ${\bar C}>0$ 
where ${\dot e}^{\mu}_I$ are RNC components of the edge tangent ${\dot e}^{a}_I$.
Note that this sort of a bound is independent of the parameter value $t_I$ and hence valid for all $e(t_I)$ whether or not  within a distance of $2\eta$ from   $e_I(\tau_{n,I})$ or $e_I({\tau_{I}})$.
Denote the set of points $e(t_I)$ within a distance of $2\eta$ from   $e_I(\tau_{n,I})$ or $e_I({\tau_{I}})$ by ${\bf U}_{I,2\eta}$ and denote the set of parameter values $t_I$ of these points by
$U_{I,2\eta, }$
From expansions of the type (\ref{4.17}) it follows that 
\be
|\int_{U_{I,2\eta}} dt_I{\dot e}^{\mu}_I| \leq  {\bar C}_1\eta
\label{u2eta}
\ee
for some ${\bar C}_1 >0$. 
Denote the open interval in parameter space with end points $\tau_{n,I}, {\tau_I}$ by $(\tau_{n,I}, {\tau_I})$.
We may write  (\ref{l4}) as:
\ba
{\hat L}^a_{\tau,\eta}(v) S
&=& 
(\frac{1}{{\frac{4}{3}}\pi})^2\frac{i}{\tau\eta}\omega_c(v)\left(\sum_I [\int_{(\tau_{n,I}, {\tau_I}) - U_{I, 2\eta}} 
dt_I{\dot e}^a_I \big(- i(j_I)(j_I+1) 2{\hat e}^c (t_I) \eta  (S_{\lambda_v})_{\lambda_v} \big) ]
+  O(\eta^2)\right)\;\;\;\;\;\;\;\;
\label{l12}\\
&=&
(\frac{1}{{\frac{4}{3}}\pi})^2\frac{i}{\tau\eta}\omega_c(v)\left(\sum_I [\int_{(\tau_{n,I}, {\tau_I})} 
dt_I{\dot e}^a_I \big(- i(j_I)(j_I+1) 2{\hat e}^c (t_I) \eta  (S_{\lambda_v})_{\lambda_v} \big) ]
+  O(\eta^2)\right)
\label{l13}\\
&=& 
(\frac{1}{{\frac{4}{3}}\pi})^2\frac{i}{\tau}\omega_c(v)\left(\sum_I [\int_{(\tau_{n,I}, {\tau_I})} 
dt_I{\dot e}^a_I \big(- i(j_I)(j_I+1) 2{\hat e}^c (t_I)   (S_{\lambda_v})_{\lambda_v} ]
+  O(\eta)\right).
\label{l14}
\ea
In (\ref{l4})  the $t_I$ integration range can be divided into $U_{I,2\eta}$ and its complement. The integral over $U_{I,2\eta}$ may be bounded using the boundedness of the RNC components $\omega_{\mu}(v)$ of $\omega_{c}(v)$ together with 
(\ref{l11}) and (\ref{u2eta}) to obtain  the 
$O(\eta^2)$ term in 
(\ref{l12}).  For the remaining range of integration  we use (\ref{l10}) to get the rest of (\ref{l12}). 
In (\ref{l13}) we increase the range of the complement of $U_{I,2\eta}$ back to its original range at the cost of an  extra integral over $U_{I,2\eta}$. Using the bound (\ref{u2eta}) 
it is straightforward to see that  this extra integral also yields a term of $O(\eta^2)$ and  we obtain (\ref{l14}).
Using expansions of the type (\ref{4.17}) in (\ref{l14}) together with $n>>1$ yields:
\ba 
{\hat L}^a_{\tau,\eta}(v) S
&=& (\frac{1}{{\frac{4}{3}}\pi})^2\frac{2}{\tau}\omega_c(v) \left( \sum_I \tau {\hat e}^c_I(t_I=0) {\hat e}^a_I(t_I=0)(j_I)(j_I+1) (S_{\lambda_v})_{\lambda_v}
+   O(\tau^2) + O(\eta) \right)\;\;\;\;\;
\label{l15}\\
&=& \left( (\frac{2}{{\frac{4}{3}}\pi})^2\omega_c(v)\sum_I {\hat e}^c_I(t_I=0) {\hat e}^a_I(t_I=0)(j_I)(j_I+1) (S_{\lambda_v})_{\lambda_v}\right)
+ O(\tau) + \frac{O(\eta)}{\tau} .
\label{l16}
\ea
Taking the limit $\eta\rightarrow 0$ followed by that of $\tau\rightarrow 0$ yields 
\be
{\hat L}^a (v) S:= (\frac{2}{{\frac{4}{3}}\pi})^2\omega_c(v) \sum_I {\hat e}^c_I(t_I=0) {\hat e}^a_I(t_I=0)(j_I)(j_I+1) (S_{\lambda_v})_{\lambda_v}
\label{l17}
\ee
Finally, {\em exactly} as for the electric shift \cite{p4} we replace ${\hat e}^a_I(t_I=0)$ by ${\hat e}^a_{I,\e}(p)$ in (\ref{l17}) to obtain the desired result:
\be
{\hat L}^a_\e (p) S:= (\frac{2}{{\frac{4}{3}}\pi})^2\omega_c(v) \sum_I {\hat e}^c_I(t_I=0) {\hat e}^a_{I,\e}(p)(j_I)(j_I+1) (S_{\lambda_v})_{\lambda_v}.
\label{l18}
\ee
Since ${\hat e}^a_{I,\e}(p)$ is a $C^{r-1}$ vector field of compact support around $v$ such that $\lim_{\e\rightarrow 0}{\hat e}^a_{I,\e}(p)$ is nonvanishing  only when $p=v$ with
$\lim_{\e\rightarrow 0}{\hat e}^a_{I,\e}(v) = {\hat e}^a_I(t_I=0) (v)$, it follows that $\lim_{\e\rightarrow 0} {\hat L}^a_\e (p) S$ is also non-trivial only at $v$ where it
equals the right hand side of (\ref{l17}). Hence, from (\ref{lpneqv}) and (\ref{l17}) it follows that (\ref{l18}) defines a regulation of the quantum metric shift operator
${\hat L}^a$.

\subsection{\label{sec5.2} Construction of  ${\hat O}_\e(M,N)$}

%Our considerations below from (\ref{51}) to (\ref{61}) apply regardless of whether $v$ is a GR or NGR. 
%In section \ref{sec5.2.1} we focus on the GR case  starting from (\ref{61}) and in section \ref{sec5.2.1} on the NGR case.

From (\ref{pbhh}) we define the operator correspondent of  ${ O}(M,N)$ at regularization parameter $\e$ to be:
\be
{\hat O_\e(M,N)} S = \int\mathrm{d}^{3}x      {\hat F}_{ab}^{k}{\hat E}_{k}^{b}{\hat L}^a_\e S
\label{51}
\ee
with ${\hat L}^a_\e S$ given by (\ref{lhate}). Substituting from (\ref{lhate}) we obtain:
\be
{\hat O_\e(M,N)} S = 
\frac{2}{ (\frac{4\pi}{3})^2}\sum_v(\omega_c(v){\hat e}^c_I(v)) 
\sum_{I_v=1}^{N_v} (j_I)(j_I+1) 
\int_{B_{2\e}(v)}\mathrm{d}^{3}x      {\hat e}^a_{I_v,\e} {\hat F}_{ab}^{k}{\hat E}_{k}^{b}
(S_{\lambda_v})_{\lambda_v} .
\label{52}
\ee
where the sum is over all nondegnerate vertices $v$ in $S$.
The operator under the integral  is exactly the diffeomorphism constraint smeared with the $C^r$ vector field  ${\hat e}^a_{I_v,\e}$ taking on the role of a shift vector field.
From \cite{diffeoconstr}, we expect that the diffeomorphism constraint operator smeared with this shift can be expressed 
%as $\e^{-1}$ times 
in terms of the difference between a small  diffeomorphism and
the identity divided by the smallness parameter.
In section \ref{sec5.2.1} we 
show that this expectation is borne out when  $v$ is a GR vertex, the small parameter being $\e$.
In order to do so we use  heuristic argumentation which closely parallels that underlying the derivation of the Hamiltonian constraint action in \cite{p4}, the 
only significant difference with \cite{p4} being that the finite transformation generated by ${\hat e}^a_{I_v,\e}$ is taken to be a semianalytic diffeomorphism rather than an electric diffeomorphism.
In section \ref{sec5.2.2} we consider the case that  $v$ is an NGR vertex. In contrast to our treatment of the GR case, here  we provide a heuristic argumentation which does not involve the area of the loops
formed by segments of the deformed and undeformed edges, but which directly seeks to write the operator under the integral as the difference between a small  diffeomorphism and the identity.
%we prefer to rely on the results of \cite{diffeoconstr}, 
%these results being applicable regardless of the GR or NGR nature of the vertex modulo certain technical
%fine points  which we discuss in sections \ref{sec5.2.2} and \ref{sec7}.

The reason we adopt different approaches for the GR and NGR cases is as follows.  Recall that in the case of the  electric diffeomorphisms which underlie the Hamiltonian constraint action, 
we transited from (\ref{mixed}) to (\ref{abaction0}) in section \ref{sec2.1.2} so as to conveniently absorb overall factors by changing the area of the small loops $l_{IJ,\e}$, these small loops
being tied to the choice of electric diffeomorphism \cite{p4}.
It turns out to be crucial,  for the satisfaction of the anomaly free condition,   to absorb overall factors in a similar manner for the action (\ref{52})  for the  GR case.  Clearly, in order to do so, we require an argumentation in which 
small loops of area  $O(\e^2)$ play a key role.  
%However, 
%, we absorb overall factors in (\ref{61}) by changing the area of the small loops $\bl_{IJ,\e}$.
%as we shall see in section \ref{sec5.2.2}, the considerations of Reference \cite{diffeoconstr}
%when straightforwadly applied, are not phrased in terms of such $\e$- sized small loops. 
This is the reason we 
rely on the argumentation of \cite{p4} for the GR case. In contrast, when $v$ is NGR, the anomaly free condition does not require such an absorption of overall factors and 
a more direct heuristic argumentation suffices for our purposes.

\subsubsection{\label{sec5.2.1} The case when $v$ is GR}

Due to the close similarity with the considerations of \cite{p4} we shall be succinct, for details the reader is urged to consult Reference \cite{p4}. 

Let us focus on the contribution from a nondegenerate vertex $v$ of $S$. To avoid notational clutter we suppress the subscript $v$ at various places.
We have:
\ba
&&\int_{B_{2\e}(v)}\mathrm{d}^{3}x      {\hat e}^a_{I_v,\e} {\hat F}_{ab}^{k}{\hat E}_{k}^{b}
(S_{\lambda})_{\lambda} 
\nonumber \\
&=& i\sum_J\int_{B_{2\e}(v)}\mathrm{d}^{3}x      {\hat e}^a_{I,\e} {\hat F}_{ab}^{k} \int_{e_J}dt_J{\dot e}^a(t_J) \delta^3(x,e_J(t_J)) 
[h_{e_J}(1,t_J) \tau_k h_{e_J}(t_J,0)]^{A_J}_{\;\;B_J} \frac{\partial{(S_{\lambda})_{\lambda} }}{\partial h_{e_J\;B_J}^{A_J}} \nonumber \\
%\int_{0}^1 dt {\dot e}^a(t) \delta^3(x,e(t)) (h_e(1,t)\tau^{(j)}_i h_e (t,0)) ^{\;A}_{\;\;B} 
%
&=& i \sum_{J }\int_{0}^{t_{\e,J,I}} dt_J  ({\hat e}^a_{I,\epsilon}{\hat F}_{ab}^k) {\dot e}^b_{J}(t_J)
                                     [ h_{e_J}(1,t_J) \tau_k h_{e_J}(t_J,0)]^{A_J}_{\;\;B_J} \frac{\partial{(S_{\lambda})_{\lambda} }}{\partial h_{e_J\;B_J}^{A_J}} \nonumber \\
&=& i  \sum_{J }\int_{0}^{t_{\e,J,I}} dt_J  ({\hat e}^a_{I,\epsilon}{\hat F}_{ab}^k) {\dot e}^b_{J}(t_J)
                                      h_{e_J}(1,0) ({\tau_k} + O(\e ))^{A_J}_{\;\;B_J}  \frac{\partial (S_{\lambda})_{\lambda}}{\partial h_{e_J\;B_J}^{A_J}} \nonumber\\
&=& i \sum_{J\neq I }\int_{0}^{t_{\e,J,I}} dt_J  ({\hat e}^a_{I,\epsilon}{ F}_{ab}^k) {\dot e}^b_{J}(t_J)
                                     ( h_{e_J}(1,0) \tau_k)^{A_J}_{\;\;B_J}  \frac{\partial (S_{\lambda})_{\lambda}}{\partial h_{e_J\;B_J}^{A_J}} .
\label{53}
\ea      
In the second line above we used the standard action  of the triad operator. The third line results from integrating the delta function in the second over $d^3x$  with  the edge $e_J$ intersecting the 
boundary of $B_{2\e}(v)$ in a single point with parameter value $t_{\e, J,I}$. In the 4th line we use the  standard holonomy expansion for smooth connections in conjunction with expansions of type (\ref{4.17}).
In the 5th line we retain only the leading order term from the 4th line and use (\ref{propeie}) to restrict the sum  over $J$.

From an analysis identical to  \cite{p4} we have that:
\be
\int_{B_{2\e}(v)}\int_{0}^{t_{\e,J,I}} dt_J   {\dot e}^b_{J} {\hat e}^a_{I,\epsilon}{ F}_{ab}^i\tau_i
= - \frac{(h_{{\bl}_{IJ,\e}} - {\bf 1})}{\e}+ O(\e^2) 
\label{54}
\ee
where the loop $l_{IJ,\e}$ is defined as: 
\be
\bl_{IJ,\e} = ({e_J}_{(t_{\e,J,I}, 0)}   )^{-1}\circ  {\bar \phi}_{I,\e} ({e_J}_{(t_{J,I,\e},0)}    ) \circ {e_I}_{ (\e_I,0) } 
\label{defblij}
\ee
Here we use an obvious notation in which the part of an edge $e$ between parameter values $t_1$ and $t_2$ is denoted by $e_{t_2,t_1}$,  the image of such an edge by a diffeomorphism $\phi$
is denoted by $\phi(e_{t_2,t_1})$ and edge composition is denoted by `$\circ$'.
%Recall from the analysis of section 3, \cite{p4} that
From Appendix \ref{sec4.0b}, similar to section 3 of Reference \cite{p4}, 
$t_I= \e_I$ is the parameter value such that the point $e_I(t_I=\e_I)$ is exactly $v_{I,\e}$ i.e.:
\be
e_I(t_I=\e_I) = v_{I,\e}.
\label{defti=ei}
\ee

We remark here that the  {\em only} significant difference with the analysis of  section 3, Reference \cite{p4} is that the {\em electric} diffeomorphism  ${ \phi}_{I,\e}$ of that work is replaced here by  the {\em semianalytic}
diffemorphism ${\bar \phi}_{I,\e}$.
\footnote{An additional but insignificant difference with the analysis of \cite{p4}  is the difference in the detailed specification of the support of 
the vector field ${\hat e}^a_{I,\epsilon}$ in Appendix \ref{sec4.0b} and in that work.}
%is supported within a coordinate sphere of 
%radius $2\e$, in Reference \cite{p4} the support was chosen to be  a coordinate sphere of radius $\e +\e^m,\;m>2$ \cite{p4}}
%
Hence in contrast to the situation with ${ \phi}_{I,\e}$,  {\em no kinks} are created by ${\bar \phi}_{I,\e}$.
In more detail, the  semianalytic diffeomorphism ${\bar \phi}_{I,\e}$ is imagined to be the finite transformation generated by  the vector field $e_{I,\e}$ and is assumed to enjoy the following properties:\\
(i) ${\bar \phi}_{I,\e}$  is the identity outside the support  $B_{2\e}(v)$ of $e_{I,\e}$.\\
(ii) Consistent with (\ref{propeie}), ${\bar \phi}_{I,\e}$ maps   $e_I$ to itself.\\
(iii) In particular  ${\bar \phi}_{I,\e}$ displaces $v$ to $v_{I,\e}$ where as in section \ref{sec2.3.1}, $v_{I,\e}$ is located at a coordinate distance $\e$ from $v$ along the edge $e_I$, the coordinates
being the RNC at $v$.\\
 ${\bar \phi}_{I,\e}$ subject to (i)-(iii) is constructed in Appendix \ref{sec4.0b} (see equation (\ref{adefphie})).
% We discuss the  technicalities connected with the assumptions (i)-(iii) in 
%section \ref{sec7}.  

%Note that the area of the loop in RN coordinates around the vertex $v$  is, similar to the case of the small loop  $ l_{IJ,e}$ of \cite{p4}, of $\O(\e^2)$.
%As in section \ref{sec2.

Substituting (\ref{54}) in (\ref{53}) we get to leading order in $\e$ that:
\be
\int_{B_{2\e}(v)}\mathrm{d}^{3}x      {\hat e}^a_{I_v,\e} {\hat F}_{ab}^{k}{\hat E}_{k}^{b}
(S_{\lambda})_{\lambda}  
= 
-\frac{i}{\e}\sum_{J\neq I}( h_{e_J}(1,0) (h_{{\bl}_{IJ,\e}} - {\bf 1}))^{A_J}_{\;\;B_J}  \frac{\partial (S_{\lambda})_{\lambda}}{\partial h_{e_J\;B_J}^{A_J}} .
\label{55}
\ee
In equation (\ref{55}) each $ h_{e_J}(1,0)(h_{{\bl}_{IJ,\e}} - {\bf 1})$ term is of $O(\e^2)$ when evaluated on smooth (i.e. $C^r$) connections.
As in \cite{p4} we re-express the sum over such contributions in (\ref{55}) in terms of products of such contributions, this re-expression agreeing with (\ref{55}) to leading order in $\e$
so as to lead to a valid approximant constraint action at regulation  parameter value $\e$. 
%Equation (\ref{55}) is of the form of a sum over the contribution of small deformations, each such deformation `$h_{e_J}(1,0) (h_{{\bl}_{IJ,\e}} - {\bf 1})$
We recall the essential steps below, for details the reader may consult Reference \cite{p4}.

Denoting  $h_{e_J}(1,0)$ by $x_{\alpha}$,  $(h_{e_J}(1,0)(h_{{\bl}_{IJ,\e}} - {\bf 1}))^{A_J}_{\;\;B_J}  $ by  $\delta x_{\alpha}$ and $(S_{\lambda})_{\lambda}$ by $F(x)$,
equation (\ref{55}) is of the form $\sum_{\alpha} \delta x_{\alpha}\frac{\partial}{\partial x_{\alpha}} F (x)$ which can be rewritten  to leading order in $\delta x_{\alpha}$ as $F(x+\delta x) - F(x)$.
The fact that   $F \sim (S_{\lambda})_{\lambda}$ is a {\em product} over edge holonomies leads to the replacement of the sum over edges of the form (\ref{55}) by a difference of products over edges.
Using the notation employed in \cite{p4} we write $(S_{\lambda})_{\lambda}$ as:
\be
 (S_{\lambda})_{\lambda} = C_{\lambda, \lambda }^{A_1..A_N} (\prod_{K=1}^N h_{e_K}(1,0)^{B_K}{}_{A_K} ) S_{rest}{}_{B_1..B_N}
\label{s=ch}
\ee
where  $C_{\lambda, \lambda }^{A_1..A_N}$ is the intertwiner at $v$ and $S_{rest}$ does {\em not} depend on the  edge holonomies over  edges which emananate from $v$.
From the discussion above, we rewrite (\ref{55}) to leading order in $\e$ as:
\ba
&&\int_{B_{2\e}(v)}\mathrm{d}^{3}x      {\hat e}^a_{I_v,\e} {\hat F}_{ab}^{k}{\hat E}_{k}^{b}
(S_{\lambda})_{\lambda}  
= -\frac{i}{\e}\left(C_{\lambda, \lambda }^{A_1..A_N} (\prod_{J\neq I} h_{e_J}(1,0)h_{{\bl}_{IJ,\e}}^{B_J}{}_{A_J} ) S_{rest}{}_{B_1..B_N}
- (S_{\lambda})_{\lambda}\right) 
\label{56}
\\
&&=
-\frac{i}{\e}\left(
C_{\lambda, \lambda }^{A_1..A_N} \prod_{J\neq I} (h_{e_J}(1,t_{\e,J,I})  h_{   {\bar \phi}_{I,\e} ({e_J})}(t_{J,I,\e},0) h_{{e_I} } (\e_I,0)) ^{B_J}{}_{A_J} h_{e_I}^{B_I}{}_{A_I} S_{rest}{}_{B_1..B_N}
- (S_{\lambda})_{\lambda} \right)\;\;\;\;\;\;\;\;\;\;
\label{57}
\\
&=& 
-\frac{i}{\e}\left(
C_{\lambda, \lambda }^{A_1..A_N} \prod_{J\neq I}( h_{   {\bar \phi}_{I,\e} ({e_J})}(1,0) h_{{e_I} } (\e_I,0)) ^{B_J}{}_{A_J} h_{e_I}(1,0)^{B_I}{}_{A_I} S_{rest}{}_{B_1..B_N}
- (S_{\lambda})_{\lambda} \right) .
\label{58}
\ea
In (\ref{56}) we have used the equivalent to leading order  `difference in products' form as discussed above. In (\ref{57}) we have used (\ref{defblij}). In (\ref{58}),
$h_{   {\bar \phi}_{I,\e} ({e_J})}(1,0)$ denotes the holonomy over the image of the edge ${e_J}$ by the semianalytic diffeomorphism ${\bar \phi}_{I,\e}$. 
In obtaining (\ref{58}) from (\ref{57}), 
we have used the fact that the diffeomorphism ${\bar \phi}_{I,\e}$ is of compact support and that the part of the $J$th edge which lies within this support is that between parameter
values $t_J=0$ and $t_J= t_{\e, J, I}$ to conclude that:
\be
{\bar \phi}_{I,\e} ({e_J}_{(1,0)}    ) = {\bar \phi}_{I,\e} ({e_J}_{(1, t_{J,I,\e})}    )\circ {\bar \phi}_{I,\e} ({e_J}_{(t_{J,I,\e},0)}    )
= {e_J}_{(1, t_{J,I,\e})} \circ {\bar \phi}_{I,\e} ({e_J}_{(t_{J,I,\e},0)}    ).
\ee
Next, recall that the invariance of the  intertwiner $C_{\lambda, \lambda }$ under a  gauge transformation $g$ implies that:
\be
(C_{\lambda})_{\lambda}^{A_1..A_N} (\prod_{J\neq I} g^{B_J}{}_{A_J})  = (C_{\lambda})_{\lambda}^{B_1.. .B_N} g^{-1}{}^{A_I}{}_{B_I}.
\label{ggeinvc}
\ee
Setting $g=  h_{{e_I} } (\e_I,0)$  it is straightforward to obtain:
\ba
&&C_{\lambda, \lambda }^{A_1..A_N} \prod_{J\neq I}( h_{   {\bar \phi}_{I,\e} ({e_J})}(1,0) h_{{e_I} } (\e_I,0)) ^{B_J}{}_{A_J} h_{e_I}(1,0)^{B_I}{}_{A_I} S_{rest}{}_{B_1..B_N}
\nonumber\\
&&= 
C_{\lambda, \lambda }^{A_1..A_N} \prod_{J\neq I}( h_{   {\bar \phi}_{I,\e} ({e_J})}(1,0) ) ^{B_J}{}_{A_J} h_{e_I}(1,\e_I)^{B_I}{}_{A_I} S_{rest}{}_{B_1..B_N}
\label{59}
\ea
From properties (i)-(iii) of ${\bar \phi}_{I,\e}$ in conjunction with  (\ref{defti=ei})
it is straightforward to see that:
\ba
&&C_{\lambda, \lambda }^{A_1..A_N} \prod_{J\neq I}( h_{   {\bar \phi}_{I,\e} ({e_J})}(1,0) ) ^{B_J}{}_{A_J} h_{e_I}(1,\e_I)^{B_I}{}_{A_I} S_{rest}{}_{B_1..B_N}\nonumber \\
&=& C_{\lambda, \lambda }^{A_1..A_N} \prod_{J\neq I}( h_{   {\bar \phi}_{I,\e} ({e_J})}(1,0) ) ^{B_J}{}_{A_J} 
h_{ {\bar \phi}_{I,\e} (e_I)}(1,0)^{B_I}{}_{A_I} S_{rest}{}_{B_1..B_N}\\
&=&  C_{\lambda, \lambda }^{A_1..A_N} \prod_{K}( h_{   {\bar \phi}_{I,\e} ({e_K})}(1,0) ) ^{B_K}{}_{A_K}  S_{rest}{}_{B_1..B_N}\\
&=& {\hat U}_{{\bar \phi}_{I,\e}} (S_{\lambda})_{\lambda}
\label{60}
\ea
where ${\hat U}_{{\bar \phi}_{I,\e}}$ is the unitary operator corresponding to the semianalytic  diffeomorphism 
${{\bar \phi}_{I,\e}}$.
Using (\ref{60}) in (\ref{58}) we have:
\be
\int_{B_{2\e}(v)}\mathrm{d}^{3}x      {\hat e}^a_{I_v,\e} {\hat F}_{ab}^{k}{\hat E}_{k}^{b}
(S_{\lambda})_{\lambda}  = -i\frac{({\hat U}_{{\bar \phi}_{I,\e}} -{\bf 1})}{\e}(S_{\lambda})_{\lambda} .
\label{61}
\ee

Similar to the transition from (\ref{mixed}) to (\ref{abaction0}) in section \ref{sec2.1.2}, we absorb overall factors in (\ref{61}) by changing the area of the small loops $\bl_{IJ,\e}$.
Specifically we assume the existence of semianalytic diffeomorphisms ${\bar \phi}_{I,c_I,\e}$ such that the loop $\bl_{IJ,, c_I,\e}$ has a coordinate area which is $c_I$ times the coordinate
area of its counterpart $\bl_{IJ,\e}$ with the loop $\bl_{IJ,, c_I,\e}$ defined as:
\be
\bl_{IJ,c_I, \e} = ({e_J}_{(t_{\e,J,I,c_I}, 0)}   )^{-1}\circ  {\bar \phi}_{I,c_I,\e} ({e_J}_{(t_{\e, J,I,c_I},0)}    ) \circ {e_I}_{ (\e_I,0) } 
\label{defblijc}
\ee
Here ${\bar \phi}_{I,c_I,\e}$  is subject to similar  properties as ${\bar \phi}_{I,\e}$ i.e.\\
(i)${}_{c_I}$ ${\bar \phi}_{I,c_I,\e}$  is the identity outside a ball $U_{c_I,\e}$ of radius ${\bf O}(\e)$ around $v$.
(ii)${}_{c_I}$ ${\bar \phi}_{I,c_i,\e}$    maps   $e_I$ to itself.\\
(iii)${}_{c_I}$ ${\bar \phi}_{I,c_I, \e}$  displaces $v$ to $v_{I,\e}$ where as in section \ref{sec2.3.1}, $v_{I,\e}$ is located at a coordinate distance $\e$ from $v$ along the edge $e_I$, the coordinates
being the RNC at $v$.
%We shall discuss techincalities associated with (i)${}_{c_I}$ - (iii)${}_{c_I}$ in section \ref{sec7}.

${\bar \phi}_{I,c_I,\e}$ subject to (i)${}_{c_I}$ - (iii)${}_{c_I}$  is constructed in Appendix \ref{sec4.0b}.

In equation (\ref{defblijc}),  the parameter value $t_J=t_{\e, J,I,c_I}$ is the parameter value at which the edge $e_J$ intersects the boundary of the ball $U_{c_I,\e}$.
%In Fig .. we show qualitatively how we visualize the deformation of the vertex structure by the action of ${\bar \phi}_{I,c_I,\e}$.
We emphasize that while the area of the small loops change by a factor of $c_I$, the position of the vertex $v_{I,\e}$ is unchanged.
With this change, we have, to leading order in $\e$, for holonomy evaluations on smooth (i.e. $C^r$ connections), that:
\be
h_{ \bl_{IJ,c_I, \e}}- {\bf 1} = c_I (h_{ \bl_{IJ, \e}}- {\bf 1}) 
\label{62}
\ee
so that (\ref{54}) takes the form
\be
\int_{0}^{t_{\e,J,I}} dt_J   {\dot e}^b_{J} {\hat e}^a_{I,\epsilon}{ F}_{ab}^i\tau_i
= - \frac{(h_{{\bl}_{IJ,\e}} - {\bf 1})}{\e}+ O(\e^2) 
= - \frac{(h_{{\bl}_{IJ,c_I\e}} - {\bf 1})}{c_I\e}+ O(\e^2).
\label{54c}
\ee
Using (\ref{54c}), (\ref{defblijc}) instead of (\ref{54}), (\ref{defblij}) in the analysis of section \ref{sec5.2} subsequent to (\ref{54})  it is straightforward to obtain:
\be
\int_{B_{2\e}(v)}\mathrm{d}^{3}x      {\hat e}^a_{I_v,\e} {\hat F}_{ab}^{k}{\hat E}_{k}^{b}
(S_{\lambda})_{\lambda}  = -i\frac{({\hat U}_{{\bar \phi}_{I,c_I,\e}} -{\bf 1})}{c_I\e}(S_{\lambda})_{\lambda} .
\label{61c}
\ee
Using (\ref{61c}) in (\ref{52}), restoring the `$v$' subscripts and denoting the contribution to the operator action ${\hat O_\e(M,N)} S$ from the nondegenerate GR vertices of $S$ by 
${\hat O_\e(M,N)} S|_{V_{GR}(S)}$
yields:
\ba
{\hat O_\e(M,N)} S|_{V_{GR}(S)} &=& 
\frac{2}{ (\frac{4\pi}{3})^2}\sum_{v\in V_{GR}(S)}\sum_{I_v=1}^{N_v}(\omega_c(v){\hat e}^c_I(v)) 
 (j_{I_v})(j_{I_v}+1) 
\int_{B_{2\e}(v)}\mathrm{d}^{3}x      {\hat e}^a_{I_v,\e} {\hat F}_{ab}^{k}{\hat E}_{k}^{b}
(S_{\lambda_v})_{\lambda_v} \nonumber\\
&=&
-i\frac{2}{ (\frac{4\pi}{3})^2}\sum_{v\in V_{GR}(S)}\sum_{I_v=1}^{N_v}(\omega_c(v){\hat e}^c_I(v)) 
 {(j_{I_v})(j_{I_v}+1)}
\frac{({\hat U}_{{\bar \phi}_{I_v,c_{I_v},\e}} -{\bf 1})}{c_{I_v}\e}(S_{\lambda_v})_{\lambda_v}.
\label{63}
\ea
Finally, we choose 
\be
c_{I_v} = \frac{128(j_{I_v})(j_{I_v}+1)}{N_v}
\label{ci=128jj}
\ee
to get:
\ba
{\hat O_\e(M,N)} S|_{V_{GR}(S)} &=& 
\frac{2}{ (\frac{4\pi}{3})^2}\sum_{v\in V_{GR}(S)}\sum_{I_v=1}^{N_v}(\omega_c(v){\hat e}^c_I(v)) 
 (j_{I_v})(j_{I_v}+1) 
\int_{B_{2\e}(v)}\mathrm{d}^{3}x      {\hat e}^a_{I_v,\e} {\hat F}_{ab}^{k}{\hat E}_{k}^{b}
(S_{\lambda_v})_{\lambda_v} \nonumber\\
&=&
-i  (\frac{3}{8\pi})^2 \frac{1}{16}   \sum_{v\in V_{GR}(S)}\sum_{I_v=1}^{N_v} (\omega_c(v){\hat e}^c_I(v)) N_v
\frac{({\hat U}_{{\bar \phi}_{I_v,c_{I_v},\e}} -{\bf 1})}{\e}(S_{\lambda_v})_{\lambda_v}.
\label{oegr}
\ea

\subsubsection{\label{sec5.2.2} The case when $v$ is NGR}

In this section we motivate the action of  the quantum operator 
$\int_{B_{2\e}(v)}\mathrm{d}^{3}x      {\hat e}^a_{I_v,\e} {\hat F}_{ab}^{k}{\hat E}_{k}^{b}$ in (\ref{52}). 
by thinking of it as a quantization of the classical expression 
$\int_{B_{2\e}(v)}\mathrm{d}^{3}x      {\hat e}^a_{I_v,\e} { F}_{ab}^{k}{ E}_{k}^{b}$.

Recall that on  the Gauss Law constraint surface we have, for any shift vector field $N^a$, the identity:
\be
\int_{B_{2\e}(v)}\mathrm{d}^{3}x      {N}^a { F}_{ab}^{k}{ E}_{k}^{b} = \int_{B_{2\e}(v)}\mathrm{d}^{3}x  {\cal L}_{\vec N} A_b^kE_k^b
\ee
where ${\cal L}_{\vec N}$ denotes the Lie derivative with respect to $N^a$. At the quantum level, we replace the triad operator by the functional derivative with respect to the connection and
obtain, heuristically, the following operator action on  a gauge invariant  wave function $\Psi (A)$ of the connection
\ba
\int_{B_{2\e}(v)}\mathrm{d}^{3}x      {N}^a {\hat F}_{ab}^{k}{\hat E}_{k}^{b} \Psi (A)& =& -i\int_{B_{2\e}(v)}\mathrm{d}^{3}x  {\cal L}_{\vec N} A_b^k \frac{\delta}{\delta A_b^k}\Psi (A) \nonumber\\
&=& -i \frac{\Psi (A_b^k +\delta {\cal L}_{\vec N} A_b^k)  - \Psi (A_b^k)}{\delta} + O(\delta).
\ea
We may then replace $A_b^k +\delta {\cal L}_{\vec N} A_b^k$ by $((\phi_{\vec N, \delta})_* A)_b^k$ where $\phi_{\vec N, \delta}$ is a diffeomorphism which translates points an affine amount $\delta$
along the orbits of $N^a$ to obtain a finite $\delta$- approximant 
to the operator:
\be
(\int_{B_{2\e}(v)}\mathrm{d}^{3}x      {N}^a {\hat F}_{ab}^{k}{\hat E}_{k}^{b})_{\delta} \Psi (A) = -\frac{{\hat U}_{\phi_{\vec N, \delta}}- {\bf 1}}{\delta} \Psi (A) 
\ee
where the $\delta$ subscript indicates that the expression is a finite $\delta$ approximant and ${\hat U}_{\phi_{\vec N, \delta}}$ is the unitary operator corresponding to the 
the diffeomorphism $\phi_{\vec N, \delta}$.
Setting $N^a\equiv {\hat e}^a_{I_v,\e}$,  $\Psi (A)\equiv (S_{\lambda_v})_{\lambda_v}$ in the above equation obtains:
\be
(\int_{B_{2\e}(v)}\mathrm{d}^{3}x      {\hat e}^a_{I_v,\e} {\hat F}_{ab}^{k}{\hat E}_{k}^{b})_{\delta}(S_{\lambda_v})_{\lambda_v}
= \frac{{\hat U}_{{\bar \phi}_{I, \delta}}- {\bf 1}}{\delta}(S_{\lambda_v})_{\lambda_v} 
\label{oedngr}
\ee
where ${\bar \phi}_{I_v, \delta}$ is constructed through Appendix \ref{sec4.0b} simply as ${\bar \phi}_{I_v, \delta} \equiv \psi_{\lambda= \delta}$ for sufficiently small $\delta$ and corresponds exactly to 
the desired finite transformation which translates points along the orbits of ${\hat e}^a_{I_v,\e}$ by an affine amount $\delta$ to leading order in $\delta$.

The $\delta\rightarrow 0$ limit dual operator action on the off shell basis state $\Psi_{B,f,h}$ then defines the $\e$ approximant dual operator action:
\ba
&&\Psi_{B,f,h} \left(\int_{B_{2\e}(v)}\mathrm{d}^{3}x      {\hat e}^a_{I_v,\e} {\hat F}_{ab}^{k}{\hat E}_{k}^{b} (S_{\lambda})_{\lambda}     \right) 
\nonumber\\
&:=&\lim_{\delta\rightarrow 0} \Psi_{B,f,h} \left(\left(\int_{B_{2\e}(v)}\mathrm{d}^{3}x      {\hat e}^a_{I_v,\e} {\hat F}_{ab}^{k}{\hat E}_{k}^{b}\right)_{\delta}
(S_{\lambda})_{\lambda} \right)
\nonumber \\
&=&
%\nonumber \\
\lim_{\delta\rightarrow 0}\Psi_{B,f,h}(\frac{({\hat U}_{{\bar \phi}_{I_v, \e, \delta}} -{\bf 1})}{\delta}(S_{\lambda_v})_{\lambda_v}) .
\label{oengr}
\ea
By assumption the 
state  $(S_{\lambda_v})_{\lambda_v}$  has an NGR vertex at $v$ and, hence, so does its diffeomorphic image by the semianalytic diffeomorphism ${\hat U}_{{\bar \phi}_{I_v, \e, \delta}}$.
It follows that the states appearing on the right hand side of (\ref{oengr}) have no overlap with $B$ which in turn implies that the right hand side vanishes. Thus for all small enough $\e$ we have that:
\be
\Psi_{B,f,h} \left(\int_{B_{2\e}(v)}\mathrm{d}^{3}x      {\hat e}^a_{I_v,\e} {\hat F}_{ab}^{k}{\hat E}_{k}^{b} (S_{\lambda})_{\lambda}     \right)  = 0.
\label{oengr1}
\ee

%We note, in conclusion, that a  further fine technical  point arises with respect to (a) above. We remark on this point briefly here and discuss it in detail in section \ref{sec7}.
%The derivation in \cite{diffeoconstr} is in the context of the {\em analytic} category. Here we work with the {\em semianalytic} category.  For the semianalytic category
%we do not know if semianalytic diffeomorphisms can be generated by appropriate vector fields. Nevertheless, as we shall argue in section \ref{sec7}, the results
%of \cite{diffeoconstr}  continue to hold for  the shift vector fields ${\hat e}^a_{I_v,\e}$
%used here. In particular equations (\ref{oedngr})-(\ref{oengr}) hold with ${\bar \phi}_{I_v,\e, \delta}$ being a semianalytic diffeomorphism. This in turn implies that (\ref{oengr1}) holds.

\subsection{\label{sec5.3} The Continuum Limit Operator ${\hat O}(M,N)$} 

In this section we evaluate the continuum limit (dual) operator action on the state $\Psi_{B,f,h}$.
Consider the case that $S$ has no overlap with $B$. It follows from (i), section \ref{sec3.2} that the
embedded colored graph underlying $S$ is distinct from the embedded colored graph underlying any
element of $B$. Since $(S_{\lambda_v})_{\lambda_v}$ lives on the same embedded colored graph as $S$,
$(S_{\lambda_v})_{\lambda_v}$ also cannot have overlap in $B$.
From the discussion around (\ref{may5.0}), (\ref{may5.1}) in section (\ref{sec3.4.3}), we have that
any diffeomorphic image of $(S_{\lambda_v})_{\lambda_v}$ also cannot have overlap in $B$.
Since the actions (\ref{oegr}), (\ref{oengr}) result in states identical to or diffeomorphic to
$(S_{\lambda_v})_{\lambda_v}$ irrespective of whether $v$ is GR or NGR, it follows that 
the dual operator action amplitude $\Psi_{B,f,h}( {\hat O}_\e(M,N)S)$ vanishes for  $S$ with no overlap
in $B$.

%Consider the case that $S$ has no overlap with $B$. Then $S$ can have exclusively GR vertices or at least one NGR vertex.
%First consider the case that $S$ (and hence $(S_{\lambda_v})_{\lambda_v}$ for any $v$) has at least one NGR vertex.
%From (\ref{52}), the contributions to the continuum limit operator action arise from vertices which are GR and vertices which are NGR. From (\ref{oengr1}) the contributions from NGR vertices
%vanish. The states which are generated by the operator action on GR vertices do not alter the vertex structure around the existing NGR vertices. Therefore all these states have at least one NGR vertex
%and cannot have overlap in $B$. Hence the dual operator action amplitude $\Psi_{B,f,h}( {\hat O}_\e(M,N)S)$ vanishes for such $S$.

%Next, let $S$ have exclusively GR vertices but no overlap in  $B$. Then $(S_{\lambda})_{\lambda}$ cannot have overlap in  $B$, as, if it did have overlap property (i), section \ref{sec3.2} implies that $S$ has overlap in $B$.  
%Since the action of ${\hat O_\e(M,N)}$  on $S$ (\ref{oegr}) results in spin networks which are diffeomorphic to $(S_{\lambda})_{\lambda}$,  property (ii), section \ref{sec3.2} implies that none of these spin networks
%can have overlap in  $B$.  Hence for such $S$, 
%we have that $\Psi_{B,f,h} ({\hat O_\e(M,N)} S)=0$ for all sufficiently small $\e$ which in turn implies that the 
%continuum limit operator amplitude for such $S$ vanishes.

%To summarise: if $S$ has no overlap with $B$, the continuum limit operator action amplitude for such $S$ vanishes.
Hence, we need consider amplitudes only for  $S$ with overlap in $B$. Accordingly let $S$ have overlap with $B$.
Two succesive actions of ${\hat \lambda}_v$ on $S$ yields $(S_{\lambda})_{\lambda}$ which, from an analysis
similar to that underlying (\ref{may1}) and (\ref{may3.0}) indicates that $(S_{\lambda})_{\lambda}$ has overlap
in $B$. The discussion around (\ref{may5.0}) and (\ref{may5.1}) then implies that all diffeomorphic
images of $(S_{\lambda})_{\lambda}$ have overlap in $B$.
%Properties (i) and  (ii) of section \ref{sec3.2}  then imply that for such $S$, $(S_{\lambda})_{\lambda}$ and its diffeomorphic images have overlap in $B$.
It then follows from (\ref{oegr}), (\ref{ci=128jj}) that:
\be
\Psi_{B,f,h}({\hat O_\e(M,N)} S) =
-i  (\frac{3}{8\pi})^2 \frac{1}{16}   \sum_{v\in V_{GR}(S)}  \sum_{I_v=1}^{N_v} (\omega_c(v){\hat e}^c_{I_v}(v)) 
 N_v
\Psi_{B,f,h}(\frac{({\hat U}_{{\bar \phi}_{I_v,c_{I_v},\e}} -{\bf 1})}{\e}(S_{\lambda_v})_{\lambda_v}).
\label{531}
\ee
Denoting 
\be
{\hat U}_{{\bar \phi}_{I_v,c_{I_v},\e}} (S_{\lambda_v})_{\lambda_v}:= ((S_{\lambda_v})_{\lambda_v})_{{\rm diff},I_v, \e}
\label{532}
\ee
we have that 
\be
\Psi_{B,f,h}({\hat U}_{{\bar \phi}_{I_v,c_{I_v},\e}} (S_{\lambda_v})_{\lambda_v})
= (\prod_{\bv\in V(((S_{\lambda_v})_{\lambda_v})_{{\rm diff},I_v, \e}   )} f(\bv)) g_{((S_{\lambda_v})_{\lambda_v})_{{\rm diff},I_v, \e}, h} \gamma_{((S_{\lambda_v})_{\lambda_v})_{{\rm diff},I_v, \e}}.
\label{533}
\ee
Diffeomorphism invariance of the RS Volume operator  together with (ii), \ref{vib} of section \ref{sec3.1} implies that the gamma factor is invariant under diffeomorphisms (more specifically, see (\ref{may5.1})) so that 
\be
\gamma_{((S_{\lambda_v})_{\lambda_v})_{{\rm diff},I_v, \e}} = \gamma_{(S_{\lambda_v})_{\lambda_v}}.
\label{534}
\ee
From the fact that no kinks are created by the action of ${\bar \phi}_{I_v,c_{I_v},\e}$ and the fact that  ${\bar \phi}_{I_v,c_{I_v},\e}$ is non-trivial only it an $\e$-vicinity of $v$, it follows that for small enough $\e$,
%and the fact that $S$ and $(S_{\lambda_v})_{\lambda_v}$ have the same kinks
we have that:
\be
g_{((S_{\lambda_v})_{\lambda_v})_{{\rm diff},I_v, \e}, h} = g_{S,h}.
\label{535}
\ee
Using (\ref{534}) and (\ref{535}) in (\ref{533}), together with the fact that $v$ is displaced to $v_{I_v,\e}$ in $((S_{\lambda_v})_{\lambda_v})_{{\rm diff},I_v, \e}$
we obtain
\be
\Psi_{B,f,h}({\hat U}_{{\bar \phi}_{I_v,c_{I_v},\e}} (S_{\lambda_v})_{\lambda_v})
= g_{S,h}  \gamma_{(S_{\lambda_v})_{\lambda_v}}(\prod_{\bv \in V(S), \bv \neq v}f(\bv)) f(v_{I_v,\e}).
\label{536}
\ee
Using (\ref{536}) in (\ref{531}) yields:
\ba
&&\Psi_{B,f,h}({\hat O_\e(M,N)} S) 
\nonumber \\
&=&
-i  (\frac{3}{8\pi})^2 \frac{1}{16} 
g_{S,h}  
 \sum_{v\in V_{GR}(S)}   \gamma_{(S_{\lambda_v})_{\lambda_v}} \sum_{I_v=1}^{N_v}  (\omega_c(v){\hat e}^c_{I_v}(v)) (\prod_{\bv \in V(S), \bv \neq v}f(\bv))N_v  
 \frac{f(v_{I_v,\e}) - f(v)}{\e} .
\label{537}
\ea
Using (\ref{4.17f}) in the above equation yields the continuum limit operator amplitude:
\ba
&&\lim_{\e\rightarrow 0}\Psi_{B,f,h}({\hat O_\e(M,N)} S) 
\nonumber \\
&=& 
-i  (\frac{3}{8\pi})^2 \frac{1}{16} g_{S,h} 
\sum_{v\in V_{GR}(S)}    \gamma_{(S_{\lambda_v})_{\lambda_v}}  \sum_{I_v=1}^{N_v}     (\omega_c(v){\hat e}^c_{I_v}(v)) (\prod_{\bv \in V(S), \bv \neq v}f(\bv))N_v
 {\hat e}_{I_v}^a\partial_a f(v).
\ea
Thus we have that:
\ba 
&&\lim_{\e\rightarrow 0}\Psi_{B,f,h}({\hat O_\e(M,N)} S) := \Psi_{B,f,h}({\hat O(M,N)} S) 
\nonumber \\
&=& 0 \;{\rm if} \; S \; {\rm has} \; {\rm no} \; {\rm overlap} \; {\rm in} \; B
\label{omnpsi0}
\\
&=&
-i  (\frac{3}{8\pi})^2 \frac{1}{16} g_{S,h} \sum_{v\in V_{GR}(S)}    \gamma_{(S_{\lambda_v})_{\lambda_v}}
\nonumber \\
&&    \sum_{I_v=1}^{N_v}      (\omega_c(v){\hat e}^c_{I_v}(v)) (\prod_{\bv \in V(S), \bv \neq v}f(\bv))N_v
 {\hat e}_I^a\partial_a f(v) \;\; {\rm if} \; S \; {\rm has} \; {\rm overlap} \; {\rm in} \; B.
\label{omnpsi1}
\ea

Comparing (\ref{comm1}), (\ref{commfinal}) with (\ref{omnpsi0}), (\ref{omnpsi1}), it follows that:
\be
\Psi_{B,f,h} ([ {\hat H}(M), {\hat H}(N)] S) = i \Psi_{B,f,h} (\widehat{\{H(M), H(N)\}}S), \;\;\;\forall S.
\label{anomfreeamp}
\ee
From  (\ref{anomfreeamp}) it follows that the commutator between a pair of Hamiltonian constraints is anomaly free on a domain consisting of the 
finite linear span of off shell basis states.

\section{\label{sec6} Action of Spatial Diffeomorphisms }

In section \ref{sec6.1} we show that the group of semianalytic diffeomorphisms is represented without anomalies on any off shell basis state, and hence on the finite span of such states.
In section  \ref{sec6.2} we show that the continuum limit (dual) action, of the Hamiltonian constraint as well as that of the  product of a pair of Hamiltonian constraints,  on any off shell basis state
is diffeomorphism covariant. As we shall see,  the key feature of our constructions which simply and directly ensures diffeomorphism covariance is the tying of  the choice of regulating coordinates (i.e. the RNC's) to the metric label
of the off shell basis state. In section \ref{sec6.3}, we discuss the conditions under  which  the results of section \ref{sec6.2} extend to the finite span of off shell basis states.

\subsection{\label{sec6.1} Action on $\Psi_{B,f,h}$}
Let the unitary operator corresponding to the semianalytic diffeomorphism $\phi$  be ${\hat U}(\phi)$.
and denote  the diffeomorphic image of the state $S$ by ${\hat U}(\phi)$  by $S_{\phi}$. 
Let $S$ have overlap in $B$. 
From the discussion around equations (\ref{may5.0}), (\ref{may5.1}), it follows that 
%(ii) section 3.2 it follows that 
$S_{\phi}$ also has overlap in $B$.
Then we have that:
\ba
\Psi_{B,f,h}( {\hat U}^{\dagger}(\phi)S) &=& \Psi_{B,f, h}(S_{\phi^{-1}})
\\
&=& g_{S_{\phi^{-1}}, h} \gamma_{S_{\phi^{-1}}}\prod_{v\in V(S_{\phi^{-1}})} f(v).
\label{dc1}
\ea
Clearly every $v \in V(S_{\phi^{-1}})$ is the image by $\phi^{-1}$ of some $\bv \in V(S)$ so that:
\be
\prod_{v\in V(S_{\phi^{-1}})} f(v) = \prod_{\bv\in V(S)} f(\phi^{-1}(v)) = \prod_{\bv\in V(S)} (\phi^*f)(\bv)
\label{fphi}
\ee
where  we use {\em Wald's notation} \cite{waldbook}  in which $(\phi^*f)$ denotes the {\em push forward} of $f$ with respect to $\phi$.
Next in order to emphasize its dependence on the metric,  let us denote the distance  $d$, defined in section \ref{sec3.3} with respect to the metric $h$, by $d_h$. 
It is straightforward to verify that for any pair of points $p,q\in \Sigma$,
\be
d_h(\phi^{-1}p, \phi^{-1}q) = d_{\phi^*h}(p,q).
\label{dphi}
\ee
Since $g_h$ is determined by the network of interkink distances with respect to $h$,  and since any kink at the position $k$ in $S$ is moved to the position $\phi^{-1}(k)$ in $S_{\phi^{-1}}$
equation (\ref{dphi}) immediately implies that:
\be
g_{S_{\phi^{-1}}, h} = g_{S, \phi^*h}.
\label{gphi}
\ee
Finally, it is straightforward to see that  the diffeomorphism invariance of the RS volume operator together with (ii),\ref{vib}, section \ref{sec3.1}  (in particular, equation (\ref{may5.1})) ensures that:
\be
\gamma_{S_{\phi^{-1}}}= \gamma_{S}.
\label{gammaphi}
\ee
Equations (\ref{dc1}), (\ref{fphi}), (\ref{gphi}), (\ref{gammaphi}) together with definition of off shell state amplitudes (\ref{psiamps}) imply that:
\be
\Psi_{B,f,h}( {\hat U}^{\dagger}(\phi)S) 
= g_{S, \phi^*h} \gamma_S\prod_{v\in V(S)} \phi^*f(v)
=\Psi_{B,\phi^*f, \phi^*h} (S).
\label{phipsi}
\ee
On the other hand if $S$ has no overlap in $B$, 
the discussion around equations (\ref{may5.0}), (\ref{may5.1})
%property (ii) of section \ref{sec3.2} 
implies that  $S_{\phi^{-1}}$ cannot have overlap in $B$ so that for such $S$
\be
\Psi_{B,f,h}( {\hat U}^{\dagger}(\phi)S) =0 = \Psi_{B,\phi^*f, \phi^*h} (S),
\label{phipsi0}
\ee
with the last equality following from the fact that $S$ has no overlap in $B$.
% that the Bra Set label remains unchanged.
\footnote{Note that (ii), \ref{vib},section \ref{sec3.1} in conjunction with the invertibility of $\phi$ implies that $\phi$ induces a bijection on $B$.}
Equations (\ref{phipsi}), (\ref{phipsi0}) together imply that:
\be
{\hat U}(\phi) \Psi_{B,f,h} = \Psi_{B, \phi^*f, \phi^*h}
\label{uphipsi}
\ee
where the equality is of two elements of the algebraic dual to the finite span of spin nets. It follows that:
\ba
{\hat U}(\phi_2) {\hat U}(\phi_1)\Psi_{B,f,h} &=& {\hat U}(\phi_2)\Psi_{B, \phi_1^*f, \phi_1^*h} 
\\
&=&   \Psi_{B, \phi_2^*\phi_1^*f, \phi_2^*\phi_1^*h} = \Psi_{B, (\phi_2\circ\phi_1)^*f, (\phi_2\circ\phi_1)^*h}\\
&=& {\hat U}(\phi_2\circ\phi_1) \Psi_{B,f,h}.
\label{diffgrp}
\ea
Equation (\ref{diffgrp}) shows that the group of semianalytic diffeomorphisms is represented without anomaly on any basis off shell state.
It is immediate to see that our considerations above extend to the finite span of such basis states and that diffeomorphisms map elements of this finite span
into the finite span so that 
the group of semianalytic diffeomorphisms is represented without anomaly on this finite span.

Finally, consider a pair of diffeomorphisms  $\phi_1,\phi_2$ such that  $\phi_1\neq \phi_2$.
From (\ref{uphipsi}), we have ${\hat U}(\phi_i)\Psi_{B,f,h} = \Psi_{B, \phi_i^*f, \phi_i^*h}, i=1,2$.
Since $h$ has no symmetries it follows that $\phi_1^*h \neq \phi_2^*h$. Then the considerations of section \ref{secl}
imply that $\Psi_{B, \phi_1^*f, \phi_1^*h} \neq \Psi_{B, \phi_2^*f, \phi_2^*h}$. Conversely, it follows that if
${\hat U}(\phi_1) \Psi_{B,f,h} = {\hat U}(\phi_2) \Psi_{B,f,h}$, it must be the case that $\phi_1= \phi_2$.
This implies that the representation of diffeomorphisms on the finite span of basis off shell states is also {\em faithful}.

\subsection{\label{sec6.2} Diffeomorphism Covariance on an off shell basis state}

Diffeomorphism covariance of the single Hamiltonian constraint action is the statement that:
\be 
{\hat U}^{\dagger}(\phi){\hat H}(N) {\hat U}(\phi)  = H(\phi_*N) , \forall \phi
\label{diffcov1}
\ee
where $\phi$ is a semianalytic $C^r$ diffeomorphism. The dual action amplitude of the left hand side operator on an off shell basis state $\Psi_{B,f,h}$ is
${\hat U}^{\dagger}(\phi){\hat H}(N) {\hat U}(\phi) \Psi_{B,f,h} (S)$. We have that:
\ba
{\hat U}^{\dagger}(\phi){\hat H}(N) {\hat U}(\phi) \Psi_{B,f,h} (S)
&=&   {\hat U}^{\dagger}(\phi){\hat H}(N)  \Psi_{B, \phi^*f, \phi^*h} (S)\nonumber\\
&=:&  \Psi_{B, \phi^*f, \phi^*h} ({\hat H}(N) {\hat U}(\phi)S)\nonumber \\
&:=& 
\lim_{\e\rightarrow 0} 
\Psi_{B,\phi^*f, \phi^*h} ( {\hat H}_{\e}(N) {\hat U}(\phi)    S).
\ea
Here we have used (\ref{uphipsi}) in the first line. In the second line we have used the definition of dual action. In doing so we have used  the flexibility, given its classical reality,  of calling 
the  Hamiltonian constraint in the first line ${\hat H}(N)$ instead of ${\hat H}^{\dagger}(N)$. In the third line we have used the definition of the continuum limit action. Hence the 
diffeomorphism covariance condition can be written in terms of amplitudes as:
\be
\Psi_{B, \phi^*f, \phi^*h} ({\hat H}(N) {\hat U}(\phi)S) = \Psi_{B,f,h} ( H(\phi_*N) S)\;\;\; \forall \;\;S.
\label{diffcov1s}
\ee
From section \ref{sec4.1}  both the left hand side and the right hand side of the above equation are known. Below, we show their equality.
As noted above, if $S$ has no overlap in $B$, neither does ${\hat U}(\phi) S$ and (\ref{first0}) implies that 
\be
\Psi_{B, \phi^*f, \phi^*h} ({\hat H}(N) {\hat U}(\phi)S) = \Psi_{B,f,h} ( H(\phi_*N) S)= 0 \;\;\; \forall \;\;S\;\;{\rm with}\;{\rm no}\;{\rm overlap}\;{\rm in}\; B.
\label{diffcov1s0}
\ee
On the other hand, 
as noted above (see the discussion around (\ref{may5.0}), (\ref{may5.1}))
%from (ii), section \ref{sec3.2},
if  $S$ has  overlap in $B$ so does ${\hat U}(\phi) S$ and we may use (\ref{first1}) to evaluate both sides of (\ref{diffcov1s}).
Below, we denote  ${\hat U}(\phi) S$ by $S_{\phi}$ for notational convenience. In what follows, it is important to keep track of the metric dependence of various quantities. Accordingly, 
we  restore the metric subscript to the interkink distance function so as to denote  the interkink distance function
for the state $S$ with respect to metric $h$ by $g_{S,h}$. 

We also append a subscript $h$ to the unit edge tangent ${\hat e}_{I_v}^{a}$ with respect to $h$ so as to denote it by ${\hat e}_{I_v, h}^{a}$.
We also denote the RNCs $\{x\}$ with respect to the metric $h$ by $\{x\}_h$ and the lapse computed at point $v$ with respect to the RNCs $\{x\}_h$ centered at  $v$ by $N(x_h(v))$.
Then for any $S$ with overlap in $B$, equation F (\ref{first1}) implies that 
the left hand side of (\ref{diffcov1s}) is:
\ba
&&\Psi_{B, \phi^*f, \phi^*h} ({\hat H}(N) {\hat U}(\phi)S)
\nonumber\\
&=& \frac{3}{8\pi}\sum_{v\in V_{GR}(S_{\phi}))}N(x_{\phi^*h}(v)) g_{S_{\phi},\phi^*h }
 \gamma_{(S_{\phi})_{\lambda_v}}(\prod_{v^{\prime}\neq v} (\phi^*f)(v^{\prime}))(\sum_{I_v=1}^{N_v}   \frac{1}{4}{\hat e}_{I_v,\phi^*h}^{a} \partial_{a} (\phi^*f)(v)) \\
&=&
\frac{3}{8\pi}\sum_{\bv\in V_{GR}(S))}N(x_{\phi^*h}(\phi(\bv))) g_{S_{\phi},\phi^*h }
 \gamma_{(S_{\phi})_{\lambda_{\phi(\bv)}}}(\prod_{\bv^{\prime}\neq \bv} (\phi^*f)(\phi(\bv^{\prime})))(\sum_{I_{\phi(\bv)}=1}^{N_{\phi(\bv)}}   \frac{1}{4}{\hat e}_{I_{\phi (\bv)},\phi^*h}^{a} \partial_{a} (\phi^*f)(\phi(\bv)).\;\;\;\;\;\;\;\;\;
\label{62.0}
 \ea
where we have used the invariance under diffeomorphisms of the nondegeneracy of vertices to replace $V_{GR}(S_{\phi})$ by $V_{GR}(S)$ in the third line and where $v^{\prime},{\bv}^{\prime}$ range over $V(S_{\phi}), V(S)$ respectively (recall that $V_{GR}(S)$ denotes  the non-degenerate, (and in the case of overlap in $B$, exclusively) GR vertices of $S$ whereas $V(S)$ refers to all (and in the case of overlap in $B$, exclusively GR,) vertices of $S$).
 
Diffeomorphism covariance of the RS Volume operator implies that  $(S_{\phi})_{\lambda_{\phi(\bv)}}$ is the image of $S_{\lambda_{\bv}}$ by $\phi$, so that:
\be
\sum_{I_{\phi(\bv)}=1}^{N_{\phi(\bv)}}   \frac{1}{4}{\hat e}_{I_{\phi (\bv)},\phi^*h}^{a} \partial_{a} (\phi^*f)(\phi(\bv))
= 
\sum_{I_{\bv}=1}^{N_{\bv}}   \frac{1}{4}{\hat e}_{I_{\bv},h}^{a} \partial_{a} f(\bv).
\label{62.1}
\ee
It also follows that the elements of the kink set of $S_{\phi}$ are diffeomorphic images by $\phi$ of elements of the kink set of $S$. It immediately follows from the definition of the interkink distance function $g_{S,k}$ 
in section \ref{sec3.4.1} that: 
\be
 g_{S_{\phi},\phi^*h } =  g_{S,h }.
\label{62.2}
\ee
%Diffeomorphism invariance of the RS volume operator together with (ii), \ref{vib}, section \ref{sec3.1} implies that  
As noted above, $(S_{\phi})_{\lambda_{\phi(\bv)}}$ is the image of $S_{\lambda_{\bv}}$ by $\phi$. Equation 
(\ref{may5.1}) then implies that:
\be
\gamma_{ (S_{\phi})_{\lambda_{\phi (\bv)}} } = \gamma_{  S_{ \lambda_{\bv} }   }.
\label{62.3}
\ee
Next, from the definition of RNCs it follows that the RNCs $\{x_h\}$ with respect to the metric $h$ around the point $v$ and the RNCs $\{x_{\phi^*h}\}$ with respect to the push forward of $h$ around the point $\phi (v)$
are related, in general, by a constant (i.e. $p$ independent) rigid rotation  so that for $p$ in a convex normal neighborhood of $v$:
\be
x^{\mu}_{\phi^*h} (\phi (p)) = R^{\mu}{}_{\nu} x^{\nu}_h (p).
\label{phirnc}
\ee
Denoting the evaluation of a lapse $N$ at $p$ in coordinates $\{x_h\}$ by $N(x_h(p))$,
\footnote{\label{fnnrncp}This notation looks similar but is different to that utilised in section \ref{sec4.2.2} where the $v$ subscript on $x$ referred to the point around which the RNCs $\{x\}$ is centered. In contrast, 
here the subscript $h$ refers to the metric with respect to which the RNCs are defined with the center point not being explicitly denoted but, rather, implicitly understood from context.}
it follows from (\ref{phirnc}) together with the fact that $\det R= 1$ that:
%\be
%M( x_{\phi^*h}( e_{I_{\phi(\bv)}}( t ))   )= \phi_*M (  x_h(e_{I_v}(t))).
%\label{lapsephit}
%\ee
%Finally, using the fact that  the pull back and push forward actions are inverses of each other, the RNCs $ \{x\}_{\phi^*h}$ around $\phi(\bv)$ pull back to $\phi_*(\{x\}_{\phi^*h})  = \{x\}_h$ 
%around $\bv$ so that:
%
\be
N(x_{\phi^*h}(\phi(\bv))) = (\phi_*N)(x_h(\bv)).
\label{62.4}
\ee
Using (\ref{62.1})-(\ref{62.4}), together with the definition of the push forward of $f$ whereby $(\phi^*f)(\phi(\bv^{\prime}))= f(\bv^{\prime})$,   in (\ref{62.0}) implies:
\ba
&&\Psi_{B, \phi^*f, \phi^*h} ({\hat H}(N) {\hat U}(\phi)S)
\nonumber\\
&=&\frac{3}{8\pi}\sum_{\bv\in V(S))}(\phi_*N)(x_h(\bv))g_{S,h }
\gamma_{S_{\lambda_{\bv}}}
(\prod_{\bv^{\prime}\neq \bv} f(\bv^{\prime}))
\sum_{I_{\bv}=1}^{N_{\bv}}   \frac{1}{4}{\hat e}_{I_{\bv},h}^{a} \partial_{a} f(\bv)\;\;\;\;\;\;\;\;\;
\label{62.5}
 \ea
It is immediate to see that the right hand side of the above equation is exactly $\Psi_{B,f,h} ( H(\phi_*N) S)$ which, together with (\ref{diffcov1s0}) completes the 
demonstration of (\ref{diffcov1s}).

Next we turn to a proof of diffeomorphism invariance of the product of two Hamiltonian constraints. Since the proof is along the same lines as for the single Hamiltonian constraint action, we shall be brief. We need to show that:
\be 
{\hat U}^{\dagger}(\phi){\hat H}(N){\hat H}(M) {\hat U}(\phi)  = H(\phi_*N)H(\phi_*M) , \forall \phi
\label{diffcov2}
\ee
where $\phi$ is a semianalytic $C^r$ diffeomorphism. The amplitude on $S$ for  the (dual) action of the left hand side operator on an off shell basis state $\Psi_{B,f,h}$ is
${\hat U}^{\dagger}(\phi){\hat H}(N) {\hat H}(M) 
{\hat U}(\phi) \Psi_{B,f,h} (S)$. From (\ref{uphipsi}),  the definition of dual action and the classical reality driven flexibility of choice between ${\hat H}(L)$ and ${\hat H}^{\dagger}(L)$ for $L=N,M$, we have that:
\ba
{\hat U}^{\dagger}(\phi){\hat H}(N){\hat H}(M)  {\hat U}(\phi) \Psi_{B,f,h} (S)
&=&   {\hat U}^{\dagger}(\phi){\hat H}(N){\hat H}(M)   \Psi_{B, \phi^*f, \phi^*h} (S)\nonumber\\
&=:&  \Psi_{B, \phi^*f, \phi^*h} ({\hat H}(M) {\hat H}(N) {\hat U}(\phi)S)\nonumber \\
&:=& 
\lim_{\e\rightarrow 0} \left(\lim_{{\bar \e}\rightarrow 0} 
\Psi_{B,\phi^*f, \phi^*h} ({\hat H}_{\bar{\e}}(M)  {\hat H}_{\e}(N) {\hat U}(\phi)    S)\right).
\ea
%Here we have used (\ref{uphipsi}) in the first line. In the second line we have used the definition of dual action. 
%In doing so we have used  the flexibility, given its classical reality,  of calling 
%the  Hamiltonian constraint in the first line ${\hat H}(N)$ instead of ${\hat H}^{\dagger}(N)$. In the third line we have used the definition of the continuum limit action. 
Hence the 
diffeomorphism covariance condition can be written in terms of amplitudes as:
\be
\Psi_{B, \phi^*f, \phi^*h} ({\hat H}(M){\hat H}(N) {\hat U}(\phi)S) = \Psi_{B,f,h} (H(\phi_*M) H(\phi_*N) S)\;\;\; \forall \;\;S.
\label{diffcov2s}
\ee
From section \ref{sec4.2}  both the left hand side and the right hand side of the above equation are known.
As above, if $S$ has no overlap in $B$, neither does ${\hat U}(\phi) S$ and section \ref{sec4.2.1}  implies that 
\be
\Psi_{B, \phi^*f, \phi^*h} ( {\hat H}(M)  {\hat H}(N) {\hat U}(\phi)S) = \Psi_{B,f,h} (H(\phi_*M)  H(\phi_*N) S)= 0 \;\;\; \forall \;\;S\;\;{\rm with}\;{\rm no}\;{\rm overlap}\;{\rm in}\; B.
\label{diffcov2s0}
\ee
Let $S$, and hence $S_{\phi}$,  have overlap in  $B$.  In the notation developed above, equation (\ref{hmhnfinal}) implies that:
\ba
&&\Psi_{B, \phi^*f, \phi^*h} ({\hat H}(M){\hat H}(N) {\hat U}(\phi)S)\nonumber\\
&=& (\frac{3}{8\pi})^2\sum_{v\in V_{GR}(S_{\phi})}N(x_{\phi^*h}(v))  \frac{1}{4}g_{S{\phi}, \phi^*h}
\nonumber\\
&& 
\Big(\;\;\sum_{v^{\prime}\in V_{GR}(S_{\phi}), v^{\prime}\neq v}    M(x_{\phi^*h}(v^{\prime})) 
 \gamma_{ ((S_{\phi})_{\lambda_v} )_{\lambda_{v^{\prime}}}}
 (\prod_{v^{\prime\prime}\in V(S_{\phi}), v^{\prime\prime}\neq v^{\prime}, v} (\phi^*f)(v^{\prime\prime})) \nonumber \\
&&
(\sum_{I_v=1}^{N_v} \frac{1}{4}{\hat e}_{I_{v},\phi^*h}^{a} \partial_{a} (\phi^*f)(v))
 (\sum_{I_{v^{\prime}}=1}^{N_{v^{\prime}}}   \frac{1}{4}{\hat e}_{I_{v^{\prime}},\phi^*h}^{a} \partial_{a} (\phi^*f)(v^{\prime}))
%\;\;\;;\;\;\;\;\;\;\;\;\;\;\;\;\;\;\;;\;\;\;\;\;\;\;\;\;\;\;\;\;
 \nonumber \\
&+&
 \gamma_{ ((S_{\phi})_{\lambda_v} )_{\lambda_{v}}}
 (\prod_{v^{\prime}\in V(S_{\phi}),v^{\prime}\neq v} (\phi^*f)(v^{\prime})) \frac{1}{4}N_v {\hat e}_{I_v,\phi^*h}^{b}  \partial_{b}(M(x_{\phi^*h}(v)) {\hat e}_{I_v,\phi^*h}^{a}  \partial_{a} (\phi^*f)(v))\;\;\Big) 
%\;\;\;;\;\;\;\;\;\;\;\;\;\;\;\;\;
 \nonumber 
\\
&=& (\frac{3}{8\pi})^2\sum_{\bv\in V_{GR}(S)}N(x_{\phi^*h}(\phi(\bv)))  \frac{1}{4}g_{S_{\phi}, \phi^*h}
\nonumber\\
& 
\Big(&\sum_{\bv^{\prime}\in V_{GR}(S),\bv^{\prime}\neq \bv}    M(x_{\phi^*h}(   \phi(\bv^{\prime})  )) 
 \gamma_{ ((S_{\phi})_{\lambda_{\phi(\bv)}  } )_{\lambda_{  \phi( \bv^{\prime})   }}}
 (\prod_{\bv^{\prime\prime}\in V(S),\bv^{\prime\prime}\neq \bv^{\prime}, \bv} (\phi^*f)(\phi(\bv^{\prime\prime}))) 
\nonumber\\
&& (\sum_{I_{\phi(\bv)}   =1}^{N_{\phi(\bv)}     } \frac{1}{4}{\hat e}_{I_{\phi(\bv)},\phi^*h   }^{a} \partial_{a} (\phi^*f)(\phi(\bv)))
 (\sum_{I_{   \phi(\bv^{\prime})  }=1}^{N_{   \phi(\bv^{\prime})    }}   \frac{1}{4}{\hat e}_{I_{  \phi( \bv^{\prime}), \phi^*h    }}^{a} \partial_{a} (\phi^*f)(\phi(\bv^{\prime})  ))
%\;\;\;\;\;\;\;\;\;\;\;\;\;\;\;\;\;\;\;\;\;\;\;\;\;\;\;\;\;\;\; \;\;\;\;\;\;\;\;\;\;\;\;\;\;\;\;\;\;\;\;\;\;\;\;\;\;\;\;\;\;\;\;\;\;\;\;\;\;\;\;\;\;\;\;\;\;\;\;\;\;\;\;\;\;\;\;\;\;\;\;\;\;
\nonumber \\
&+&
 \gamma_{ ((S_{\phi})_{\lambda_{\phi(\bv)}} )_{\lambda_{  \phi(\bv)   }}}
 (\prod_{ \bv^{\prime}\in V(S), \bv^{\prime}\neq \bv} (\phi^*f)(   \phi(\bv^{\prime} )  )) \frac{1}{4}N_{\phi (\bv)} {\hat e}_{I_{\phi(\bv)}, \phi^*h   }^{b}  \partial_{b}(M(x_{\phi^*h}( \phi(\bv)  ))
 {\hat e}_{I_{\phi(\bv)}, \phi^*h  }^{a}  \partial_{a} (\phi^*f)(\phi(\bv)))\;\;\Big). \;\;\;\;\;\;\;\;\;\;\;
\label{62.6} 
 \ea 
%Using the diffeomorphism invariance of the RS volume operator in conjunction with (ii),\ref{vib},section \ref{sec3.1}  to simplify the $\gamma$ factors above, 
Diffeomorphism covariance of the RS volume operator implies that 
$((S_{\phi})_{\lambda_{\phi(\bv)}  } )_{\lambda_{  \phi( \bv^{\prime})   }}$ is the image by the diffeomorphism 
$\phi$ of $(S_{\lambda_{\bv}  } )_{\lambda_{   \bv^{\prime}   }}$. Using this fact  with (\ref{may5.1})
to simply the gamma factors above, 
together with (\ref{62.1}), (\ref{62.2}) and (\ref{62.4}), 
it follows that (\ref{62.6}) simplifies to:
\ba
&&\Psi_{B, \phi^*f, \phi^*h} ({\hat H}(M){\hat H}(N) {\hat U}(\phi)S)\nonumber\\
&=& (\frac{3}{8\pi})^2\sum_{\bv\in V(S)}(\phi_*N)(x_{h}(\bv))  \frac{1}{4}g_{S, h}
\nonumber\\
&& 
\Big(\;\;\sum_{\bv^{\prime}\neq \bv}    (\phi_*M)(x_{h}( \bv^{\prime}  )) 
 \gamma_{ (S_{\lambda_{\bv}  } )_{\lambda_{   \bv^{\prime}   }}}
 (\prod_{\bv^{\prime\prime}\neq \bv^{\prime}, \bv} f(\bv^{\prime\prime})) 
 (\sum_{I_{\bv}   =1}^{N_{\bv}     } \frac{1}{4}{\hat e}_{I_{\bv} ,h  }^{a} \partial_{a} f(\bv))
 (\sum_{I_{   \bv^{\prime}  }=1}^{N_{   \bv^{\prime}   }}   \frac{1}{4}{\hat e}_{I_{   \bv^{\prime},h  }}^{a} \partial_{a} f(\bv^{\prime})  )\nonumber \\
&+&
 \gamma_{ (S_{\lambda_{\bv}} )_{\lambda_{  \bv  }}}
 (\prod_{  \bv^{\prime}\neq \bv} f(\bv^{\prime} )  ) \frac{1}{4}N_{\phi (\bv)} 
 {\hat e}_{I_{\phi(\bv)}, \phi^*h    }^{b}  \partial_{b}(M(x_{\phi^*h}( \phi(\bv)  )) {\hat e}_{I_{\phi(\bv)}, \phi^*h  }^{a}  \partial_{a} (\phi^*f)(\phi(\bv)))\;\Big). \;\;\;
\label{62.7} 
 \ea 
Next, note that  any parameterization $e_{I_{\bv}}(t)$ for the $I_{\bv}$th edge $e_{I_{\bv}}$ emanating from  $\bv$ in $S$, 
defines a natural parameterization by the same parameter $t$  for its diffeomorphic image $e_{ I_{\phi(\bv)}} := \phi (e_{I_{\bv}})$  emanating from  $\phi(\bv)$ in $S_{\phi}$ through
\be
e_{I_{\phi(\bv)}} (t): = \phi (e_{I_{\bv}}(t)).
\label{natpar}
\ee
In this parameterization the following results hold. 

First, from the defining  properties of push forward by $\phi$, we have that 
\be
{\dot e}^a_{ I_{\phi({\bv})  }}(t)= \phi^*{\dot e}^a_{I_{\bv}}(t)
\label{dotephi}
\ee
where similar to  (\ref{A2.43a}), the  tangent vector to an edge $e(t) $  with respect to its parameter $t$ is denoted by ${\dot e}^a (t)$,  
It then follows, again from the propeprties of the pushforward operation that:
\be
{\dot e}^a_{ I_{\phi({\bv})  }}(t){\dot e}^b_{ I_{\phi({\bv})  }}(t) \phi^*h_{ab}(    e_{I_{\phi(\bv)}} (t))
= \phi^*{\dot e}^a_{I_{\bv}}(t)\phi^*{\dot e}^b_{I_{\bv}}(t)\phi^*h_{ab}(   \phi( e_{I_{\bv}} (t)) )
=  {\dot e}^a_{I_{\bv}}(t){\dot e}^b_{I_{\bv}}(t)h_{ab}(  e_{I_{\bv}} (t)) 
\label{norm2phie}
\ee
from which the equality of the metric norms of these edge tangents follow i.e.
\be 
|{\vec{\dot e}}_{ I_{\phi({\bv})  }}(t)|_{\phi^*h}= |{\vec {\dot e}}_{I_{\bv}}(t)|_h
\label{normphie}
\ee
where the metric subscripts indicate which metric is being used to calculate the norm.
Second, reiterating the discussion around (\ref{phirnc}), 
from the definition of RNCs it follows that the RNCs $\{x_h\}$ with respect to the metric $h$ around the point $v$ and the RNCs $\{x_{\phi^*h}\}$ with respect to the push forward of $h$ aroud the point $\phi (v)$
are related, in general, by a constant rigid rotation so that for $p$ in a convex normal neighborhood of $v$.
%\be
%x^{\mu}_{\phi^*h} (\phi (p)) = R^{\mu}{}_{\nu} x^{\nu}_h (p).
%\label{phirnc1}
%\ee
Similar to (\ref{62.4}), denoting the evaluation of a lapse $M$ at $p$ in coordinates $\{x_h\}$ by $M(x_h(p))$,
%\footnote{This notation looks similar but is different to that utilised in section \ref{sec4.2.2} where the $v$ subscript on $x$ referred to the point around which the RNCs $\{x\}$ is centered. In contrast, 
%here the subscript $h$ refers to the metric with respect to which the RNCs are defined with the center point not being explicitly denoted but, rather, implicitly understood from context.}
it then follows from (\ref{phirnc}) together with the fact that $\det R= 1$ therein, that:
\be
M( x_{\phi^*h}( e_{I_{\phi(\bv)}}( t ))   )= \phi_*M (  x_h(e_{I_v}(t))).
\label{lapsephit}
\ee
Third, let $F: e_{I_v}\rightarrow {\bf C}$ be a function of points on the $I_v$th edge at $v$.  Defining its pushforward $F_{\phi}: e_{I_{\phi(\bv)}}= \phi (e_{I_v})\rightarrow {\bf C}$ by 
$F_{\phi}(e_{I_{\phi(\bv)}}(t)): =F (e_{I_v}(t))$, it immediately follows that:
\be
\frac{dF_{\phi}(e_{I_{\phi(\bv)}}(t))}{dt}= \frac{dF (e_{I_v}(t))}{dt}.
\label{dfphidt}
\ee

Next, recalling the notation (\ref{A2.43a}) employed in section \ref{sec4.2.2}, we have that:
\ba
&&{\hat e}_{I_{\phi(\bv)}, \phi^*h    }^{b}  \partial_{b}(M(x_{\phi^*h}( \phi(\bv)  )) {\hat e}_{I_{\phi(\bv)}, \phi^*h  }^{a}  \partial_{a} (\phi^*f)(\phi(\bv)))\nonumber\\
&:=& \frac{1}{ |{\vec{{\dot e}}}_{I_{\phi({\bv})  }}(t)|_{\phi^*h}}\frac{d}{dt} \frac{   M( x_{\phi^*h}( e_{I_{\phi(\bv)}}( t ))    )        }{|{\vec{{\dot e}}}_{I_{\phi(\bv)}  }(t)|_{\phi^*h}}
\frac{d(\phi^*f)( e_{I_{\phi(\bv)}} (t)          )}{dt}|_{t=0}
\\
%&=& 
% \frac{1}{|{\vec{{\dot e}}}_{I_{\phi({\bv}) } }(t_{\phi})|_{\phi^*h}}\frac{d}{dt_{\phi}} (\frac{M(t_{\phi})}{|{\vec{{\dot e}}}_{I_{\phi(\bv)}  }(t_{\phi})|_{\phi^*h}}\frac{d(\phi^*f)(t_{\phi})}{dt_{\phi}}|_{t_{\phi}=0}\\
&=&
 \frac{1}{|{\vec{{\dot e}}}_{I_{{\bv} } }(t)|_{h}}\frac{d}{dt} \frac{\phi_*M( x_h(e_{I_v}(t))      )       }{|{\vec{{\dot e}}}_{I_{\bv}  }(t)|_{h}}\frac{df(   e_{I_{\bv}}(t)          )}{dt}|_{t=0}\\
&=& 
{\hat e}_{I_{\bv}, h    }^{b}  \partial_{b}(\phi_*M(x_{h}( \bv  )) {\hat e}_{I_{\bv}, h  }^{a}  \partial_{a} f(\bv)).
\label{phimeqn}
\ea
%Above,  in the second line we have added the subscript $\phi^*h$ to the norm to indicate that the norm is taken with respect to the metric $\phi^*h$.
In the third line we used (\ref{normphie}),  (\ref{lapsephit}) and  (\ref{dfphidt}). In the fourth line we have  again used  notation of the type (\ref{A2.43a}) to revert to the  parameterization independent 
form (\ref{phimeqn}). Using (\ref{phimeqn}) in 
(\ref{62.7})  together with the fact that vertex valence is invariant under diffeomorphisms (so that $N_{\phi(\bv)}= N_{\bv}$)
\footnote{Please note that although the same letter $N$ is involved in the notation of one of the lapses as well as vertex valence, the latter is distinguished from the former by the presence of the vertex
subscript to $N$}
yields:
\ba
&&\Psi_{B, \phi^*f, \phi^*h} ({\hat H}(M){\hat H}(N) {\hat U}(\phi)S)\nonumber\\
&=& (\frac{3}{8\pi})^2\sum_{\bv\in V_{GR}(S)}(\phi_*N)(x_{h}(\bv))  \frac{1}{4}g_{S, h}
\nonumber\\
&& 
\Big(\;\;\sum_{\bv^{\prime}\in V_{GR}(S), \bv^{\prime}\neq \bv}    (\phi_*M)(x_{h}( \bv^{\prime}  )) 
 \gamma_{ (S_{\lambda_{\bv}  } )_{\lambda_{   \bv^{\prime}   }}}
 (\prod_{\bv^{\prime\prime}\in V(S),\bv^{\prime\prime}\neq \bv^{\prime}, \bv} f(\bv^{\prime\prime})) 
 (\sum_{I_{\bv}   =1}^{N_{\bv}     } \frac{1}{4}{\hat e}_{I_{\bv} ,h  }^{a} \partial_{a} f(\bv))
 (\sum_{I_{   \bv^{\prime}  }=1}^{N_{   \bv^{\prime}   }}   \frac{1}{4}{\hat e}_{I_{   \bv^{\prime},h  }}^{a} \partial_{a} f(\bv^{\prime})  )\nonumber \\
&+&
 \gamma_{ (S_{\lambda_{\bv}} )_{\lambda_{  \bv  }}}
 (\prod_{ \bv^{\prime}\in V(S) \bv^{\prime}\neq \bv} f(\bv^{\prime} )  ) \frac{1}{4}N_{\bv} 
{\hat e}_{I_{\bv}, h    }^{b}  \partial_{b}(\phi_*M(x_{h}( \bv  )) {\hat e}_{I_{\bv}, h  }^{a}  \partial_{a} f(\bv))\Big).
\label{finalphimeqn}
\ea
It is straightforward to verify, from (\ref{hmhnfinal}) that the right hand side of (\ref{finalphimeqn}) is exactly 
$\Psi_{B,f,h} (H(\phi_*M) H(\phi_*N) S)$. 
This completes our proof of (\ref{diffcov2s}).

\subsection{\label{sec6.3} Diffeomorphism Covariance on finite span of off shell basis states}

The exposition of the previous two sections highlights the fact that the key ingredient which ensures diffeomorphism covariance is the metric label dependence of the 
regulating coordinate structure in the constraint action.
Thus, given an off shell state with metric label $h$, the regulation prescription is to use,  as regulating coordinates, the  RNCs associated with $h$.
In the case of a finite superposition of such off shell states with  distinct metric labels, the prescription is well defined iff any such finite superposition 
is {\em uniquely} decomposable into its constituent basis states so that for the constraint action on each basis state,  we use the RNCs associated with its metric label and then superpose the results.
 
If this is not true,  there exists an element in the finite span which is  expressible as two distinct superpositions of basis states,
Since the prescription underlying the constraint action  depends on the metric labels of the constituent basis elements, the constraint action is then  no longer uniquely defined.
To illustrate what we mean let us suppose, just for the sake of argument,  that the  same element $\Psi$ in the finite span  could be written in two different ways as:
\be
\Psi = a_1\Psi_{B,f,h_1} + a_2\Psi_{B,f ,h_2} = b_1 \Psi_{B,f ,h_1} + b_2 \Psi_{B,f,h_2}
\ee
with $a_i\neq b_i, i=1,2$ and $h_1\neq h_2$. Since the constraint action on each basis state is different the amplitudes $A_i(N,S) = \Psi_{B,f,h_i}({\hat H(N)}S)$ would be different for $i=1$ and $i=2$.
The amplitude $\Psi ({\hat H(N)}S)$ would then evaluate to  $a_1A_1 + a_2 A_2$ when using the first superposition and $b_1A_1+ b_2A_2$ when using the second and there is no reason to expect that these
two combinations coincide. Consequently a weak formulation of diffeomorphism covariant anomaly free action results which applies to each specific basis expansion representation of the state $\Psi$
separately. In particular, anomaly freedom holds in the sense of the equality of $[{\hat H}(M, {\hat H}((N)]$ and $i\widehat{\{H(M),H(N)\}}$ on identical choices of basis expansion of $\Psi$
and a similarly weak statement holds for diffeomorphism covariance so the left and right hand sides of   (\ref{diffcov1}), (\ref{diffcov2}) are equal when evaluated on identical choices of basis expansion of $\Psi$.
\footnote{In the absence of a proof of uniqueness of decomposition in \cite{p3}, a similarly weak statement (see section 13\cite{p3} for its articulation) holds for the results derived in that work.}

On the other hand if the uniqueness of decomposition is true, the prescription is well defined (in the context of the simple example above this would imply  that $a_i=b_i, i=1,2$).
In section \ref{secl} we prove certain properties of linear independence of off shell states which ensure the required uniqueness of decomposition so as to render the metric label dependent 
regularization prescription for the constraint action well defined.  The proof requires that the Bra Set labels satisfy an additional property (iv) beyond properties (i)-(iii) of section \ref{sec3.2}.
%In section \ref{secl} we show how to augment any Bra Set satisfying (i)-(iii) with an appropriate set of elements so that the augmented Bra Set satisfies properties (i)-(iv).
%It follows that with these augmented Bra Set labels, the results of sections \ref{sec6.1}- \ref{sec6.2} showing diffeomorphism covariance of the single and product constraint actions
%can be extended to the finite span of basis off shell states. While the augmented Bra Sets explicitly satisfy property (iv) by virtue of the augmentation, it seems plausible to us that this 
%augmentation may be unnecessary and 
While it seems plausible to us that property (iv) may follow directly from properties (i)-(iii) a putative proof is beyond the scope of this work. Instead, as discussed in the last paragaraph of  section \ref{sec3.2}
we shall provide an explicit example of a rich family of Bra Sets which satisfy properties (i)- (iv).
These matters are discussed further  in section \ref{secl}.

Finally, we note that we have phrased our results in section \ref{sec4.2} and in  section \ref{sec6.2}  for the constraint operator  product,   in terms of a definition of this product as 
%in terms of the continuum limit of the single  approximant action and and
 the continuum limit of the product of single approximant actions  on the finite span of basis off shell states. 
In this definition the action of the {\em product} of approximants is evaluated {\em first} and  the {\em continuum limit} is taken {\em second}.
Instead, these results can also be phrased in terms of a definition of this product as that of two continuum limit single constraint actions each on a suitable domain of states in the algebraic dual. 
In this definition, one can think of the {\em continuum limit}  being taken first and the {\em product} second.
%These results can also be phrased in terms of the the product of two such continuum limit actions on suitable domains of states in the algebraic dual. 
In section \ref{secl} these results are rephrased, 
%under the same technical assumptions for which  properties of linear independence hold,
as precise statements of anomaly free Hamiltonian 
%in terms of  continuum limit dual operator action on  domains in the algebraic dual space.
%diffeomorphism covariant 
constraint commutators in terms of  continuum limit dual operator action on  domains in the algebraic dual space.
%are formulated,  discussed
%and proved 

%in section \ref{secl}.

The next section, section \ref{secl}, is quite technical  and may  be skipped  on a first reading

\section{\label{secl}Linear Independence of Off Shell States and Anomaly Free Domains}
In section \ref{secl1} we consider a set ${\bf K}$ consisting of 3$m$ points in the vicinity of a point $p \in \Sigma$. 
%We fix RNCs centered at $p$ with respect to a metric $h$. 
The elements of ${\bf K}$ are
close enough to $p$ that ${\bf K}$ is contained in a single 
%semianalytic 
coordinate patch $\{x\}$ around $p$ with $\{x\}$ being some fixed coordinates which are not necessarily RNCs.
%in a convex normal neighborhood of $p$. 
The elements of ${\bf K}$  are  placed in such a manner as to afford ${\bf K}$ a segregation structure
in terms of  $m$ 3 point subsets, $K_3^{(i)}$ as described in section \ref{sec3.3} except that the distance used here is the coordinate distance in contrast to the geodesic distance used in section \ref{sec3.3}. 
The detailed placement of points is dictated by the choice of 2 tangent vectors $u,v$ at $p$. 
In section \ref{secl2}  we show that the coordinate distance based segregation structure  of section \ref{secl1} implies the same segregation structure in terms of geodesic distances.
We also  determine $g({\bf K}, h)$ as defined in section \ref{sec3.3}. In section \ref{secl3} we generalise the construction of ${\bf K}$ with the `single nested structure' around $p$ displayed in 
section \ref{secl1.0} to that with multiple nests and determine $g({\bf K}, h)$  for this multinested set ${\bf K}$.
In section \ref{secl4} we prove a key Lemma. In section \ref{secl5} we use  the results of sections \ref{secl1.0} to \ref{secl4} to construct the desired  proof of linear independence for the case of states
with a fixed Bra Set label $B$ subject to an additional technical property (property (iv) of section \ref{secl5}). 
In section \ref{secl6} we remove the restriction of fixed Bra Set label.
We show how the obtained linear independence ensures uniqueness of decomposability
of states in the finite span of off shell basis states and rephrase the property of 
%diffeomorphism covariant
anomaly free action proved in sections \ref{sec4} and \ref{sec5} in terms of continuum limit operator actions on domains.
In section \ref{secl61} we extend the results of section \ref{secl6} to the vector space  sum of the finite span of offshell basis states and the space of physical (`onshell') states.
In section \ref{secl7}  we discuss how  property (iv) could  be satisfied.
%In section \ref{secl6} weLet the valen it presents itself as a bivalent kink in such 
%comment on possible generalizations of the proof of linear independence under appropriate technical assumptions.
%Finally in section \ref{secl7} we rephrase the property of diffeomorphism covariant anomaly free action proved hitherto in terms of continuum limit operator actions on domains.

\subsection{\label{secl1.0}Single Nest} 

%with respect to  Coordinate Distances}

Fix a point $p\in \Sigma$ and a  coordinate patch $\{x\}$ in a neighborhood $U_p$ of $p$.
%Let the Riemann Normal Coordinates  centered at $p$ and defined with respect to a metric $h$  be $\{x\}_{h,p}$
Let $q \in U_p$ 
%be a point in a Convex Normal Neighborhood of $p$. 
Denote the coordinates of $q$ in $\{x\}$ by $x^{\mu}(q), \mu=1,2,3$.
In what follows we shall omit the subscripts $p,h$ whenever convenient to reduce notational clutter.
Let $u,v$ be a  pair of (non-trivial) tangent vectors at $p$ with coordinates $u^{\mu}, v^{\mu}$.

We proceed as follows.
First we specify the positions of the $3m$ elements of ${\bf K}\equiv {\bf K}_{m, p,\e,\delta} $ relative to $p$ in terms of $\{x\}, u^{\mu}, v^{\mu}$ and the parameters $\e, \delta$.
Here $\delta$ is independent of $\e$ and subject only to the condition:
\be
1-m \delta >0 .
\label{delta}
\ee
Next, we show that in terms of coordinate distances, the placement of points endows ${\bf K}$ with the desired nested 
structure. 
%We prove a lemma correlating (small enough) coordinate distances with 
%geodesic distances $d$. 
Finally we 
%use the lemma to 
show that coordinate nesting of ${\bf K}$ implies the desired nesting in terms of $d$.

\subsubsection{\label{secl1}Single Nest with respect to Coordinate Distances}

The $3m$ points are divided into triples. Elements of $j$th triple are denoted by $p^{(j)}_q, q=1,2,3$ and the set of this triple by $K^{(j)}_3$. 
\footnote{We have used the notation $K^{(j)}_3 \subset {\bf K}$  in section \ref{sec3.3}  for the set of  3 points with specific geodesic distance properties in relation to elements of ${\bf K}$. While we abuse this notation
here, in the next section we shall show that  $K^{(j)}_3 $ does indeed satisfy its defining properties detailed in section \ref{sec3.3}.}
With this notation in place, we specify the positions of the elements of ${\bf K}$ as follows:\\
\ba
x^{\mu}(p_1^{(1)}) - x^{\mu} (p) &=& O(\e^{2m(1+2\delta)}) \label{l1.1}\\
x^{\mu}(p_2^{(1)}) - x^{\mu} (p_1^{(1)}) &=&       \e^{m(1+\delta)} (u^{\mu} + O(\e^{m\delta})) \label{l1.2}\\
x^{\mu}(p_3^{(1)}) - x^{\mu} (p_1^{(1)}) &=&       \e^{m} (v^{\mu} + O(\e^{m\delta})) \label{l1.3}
\ea
From (\ref{l1.2}), (\ref{l1.3}) it is straightforward to show that:
\be
x^{\mu}(p_3^{(1)}) - x^{\mu} (p_2^{(1)}) =       \e^{m} (v^{\mu} + O(\e^{m\delta})) .\label{l1.4}\\
\ee
This places the points in the first triple. We specify the placement of elements of the remaining triples as follows.

Let $m\geq j \geq 2$. Then the placements satisfy:
\ba
x^{\mu}(p_q^{(j)}) - x^{\mu} (p^{(k)}_r) &=& \O(\e^{(1-\delta)(m-j+2)})\;\;j>k, k=1,..,m-1 \;\; q=1,2\;\; r=1,2,3 \label{l2.1}\\
%x^{\mu}(p_2^{(1)}) - x^{\mu} (p_1^{(1)}) &=&       \e^{m(1+\delta)} (u^{\mu} + O(\e^{m\delta})) \label{l1.2}\\
x^{\mu}(p^{(j)}_{3}) - x^{\mu} (p_r^{(k)}) &=&      \O( \e^{m-j+1}) \;\;j>k, k=1,..,m-1\;\;\;r=1,2,3 \label{l2.2}
\ea
and 
\ba
x^{\mu}(p_2^{(j)}) - x^{\mu} (p_1^{(j)}) &=&  \e^{(1-\frac{\delta}{2})(m-j+2)} (u^{\mu} + O(\e^{(1-\frac{m\delta}{2})})) \label{l2.3}\\
x^{\mu}(p_3^{(j)}) - x^{\mu} (p_1^{(j)}) &=&    \e^{m-j+1} (v^{\mu} +O(\e^{(1-\frac{m\delta}{2})})). \label{l2.4}
\ea
From (\ref{l2.3}), (\ref{l2.4}) it is straightforward to show that:
\be
x^{\mu}(p_3^{(j)}) - x^{\mu} (p_2^{(j)}) =      \e^{m-j+1} (v^{\mu} +O(\e^{(1-\frac{m\delta}{2})})). \label{l2.5}
\ee
It is a straightforward exercise  to verify that for all small enough $\e$ these placements endow ${\bf K}$ with a nested structure,  
in terms of {\em coordinate}  distances, of the type described in  section \ref{sec3.3} (we emphasize that in contrast to the coordinate distance used here, the nested structure in section \ref{sec3.3} is defined with respect
to geodesic distances). This structure is as follows. If $m=1$ we have a single triple of points so that ${\bf K}= K^{(1)}_3$. If $m>1$ then
for any $k<m$ the inter point coordinate distances in $K^{(k)}_3$ are smaller than any of the coordinate distances between any of these points and any point in ${\bf K}- \cup_{k^{\prime}=1}^{k} K^{(k^{\prime})}_3$
as well as the interpoint coordinate distances between elements of ${\bf K}- \cup_{k^{\prime}=1}^{k} K^{(k^{\prime})}_3$.

\subsubsection{\label{secl2}Single Nest with respect to  Geodesic Distances}

Fix a  point 
%$p_\e \in \Sigma$ 
$p_\e \in U_p$ 
and a tangent vector $v_\e^a$ at $p_\e$. Let the metric norm $|{\vec v}_\e|$ of $v_\e^a$ with respect to $h$ be of $\O(1)$.
Consider a point $q_\e \in U_p$ 
%such that $p_\e, q_\e$ are in a single coordinate patch 
with  $x^{\mu}(q_{\e}) - x^{\mu}(p_\e)= \e v_\e^{\mu}$ for small enough $\e$.  
Let the geodesic connecting them be parameterized with affine parameter $\lambda$ such that the geodesic tangent  $(\frac{d}{d\lambda})^a$ is of unit metric norm so $h_{\mu \nu} \frac{dx^{\mu}}{d\lambda}\frac{dx^{\nu}}{d\lambda}=1$.
Let $p_\e$ be located at $\lambda=0$ and $q_\e$ at $\lambda_\e$.
Using the fact that the $C^{r-1}, r>>1$ differentiability of the metric $h_{\mu \nu}$ implies sufficient differentiability of geodesics for Taylor expansion to first order in the geodesic affine parameter, we have that:
\be
x^{\mu}(q_\e) - x^{\mu}(p_\e) =  \e v_\e^{\mu}=  \lambda_\e \frac{dx^{\mu}}{d\lambda}|_{\lambda=0} + O(\lambda_\e^2)
\ee
which implies 
\be
|{\vec v}_\e(p_\e)| \e= \lambda_\e + O(\lambda_\e^2)
\ee
where $|{\vec v}_\e(p_\e)|$ is the metric norm of $v^a_\e$ at $p_\e$.
On the other hand by virtue of the choice of affine parameterization it immediately follows that the geodesic distance $d(q_\e, p_\e)$ equals the affine distance $\lambda_\e$.
This in conjunction with the above equation implies that:
\be 
d(q_\e, p_\e) = |{\vec v}_\e| \e + O(\e^2).
\label{geodiste}
\ee
In particular if for some $\eta >0$ and for some tangent vector $v^{a}\neq 0 $ at $p_\e$ with coordinate components $v^{\mu}$ independent of $\e$,  $v^{a}_\e \equiv v^{a}_{\e,\eta}$ at $p_\e$ is such that
\be
v^{a}_{\e,\eta} = v^{a} + O(\e^{\eta}),
\label{veta}
\ee
it follows from (\ref{geodiste}) that:
\be 
d(q_\e, p_\e) = |{\vec v}(p_\e)| \e + O(\e^2) + O(\e^{\eta+1}), \;\;\;|{\vec v}(p_\e)| := (h_{\mu \nu}(p_\e) v^{\mu}v^{\nu})^{\frac{1}{2}}.
\label{geodisten}
\ee
Finally, let $p_\e$ be located at a  coordinate distance of $O(\e^{\tau}), \tau>0$ from the fixed  point  $p\in U_p$ so that:
%independent location $p$ which is also in the same coordinate patch so that:
\be
x^{\mu}(p_\e)- x^{\mu}(p) = O(\e^{\tau}).
\ee
Then defining $v^a$ at every point in $U_p$  to have the $\e$ independent coordinates $v^{\mu}$ so that the norm of $v^a$ at $p$ is $|{\vec v}(p)|= (h_{\mu \nu}(p) v^{\mu}v^{\nu})^{\frac{1}{2}}$, we have via Taylor
expansion of $h$ around $p$ that:
\be
|{\vec v}(p_\e)| = |{\vec v}(p)| + O(\e^{\tau})
\ee
which together with (\ref{geodisten}) implies that:
\be 
d(q_\e, p_\e) = |{\vec v}(p)| \e + O(\e^2) + O(\e^{\eta + 1}) + O(\e^{\tau + 1 }) , \;\;\;|{\vec v}(p)| := (h_{\mu \nu}(p) v^{\mu}v^{\nu})^{\frac{1}{2}}.
\label{geodistent}
\ee
Using (\ref{geodistent}) in conjunction with (\ref{delta}) and equations (\ref{l1.1}) - (\ref{l2.5}), a tedious but straightforward analysis implies that the network of geodesic distances between elements of ${\bf K}$
satisfy the following equations.
\ba 
d(p^{(1)}_1, p) &=  & O(\e^{2m(1+\delta)}) \label{d1.1}\\
d(p^{(1)}_2, p^{(1)}_1) &=&    \e^{m(1+\delta)} (|u(p)|+ O(\e^{m\delta})) \label{d1.2}\\
d(p^{(1)}_3, p^{(1)}_1) &=&   \e^{m}(|v(p)|+ O(\e^{m\delta})). \label{d1.3}
\ea
From (\ref{d1.2}), (\ref{d1.3}) it is straightforward to show that:
\be
d(p^{(1)}_3, p^{(1)}_2) =  \e^{m}(|v(p)|+ O(\e^{m\delta})). \label{d1.4}
\ee
For  $m\geq j \geq 2$ we have that:
\ba
d(p_q^{(j)}, p^{(k)}_r) &=& \O(\e^{(1-\delta)(m-j+2)})\;\;j>k, k=1,..,m-1 \;\; q=1,2\; r=1,2,3 \label{d2.1}\\
%x^{\mu}(p_2^{(1)}) - x^{\mu} (p_1^{(1)}) &=&       \e^{m(1+\delta)} (u^{\mu} + O(\e^{m\delta})) \label{l1.2}\\
d(p_3^{(1)}, p_1^{(r)}) &=&      \O( \e^{m-j+1}) \;\;j>k, k=1,..,m-1\;\;r=1,2,3 \label{d2.2}
\ea
and 
\ba
d(p_2^{(j)}, p_1^{(j)}) &=&  \e^{(1-\frac{\delta}{2})(m-j+2)} (|u(p)| + O(\e^{(1-\frac{m\delta}{2})})) \label{d2.3}\\
d(p_3^{(j)}, p_1^{(j)}) &=&    \e^{m-j+1} (|v(p)| +O(\e^{(1-\frac{m\delta}{2})})) \label{d2.4}
\ea
From (\ref{d2.3}), (\ref{d2.4}) it is straightforward to show that:
\be
d(p_3^{(j)}, p_2^{(j)}) =      \e^{m-j+1} (|v(p)|+O(\e^{(1-\frac{m\delta}{2})})). \label{d2.5}
\ee
It is again a  straightforward exercise  to verify that for all small enough $\e$ these placements endow ${\bf K}$ with a nested structure,  
in terms of geodesic distances, of the type described in  section \ref{sec3.3}. 
This structure is as follows. If $m=1$ we have a single triple of points so that ${\bf K}= K^{(1)}_3$. If $m>1$ then
for any $k<m$ the inter point coordinate distances in $K^{(k)}_3$ are smaller than any of the coordinate distances between any of these points and any point in ${\bf K}- \cup_{k^{\prime}=1}^{k} K^{(k^{\prime})}_3$
as well as the interpoint coordinate distances between elements of ${\bf K}- \cup_{k^{\prime}=1}^{k} K^{(k^{\prime})}_3$.

A brief account of how this structure is implied by (\ref{delta}) and (\ref{d1.1})-(\ref{d2.5}) is as follows.

First note that for $m\geq j\geq 2$, 
\ba
(m-j+2)(1-\frac{\delta}{2})- (m-j+1) &=& 1 - \frac{\delta}{2} (m-j+2)  \geq (1- \frac{\delta}{2} m)  >0
\label{mdeltahalf}
\\
(m-j+2)(1-\delta)- (m-j+1) &=& 1 - \;\delta(m-j+2)  \geq (1- \;\delta m)  >0
\label{mdelta}
\ea
where the last inequality in both equations  follows from (\ref{delta}). 

Next note that (\ref{d1.2})-(\ref{d1.4}) imply that largest inter point distance in $K^{(1)}_3$ is $\O (\e^m)$. If $m>1$, 
equations (\ref{d2.1}), (\ref{d2.2}) and (\ref{mdelta})  imply that the shortest distance between any element of $K^{(1)}_3$ and any element of $K^{(j\neq 1)}_3$
is $\O (\e^{(1-\delta) (m-j+2)})$, which is smallest for $j=2$ so that the shortest distance between any element of $K^{(1)}_3$ and any element of $\cup_{j\neq 1}K^{(j)}_3$
is $\O (\e^{(1-{m\delta})})$, which is clearly larger than $\O (\e^m)$ for $m>1$.
%(Note also  that  $1-2\delta >0$ for $m>1$ from (\ref{delta})).
Note also that the shortest interpoint distance between elements of $K^{(j\neq 1)}_3$ is, from (\ref{d2.3})-(\ref{d2.5}) and (\ref{mdelta}), 
$\O(\e^{(1-\frac{\delta}{2})(m-j+2)} )$ which is smallest at $j=2$ in which case this distance is $\O(\e^{(1- \frac{\delta}{2})m} )$ which is also larger than $\O(\e^m)$.
Hence $K^{(1)}_3$ exhibits the segregation structure alluded to above.

Next fix $k$ such that  $m>k>1$. Then  from (\ref{d2.3})- (\ref{d2.5}) and (\ref{mdeltahalf})  the largest interpoint distance in $K^{(k)}_3$ is $\O(\e^{m-k+1})$.
From (\ref{d2.1}), (\ref{d2.2}) and (\ref{mdelta}), the 
shortest distance between any element of $K^{(j>k)}_3$ and any element of  $K^{(k)}_3$  is $\O(\e^{(1-\delta)(m-j+2)})$ which is smallest at $j=k+1$. Thus the 
the shortest distance between any element of $K^{(k)}_3$ and any element of $\cup_{j>k}K^{(j)}_3$
is $\O(\e^{(1-\delta)(m-k +1)})$  which is clearly larger than $\O(\e^{m-k+1})$.
Note also that the shortest interpoint distance betweem elements of $K^{(j\neq k)}_3$ is, from (\ref{d2.3})- (\ref{d2.5}) and (\ref{mdelta}), 
$\O(\e^{(1-\frac{\delta}{2})(m-j+2)} )$ which is smallest for $j=k+1$ in which case this distance is $\O(\e^{(1-\frac{\delta}{2})(m-k+1)} )$, which is again
larger than $\O(\e^{m-k+1})$. 

This immediately implies that ${\bf K}$ exhibits the segregation structure alluded to above.

We now evaluate $g({\bf K}, h)$. To do so note that in $K^{(1)}_3$  we have to leading order in $\e$ that 
\be
(\frac{d^{(1)}_{\min}}{d^{(1)}_{\max}}) = (\frac{\e^{m(1+\delta)}}{ \e^{m}})(\frac{|u(p)|}{|v(p)|}) = \e^{m\delta}\frac{|u(p)|}{|v(p)|}
\label{gk1}
\ee
and that in $K^{(j>1)}_3$ to leading order in $\e$ we have that:
\be
(\frac{d^{(j)}_{\min}}{d^{(j)}_{\max}}) =  \frac{ \e^{(1-\frac{\delta}{2})(m-j+2)} }{  \e^{m-j+1}}\frac{|u(p)|}{|v(p)|}
 =  \e^{(1- \frac{\delta}{2} (m-j-2))       }\frac{|u(p)|}{|v(p)|}.
\label{gkj}
 \ee
From (\ref{gk1}) and (\ref{gkj}) and the definition of $g$ (\ref{defgcase2}) it follows that to leading order in $\e$ we have that:
\be
g({\bf K},h) = \e^{2m\delta}(\prod_{j=2}^m \e^{2(1- \frac{\delta}{2} (m-j+2))       })\left(\frac{|u(p)|}{|v(p)|}\right)^{2m}
=: \e^{\alpha(m,\delta)}\left(\frac{|u(p)|}{|v(p)|}\right)^{2m}.
\label{gke1}
\ee
where we have defined the positive exponent $\alpha (m,\d)$ as:
\be
\alpha (m,\d):=  2m\delta +  \sum_{j=2}^m  2(1- \frac{\delta}{2} (m-j+2))    
\ee

\subsection{\label{secl3} Multiple Nests}

Given $n$ distinct  points $p_i, i=1,..n$,  a coordinate patch $\{x\}_i$ around each $p_i$, parameters $\e_i$  with 
\be
\e_i>>\e_j\;\; {\rm iff}\;\; i>j, 
\label{esize}
\ee
  positive integers $m_i$ and  the $\e_i$-independent (but $m_i$ dependent) parameters $\delta_i = \delta_i (m_i) $,
with 
\be
1- \delta_i m_i > 0, 
\ee
and pairs of tangent vectors $u_i,v_i$ at $p_i$, 
we can repeat  the considerations of sections \ref{secl1} for each $i$ by setting  $(p,\e, m, \delta, u, v ):= (p_i,\e_i,m_i, \delta_i, u_i, v_i)$ and placing $3m_i$ points around $p_i$
in the manner described in section \ref{secl1}. For small enough $\{\e_i\}$, the placement around any $p_i$ is not influenced by the placement around $p_{j\neq i}$. 
In this way for small enough $\{\e_i\}$ we obtain $n$ nested sets ${\bf K}_i$ with their union ${\bf K}= \cup_{i=1}^n {\bf K_i}$ also exhibiting a segregation structure  in terms of triples of points by virtue of the 
relative size (\ref{esize}) of the small parameters $\{\e_i\}$.

Setting 
\be
{\bf K}\equiv {\bf K}_{\{ \{x\}_i,  \e_i, p_i, u_i, v_i, m_i, \delta_i (m_i) \}},
\label{kembroid}
\ee
it immediately follows that to leading order in these small parameters we have that:
\be
g({\bf K}_{ \{ \{x\}_i,\e_i, p_i, u_i, v_i, m_i, \delta_i (m_i) \}},h) = (\prod_{i=1}^n \e_i^{\alpha (m_i,\delta_i})  \prod_{j=1}^n\left(\frac{|u_j(p_j)|}{|v_j(p_j)|}\right)^{2m_j}
\label{gken}
\ee
which in turn implies that 
\be
\lim_{\e_n\rightarrow 0}(...(\lim_{\e_1\rightarrow 0})...) (\prod_{i=1}^n \e_i^{-\alpha (m_i\d_i)}) g({\bf K}_{\{ \{x\}_i,\e_i, p_i, u_i, v_i, m_i, \delta_i (m_i) \}},h)
=  \prod_{i=1}^n\left(\frac{|u_i(p_i)|_h}{|v_i(p_i)|_h}\right)^{2m_i}
\label{normextr}
\ee
where we have appended the subscript $h$ to the norm to emphasize that the metric used is $h$. For future purposes it is convenient to denote the $\e_i$ dependent prefactor as follows:
\be
(\prod_{i=1}^n \e_i^{-\alpha (m_i,\delta_i)})=: E(\{ \e_i\}, \{m_i\}) \equiv E({\vec \e}, {\vec m})
\label{pref}
\ee
where we have ommitted the explicit dependence of $E$ on $\{\delta_i (m_i)\}$ to avoid notational clutter. In this notation we write (\ref{normextr}) as:
\be
\lim_{{\vec \e} \rightarrow 0} E({\vec \e}, {\vec m})g({\bf K}_{\{\{ \{x\}_i, \e_i, p_i, u_i, v_i, m_i, \delta_i (m_i) \}},h)= \prod_{i=1}^n\left(\frac{|u_i(p_i)|_h}{|v_i(p_i)|_h}\right)^{2m_i}.
\label{extract}
\ee

\subsection{\label{secl4} An Important Lemma}

The Lemma which we prove in this section concerns the following set up.  We are given $n$ distinct elements  $h_i, i=1,..n$ of ${\cal H}_{h_0}$. Clearly, no pair of these metrics are conformal to each other.
Hence, given $h_1, h_2$ we can always find a point $p_1$ and vectors $u_1,v_1$ at $p_1$ such that the ratio of their norms with respect to $h_1,h_2$ are unequal.
Denote the ratio of these norms with respect to the $i$th metric by $r_i^{(1)}$ and look for $i$ such that $r_i^{(1)} \neq r_1^{(1)}$ i.e. look for all metrics in addition to $h_2$ for which the ratio of metric norms
of $u_1,v_1$ are unequal to that computed with respect to $h_1$. Renumber the set of metrics so these metrics with norm ratios different  from $r_1^{(1)}$  at $p_1$  are $\{h_{I_1}, I_1=1,..n_1\}$ . For the remaining metrics these ratios must be equal to $r_1^{(1)}$.

Next, consider $h_{n_1+1}$. There exists a point $p_2$ and vectors $u_2,v_2$ at $p_2$ such that the ratio of their norms with respect to this metric and with respect to $h_1$ are unequal.
If this point happens to be $p_1$, due to the $C^{r-1}$ nature of these metrics, we can go to a  nearby point and choose a `nearby' pair of vectors  for which the desired inequality holds.
More in detail, fix a neighborhood of $p_2$ and a chart thereon. Then it is straightforward to see that we can  choose a point close enough to $p_2$ in this chart and vectors with components close enough to
those at $p_2$ such that the inequality continues to hold.  Rename this point as $p_2$ and the vectors as $u_2,v_2$.
Denote the ratio of  the norms of $u_2,v_2$ with respect to $h_i$ by $r_i^{(2)}$. 
Look for all metrics in addition to $h_{n_1}+1$ in the set $\{h_{n_1+1}, ..,h_n\}$ for which the norm ratios are unequal to the ratio $r_1^{(2)}$ computed with respect to $h_1$. 
Clearly we have $r_{I_2}^{(2)} \neq r_1^{(2)}$ with   $r_{I_2}^{(1)}= r_1^{(1)} $ i.e. disagreement at the second point, agreement at the first.
Renumber this set of remaining metrics so that these metrics are numbered  $h_{I_2}, I_2= n_1+1,..n_2$.
 
Iterate this procedure so as to get $p_3$ distinct from $p_1,p_2$ and $u_3,v_3$ such that in obvious notation $r_{I_3}^{(3)}  \neq r_1^{(3)}$, but  $r_{I_3}^{(i^{\prime})}= r_1^{(i^{\prime})}, i^{\prime}=1,2$
where $I_3= n_2+1,.., n_3$. Continue this iterative procedure until it ends  after some finite number  $m^{\prime} \leq n$ steps and provides the setting for the Lemma below.
\\

\noindent{\bf Lemma}: Let there exist  points $p_{ i^{\prime}} \in \Sigma , {\di} =1,..,m^{\prime}$, pairs of vectors $u_{\di}, v_{\di}$ at $p_{\di}$,  metrics $h_i, i=1,..,n$,
positive integers $n_{\di}$ with $n_{\di}< n_{\dj}$ for $\di < \dj$ with $n_{\dm}=n$, indices $I_{\di}$ with range $n_{\di -1}+1,.., n_{\di}$ (where $n_0:=0$)
 such that for any fixed $\di$ and all $I_{\di}$ we have that:
\ba
&& r_{I_{\di}}^{(\dj)} = r_{1}^{(\dj)}\;\;{\rm if}\;\; \dj < \di,  \nonumber \\
&& r_{I_{\di}}^{(\di)} \neq  r_{1}^{(\di)}. 
\label{ri=r1}
\ea
where 
\be
r_k^{(\dj)}:= \left(\frac{|u_{\dj}(p_{\dj})|_{h_k}   }{|v_{\dj}(p_{\dj})|_{h_k}}\right)^2 \;\;\;k=1,..,n.
\ee
%\ba
%&&  \nexists \lambda_{p_{\di}}>0 \;{\rm s.t.}\;\;  h_{I_{\di}}({(p_{\di})}) = \lambda_{p_{\di}}   h_1({(p_{\di})}),   \nonumber\\
%&&\exists \alpha_{p_{\dj}}>0  \;{\rm s.t.}\;\;  h_{I_{\di}}({(p_{\dj<\di})}) = \alpha_{p_{\dj}} h_1({(p_{\dj})}), \nonumber\\
%&&I_{\di}= n_{\di -1}+1, ...,n_{\di}, \;\;\; n_0:=1 \;\;\;n_{\dm}= n.
%\label{hi=hj}
%\ea
Then there exist positive integers $N_{\dj}$,  
%and pairs of tangent vectors $u_{\dj}, v_{\dj}$ at $p_{\dj}$ for  
$\dj = 1,.., \dm$  such that:
\ba
\prod_{\dj=1}^{\dm} (r_1^{(\dj)})^{N_{\dj}} \neq \prod_{\dj=1}^{\dm} (r_k^{(\dj)})^{N_{\dj}}, \;\;\; k=2,..,n
\label{rneqr}
\ea
\\

\noindent{\bf Proof}: 
%First note that  (\ref{hi=hj}) implies that $\forall \dj = 1,.., \dm$ we can choose pairs $u_{\dj}, v_{\dj}$ at $p_{\dj}$ such that:
%\be
%r^{(\di)}_{I_{\di}} \neq r^{(\di)}_1 , \;\;\; r^{(\dj<\di)}_{I_{\di}} =  r^{(\dj)}_1 .
%\label{ri=r1}
%\ee
%Hereon we choose $u_{\dj}, v_{\dj}$ at $p_{\dj}$ so that (\ref{ri=r1}) holds.
Choose $N_1=1$. From (\ref{ri=r1}), we have 
\be
r^{(1)}_{I_1} \neq r^{(1)}_1.
\label{p1}
\ee
From (\ref{ri=r1}), $r^{(2)}_{I_1} \neq r^{(2)}_1$ but $r^{(1)}_{I_2} =  r^{(1)}_1$. It follows that:
\be
(r^{(2)}_{I_2})^{N_2}  r^{(1)}_{I_2}\neq     (r^{(2)}_1)^{N_2}                r^{(1)}_1    \;\;\;{\rm for}\;{\rm any}\;N_2.
\label{p2}
\ee
If $I_1$ is such that $r^{(2)}_{I_1} = r^{(2)}_1$, then again  
\be
(r^{(2)}_{I_1})^{N_2}  r^{(1)}_{I_1}\neq     (r^{(2)}_1)^{N_2}                r^{(1)}_1    \;\;\;{\rm for}\;{\rm any}\;N_2.
\label{p3}
\ee
If  $I_1$ is such that $r^{(2)}_{I_1} \neq r^{(2)}_1$, then there exists $N_{I_1}$ such that:
\be
(r^{(2)}_{I_1})^{N}  r^{(1)}_{I_1}\neq     (r^{(2)}_1)^{N}                r^{(1)}_1    \;\;\forall N\geq N_{I_1}.
\label{p4}
\ee
To see this, first suppose that $(r^{(2)}_{I_1})^{N}  r^{(1)}_{I_1}\neq     (r^{(2)}_1)^{N} r^{(1)}_1        \forall N \geq 1$. In this case $N_{I_1}=1$. Suppose  that
it is not the case that  $(r^{(2)}_{I_1})^{N}  r^{(1)}_{I_1}\neq     (r^{(2)}_1)^{N}r^{(1)}_1 \forall N \geq 1$. Then there exists some $N_0\geq 1$ for which
$(r^{(2)}_{I_1})^{N_0}  r^{(1)}_{I_1}=   (r^{(2)}_1)^{N_0}r^{(1)}_1$. 
Then $\forall N>N_0$, since $r_{I_1}^{(2)}\neq r_1^{(2)}$ we have that 
$(r^{(2)}_{I_1})^{N- N_0} (r^{(2)}_{I_1})^{N_0}  r^{(1)}_{I_1}) \neq  (r^{(2)}_1)^{N-N_0})  ((r^{(2)}_1)^{N_0}r^{(1)}_1)$ so that we can set $N_{I_1} = N_0+1$,
\footnote{Note that it must be the case that for all $N<N_0$, we have that $(r^{(2)}_{I_1})^{N}  r^{(1)}_{I_1}\neq     (r^{(2)}_1)^{N}r^{(1)}_1$
else there would exist $N_1<N_0$ for which equality holds, in which case our argument above shows that the inequality must hold for all $N>N_1$ including $N=N_0$ which is 
a contradiction. Hence $N_0$ is unique.}
which completes the proof of (\ref{p4}).

In the case that there exists $I_1$ such that  $r^{(2)}_{I_1} \neq  r^{(2)}_1$ we set $N_2 = \max_{I_1 \;{\rm s.t}\; r^{(2)}_{I_1}\neq r^{(2)}_1} N_{I_1}$ else we choose $N_2=1$.
Next, define:
\be
\prod_{\di=1}^2 (r_k^{(\di)})^{N_{\di}}  :=  \alpha^{(2)}_k.
\label{defalpha2}
\ee
Our choice of $N_1, N_2$ implies that 
\be
 \alpha^{(2)}_{I_{\dk}} \neq   \alpha^{(2)}_1 \;\; \forall \dk \leq 2.
 \label{alpha2neq}
 \ee
Now  $r^{(3)}_{I_3}\neq r_1^{(3)}$ but $r^{\di<3}_{I_3}= r^{(\di)}_1$ which implies that $ \alpha^{(2)}_{I_{3}} =  \alpha^{(2)}_1$  (Indeed, $ \alpha^{(2)}_{I_{\dk}} =  \alpha^{(2)}_1, \; \forall \dk\geq 3$).
This implies that:
\be
(r^{(3)}_{I_3})^N \alpha^{(2)}_{I_{3}}\neq (r_1^{(3)})^N  \alpha^{(2)}_1    \;\;\forall N.
\label{r3i3}
\ee
Next, if there exists any $I_{\di <3}$ such that $r^{(3)}_{I_{\di}}= r_1^{(3)}$, we have that 
\be
(r^{(3)}_{I_{\di}})^N \alpha^{(2)}_{I_{3}}\neq (r_1^{(3)})^N  \alpha^{(2)}_1    \;\;\forall N.
\label{r3i}
\ee
If there exists $I_{\di <3}$ such that $r^{(3)}_{I_{\di}} \neq r_1^{(3)}$, then by an argument identical to that used to prove (\ref{p4}),  there exists $N_{I_{\di}}$ such that:
\be
(r^{(3)}_{I_{\di}})^N \alpha^{(2)}_{I_{3}}\neq (r_1^{(3)})^N  \alpha^{(2)}_1    \;\;\forall N \geq N_{I_{\di}}.
\label{r3ineq}
\ee
If there exist $I_{\di <3}$ such that $r^{(3)}_{I_{\di}} \neq r_1^{(3)}$, we set $N_3 = \max_{\di <3 \;{\rm s.t}\; r^{(3)}_{I_{\di}}\neq r^{(3)}_1} N_{I_{\di}}$, else we set $N_3=1$,
Defining  $\prod_{\di=1}^3 (r_k^{(\di)})^{N_{\di}}  :=  \alpha^{(3)}_k$, our choice for $N_1,N_2,N_3$ implies that:
\ba
&&\alpha^{(3)}_{I_{\dk}} \neq \alpha^{(3)}_1, \;\; \forall \dk \leq 3,
\label{alpha3neq}\\
&&\alpha^{(3)}_{I_{\dk}} =  \alpha^{(3)}_1, \;\; \forall \dk >3.
\ea
Continuing on in this manner, at the ${ f}^{\prime}$th step a choice of $N_{\di}, \di=1,.., {f^{\prime}}$ exists such that:
\ba
&&\alpha^{({ f^{\prime}})}_{I_{\dk}} \neq \alpha^{({ f^{\prime}})}_1, \;\; \forall \dk \leq { f^{\prime}},
\label{alphafneq}\\
&&\alpha^{({ f^{\prime}})}_{I_{\dk}} =  \alpha^{({ f^{\prime}} )}_1, \;\; \forall \dk > { f^{\prime}}.
\ea
This procedure terminates at the $\dm$th step and completes the proof.

\subsection{\label{secl5} Proof of Linear Independence: The case of a single Bra Set Label}
As indicated towards the end of section \ref{sec6.3}, it is necessary for  admissible Bra Set labels for basis states to satisfy additional properties beyond properties (i)-(iii) of  section \ref{sec3.2}, for
the proof of linear independence below to go through. We specify the additional properties as  (iv) below. In order to do so we introduce some preliminary structures.

%Let $\gamma (S)$ be the coarsest graph underlying the (embedded) spin network $S$. Let the edges of this graph be colored with its colors in $S$. Call the resulting colored graph $\gamma_c$.
%Thus, $\gamma_c$ has all the information which $S$ has except for vertex intertwiner information.

First,define the relation  $\approx$ between elements of $B$ as  follows.  Let $S_1,S_2 \in B$ . Then $S_1\approx S_2$ iff:\\
\noindent (a) $V(S_1) = V(S_2)$ where $V(S)$ is the set of  vertices of $S$.\\
\noindent (b) For every $v_{\alpha} \in  V(S_1) \exists \e^{\alpha}_0$ such that $\forall \e<\e^{\alpha}_0$, $B_{\e}(v_{\alpha})\cap S_1= B_{\e}(v_{\alpha})\cap S_2$
where $B_{\e}(v)\cap S$ is the (in general) open spin network  obtained by restricting $S$ to the coordinate ball  $B_{\e}(v)$ of size $\e$ about $v$. 
%Clearly, this definition
%does not depend on the choice of coordinates as $\e^{\alpha}_0$ can be arbitrarily small.
%\noindent (b)  For every $v_{\alpha} \in  V (S_1) \exists \e^{\alpha}_0(S_1,S_2)$ such that $\forall \e< \e^{\alpha}_0 (S_1,S_2)$, $B_{\e}(v_{\alpha})\cap \gamma_c (S_1)= B_{\e}(v_{\alpha})\cap \gamma_c (S_2)$
%where $B_{\e}(v)\cap \gamma_c (S)$ is the (in general) open graph  obtained by restricting the colored graph $\gamma_c (S)$ underying $S$ to the coordinate ball  $B_{\e}(v)$ of size $\e$ about $v$. 
\\

Clearly, this definition
does not depend on the choice of coordinates as $\e^{\alpha}_0 $ can be arbitrarily small. It is straightforward to see that the fact that  $\e^{\alpha}_0 $ can be chosen arbitrarily small also implies that $\approx$
is an equivalence relation.
Denote the equivalence class of spin nets equivalent to $S$ under $\approx$ by  $[S]_B$ and call the set of these equivalence classes for all $S$ as ${\cal S}_B$. Denote complex valued functions on this 
${\cal S}_B$ by calligraphic alphabets, ${\cal F}: {\cal S}_B \rightarrow {\bf C}$,  with the evaluation of $\F$ on $[S]_B\in {\cal S}_B$ denoted by  $\F([S]_B)$. Similar to the off shell basis state $\Psi_{B,f,h}$, we define
the state $\Psi_{B, {\F}, h}$  (where as before, the bra set label  $B$ satisfies (i)-(iii) of section \ref{sec3.2} and the metric label $h$ is such that $h\in {\cal H}_{h_0}$) through its amplitudes on elements of $B$.
Let $B_{\perp}$ be set of spin nets which are orthogonal to every element of $B$. Then:
\ba
\Psi_{B, {\F} ,h}(|s\ket) &:=& 0 \;\;\;\;\;{\rm if} \;\; s\in B_{\perp},   \label{psif0}\\
&=& {\F} ([s]_B) g_{s,h}    \;\;\;\;\;{\rm if}  s  \in B,   \label{psifs}.
\ea
%\be
%\gamma_{[s]}:= \gamma_s  
%=   \sum_i { c^s_i}, 
%\label{defpsiff}
%\ee
%where in the last equation, we have used the fact that the $\gamma$ factor depends only on the vertex structure at vertices of $s$ and hence only on $[s]$.
The action of the complex map  can be  extended  to the finite linear span of spin nets by  linearity to yield an element of the algebraic dual space and this map so extended
defines the action of the distribution $\Psi_{B, {\F} ,h}$ on the finite linear span of spin nets. 
We note here that $\F$ simply denotes a complex map on ${\cal S}_B$. Concrete examples of ${\F}$  may require auxilliary constructs such as  $h$ dependent coordinate patches around vertices
for its definition. Our considerations in this section are independent of the detailed choice of $\F$; it is only necessary that $\F$ be a well defined function on the space ${\cal S}_B$ of equivalence classes $[S]_B$.
For these considerations to be applicable to the kind of distributions encountered in sections \ref{sec3}, \ref{sec4} and \ref{sec6}, this definition of $\F$ must be general enough that those distributions are of the type
$\Psi_{B, \F, h}$. We digress briefly to indicate that this is indeed the case.
%Since the only non-trivial linear relations between 
%spin nets are amongst those which share the same colored graph, it is straightforward to check that this extension is well defined so that $\Psi_{B, {\F}, h}$ is  a distribution which lies in the algebraic dual to
%the finite span of spin net states.
%\footnote{\label{fnlocal}
%Note that the off shell basis states of section \ref{sec3} as well as the states obtained by the single and product action of the Hamiltonian constraint on these off shell basis states
%are all of the type $\Psi_{B,{\F}, h}$. 

The off shell basis state $\Psi_{B,f,h}$ of section \ref{sec3} has vanishing amplitudes for states in $B_{\perp}$. Its amplitude on a state $S$ may be obtained by setting ${\cal F}([S]_B)= \prod_{v\in V(S)} f(v)$; 
${\cal F}$ clearly only depends on the values of $f$ at vertices of $S$ 
and hence only on $[S]_B$. For an off shell  state  obtained by the first action of the Hamiltonian constraint (\ref{first1}) on $\Psi_{B,f,h}$, amplitudes on states in $B_{\perp}$ vanish by virtue of (\ref{first0}) and 
the amplitude on a state in $B$ is given by (\ref{first1})
with the $\gamma$ factor set to the appropriate eigenvalue of the ${\hat \lambda}_v$ operator (see section \ref{sec3.4}). ${\cal F}$ is then obtained as the product over vertex contributions. Each such contribution is either an evaluation of $f$ 
at a vertex of $S$
or a sum over evaluations of directional derivatives of $f$ along edge tangents at nondegenerate vertices of $S$ weighted with the ${\hat \lambda}_v$ eigenvalue at that vertex.
%certain powers of the volume operator eigen values at that vertex. 
The latter contributions  clearly only
depend on the coloring of edges at the vertex, the vertex intertwiner,
\footnote{Note the property of non-degeneracy depends on these edge colorings and this vertex intertwiner.}
the behavior of $f$ in an arbitrarily small neighborhood of this vertex and the unit edge tangents at the vertex 
normalized with respect to certain coordinates in this arbitrarily small neighborhood of the vertex (these coordinates happen to be the RNC's
with respect to the metric $h$). Thus this contribution too only depends on $[S]_B$. Similar comments hold for the action of the constraint product (\ref{hmhnfinal}). In this case in the expression for the amplitude on a state $S\in B$ there are 
also `2nd derivative' contributions but clearly these contributions also   depend only $[S]_B$.
%}

Next, for the purposes of section \ref{secl61}, we consider the action of diffeomorphisms on $\Psi_{B, {\F}, h}$. To this end note  that  
given  any diffeomorphism $\phi$ and a pair of states $S_1,S_2\in B$, we have that   $S_1\approx S_2$ iff $(S_{1})_{\phi}\approx (S_{2})_{\phi}$ where we have used $S_{\phi}$ to denote the diffeomorphic image of $S$ obtained by the action of the unitary operator ${\hat U}(\phi)$ on $S$  
i.e. $S_{\phi}:= {\hat U}(\phi) S$.
\footnote{ Property (ii), Section \ref{sec3.4.3} implies closure of $B$ under the action of diffeomorphisms.}
This follows from  the fact that:\\
\noindent $\rm{(a)}_{\phi}$ $V((S_{1})_{\phi})=  V((S_{2})_{\phi})$ iff (a) holds.
  \\
\noindent $\rm{(b)}_{\phi}$ The phrasing of (b) does not depend on choice of coordinates so that properties of $S_1,S_2$ with respect to coordinate spheres in coordinates $\{x\}$
are replicated by $(S_{1})_{\phi}, (S_{2})_{\phi}$ with respect to the images of these spheres which are then coordinate spheres  with respect to the push forward coordinates $\phi^*\{x\}$.
\\

It follows that 
%from the invertibility of  $\phi$ that 
$\phi$ induces a well defined  map on the space ${\cal S}_B$ of these equivalence classes of elements of $B$ through
\be
\phi ([S]_B):= [S_{\phi}]_B
\label{phiequivs}
\ee
We have that:
\be
(\phi_1  (\phi_2 ([S]_B)) = \phi_1 ([{\hat U} (\phi_2)S ]_B )= [{\hat U} (\phi_1) {\hat U}(\phi_2) S]_B= [S_{\phi_1\circ \phi_2}]_B
\label{phi1phi2equivs}
\ee
so that the map provides a representation of the diffeomorphism group on ${\cal S}_B$.
Equation (\ref{phiequivs}) implies that  we may define the function $\F_{\phi}$ on ${\cal S}_B$ by:
\be
\F_{\phi}([S]_B) := \F ([S_{\phi}]_B)
\label{defcalfphi}
\ee
Next, recall that the dual action of ${\hat U}(\phi)$  on  $\Psi_{B,{\F}, h}$ is defined through:
\ba
{\hat U}(\phi) \Psi_{B, {\F} ,h} (S) &=&  \F ([S_{\phi^{-1}}]) g_h(S_{\phi^{-1}}),\;\;\;\;{\rm for}\; S\in B \\
%\gamma_{S_{\phi^{-1}}}
&=& 0 \;\;\;\;{\rm for}\; S\in B_{\perp} \\
\label{dualphiff}
\ea
%for any spin net state $S$.
From (\ref{dualphiff}), (\ref{defcalfphi}),  (\ref{gphi}),(\ref{gammaphi}), (\ref{phipsi}),  (\ref{phiequivs}) and (\ref{phi1phi2equivs}),  it follows that:
\ba
{\hat U}(\phi) \Psi_{B, {\F} ,h} &=& \Psi_{B, {\F_{\phi^{-1}}}, \phi^*{h}} \label{uphipsiff} \\
{\hat U}(\phi_1){\hat U}(\phi_2) \Psi_{B, {\F} ,h} &=& {\hat U}(\phi_1\circ \phi_2) \Psi_{B, {\F} ,h} \label{u1u2psiff}
\ea
This concludes our analysis of the action of diffeomorphisms on states of the form $\Psi_{B, {\F} ,h}$.

Next, we define some useful notation.
Recall the set ${\bf K}_{\{\{x\}_i, \e_i, p_i, u_i, v_i, m_i, \delta_i (m_i),  \}}$ from equation (\ref{kembroid}). 
Let ${\bf K}_0$ be a fixed set of points such that no $p_i\in {\bf K}_0$. Fix an equivalence class $[S]_B$. 
%For every 
%${\bf K}_{\{\{x\}_i,\e_i, p_i, u_i, v_i, m_i, \delta_i (m_i),  \}}$ with small enough $\{\e_i\}$ let 
 Let ${\bar S}_{{\bf K}_0, [S]_B}({\bf K}_{\{ \{x\}_i,\e_i, p_i, u_i, v_i, m_i, \delta_i (m_i) \}})$  denote a spin net in $[S]_B$ (assuming such a spin net exists)  with Kink set
 ${\bf K}_{\{ \{x\}_i,\e_i, p_i, u_i, v_i, m_i, \delta_i (m_i)  \}} \cup {\bf K}_0$ such that for 
 for fixed $h$ and  sufficiently small  $\{\e_i\}$, the elements of 
 ${\bf K}_{\{ \{x\}_i,\e_i, p_i, u_i, v_i, m_i, \delta_i (m_i) \}}$ comprise the `closest kinks'  (see section \ref{sec3.3})
\footnote{Note that even if some or all of the $p_i$ are in $V(S)$, ${\bf K}_{\{ \{x\}_i,\e_i, p_i, u_i, v_i, m_i, \delta_i (m_i) \}}$ can be chosen so that  its elements
are non-coincident with elements of $V(S)$ by virtue of the fact that the locations of the elements of ${\bf K}_{\{ \{x\}_i,\e_i, p_i, u_i, v_i, m_i, \delta_i (m_i) \}}$
are specified only upto certain orders of $\e$ (see section \ref{secl1}).} 
 so that:
\be
g_{{\bar S}_{{\bf K}_0, [S]_B}({\bf K}_{\{ \{x\}_i,\e_i, p_i, u_i, v_i, m_i, \delta_i (m_i)  \}}), h} = g ({\bf K}_{\{ \{x\}_i,\e_i, p_i, u_i, v_i, m_i, \delta_i (m_i)  \}}, h)
g ({\bf K}_0, h).
\label{small}
\ee
With these preliminaries in place we require that any Bra Set subject to (i)-(iii), section \ref{sec3.2} is required to satisfy
an extra condition which guarantees the existence of states of the type ${\bar S}_{{\bf K}_0, [S]_B}({\bf K}_{\{ \{x\}_i,\e_i, p_i, u_i, v_i, m_i, \delta_i (m_i) \}})$.
The behavior of the interkink distance function for these states (see (\ref{small})) for sufficiently small $\{\e_i\}$ allows us to connect a statement of linear independence
of basis off shell states with different metric labels to the Lemma proved in section \ref{secl4} and thereby prove such a statement.
The extra condition (iv)(a) is as follows.\\

%specify the additional condition (iv) as follows:
\noindent{\bf Condition (iv)(a)}: 
%Any Bra Set subject to (i)-(iii), section \ref{sec3.2} is required to satisfy the following additional properties:\\
%\noindent 1. For any $S\in B$, $\exists S_0 \in B$ such that $S_0 \in [S]$ and $S_0$ has no kinks.\\
%Recall the set ${\bf K}_{\{\{x\}_i, \e_i, p_i, u_i, v_i, m_i, \delta_i (m_i),  \}}$ from equation (\ref{kembroid})). 
%Let ${\bf K}_0$ be a fixed set of points such that no $p_i\in {\bf K}_0$. Fix an equivalence class $[S]$. 
%For every 
%${\bf K}_{\{\{x\}_i,\e_i, p_i, u_i, v_i, m_i, \delta_i (m_i),  \}}$ with small enough $\{\e_i\}$ let 
% ${\bar S}_{{\bf K}_0, [S]}({\bf K}_{\{ \{x\}_i,\e_i, p_i, u_i, v_i, m_i, \delta_i (m_i),  \}})$  be a spin net in $[S]$  with Kink set
% ${\bf K}_{\{ \{x\}_i,\e_i, p_i, u_i, v_i, m_i, \delta_i (m_i),  \}} \cup {\bf K}_0$ such that ${\bf K}_{\{ \{x\}_i,\e_i, p_i, u_i, v_i, m_i, \delta_i (m_i),  \}}$ comprise the `closest kinks' so that 
% for fixed $h$ and  sufficiently small  $\{\e_i\}$:
%\be
%g_{{\bar S}_{{\bf K}_0, [S]}({\bf K}_{\{ \{x\}_i,\e_i, p_i, u_i, v_i, m_i, \delta_i (m_i),  \}}), h} = g ({\bf K}_{\{ \{x\}_i,\e_i, p_i, u_i, v_i, m_i, \delta_i (m_i),  \}}, h)
%g ({\bf K}_0, h).
%\label{small}
%\ee
For any $S\in B$,   and for any choice of $p_i, u_i, v_i, m_i$ and  $\delta_i >0$ (such that $1- m_i\delta_i>0$), there 
exists  ${\bf K}_0$ and  a choice of $\{x\}_i$ at $p_i$, such that for every  small enough 
$\{\e_i\}$, $B$ contains 
a state   $S_{{\bf K}_0, [S]_B}({\bf K}_{\{ \{x\}_i,\e_i, p_i, u_i, v_i, m_i, \delta_i (m_i)  \}}) \in [S]_B$
where we have used the notation developed in the previous paragraph and 
%$\exists S_{{\bf K}_0, [S]}({\bf K}_{\{ \{x\}_i,\e_i, p_i, u_i, v_i, m_i, \delta_i (m_i)  \}}) \in B$ 
%with 
% $S_{{\bf K}_0, [S]}({\bf K}_{\{\{x\}_i, \e_i, p_i, u_i, v_i, m_i, \delta_i (m_i),  \}})  \in [S]$ 
%for all  small enough $\{\e_i\}$,   
%Here $S_0({\bf K}_{\{\{x\}_i, \e_i, p_i, u_i, v_i, m_i, \delta_i (m_i),  \}})$
% is a spin network with kink set ${\bf K}_{\{ \e_i, p_i, u_i, v_i, m_i, \delta_i (m_i),  \}}$. 
where by small enough $\{\e_i\}$ we mean small enough that (\ref{small}) holds.
%\end{document}
%.  Consider the fixed set of labels  $\{ p_i, u_i, v_i, m_i, \delta_i (m_i), i=1,..,n \}$.  We require that there exist strictly positive
%$\e^0_i \equiv \e^0_i (\{ p_j, u_j, v_j, m_j, \delta_j (m_j), i=1,..,n \}),  i =1,..n$ such that $B$ contains a spin network
% $S_0({\bf K}_{\{\{x\}_i \e_i, p_i, u_i, v_i, m_i, \delta_i (m_i),  \}}) $ for all  $\e_i$ such that $0 < \e_i <\e^0_i$. 
%This completes the specification of property (iv).
\\
%Consider any set of {\em disjoint}  Bra Sets ${\cal B}= \{B_{\alpha}\}$ subject to properties (i)- (iv).
%Consider the finite linear span ${\cal L}_{\cal B}$ of off shell basis states with bra set labels which are restricted to be elements of ${\cal B}$. Then our proof of linear independence
%holds for any state in ${\cal L}_{\cal B}$ so that  for any such state, the metric label dependent constraint action is well defined, diffeomorphism covariant and anomaly free.
%While ${\cal L}_{\cal B}$ would seem to be a very restricted space, we find it plausible that this is not the case for the following reason.

\noindent In what follows we refer to (i)-(iii), section \ref{sec3.2} and Condition (iv)(a) above simply as properties (i)-(iv)(a). We proceed to a  statement of linear independence and its proof.\\

\noindent{\bf Statement} Let $\Psi_{B, \F_i,h_i}\neq 0$ be as defined in (\ref{psif0}), (\ref{psifs}) with $B$ subject to properties (i)-(iv)(a),  $\{h_i, i=1,..,n\}$ being distinct metric labels.
Let $a_i, i=1,..,n$ denote complex numbers. Then 
\be
\sum_{i=1}^n a_i \Psi_{B, \F_i,h_i}= 0 \Rightarrow a_i= 0, i=1,..,n  .
\label{linindep}
\ee

\noindent{\bf Proof}: Let $S\in B$ such that 
\be
\F_1([S]_B)\neq 0.
%\gamma_S\neq 0
\label{fneq0}
\ee
(if no such $S$ exists $\Psi_{B, \F_1,h_1} =0$). Consider ${\bf K}_{\{ \{x\}_{\di},\e_{\di}, p_{\di}, u_{\di}, v_{\di}, N_{\di}, \delta_{\di} (N_{\di}), \di=1,..,\dm  \}})$
where $\{N_{\di}\}$ are chosen exactly as in section \ref{secl4}. In particular we can choose $p_{\di}, u_{\di},v_{\di}$ such that (\ref{ri=r1}) is satisfied by virtue of the fact that $h_i \in {\cal H}_{h_0}$ and $h_0$ has no 
conformal isometries.
Define $\alpha_k$ by 
\be
\alpha_k := \prod_{\dj=1}^{\dm} (r_k^{(\dj)})^{N_{\dj}} \;\;k=1,..,n .
\label{defalpha}
\ee
%Set $\e_{\di} = \frac{1}{\lambda^{\dm -\di}, \lambda >>1 $.
Property (iv)(a), guarantees the existence of $S_{{\bf K}_0, [S]_B}({\bf K}_{\{ \{x\}_{\di},\e_{\di}, p_{\di}, u_{\di}, v_{\di}, N_{\di}, \delta_{\di} (N_{\di}), \di=1,..,\dm  \}})$. 
%Moreover
%property (i) allows us to choose $S_{{\bf K}_0, [S]_B}({\bf K}_{\{ \{x\}_{\di},\e_{\di}, p_{\di}, u_{\di}, v_{\di}, N_{\di}, \delta_{\di} (N_{\di}), \di=1,..,\dm  \}})$ such that its intertwiners
%are basis intertwiners thereby guaranteeing that
%\be
%\gamma_{S_{{\bf K}_0, [S]}({\bf K}_{\{ \{x\}_{\di},\e_{\di}, p_{\di}, u_{\di}, v_{\di}, N_{\di}, \delta_{\di} (N_{\di}), \di=1,..,\dm  \}})} \neq 0.
%\label{71a}
%\ee
We have that: 
\ba
\lim_{{\vec \e}\rightarrow 0} E({\vec \e}, {\vec N})&&
\Psi_{B, \F_i,h_i} ( S_{{\bf K}_0, [S]_B}({\bf K}_{\{ \{x\}_{\di},\e_{\di}, p_{\di}, u_{\di}, v_{\di}, N_{\di}, \delta_{\di} (N_{\di}), \di=1,..,\dm  \}}))
\nonumber \\
&=& \F_i([S]_B) g({\bf K}_0, h) 
%\gamma_{ S_{{\bf K}_0, [S]}({\bf K}_{\{ \{x\}_{\di},\e_{\di}, p_{\di}, u_{\di}, v_{\di}, N_{\di}, \delta_{\di} (N_{\di}), \di=1,..,\dm  \}})           }   
\alpha_i
\label{71}
\ea 
where we used  (\ref{extract})  in (\ref{71}). Next we impose the left hand side of (\ref{linindep}):
\be
\sum_{i=1}^n a_i \Psi_{B, \F_i,h_i}= 0 .
\label{72}
\ee
Evaluating the amplitude of  (\ref{72}) for  $S_{{\bf K}_0, [S]_B}({\bf K}_{\{ \{x\}_{\di},\e_{\di}, p_{\di}, u_{\di}, v_{\di}, N_{\di}, \delta_{\di} (N_{\di}), \di=1,..,\dm  \}})$
implies that:
\be
 \sum_i a^{\prime}_i \alpha_i = 0, \;\;\;; a^{\prime}_i: = a_i  \F_i([S]_B) g({\bf K}_0, h) 
 %\gamma_{S_{{\bf K}_0, [S]}({\bf K}_{\{ \{x\}_{\di},\e_{\di}, p_{\di}, u_{\di}, v_{\di}, N_{\di}, \delta_{\di} (N_{\di}), \di=1,..,\dm  \}})}.
\label{73}
\ee
%where we have used  (\ref{fneq0}) in the last part of (\ref{72}).
Replacing $N_{\di}$ by $\Lambda N_{\di}$ with $\Lambda$ a positive integer and repeating these considerations then  yields:
\be
\sum_i a^{\prime}_i (\alpha_i)^{\Lambda} = 0 .
\label{74}
\ee
Let the set of  $\bn$ distinct   values of $\alpha_i$  be $\{ \beta_{\bi}, \bi =1,..,{\bar n} \}$ and let $\bI_{\bi}$ run over the index values   in $\{1,..,n\}$ for which $\alpha_{\bI_{\bi}}= \beta_{\bi}$.
Setting, for fixed $\bi$, 
\be
\sum_{\bI_{\bi}} a^{\prime}_{\bI_{\bi}}=: b_{\bi},
\label{75}
\ee
equation (\ref{74}) can be written as:
\be
\sum_{\bi=1}^{{\bar n}}b_{\bi} (\beta_{\bi})^{\Lambda}=0.
\label{76}
\ee
Renumbering the summands so that  $b_{\bi} =: b^{\prime}_{\bar{\dj}}, \beta_{\bi}= \beta^{\prime}_{\bar{\dj}}$ with $\beta^{\prime}_{{\bar \dj}} > \beta^{\prime}_{\bar{\di}}$ iff ${\bar \dj} < {\bar \di}$, 
we have that $\beta^{\prime}_1$ is the largest value of $\alpha_i, i=1,..,n$. Suppose that $\beta_{\bar {\dm}}= \alpha_1$ for some ${\bar \dm}$. Then (\ref{rneqr}) implies that  $b^{\prime}_{{\bar \dm}} = a^{\prime}_1$. 
Equation (\ref{76}) takes the form
\be
\sum_{{\bar \di} =1}^{{\bar n}}b^{\prime}_{{\bar \di}} (\beta^{\prime}_{{\bar \di}})^{\Lambda}=0.
\label{77}
\ee
Dividing (\ref{77}) by $\beta^{\prime}_1$ and taking the $\Lambda \rightarrow \infty$ limits yields $b_1^{\prime}=0$. Substituting this into (\ref{77})  yields 
\be
\sum_{{\bar \di}=2}^{{\bar n}}b^{\prime}_{{\bar \di}} (\beta^{\prime}_{{\bar \di}})^{\Lambda}=0.
\label{78}
\ee
Iterating this procedure ${\bar \dm}$ times implies $b^{\prime}_{{\bar \dm}} = a^{\prime}_1 =0$, 
which together with (\ref{fneq0}),(\ref{73}) implies that 
$a_1=0$. 

Iterating this whole procedure starting (\ref{72}) with $a_1=0$, we obtain $a_2=0$ and after $n$ iterations, we have $a_i= 0, i=1,..,n$ which proves the statement.
\\

\subsection{\label{secl6} Proof of Linear Independence: The case of  arbitrary Bra Set Labels}

Let $\Psi_{h_{{\hat \rho}}}$ be a finite linear combination of (non-trivial) off shell basis states with, in general, distinct Bra Set labels but with the same metric label $h_{\hat \rho}$ i.e. 
\be
\Psi_{h_{{\hat \rho}}} := \sum_{i_{\hat \rho}= 1}^{{\hat n}_{{\hat \rho}}} c^{({\hat \rho})}_{i_{{\hat \rho}}} \Psi_{B^{(\hat \rho)}_{i_{{\hat \rho}}}, \F^{(\hat \rho)}_{i_{{\hat \rho}}},h_{\hat \rho}}
\label{l61}
\ee
where $c^{({\hat \rho})}_{i_{{\hat \rho}}}$ are fixed complex coefficients. 

In order to prove the  statement of linear independence below  for superpositions of states with distinct Bra Set labels we need  a further property (iv)(b) of permissible Bra Set labels beyond 
(iv)(a), section \ref{secl5}:\\

\noindent{\bf Condition (iv)(b)}: Consider two distinct Bra Sets $B_1,B_2$ each satisfying (i)-(iv)(a).  
Let $S \in B_1$  and  let the family of states $S_{{\bf K}_0, [S]_{B_1}}({\bf K}_{\{ \{x\}_i,\e_i, p_i, u_i, v_i, m_i, \delta_i (m_i)  \}}) \in B_1$  for all  parameter specifications $\{ \{x\}_i,\e_i, p_i, u_i, v_i, m_i, \delta_i (m_i)  \}$ 
described in (iv)(a). Then for sufficiently small $\{\e_i\}$, either all of $S,S_{{\bf K}_0, [S]_{B_1}}({\bf K}_{\{ \{x\}_i,\e_i, p_i, u_i, v_i, m_i, \delta_i (m_i)  \}})$ are elements of $B_2$ or none of 
$S,S_{{\bf K}_0, [S]_{B_1}}({\bf K}_{\{ \{x\}_i,\e_i, p_i, u_i, v_i, m_i, \delta_i (m_i)  \}})$ are in $B_2$.
\\

\noindent{\bf Statement} Let $\Psi_{h_{{\hat \rho}}}\neq 0$ be as defined in (\ref{l61}) with  $\{h_{{\hat \rho}}, {\hat \rho}=1,..,n\}$ being distinct metric labels.
Let $a_{{\hat \rho}} , {\hat \rho}=1,..,n$ denote complex numbers. Then 
\be
\sum_{{\hat \rho} =1}^n a_{\hat \rho} \Psi_{h_{{\hat \rho}}}= 0 \Rightarrow a_{\hat \rho}= 0, {\hat \rho}=1,..,n .
\label{linindep1}
\ee

\noindent{\bf Proof}: Since  we are given the distinct metrics $h_1,h_2..,h_n$, we can fix the data\\
$\{ \{x\}_{\di},\e_{\di}, p_{\di}, u_{\di}, v_{\di}, N_{\di}, \delta_{\di} (N_{\di}), \di=1,..,\dm  \}$ and for each datum  we have the  Kink Set\\
${\bf K}_{\{ \{x\}_{\di},\e_{\di}, p_{\di}, u_{\di}, v_{\di}, N_{\di}, \delta_{\di} (N_{\di}), \di=1,..,\dm  \}}$
exactly as in section \ref{secl3}.
Let $S \in \cup_{i_1 }B^{(1)}_{i_{{1}}} $ be such that $\Psi_{h_1} (S) \neq 0$; if  such an (asymmetric spin net basis element (see (i),section \ref{sec3.2})) $S$ did not exist then $\Psi_{h_1}$ would be trivial. 
This means that there exists some fixed value of $i_1=j$ such that $S\in B^{(1)}_j$. 

Property (iv)(a) then implies that 
$S_{{\bf K}_0, [S]_{B^{(1)}_j}}(  {{\bf K}_{\{ \{x\}_{\di},\e_{\di}, p_{\di}, u_{\di}, v_{\di}, N_{\di}, \delta_{\di} (N_{\di}), \di=1,..,\dm  \}}}) \in B^{(1)}_j$ for all parameter specifications detailed in (iv)(a).

Define the set ${\cal I}_{\bar S}^{({\hat \rho})}$ to be the set of values of the index $i_{\hat \rho}$ for which ${\bar S} \in B^{({\hat \rho})}_{i_{\hat \rho}}$.
From (iv)(a,b) we have that:
\be
{\cal I}_{ S}^{({\hat \rho})}= {\cal I}_{   S_{{\bf K}_0, [S]_{B^{(1)}_j}}(  {{\bf K}_{\{ \{x\}_{\di},\e_{\di}, p_{\di}, u_{\di}, v_{\di}, N_{\di}, \delta_{\di} (N_{\di}), \di=1,..,\dm  \}}})                                   }^{({\hat \rho})}
\label{caliiv}
\ee
%$\Theta (S, B^{({\hat \rho})}_{i_{\hat \rho}}) =1$.
From (\ref{l61}) we have that:
\be
\Psi^{}_{h_1} (S)=
(\sum_{i_{1}\in {\cal I_S}^{(1)} } c^{({1})}_{i_{1}} \F^{(1)}_{i_{1}}([S]_{B^{(1)}_{i_1}})) g_{S,h_1}, 
\label{difb1}
\ee
From (\ref{l61}) and (\ref{caliiv}) we have that:
\ba
&\Psi^{}_{h_1} (S_{{\bf K}_0, [S]_{B^{(1)}_j}}(  {{\bf K}_{\{ \{x\}_{\di},\e_{\di}, p_{\di}, u_{\di}, v_{\di}, N_{\di}, \delta_{\di} (N_{\di}), \di=1,..,\dm  \}}})) 
& \nonumber\\
&=\big( \sum{}_{i_1\in
{\cal I}_{       S_{{\bf K}_0, [S]_{B^{(1)}_j}} (  {{\bf K}_{\{ \{x\}_{\di},\e_{\di}, p_{\di}, u_{\di}, v_{\di}, N_{\di}, \delta_{\di} (N_{\di}), \di=1,..,\dm  \}}})         }^{(1)}         }  &\nonumber\\            
&c^{({1})}_{i_{1}} \F^{(1)}_{i_{1}}([ S_{{\bf K}_0, [S]_{B^{(1)}_j}} (  {{\bf K}_{\{ \{x\}_{\di},\e_{\di}, p_{\di}, u_{\di}, v_{\di}, N_{\di}, \delta_{\di} (N_{\di}), \di=1,..,\dm  \}}})]_{B^{(1)}_{i_1}}    )\big)&\nonumber\\    
&g({\bf K}_0, h_1)  g({{\bf K}_{\{ \{x\}_{\di},\e_{\di}, p_{\di}, u_{\di}, v_{\di}, N_{\di}, \delta_{\di} (N_{\di}), \di=1,..,\dm  \}}},h_1) &\nonumber \\
&=
(\sum_{i_{1}\in {\cal I_S}^{(1)} } c^{({1})}_{i_{1}} \F^{(1)}_{i_{1}}([S]_{B^{(1)}_{i_1}}))
%&\nonumber\\
g({\bf K}_0, h_1)  g({{\bf K}_{\{ \{x\}_{\di},\e_{\di}, p_{\di}, u_{\di}, v_{\di}, N_{\di}, \delta_{\di} (N_{\di}), \di=1,..,\dm  \}}},h_1)&
\label{difb2}
\ea
where in the last line we have used the fact that since  $S,S_{   {\bf K}_0, [S]_{B^{(1)}_j}} (  {{\bf K}_{\{ \{x\}_{\di},\e_{\di}, p_{\di}, u_{\di}, v_{\di}, N_{\di}, \delta_{\di} (N_{\di}), \di=1,..,\dm  \}}}) \in [S]_{B^{(1)}_j}$,
it must be the case from (\ref{caliiv})  that for all $i_1 \in {\cal I_S}^{(1)}$, we have that 
\\
$S,S_{{\bf K}_0, [S]_{B^{(1)}_j}} (  {{\bf K}_{\{ \{x\}_{\di},\e_{\di}, p_{\di}, u_{\di}, v_{\di}, N_{\di}, \delta_{\di} (N_{\di}), \di=1,..,\dm  \}}}) \in [S]_{B^{(1)}_{i_1}}$.

Equations (\ref{difb1}), (\ref{difb2}) together with the fact that $\Psi_1(S)\neq 0$ implies that:
%It is straightforward to see that this,  together with fact that we have a single metric label $h_1$ for all the states 
%$\Psi_{B^{(1)}_{i_{{1}}, \F^{(1)}_{i_{1}},h_{1}}}$, 
%ensures that:
\be
\Psi_{h_1}   ( S_{{\bf K}_0, [S]_{B^{(1)}_j}}(  {{\bf K}_{\{ \{x\}_{\di},\e_{\di}, p_{\di}, u_{\di}, v_{\di}, N_{\di}, \delta_{\di} (N_{\di}), \di=1,..,\dm  \}}})   ) \neq 0.
\ee
It is then straightforward to check that  the  proof proceeds identically to that in section \ref{secl5} above with the substitutions in that proof of:
\ba
& i\rightarrow {\hat \rho} \;\;\; \;\;\;\;\;\;\;\;\; \\
&  \Psi_{B,\F_i,h_i}(S)                               )
 \rightarrow \Psi_{h_{{\hat \rho}}}(S). \\
& \lim_{{\vec \e}\rightarrow 0}E({\vec \e},{\vec N}) 
\Psi_{B,\F_i,h_i}(S_{{\bf K}_0, [S]_B}({\bf K}_{\{ \{x\}_{\di},\e_{\di}, p_{\di}, u_{\di}, v_{\di}, N_{\di}, \delta_{\di} (N_{\di}), \di=1,..,\dm  \}}) \nonumber\\
&= \F_i([S]_B) g({\bf K}_0, h_i) \alpha_i 
%\gamma_{S_{{\bf K}_0, [S]}({\bf K}_{\{ \{x\}_{\di},\e_{\di}, p_{\di}, u_{\di}, v_{\di}, N_{\di}, \delta_{\di} (N_{\di}), \di=1,..,\dm  \}}) } 
\nonumber \\
&\rightarrow
\lim_{{\vec \e}\rightarrow 0}E({\vec \e},{\vec N}) 
\Psi_{h_{{\hat \rho}}} (S_{{\bf K}_0, [S]_{B^{(1)}_j}}({\bf K}_{\{ \{x\}_{\di},\e_{\di}, p_{\di}, u_{\di}, v_{\di}, N_{\di}, \delta_{\di} (N_{\di}), \di=1,..,\dm  \}}))\\
&= (\sum_{i_{\hat \rho}\in {\cal I}_{ S}^{({\hat \rho})}    } c^{({\hat \rho})}_{i_{{\hat \rho}}}\F^{(\hat \rho)}_{i_{{\hat \rho}}}([S]_{B^{({\hat \rho})}_{i_{\hat \rho}}}) g({\bf K}_0, h_{\hat \rho}) \alpha_{\hat \rho} 
\label{782}
\ea
where as indicated above, the kink set ${\bf K}_{\{ \{x\}_{\di},\e_{\di}, p_{\di}, u_{\di}, v_{\di}, N_{\di}, \delta_{\di} (N_{\di}), \di=1,..,\dm  \}})$ 
%in the last line of (\ref{782}) above
is defined so that  the Lemma of section \ref{secl4} applies with respect to the set of metric labels $\{h_{\hat \rho}\}$.
In particular, we first show that $a_{{\hat \rho}=1}$ vanishes and then through reiteration that $\{a_{{\hat \rho}}=0 , {\hat \rho}=2,..,n\}$. \\
Q.E.D
\\

Let ${\cal L}_h$ be the finite linear span of states, each of the form  $\Psi_{B_i, \F_i, h}$, for the same  metric label $h$.
Let ${\cal L}$ be the finite linear span of states each of the form  $\Psi_{B_i, \F_i, h_i}$ where the metric labels $h_i$ are not necessarily the same for all $i$.
It follows from the above statement of linear independence that any element  
%of the linear span  
of ${\cal L}_h$ 
% off shell basis states with each such basis state labelled 
%by the {\em same} metric label $h$ 
cannot admit a decomposition into a finite set ${\cal S}$ of basis states in which any element of the set ${\cal S}$  which is labelled by 
a metric label distinct from $h$ contributes non-trivially. This implies that any element of the finite span ${\cal L}$ can be {\em uniquely} written as a finite linear combination of states $\Psi_{h_{\alpha}} \in {\cal L}_{h_{\alpha}}$.
Since the dual constraint action on each such $\Psi_{h_{\alpha}}$ employs the RNC's associated with $h_{\alpha}$ it follows that this dual constraint action is unambigous on 
${\cal L}$.
%It is then straightforward to see from this version of the statement that any element of the linear span ${\cal L}_{B}$ of off shell basis states  of the form $\Psi_{B,\F,h}$ for fixed $B$ but 
%arbitrary $\F$ and arbritrary $h$, $h\in {\cal H}_{h_0}$ is uniquely decomposable in terms of these basis states. Since the distributions obtained by the action of the single and product actions of the 
%constraint on such basis states are also of the type $\Psi_{B,\F_i, h_i}$, the unique decomposability extends to the space of states obtained by these actions as well.

It then follows that we can express the contents of section 4 in terms of continuum limit operator actions on domains as follows.
Consider the linear span ${\cal L}$ defined above. Within this define the finite linear span ${\cal L}^{0}$ of basis states for which $\F$ is determined by a $C^r$ function $f$ through (\ref{psiamps}) on $\Sigma$, 
and the finite linear span, ${\cal L}^{I}$,  of states for  which $\F$ is of the form corresponding to the action of a single continuum limit Hamiltonian constraint (\ref{first1}) (see  section \ref{secl5} for 
a discussion on how the constraint action yields functions of the form $\F$).
Then referring to the continuum limit Hamiltonian constraint with lapse $M$ as ${\hat H}(M)$ we may rephrase the results of sections \ref{sec4}, \ref{sec5} and \ref{sec6} as follows.
Define ${\cal D} ={\cal L}^{0} \cup {\cal L}^{I}$ and ${\cal D}_{0} = {\cal L}^{0}$. Then 
${\hat H}(M): {\cal D} \subset {\cal L} \rightarrow {\cal L}$ and ${\hat H}(M) {\hat H}(N): {\cal D}_0 \subset {\cal D} \rightarrow {\cal L}$,
the latter action leading to anomaly free commutators.

\subsection{\label{secl61} Inclusion of Physical states}

The results of the previous section pertained to the finite span of off shell states. Here we extend these results to the vector space sum of this finite span and the vector space of physical states.
The following Lemma plays a key role in this extension.\\

\noindent{\bf Lemma}: Given a finite set of metrics $M= \{h_{\alpha}, \alpha= 1,..,n\}, M \subset {\cal H}_{h_0}$, there exists a diffeomorphism $\psi$ such that 
$\psi (M) \cap M =\emptyset$ where $\psi(M)=\{\psi^*h_{\alpha},\alpha =1,..,n\}$.\\

\noindent{\bf Proof}: Recall that the elements of ${\cal H}_{h_0}$ have no (conformal) symmetries and are diffeomorphic images of each other.
It follows straightforwardly from this fact that given $\alpha, \beta \in \{1,..,n\}$,  there exists a unique diffeomorphism 
$\phi^{(\beta)}_{\alpha}$ such that $h_{\alpha}= (\phi^{(\beta)}_{\alpha})^*h_{\beta}$.
Let $L_{\beta} = \{\phi^{(\beta)}_{\alpha}, \alpha=1,..n \}$. It then also follows from this uniqueness that for fixed $\beta$, any  diffeomorphism $\psi_{\beta} \notin L_{\beta}$ we have that 
\be
\psi^*_{\beta} h_{\beta} \notin M
\label{psinotinlbeta}
\ee
Let $L:= \cup_{\beta=1}^n L_{\beta}$. Equation (\ref{psinotinlbeta}) implies that for any diffeomorphism $\psi \notin L$, we have that:
\be
\psi^*h_{\alpha} \notin M  \forall \alpha \in \{1,..,n\},
\label{psinotinl}
\ee
which proves the Lemma.
\\

%Next consider states $\Psi_[\alpha}$ of the form (\ref{l61}) with the fo

Next consider a state $\Psi_{h_{\alpha}}$ of the form (\ref{l61}).
%\be
%\Psi_{h_{\alpha}} := \sum_{i_\alpha= 1}^{n_{\alpha}} c_{i_{\alpha}} \Psi_{B_{i_{\alpha}}, \F_{i_{\alpha}},h_\alpha}
%\label{l61a}
%\ee
From the definition of the dual action of ${\hat U} (\phi )$ it follows that the amplitude of the state ${\hat U}(\phi) \Psi_{h_{\alpha}}$ on a spin net $S$ 
is the same as the amplitide of the state $\Psi_{h_{\alpha}}$ on $S_{\phi^{-1}}$ so that non-triviality of the former implies non-triviality of the latter.
This, in turn, yields the implication:
\be
\Psi_{h_{\alpha}} \neq 0 \Rightarrow
{\hat U}(\phi) \Psi_{h_{\alpha}} \neq 0
\label{nontrivialuphi}
\ee
The explicit action of ${\hat U}_{\phi}$ on $\Psi_{h_{\alpha}}$ for any diffeomorphism $\phi$ may be computed from  
 (\ref{l61}) and (\ref{uphipsiff}). We obtain:
\be
{\hat U}(\phi) \Psi_{h_{\alpha}} = \sum_{i_\alpha= 1}^{n_{\alpha}} c_{i_{\alpha}} \Psi_{B_{i_{\alpha}}, (\F_{i_{\alpha}})_{\phi^{-1}}, \phi^*(h_\alpha)}.
\ee
The form of above expression implies that 
the metric label of the state ${\hat U}(\phi) \Psi_{h_{\alpha}}$ is $\phi^*(h_\alpha)$. To emphasise this fact we  denote
${\hat U}(\phi) \Psi_{h_{\alpha}}$ by $\Psi^{(\phi)}_{ \phi^*(h_\alpha)}$; this notation will prove useful below.

Next we prove the following statement:\\

\noindent{\bf Statement} Let $\Psi_{h_{\alpha}}\neq 0$ be as defined in (\ref{l61})  with  $\{h_{\alpha}, \alpha=1,..,n\}$ being distinct metric labels.
Let $a_{\alpha} , \alpha=1,..,n, b$ denote complex numbers. Let $\Psi$ be a diffeomorphism invariant state. Then 
\be
\sum_{\alpha =1}^n a_\alpha \Psi_{h_{\alpha}} + b \Psi = 0 \Rightarrow a_\alpha= 0, i=1,..,n,\; b=0 .
\label{linindepphys}
\ee

\noindent{\bf Proof}: Let the diffeomorphism  $\psi$ be chosen as in the Lemma above so that (\ref{psinotinl}) holds. Denote the unitary operator corresponding to $\psi$ by ${\hat U}_{\psi}$.
The  action of  ${\hat U}_{\psi}$ on the condition
\be 
\sum_{\alpha =1}^n a_\alpha \Psi_{h_{\alpha}} + b \Psi = 0 
\label{lip1}
\ee
yields:
\be 
\sum_{\alpha =1}^n a_\alpha \Psi^{(\psi)}_{\psi^*h_{\alpha}} + b \Psi = 0 
\label{lip2}
\ee
where we have used the diffeomorphism invariance of $\Psi$ and the notation introduced above.
Subtracting (\ref{lip2}) from (\ref{lip1}) yields:
\be
\sum_{\alpha =1}^n a_\alpha \Psi_{h_{\alpha}}  - \sum_{\alpha =1}^n a_\alpha \Psi^{(\psi)}_{\psi^*h_{\alpha}}= 0.
\label{lip3}
\ee

Equations (\ref{psinotinl}), (\ref{nontrivialuphi}) together with the results of section \ref{secl6} ( see the Statement therein) imply that 
\be
a_{\alpha}= 0, \;{\rm for}\; \alpha=1,..,n.
\label{lip4}
\ee
Equation (\ref{lip4}) then implies that $b=0$ in the linear independence condition in (\ref{linindepphys}), thus proving the Statement.
\\

Next, note that any physical state must be diffeomorphism invariant so that the (\ref{linindepphys}) holds. Further any physical state is annihilated by the Hamiltonian constraint.
From these facts together with the operator domain based rephrasing  of constraint actions at the end of  section \ref{l6}, it is immediate that this rephrasing can be extended to 
accomodate physical states as follows.

Let ${\cal L}_h$, ${\cal L}, {\cal L}^0,{\cal L}^I$  be as in section \ref{l6}. Let the vector space of physical states be $V_{phys}$ with $V_{phys}$ being a subset of the algebraic dual space to the finite 
span of spin network states. Let ${\cal L}^{\prime} = {\cal L}\oplus V_{phys}$.
Any element of ${\cal L}^{\prime}$ can be {\em uniquely} written as a finite linear combination of states $\Psi_{h_{\alpha}} \in {\cal L}_{h_{\alpha}}$ together with an element of $V_{phys}$.
Since the dual constraint action on each such $\Psi_{h_{\alpha}}$ employs the RNC's associated with $h_{\alpha}$ and since the dual constraint action kills elements of $V_{phys}$,  it follows that this dual constraint action is 
unambiguous on 
${\cal L}^{\prime}$. Define 
${\cal D}^{\prime} ={\cal L}^0\oplus {\cal L}^I \oplus V_{phys}$ and 
${\cal D}^{\prime}_0:= {\cal L}^{0} \oplus V_{phys}$. Then
${\hat H}(M): {\cal D}^{\prime} \subset {\cal L}^{\prime} \rightarrow {\cal L}^{\prime}$ and ${\hat H}(M) {\hat H}(N): {\cal D}^{\prime}_0 \subset {\cal D}^{\prime} \rightarrow {\cal L}^{\prime}$,
the latter action leading to anomaly free commutators.

Note that in the above phrasing we have defined physical states as diffeomorphism invariant states annhiliated by the constraint action. Since the constraint action is defined through a continuum limit 
of finite approximants and since each of these approximants rely on RNCs associated with some metric $h$, $h\in {\cal H}_{h_0}$, the question arises as to which $h$ to use in defining the 
action of the constraint on physical states. We could arbitrarily choose some such $h$, fix it once and for all and demand that physical states be diffeomorphism invariant and killed by the
constraint action regulated with respect to this fixed $h$ without affecting any of the statements made hitherto. However, the physical state space would then be expected to 
carry an imprint of this choice of $h$. Hence
as in \cite{p4} we shall define  a physical state $\Psi$ to be a diffeomorphism state which is annihilated by the continuum limit actions of 
the $h$- dependent regulated Hamiltonian constraint {\em for all $h\in {\cal H}_{h_0}$} i.e.  $\Psi \in V_{phys}$ iff:
\ba
{\hat U}(\phi) \Psi &=& \Psi  \;{\rm for}\;{\rm  all}\; {\rm semianalytic} \; {\rm diffeomorphisms}\;\phi \label{physdef0}\\
\lim_{\e\rightarrow 0} \Psi ({\hat H}_\e(M) |s\ket) &=& 0 \;\forall\; M, \;|s\ket \label{physdef}
\ea
where  (\ref{physdef})  holds for the regulated actions  ${\hat H}_\e(M)$ with respect to all choices of regulating metrics  $h\in {\cal H}_{h_0}$.
We discuss this definition and related issues in section \ref{sec8.0}.

%Note that we could also demand annihilation with respect to the continuum limit of a regulated constraint with regulation
%with respect to a particular element of ${\cal H}_{h_0}$ and then demand that all members of the equivalenc

\subsection{\label{secl7} Discussion of Property (iv)}

%In particular, if we restrict attention to the finite span of states with a single Bra Set label $B$ which satisfies a further technical property beyond those specified in section \ref{sec3.2},
%we obtain linear independence and uniqueness of decomposition. The technical property in question can be satisfied so such Bra Sets do exist.
%It seems us that this property could follow from properties (i)- (iii), section \ref{sec3.2} if certain structural properties of diffeomorphism type children hold in relation to structural properties of their
%parents. If these structural properties hold, a statement of linear independence of off shell basis states  with distinct Bra Set labels  follows.
%While these structural properties (or some slight variation thereof) seem  to follow plausibly from the nature of the diffeomorphism type deformations together with the results on compactly supported diffeomorphisms
%discussed in Appendix, we leave a comprehensive proof for future work.  
With regard to the satisfaction of property (iv) there are two possibilities. The first is that property (iv) is a consequence of the properties detailed in section \ref{sec3.2}. The second possibility is that 
property (iv) is independent of these properties so that permissible Bra Sets need to be suitably enlarged so as to satisfy property (iv) in addition to (i)-(iii) of section \ref{sec3.2}. We shall first 
discuss how property (iv) may follow from  (i)-(iii) and then discuss a specific enlargement  of permissible Bra Sets which accomodates property (iv).

%While (iv) may seem unduly restrictive, we find it plausible that this is not the case for the following reason.
%We discuss our reasons for believing in this plausibility after our proof of linear independence  below.
Let $ B$ be  subject to (i)-(iii) section \ref{sec3.2} and let $S\in B$. 
%Given the set ${\bf K}_{\{ \{x\}_i,\e_i, p_i, u_i, v_i, m_i, \delta_i (m_i)  \}}$, 
Property (ii) section \ref{sec3.2}
implies that we can move the kinks in $S$ ever so slightly so as to avoid any given set of  points $\{p_i\}$ and still remain in $B$. Call the resulting spin network  $S^{\prime} \in B$.  Let ${\bf K}_0$  be
the kink set of $S^{\prime}$. Repeated action of the constraint at the vertices of  $S^{\prime}$ creates `diffeomorphism' type children 
with a nest around each vertex of an  ever increasing number of kink triplets. 
Using diffeomorphisms of the type discussed in Appendix \ref{seca4}, together with path connectedness of $\Sigma$ we conjecture that it should be possible to  move these kinks around to other locations in $\Sigma$ without moving 
the vertices of  $S^{\prime}$. 
This together with any other deformations which preserve property  (ii), section \ref{sec3.2}
suggests that we may be able move 
these triples to appropriate  locations and deform the graph structure  while remaining in $B$  so as to yield the kink set ${\bf K}_{\{\{x\}_i, \e_i, p_i, u_i, v_i, m_i, \delta_i (m_i)  \}}$  and  satisfy property (iv)(a)
with property (iii) section \ref{sec3.2} enforcing property (iv)(b).
%This together with 
%property  (ii), section \ref{sec3.2} leads us to believe that it is plausible 
%In this manner it is plausible that property (iv)  is satisfied by $B$. 
We leave the fleshing out of these ideas into a putative proof that (iv) follows from (i)-(iii) for future work.

In the event that (iv) does not follow from (i)-(iii), a detailed specification of a rich collection of Bra Sets which satisfy properties (i) to (iv) is as follows.
The arguments  at the end of section \ref{sec3.2} show that  
properties  (i)-(iii) hold if we choose as our Bra Set, the set 
${\cal E}_{\gamma_{0c,d}}$ defined in \ref{un}, section \ref{sec3.1},    with ${\gamma_{0c}, d}$ chosen so that:\\
\noindent (a) All vertices of valence greater than 3 are embedded as GR vertices.\\
\noindent (b) Any vertex of valence 3 is embedded either as GR vertex or as a trivalent kink.\\

Consider as Bra Sets,  the family of sets ${\cal E}=    \{{\cal E}_{\gamma_{0c,d}}\}$ for all
 for all 
(i.e. non-isomorphic) choices of ${\gamma_{0c}}$ and, for each fixed $\gamma_{0c}$, all distinct choices of $d$  subject to (a),(b).
We now show that this family of sets satisfies
property (iv). 

First recall  that the elements of each ${\cal E}_{\gamma_{0c,d}}$ with fixed $\gamma_{0c},d$ are orthogonal to each other. Second note that any element of 
${\cal E}_{\gamma_{0c,d_1}}$ is orthogonal to any element of ${\cal E}_{\gamma_{0c,d_2}}$ if the decorations $d_1$ and $d_2$ are distinct. Finally, note that 
the elements of the  sets ${\cal E}_{\gamma^{(1)}_{0c,d_1}},{\cal E}_{\gamma^{(2)}_{0c,d_2}} $ based on non-isomorphic asymmetric colored graphs $\gamma^{(1)}_{0c}, \gamma^{(2)}_{0c}$
are also orthogonal to each other.
%are mutually orthogonal in the sense that the elements of one set in the family are orthogonal to every element of a distinct set of the family
%by virtue of the fact that any embedding of any abstract decorated spin network on $\gamma^{(1)}_{0c}, d^{(1)}$ is orthogonal to any embedding of any abstract decorated spin network $\gamma^{(2)}_{0c}, d^{(2)}$
%if either (1)  $\gamma^{(1)}_{0c}$ is not isomorphic to $\gamma^{(2)}_{0c}$ or (2) $\gamma^{(1)}_{0c}$ is isomorphic to $\gamma^{(2)}_{0c}$ but the decorations $d^{(1)}$ and $d^{(2)}$ are not images of each
%other by the isomorphism which maps $\gamma^{(1)}_{0c}$  to $\gamma^{(2)}_{0c}$ (by which we mean that the decorated vertices of $\gamma^{(1)}_{0c}$ are mapped to vertices with distinct decorations
%in $\gamma^{(2)}_{0c}%\footnote{Note that any such isomorphism is unique by virtue of the asymmetry of  $\gamma^{(1)}_{0c}$, $\gamma^{(2)}_{0c}$.}
This orthogonality implies that  (iv)(b) is automatically satisfied as  Bra Sets are either identical or  have no overlap. 
It remains to show that (iv)(a) holds. 

Consider one of the elements ${\cal E}_{\gamma_{0c,d}}$ of the family of sets ${\cal E}$. Clearly for any $S \in {\cal E}_{\gamma_{0c,d}}$  there exists a spin net $ S^{\prime}\in {\cal E}_{\gamma_{0c,d}}$ such 
that ${S}^{\prime}$ is obtained by:\\
\noindent (1) removing all $C^0$ loops from $S$,
\noindent (2) moving the kinks in the remaining non-loop component of $S$ ever so slightly so as to avoid the coincidence with any element of a given  set of a finite number of points $\{p_i\}$ without changing the embeddable abstract  spin network
which embeds to $S$.
\footnote{ This can be done  by the action of diffeomorphisms each of  which is  identity outside an arbitrarily small neighborhood of the kink to be moved (see Appendix \ref{seca4} for the construction of such diffeomorphisms).}
%or by moving the contents of a small neighborhood of the kink ever so slightly and  reconnecting the edges outside this neighborhood with the contents of the neighborhood of the kink so moved smoothly using the techniques
%of $C^r$ joins described in Appendix \ref{seca2}
%and bivalent kinks from the non-loop component of $S$ i.e. ${S}^{\prime}$ is the embedded image of an abstract decorated  spin network on the decorated colored graph $\gamma_{0c,d}$.
Next consider the set ${\bf K}_{\{ \{x\}_i,\e_i, p_i, u_i, v_i, m_i, \delta_i (m_i) \}}$;
${\bf K}_{\{ \{x\}_i,\e_i, p_i, u_i, v_i, m_i, \delta_i (m_i) \}}$  is a finite set of distinct  points with locations specified approximately (i.e. upto a certain order in the small parameters $\{\e_i \}$ (see section \ref{secl}))
and which can therefore  be arranged not to intersect $S^{\prime}$.
Next, from the specification of the positions of the elements of this set, 
it is straightforward to see that for small enough $\{\e_i\}$ the 3 point sets in the $i$th nest each lie in  distinct (coordinate) spherical shells around $p_i$. The larger the interpoint distance the larger the radii of the inner and outer
boundaries of the shell containing the 3 point set.  It is clearly possible to connect the 3 points in the shell  by a single piecewise analytic (with respect to $\{x\}_i$)  loop with bivalent $C^0$ kinks at these points such that this loop 
does not intersect $S^{\prime}$ and such that this loop is confined to the shell.
Using the techniques of $C^k$ joins   we may smoothen this curve in a $C^r$ manner at all points of non-analyticity other than these 3. Finally, it is straightforward to see that  with a slight modification of the smoothening procedure
outlined in  Appendix \ref{seca2.2} we can  smoothen one of these points to a $C^1$ kink and the other to $C^2$ kink.
%md  smoothen the curve at these  3 points 
%so that these 3 points appear as $C^q, q>2$ bivalent kinks. 
In this manner for all small enough $\{\e_i\}$ we obtain a set of non-intersecting loops one for each 3 point set in each nest. Let there be a total of  $m$ such loops. Number them arbitrarily from $1,..m$,  color the $j$th 
loop so numbered by the spin $j$ and adjoin them to $S^{\prime}$ to obtain a spin network which is an embedding of its abstract decorated counterpart on $\gamma_{m,c,d}$ (see \ref{un}, section \ref{sec3.1}).
The state so obtained is clearly in ${\cal E}_{\gamma_{0c,d}}$ and is the desired state specified by (iv)(a) so that (iv)(a) holds.

\section{\label{sec7} Assorted Technicalities} 

\noindent (i) Loop Areas in the GR case:  As noted in section \ref{sec2.1}, control over areas of the loops $l_{IJ,\e}$ is required to obtain the desired numerical factors for various terms in the 
constraint action.
As outlined in section \ref{sec2.3.1}, the desired areas for the Hamiltonian constraint action can be obtained by adjusting the initial kink positions prior to their movement and/or smoothening. As can be 
seen in Appendix \ref{a2}, kink movement, kink  smoothening and upward conicality can be implemented with arbitrarily small change in these areas by choosing the supports of the various 
compactly supported functions encountered to be arbitrarily small.
As outlined in the latter part of Appendix \ref{sec4.0b}, the desired loop areas generated by the action of the quantum correspondent ${\hat O}(M,N)$ of the Hamiltonian constraint Poisson bracket can be obtained through the action of the 
diffeomorphisms constructed therein by a suitable choice of supports of the compactly supported functions and the iteration numbers $N_{\lambda}$ which define these diffeomorphisms.
The arguments provided, in our opinion, suffice at a physicists level of rigor to show that that prescribed areas can be plausibly obtained. However it would be desireable to 
develop an explicit demonstration that this possible.
%
%We note here that the arguments in Appendix \ref{sec4.0b} pertaining to obtaining desired areas through  the action of diffeomorphisms could be made more watertight. In their present form, while they 
%show conclusively that loop areas can be changed and while the constructions have enough freedom to make it quite plausible that these changes can obtain prescribed areas, 
%an explicit  demonstration that a prescribed area of $\O (\e^2)$ can be obtained would be desireable.
On the other hand as discussed above, prescribed loop areas {\em can} be obtained for the Hamiltonian constraint  without recourse to the constructions of section \ref{seca4}.
Hence, it is possible to leave the loop areas in section \ref{sec5}  unchanged and instead alter those associated with the Hamiltonian constraint so as to achieve anomaly freedom.
More in detail, let us alter the loop areas in the constraint action (\ref{abaction}) by choosing $a_{I_v}, b_{I_v}$ in (\ref{abaction0}) as follows (here we have used obvious notation in  augmenting the 
edge index by the vertex subscript $v$). We retain the same relation between $a_{I_v}, b_{I_v}$ as before so that  $b_{I_v}= \frac{4}{3}(N-1)a_{I_v}$ 
and set $a_{I_v}= d_{I_v}j_{I_v}(j_{I_v}+1)$  in (\ref{abaction0})  (instead of $a_{I_v}=j_{I_v}(j_{I_v}+1)$), with $d_{I_v}$ some constant. It is then straightforward to verify that 
the single and product constraint actions are still well defined in the continuum limit and that we obtain  anomaly free commutators {\em without} changing the loop areas in section \ref{sec5}
(i.e. we set $c_{I_v}=1$ there instead of as in (\ref{ci=128jj})) if we choose $d_{I_v}$ as
\be
d_{I_v}:= \frac{ \sqrt{128 \sum_{J_v=1}^{N_v}j_{J_v}(j_{J_v}+1)}}{128 j_{I_v}(j_{I_v}+1)}
\ee

Finally, modulo our comments above, note that even for the Hamiltonian constraint action, deformations (including those of displaced vertex movement, kink movement and cone stiffening) at different values of regulator parameter
consistent with desired loop areas can be related by the action of diffeomorphisms as discussed in Appendix \ref{seca4}.\\

%\noindent (ii) Permissible Bra Sets: As indicated at the end of section \ref{sec3.2}, conditions (i)-(iii) suffice for our demonstration of anomaly freedom in sections \ref{sec4}-\ref{sec6}.
%For the purposes of this demonstration, it seems plausible to us that condition (iii) can be weakened so as to demand closure with respect to possible parents and children related solely through diffeormorphism 
%type deformations generated by the constraint; propagation deformation  mediated heredity does  not seem to be a necessary property because the contribution of propagation type children vanishes
%in the continuum limit. We have retained the stronger condition (iii) in section \ref{sec3.2}
%as it is easier to state and suffices for our purposes but it would be of interest to confirm that the weakening above also suffices.
%\\

\noindent (ii) Choice of Constraint Action in the NGR case: \\
\noindent (a) The constraint action is assumed, without derivation, to be of the general form (\ref{ngr}). Here we show how a particular case of this general form can be derived using the considerations of 
\cite{p4}.
%in conjunction with the routings of deformed edges specified by  \cite{qsd}. 
The derivation of the `one edge at a time' action of section 3 of \cite{p4} holds for NGR vertices $v$.
In that derivation the curvature contribution from  an edge $e_J$ is taken to be proportional to the area of the small loop $l_{IJ,\e}$ \cite{p4}, and,  
with a natural visualization of the map  $\phi_{I,\e}$, this area is of $O(\e^3)$ for edges whose tangents at $v$ are collinear with the tangent of the $I$th edge at $v$.
This is a subleading contribution and so in the heuristic argumentation underlying the derivation of \cite{p4}, we may drop such contributions.
Thus we need to deform only those edges `one at time'
%only those edges $\{e_J\}$  
which have  tangents at $v$  which are non-collinear with tangent of the $I$th edge at $v$.
%are deformed by the finite deformation $\phi_{I,\e}$.

It remains to provide a `routing' for such a deformation such that the deformed edge does not intersect any of the other undeformed edges except the $I$th one at $v_{I,\e}$.
Given such a routing prescription, we shall take the view that $\phi_{I,\e}$ can be identified {\em a posteriori} as a deformation map which implements this routing.
%One possibility is to attempt to deform each such edge can then be deformed one at a time
%according to 
One possibility may be to take recourse to
the routing prescription for `extraordinary' edges developed in  Lemma 3.1 of  \cite{qsd}; however the detailed  prescription relies on {\em analyticity} of edges whereas the edges
of interest here are semianalytic. Instead we adopt the following routing procedure, which suitably adapts the procedure for the GR case described in Appendix \ref{seca2.0}.
In a small enough neighborhood $U_{I,v}$  of  $v$, we use the coordinate patch $\{x\}_I$ and identify the kink locations ${\tilde v}_J$ as desired on the edges of interest.
Note that for small enough  $U_{I,v}$ the edges emanating from $v$ outside of $v$ are {\em analytic} in $\{x\}_I$.

An explicit 
way to locate the kinks ${\tilde v}_J$ based on a visualization of $\phi_{I,\e}$ close to that of the semianalytic diffeomorphism ${\bar \phi}_{I,\e}$ of Appendix \ref{sec4.0b}
is to locate them at the intersection of $e_J$ with the boundary of a coordinate sphere of radius of $\O(\e)$ (with $\e$ small enough that this ball lies well within $U_{I,v}$), this radius being 
such that $v_{I,\e}$  is located at the intersection of the boundary of this sphere with $e_I$ at a geodesic distance $\e$ from $v$.
Connect these kinks so located to 
 $v_{I,\e}$ exactly as in  Appendix \ref{seca2.0} through the straight lines $\{l_J\}$. 
Since the lines $\{l_J\}$ are analytic in this chart and since the undeformed edges are also analytic outside $v$ (see above), it is straightforward to see that  the geometry of this configuration of lines and edges
is such that:\\
\noindent (1) No 2 lines intersect each other.\\
\noindent (2) Since the line $l_J$ joins ${\tilde v}_J$ to $v_{I,\e}$ only for those edges $e_J$  which have  tangents at $v$ which are non-collinear with that of $e_I$, there exists a small enough open ball $B$ around $p$ such that these
lines do not pass through $B$ \\
\noindent (3) Analyticity of the edges away from $v$ together with analyticity of the lines imply that  the collection of intersections of $\{l_{K}\}$ with  $\{e_{J}\}$ are isolated points away from $v$.

In the neighborhood of any such point we can slightly deform the relevant straight lines by `lifting' them off the point of intersection one at a time 
either (1) by  the action of a suitable diffeomorphism on the line to be `lifted', of the type constructed in Appendix \ref{sec4.0a}
(which is identity outside an arbitrarily small compact set around this point), or,  (2) through the replacement of a small segment of this line containing the  intersection point by a pair of segments
which meet each other away from the intersection point in a kink, and with each segment meeting the remaining part of the line also at a kink, followed by smoothening these 3 kinks using  the techniques of Appendix \ref{seca2.0}.

Finally, we move the kinks ${\tilde v}_J$ to their desired positions $O(\e^q)$ away from $v$ using either appropriate diffeomorphisms supported in small cylindrical neighborhoods around these edges
as in Appendix \ref{seca4.1} or through the $C^r$ join  techniques of Appendix \ref{seca2.1}. Note that for small enough cylindrical neighborhoods in the former approach or small enough $\delta_1$ in the latter,
no other edge presents itself in the region of interest
\footnote{ In a small enough neighborhood $U_{J,v}$ of $v$, we may choose coordinates $\{x\}_J$ so that $e_J$ is along the $z$  axis. 
%For $U_{J,v}$ small enough, by virtues of their semianalyticity all the 
%undeformed edges within this neighborhood are {\em analytic} in these coordinates. 
The cylindrical  neighborhood $C_{\delta^m}$of interest is of radius $\delta^m, m\geq 1, \; \delta <<\e$ around the $z$ axis, $v \notin  {\bar C}_{\delta^m}$.
Clearly ${\bar C}_{\delta^m} \subset {\bar C_{\delta}} \;\forall m>1$.
The $x,y$ coordinates of any edge $e_{K\neq J}$ are $C^r$ functions $x(t_K), y(t_K)$ of the parameter $t_K$  along  $e_K$  
Let   $e_K$  intersect the cylindrical neighborhood  $C_{\delta^m}$ of interest for arbitrary large  $m$, at parameter $t^m_K$.
Since $ d(t_K)= (x(t_K))^2 +  (y(t_K))^2$ is a  continuous function of $t_K$, compactness of $e_K\cap {\bar C_{\delta}}$ and the convergence of the sequence $d(t^m_K)$ to zero implies
that  $d(t_K)$ attains its infimum of zero on $e_K\cap {\bar C_{\delta}}$  which in turn implies that $e_K$ intersects $e_J$ outside of $v$ which is not the case.
}
so that there is no obstruction to the application of these techniques. 

To summarise $\phi_{I,\e}$ is chosen to be identity on edges which have collinear tangents with that of $e_I$ at $v$ and deforms the remaining edges as above and thereby defines the `one edge at a time' deformation.
 Accordingly in (\ref{ngr}) we set $B=A_I= 0, I=1,..,N$ and define $P_{\alpha_{IJ}}$ for the  edges $\{e_J\}$ through equation (3.24) of \cite{p4} with $P_{\alpha_{IK}}=0 $
for those edges $\{e_K\}$ which have  edge tangents collinear with that of $e_I$ at $v$.\\

\noindent (b) Based on preliminary results we anticipate that with further work it should be possible to also obtain an  NGR vertex analog of the `Mixed Action' for GR vertices used in this work
with deformations of edges $\{e_K\}$ with tangents collinear with that of $e_I$ at $v$ involving loops $l_{IK,\e}$ of areas subleading to $\O(\e^2)$.
We leave this for future work.\\

\noindent (iii) An Important Technicality with regard to Semianalytic Structures:  Semianalytic structures (manifolds, diffeomorphisms, hypersurfaces) play a key role in the present foundations of 
LQG and ensure the uniqueness of its kinematics \cite{lost}.  It is not clear to us if the notion of semianalyticity has a natural and useful extension to {\em differential} geometric structures 
such as vector fields and metrics, the reason being that we are unable to demonstrate that semianalyticity of components in a semianalytic chart is invariant under change of chart due to the 
occurrence of Jacobian factors. Related to this, is our inability to articulate `the generation of semianalytic diffeomorphisms from semianalytic vector fields' as a well defined statement
due to the possible ill definedness of the semianalyticity of vector fields. Indeed, it is precisely in  order to circumvent this putative obstruction that we have taken recourse to the constructions of Appendix \ref{seca4}.
An investigation into a possible natural and useful extension of the concept of semianalyticity to differential geometric structures and the establishing of the precise nature of vector field generators of 
semianalytic diffeomorphisms constitutes an important task for the future.

\section{\label{sec8} Concluding Remarks}

\subsection{\label{sec8.0} Physical States and the issue of Regulator Dependence}

As noted in section \ref{secl61},  the constraint action is defined through a continuum limit 
of finite approximants which rely on their construction on some $h\in {\cal H}_{h_0}$, raising the question as to which particular $h$ to use in defining 
%each of these approximants rely on RNCs associated with some metric $h$, $h\in {\cal H}_{h_0}$, the question arises as to which $h$ to use in defining the 
action of the constraint on physical states. While an arbitrary choice does not create any inconsistency, it does raise the issue of whether the physical state
space then carries an imprint of this choice.  It is quite plausible to us that with enough care, we could arrange for the state deformations, generated by constraint actions defined using different elements of 
${\cal H}_{h_0}$, to be related by diffeomorphisms given that ${\cal H}_{h_0}$ is the space of metrics diffeomorphic to $h_0$. In such a case, while the diffeomorphism invariance of physical states may
wash away most imprints of any individual choice of $h$, it may still be the case that the physical state space carries an imprint of the diffeomorphism class of metrics ${\cal H}_{h_0}$ and one may enquire
as to whether the physical state space would change  if we employed a different diffeomorphism class of metrics. Such a dependence on regulating structure conflicts with the spirit of background independence
and may conceivably result in a similar unphysical dependence for the emergent classical theory. Hence we propose that physical states should be defined to be those states which are annihilated by 
the constraint for all choices of metric regulating structure. 

In this regard note  that the Bra Sets underlying the {\em off shell} (as opposed to physical) states do not change with change in regulating metric due to the role of 
abstract structures in their definition (see section \ref{sec4.3} for related discussion). An obvious question is whether  an appropriate recourse to abstract structures may remove any dependence of physical states on regulating metrics.
For example, given the space of  states annhilated by the constraint action defined with respect to some metric $h$, one may restrict attention to only those 
states for which   the coefficients of  their bra summands  have the same value for {\em all} bra states which are  embeddings of the same decorated abstract spin network.
Note that the decorations we have defined in section \ref{sec3.1} correspond to a specification of differential structure properties of edges at vertices; no global features such as knotting 
are specified. As a result, any solution of interest with a certain bra summand $ s$ would have to admit all possible knotted embeddings of the decorated abstract spin network which embeds to $s$.
Since the constraint action, at least on GR vertices, does not create non-trvially knotted deformation structures, 
it could be the case that the  restricted class of solutions is then too small. 
In this regard, note that the absence of non-trivial knotting of deformation structures at GR vertices remains true {\em irrespective} of the specific metric used;
%the constraint actions do not  create ``knotted'' deformation structures at least at GR vertices
\footnote{As we shall see in section \ref{sec8.1}, it may be possible to define solutions based only on deformations of GR vertices through an appropriate choice of kink differentiability.} hence it may be appropriate to 
enlarge the kind of decorations considered to include more global features such as knotting (and lack thereof) and then look for an appropriately restricted class of solutions which is  specified by 
unknotted deformations at GR vertices.
Indeed in practice, for example see \cite{ttme} for a class of solutions to Thiemann's QSD constraint \cite{qsd}, we expect that  solutions are built mainly with regard to their abstract connectivity and absence of knotted deformation structure.
In this regard a  more `abstract' graph based physical solution space seems appealing from the viewpoint (and hope)  that the fundamental quantum gravitational  structures are not embedded and that their appearance as embedded structures
in a manifold is emergent. We leave a detailed  investigation of the structure of physical states and their possible independence from regulating metrics for future work.

\subsection{\label{sec8.1}  Choice of Anomaly Free Action, Physical States and Propagation}

In this work we have provided a detailed demonstration of nontrivial anomaly free constraint commutators in vacuum Euclidean LQG
%\footnote{The case of a cosmological constant and of matter couplings represent important open problems for future research.} 
for the specific choice of constraint action (\ref{action}). 
This action is completely specified on GR vertices by the first two lines of (\ref{action}) and is left partially unspecified on NGR vertices in the third line.
We anticipate that similar demonstrations exist for a more general and equally valid choice of  constraint actions for Euclidean LQG,
the common feature of these choices being that the  graph deformations they generate have at their core the finite transformations $\phi_{I,\e}$ generated by the Electric Shift.

These generalizations rely on changes 
in the size of the loops $l_{IJ,\e}$ and the placement and choice of type of kinks as follows.
First note that the factor of $\frac{1}{4}$ in the first line of (\ref{action}) arises from the specific choice of $a_I, b_I$ outlined immediately after (\ref{abaction0}), the freedom in this 
choice arising from the choice of size of the loops $l_{IJ,\e}$.  We anticipate that alternate choices of $a_I, b_I$ which result in a replacement of this factor of 
$\frac{1}{4}$ by any factor less than unity would also provide an anomaly free action. To see this, note that the specific placement of kinks outlined in Appendix \ref{seca3}
yields exactly  a factor of $\frac{1}{4}$ from the kink contraction behavior of the interkink distance function $g$. This conspires with the factor of 
$\frac{1}{4}$ in (\ref{action}) to yield the derivative terms in the continuum limit action (\ref{derivative}), any mismatch of these factors leading to an ill defined continuum limit.
Hence an alternate choice of factor in (\ref{action}) could presumably be compensated with a corresponding choice of placement of kinks so as to yield an identical compensating factor
under kink contraction from $g$, thereby rendering the continuum limit constraint action well defined and resulting in anomaly free commutators.
A second set of generalizations consists in the choice of the differentiability level of these  kinks. 
%Clearly, 
We believe that the current demonstration of anomaly free action appropriate for the generation of $C^0$ kinks
  by electric diffeormorphism type deformations can
be suitably modified so as to go through 
%a demonstration of anomaly free action independent of  kink differentiability type is 
%possible through a 
%would go through 
for $C^k$ kinks,  for any choice of $k<r$. One may even choose distinct kink types for the `diffeomorphism' and `propagation' deformations, as well as for the GR and non GR cases.

This second class of generalizations is expected to have profound effects on the nature of physical states, a discussion of which we turn to next.
Our discussion will be qualitative, the intention being to convey a flavor of the issues involved and set the stage for a more detailed analysis in the future, such an analysis being out of the 
scope of this paper. In what follows, we shall assume some level of familiarity with References \cite{leeprop,pftprop,propu13,ttme}.
The particular property of physical states which we shall focus on is that of {\em propagation}. The issue of propagation in the context of LQG quantizations was first noted in \cite{leeprop}
and a mechanism for propagation in this context was first uncovered in \cite{pftprop}.  We refer the reader to these and subsequent discussions of propagation \cite{propu13} and, especially, \cite{ttme}
which discusses propagation in the context of Thiemann's QSD Hamiltonian constraint \cite{qsd}. Here we recall that a {\em key  necessary} condition for propagation, first noted in \cite{pftprop}
is that of {\em non-unique parentage}. Referring to the (diffeomorphism class of) spin net being acted upon the by the constraint as a `parent' and the (diffeomorphism classes of) deformed states
generated by this action as `children', non-unique parentage refers to the phenomenon of children having more than one (diffeomorphically) distinct parent.  
We shall say that (the diffeomorphically) distinct states $c_1,c_2$ are siblings if  $c_2$ is the offspring of  a (not necessarily unique) parent of $c_1$. If the parentage of $c_1$ is not unique, we shall say that the collection
of ancestors and off spring of $c_1,c_2$ are all linked by non-unique parentage.

Recall that physical states are obtained as a formal sum over kinematic states. It is expected that the summands in a physical state are related by heredity and that, generically, given one such summand
all summands related to it through parent-child relations are also present in the sum. Clearly the number of such relations increases with non-unique parentage and finding the coefficients of this 
increasing number of summands so that the sum is a physical state becomes more complicated. Thus the structural richness of  physical states as well as the difficulty of solving the constraint equations
are expected to be proportional to the extent of non-unique parentage.  This structural richness is directly responsible for the encoding of propagation by physical states.
From this point of view, more non-unique parentage leads to vigorous propagation. On the other hand too much non-unique parentage carries the risk that almost all states are related by 
such parentage leading to  a possible undesireable decrease in the size of the physical state space. As a result it seems crucial to have a constraint action which leads to an optimal level of non-unique
parentage.

In the action (\ref{action}) one vigorous channel of non-unique parentage is through the state deformations of the type (\ref{sprop1b}). 
\footnote{Non-unique parentage through state deformations of the type (\ref{sprop2b}) also seems possible; preliminary analysis indicates the strong possibility of a mechanism of propagation similar to 
that encountered in \cite{ttme}.}
We provide a pictorial demonstration of this fact in Figure \ref{Figpropanom}.
\begin{figure}[H]
\centering
%  \begin{subfigure}[H]{0.3\textwidth}
    \centering
    \includegraphics[width=0.5\textwidth]{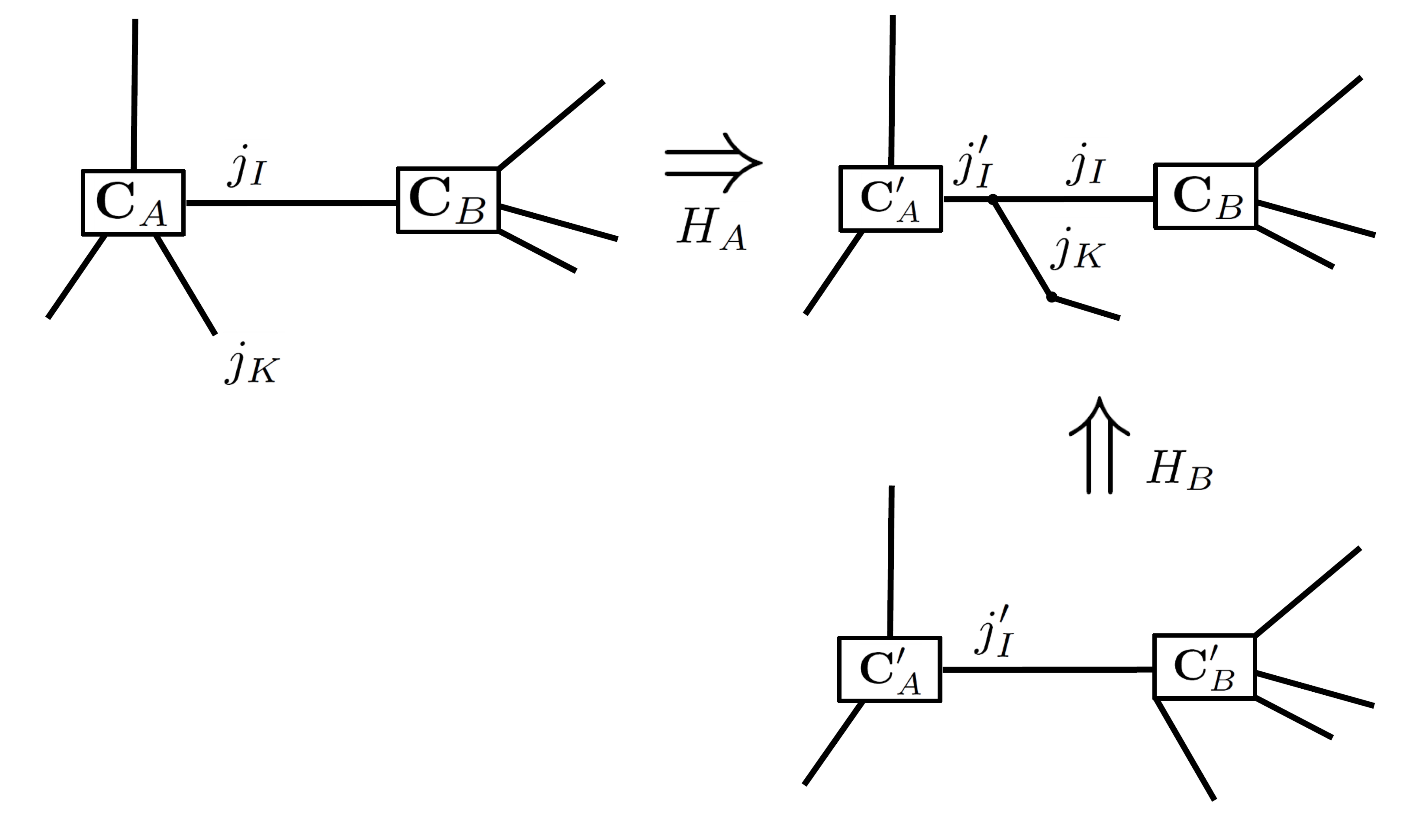}
    \caption{The figure depicts the phenomenon of non-unique parentage via state deformations of the type  (\ref{sprop1b}).  The vertex on the left of graphs depicted is understood to be the vertex $A$ and the one on the right, vertex $B$.
    $H_A$ signifies the action of the Hamiltonian constraint on vertex $A$  of the (part of a)
 parent spin net on the top left of the figure  to yield the child on its  right. Upto a diffeomporphism this child is also generated by the action of the constraint $H_B$ on the right vertex, $B$,  of the (part of the) spin network depicted 
 at the bottom right of the Figure.}
 \label{Figpropanom}
%  \end{subfigure} \quad
\end{figure}
For more details on this type of non-unique parentage, the interested reader may consult Reference \cite{propu13}.
The extent of non-unique parentage, with all its attendant repercussions, can also be seen to be closely related to the choice of kink type alluded to above. For example, if different kink types are
specified for  deformations of GR and NGR vertices, it is straightforward to see that we can block the possibility of NGR vertices creating children of GR ones i.e. of non-unique parentage involving GR and NGR vertices.
Even within the context of only GR parental vertices, different kink type choices for `diffeomorphism' and `propagation' type children can cut down on conceivable non-unique parentage.
To see this, consider the action of a `propagation' deformation of the type displayed in Figure \ref{figmixed1}  in the case in which the $I$th and $K$th edges carry the same spin label $j$.
The deformation drags part of the $K$th edge along the $I$th one between $v$ and $v_{I,\e}$ so that the Clebsch-Gordon decomposition of the product of two identical spin $j$ edges, namely the $I$th edge and this  draggged part of the $K$th edge
admits a vanishing spin representation. If this is consistent with the intertwiner at $v$ as depicted in Figure \ref{figmixed1},
\footnote{\label{fnmerger}Checking this consistency involves an analysis involving edge colorings and intertwiners
which is beyond the scope of this paper; here we simply assume such consistency as our aim is only to provide the reader with a flavor of the issues involved.}
then we have the situation depicted in Figure \ref{Figkinkmerge} in the case when $v$ is 4 valent and 
the deformation generates the pair of kinked edges with  total of 3  kinks. 

\begin{figure}[H]
\centering
%  \begin{subfigure}[H]{0.3\textwidth}
    \centering
    \includegraphics[width=0.3\textwidth]{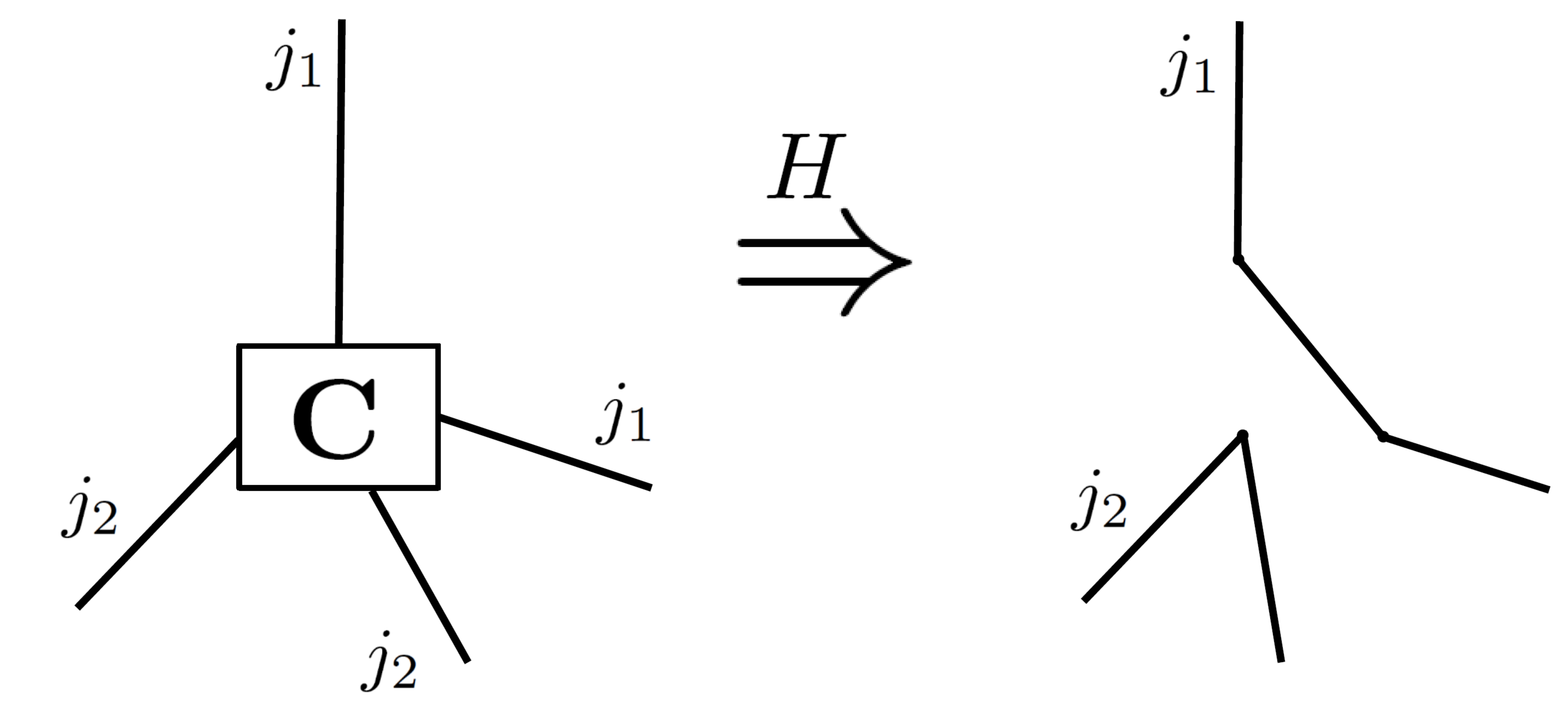}
    \caption{The figure depicts the generation of a child with 3 kinks on the right from the 4 valent parental vertex $v$ on the left through the action of the Hamiltonian constraint $H$ on this parental vertex. 
    We depict only the relevant part of the parent and child states.
    The deformation is
    of type (\ref{sprop1b}). Modulo the caveats mentioned in Footnote \ref{fnmerger}, the vertex on the left is annhilated with the creation of the triplet of kinks on the right.
    %Let the edge depicted with two kinks be $e_1$ and the edge with a single kink be $e_2$. 
    %It is assumed the extension of $e_2$ into the larger graph has a second kink (not depicted).
    As indicated in the main text, the (part of the) child state on the right
    could also be created in the course of a `diffeomorphism' type deformation  of an appropriate parent. }
    %The (relevant part of the)  parental state would have a vertex at $v^{\prime}$, 
    %from which the (undeformed versions of the) edges $e_1,e_2$ emanate, with these versions having a single  kink and no kink respectively. }
    %The two successive diffeomorphism type deformations create a pair of kinks on these edges}
 \label{Figkinkmerge}
%  \end{subfigure} \quad
\end{figure}

It is straightforward to see that if the parental structure on the left is part of an appropriately defined parental state $p$ this  deformation structure consisting of a pair of kinked edges can also be seen as a subset of the deformation structure 
which arises in the process of a `diffeomorphism' type deformation on a distinct parental vertex $v^{\prime}$ in a distinct parental state $p^{\prime}$.
Let the vertex $v^{\prime}$ be an undepicted part of the parental graph $p$  with one of the two edges colored with $j_1$ together with  one of the two edges colored with $j_2$  ending at $v^{\prime}$.
Call these edges $e_{1a},e_{2a}$ respectively and call the remaining pair of edges at $v$ in $p$ as $e_{1b},e_{2b}$ with $e_{1b},e_{2b}$ being colored with $j_1,j_2$ respectively. Further, let
$e_{1a}$ contain a single kink and $e_{2a}$, no kink in $p$.
Consider a distinct parent $p^{\prime}$,  the only difference between $p,p^{\prime}$ being  that in the latter the 
4 valent vertex $v$ of $p$ is replaced by a smoooth join between $e_{1a},e_{1b}$ to yield a single edge $e^{\prime}_1$ and  a distinct smooth join between $e_{2a},e_{2b}$ to yield $e_2^{\prime}$.
% two edges $e_{1a},e_{2a}$  in $p$ with colors $j_1$ joined smooothly together to yield an edge $e^{\prime}_1$ and the two edges colored with $j_2$ joined in an edge $e^{\prime}_2$.
%Amongst the edges emanating from this vertex would be 
%the undeformed versions of the kinked edges, so that one of the undeformed edges would have a single kink and the other no kink. 
Assuming an appropriately upward conical structure at $v^{\prime}$ in $p,p^{\prime}$, a  diffeomorphism type deformation of $v^{\prime}$ in $p^{\prime}$ would then 
deform these edges by adding an extra kink to each of them, thereby producing the desired child graph structure upto the action of diffeomorphisms.
Clearly the deformation structure in the child has this non-unique parentage under the assumption that 
%leading once again to a situation of non-unique parentage provided that 
the kinks created by
 propagation type deformations and those created by diffeomorphism type deformations are {\em identical}  $C^0$ type.
If the kink type for propagation and diffeomorphism type deformations was distinct, this type of `kink merger' based non-unique parentage would be blocked.

%on the  4 valent vertex shown in Figure \ref{kinkmerge} 
%For example , if these kinks were all $C^0$ for both types of children,  a triple of bivalent  kinks  in a diffeomorphism type child could conceivably also be generated by a propagation deformation of a parental vertex
%in which these two kinks are joined to form a 4 valent parental vertex as shown in Figure ... (whether this is actually permitted by the constraint action depends on an analysis involving edge colorings and intertwiners
%which is beyond the scope of this paper). 
The example of Figure \ref{Figkinkmerge} also touches on the proposal to demand invariance of physical states under
`extended diffeomorphisms' \cite{r-f}. Since such extended diffeomorphisms generate bivalent kinks in  edge interiors there is a danger of `too much' non-unique parentage due to the possible existence of 
parental vertices obtained by the merger of such kinks.  Perhaps a more appropriate choice would be to demand invariance under the kind of extended diffeomorphisms generated by the constraint action
in this work in conjunction with a specification of distinct kink types for propagation and diffeomorphism type deformations so as to reduce excess non-unique parentage. 
Indeed it may be that this restricted set of extended diffeomorphisms, through the mechanism of appropriate non-unique parentage, serve to render the continuous moduli of Reference \cite{g-r} irrelevant, thereby 
restoring separability of the physical state space.

We leave a detailed analyis of these and other matters for future work, our purpose here only being to point out the key role played by seemingly insignificant choices of deformation `microstructure'
generated by the constraints.

\subsection{\label{sec8.2} Other avenues for future work and improvement}

\noindent (1) $U(1)^3$ model: This toy model was introduced by Smolin \cite{leeg0} and its classical dynamics has been studied further in \cite{ttu13}. Recently the work of Barbero and collaborators \cite{feru13}
suggests the possibility, as originally envisaged by Smolin, of making contact with Euclidean gravity through a perturbation expansion about the $U(1)^3$ model. The model was studied in an LQG type quantization 
in 2+1 \cite{2+1u13} and 3+1 dimensions \cite{p1,p2,p3} with a view towards constructing an anomaly free quantum constraint algebra. The construction in Reference \cite{p3} is quite baroque in its use of 
complicated networks of coordinate patches as well as  certain not very natural structures called `interventions'. In contrast, in this work we use Riemann Normal Coordinates and avoid the use of interventions.
The technical modification relative to \cite{p3} which allows us to avoid interventions is the implementation of upward conicality (see section \ref{sec2.3.3}).
Another  significant breakthrough in this work is the ability to handle multivertex off shell states; this is in contrast to \cite{p3} which restricts attention to single vertex off shell states.
It is therefore of interest to revisit the $U(1)^3$ model and construct its quantization in the simpler and more powerful manner outlined here.
\\

\noindent (2) Better Off Shell States: The off shell states in this work are designed to trivialize the contribution of the propagation part of the action. Intuitively this choice corresponds to off shell
states which are not sufficiently `dynamical'. It would be of interest to search for off shell states which support an anomaly free action but which do not trivialise the  `quantum propagating degrees of freedom'
so brutally. Indeed, this issue seems to be of crucial relevance to (3) below.
\\

\noindent (3) Lorentzian General Relativity: As envisaged by Thiemann \cite{ttcomplexifier}  (see also \cite{aacomplexifier,mvcomplexifier}), vacuum Lorentzian gravity can be related to vacuum Euclidean gravity 
by a phase space dependent Wick rotation generated by a `complexifier' function on phase space. Preliminary calculations suggest that the quantum complexifier has a trivial action on the off shell (but {\em not} on shell) states presented in this work and this putative
triviality is intimately connected with the trivialization of propagation by  off shell states indicated in (2).
If these calculations hold up, perhaps the implication is that these off shell states should be improved  upon as suggested in (2). 
In this regard we are in the initial stages of exploring the viability of a strategy based on the use of density weight one constraints, modifications of both the constraint action and the  off shell states, and   the implementation of the `right hand side' operator ${\hat O}(N,M)$ as a generator of electric,   as opposed to purely semianalytic, diffeomorphisms.
%A search for such improvements involving density
%weight one constraints, modifications of both the constraint action and the  off shell states and  the implementation of the `right hand side' operator ${\hat O}(N,M)$ as a generator of electric,   as opposed to purely semianalytic, diffeomorphisms
In any case these issues are clearly in need of clarification.
\\

\noindent (4) Inclusion of Cosmological Constant: It is straightforward to check that the higher density weight of the constraint used in this work precludes a continuum limit action of the cosmological constant 
term. As indicated in (3), we are exploring a possible return to the use of density weight one constraints which, if successful,  could then potentially allow for a generalization to the case of a cosmological constant.\\

Notwithstanding the desireability of the improvements suggested in (2), (3),  we believe that the work in this paper provides significant support for  the physical relevance of the class of 
constraint actions developed in \cite{p4} as well as herein through a detailed demonstration of anomaly freedom, this demonstration being intimately tied to the incorporation, into
the quantum dynamics, of the Electric Shift mediated transformations underlying the classical dynamics of gravity \cite{aame}.
\\

%Notwithstanding this, we believe that the work in this paper does serve to highlight the physical relevance of the class of constraint actions developed in \cite{p4} as well as herein
%through a detailed demonstration of anomaly freedom, this demonstration being intimately tied to the incorporation, into the quantum dynamics,  of the Electric Shift mediated transformations underlying the classical dynamics of gravity \cite{aame}.\\

%\section{\label{sec9} Concluding Remarks}

%Other physical systems :U(1)3, vacuum lorentzian, cosmological constant. matter open.

\noindent{\bf Acknowledgements}: I am very grateful to Fernando Barbero for his detailed comments on a draft version of this paper and for his kind help with the figures
and to Abhay Ashtekar for his comments on the initial part of this manuscript.
I thank Abhay Ashtekar, Fernando Barbero and Jorge Pullin for their constant encouragement and support through the years.
%I am very grateful to Jorge PullinAbhay Ashtekar for his 
%constant encouragement.

\section*{Appendix}
\appendix

\section{\label{seca1} Determinant of the Jacobian between RNCs at nearby points}

Let the RNCs centered at the point $p$ be $\{x\}$. Consider a point $p_\e$ which is a coordinate distance  $\e$  from $p$ (as measured by the $\{x\}$ coordinates).
Let the RNCs centered at the point $p_\e$ be  $\{\bx\}$. We are interested in computing the determinant of the Jacobian $J(q)$ between the  unbarred and the barred coordinates
 at the point $q=p_\e$. Here the matrix components of the matrix $J(q)$ are:
\be
J^i{}_j(q):= \frac{\partial x^i}{\partial \bx^j} (q).
\label{rnc1}
\ee
Denoting the metric components in the barred and unbarred coordinates  by  $\bh_{ij}, h_{ij}$ we have that:
\be
\bh_{ij} (q)=  J^k{}_i(q)J^l{}_j(q)h_{kl}(q)
\label{rnc2}
\ee
which, in matrix notation reads:
\be
\bh (q) = J^{\tau} (q)h(q) J(q)
\label{rnc3}
\ee
where $J^{\tau}$ denotes the transpose of $J$. Next recall that   the metric  is $C^{r-1}$ as are the RNCs so that the metric components in the RNCs are $C^{r-2}$ functions of their arguments \cite{deturck-kazdan}.
For $r$ large (as assumed in this work), the metric components can be Taylor expanded to leading order in $\e^2$. Recall that the RNCs centered at a given point  are designed so as to render the 
Christofell symbols to vanish at that point.  It is straightforward to see that the vanishing of the Christofells   implies the vanishing of all first derivatives of metric components at that point.
Taylor expansion of $h_{ij}$ around $p$ then yields:
\be
h_{ij}(p_\e) = \delta_{ij}  + O(\e^2),
\label{rnc4}
\ee
where we have used $h_{ij}(p) =\delta_{ij}$ by virtue of the defining property of RNCs.
On the other hand, by this very same property applied to the barred RNCs centered at $p_\e$ we have that 
\be
\bh_{ij}(p_\e) = \delta_{ij} .
\label{rnc5}
\ee
Setting $q=p_\e$ in (\ref{rnc3}) and using (\ref{rnc4}), (\ref{rnc5}) yields:
\ba
{\bf 1} &=& J^{\tau}({\bf 1} + O(\e^2)) J \label{rnc6}\\
\Rightarrow 1 &=& (\det J)^2 (1 + O(\e^2)) \label{rnc7}\\
\Rightarrow \det J &=& 1 + O(\e^2) \label{rnc8}.
\ea
Here, we obtain (\ref{rnc7}) by taking the determinant of both sides of (\ref{rnc6}).  The desired result (\ref{rnc8}) follows straightforwardly from (\ref{rnc7}) together with an implicit choice of handedness of RNCs so that they are 
{\em righthanded} with respect to the orientation of the manifold.

\section{\label{a2} Polynomial $C^r$ joins}
\subsection{\label{seca2.0} Setup}
Let $v, v_{I,\e}, {\tilde v}_J$ be as in section \ref{sec2.3.2} so that $v_{I,\e}$ is at an RN coordinate distance of $\e$ 
from
the GR vertex $v$ and the kink ${\tilde v}_J$ is at coordinate distance of $\O(\e)$ from $v$ along $e_J$.  
In section \ref{seca2.1} we show
how to shift the position of  ${\tilde v}_J$ as close to $v$ as desired without changing the area of the loop to leading
order in $\e$ through the procedure of polynomial $C^r$ joins. The same procedure will be seen to be applicable to kink smoothening as desired 
in section \ref{sec2.3.2}.  Before we describe the procedure, it is essential to specify the 
construction of the deformed edge ${\tilde e}_J$ which connects ${\tilde v}_J$ to $v_{I,\e}$
a bit more precisely than in section \ref{sec2.3.2}.  We proceed as follows.

For small enough $\e$ fix a semianalytic coordinate system $\{ x_I \}$ around $v$ such that
the $I$th edge is a coordinate straight line along the $z_I$ axis (the existence of such coordinates
follows from the definition of a semianalytic edge \cite{lost}). Join $v_{I,\e}$ to ${\tilde v}_J$ by a coordinate
straight line $l_J$ in these coordinates. Note that  $e_J$ is not necessarily a coordinate straight line in these $x_I$ coordinates so that $l_J$ may intersect $e_J$ in more than one isolated point. The intersections are  
isolated as  $l_J$ cannot overlap
with $e_J$ because  the GR nature of $v$ and the $C^r$ nature of the edges implies that their tangents at intersection points are distinct. 
Moreover, it is straightforward to see that the GR nature of $v$ and the $C^r$ nature of the edges also imply that the coordinate distance of any such intersection point 
from ${\tilde v}_J$ is of $O(\e^2)$.

Redefine ${\tilde v}_J$ to be the first point of intersection 
between $l_J$ and $e_J$ as we proceed from $v_{I,\e}$ towards $e_J$. Define ${\tilde e}_J$ to be 
the part  of $l_I$
connecting $v_{I,\e}$ to  ${\tilde v}_J$ so redefined. 
By virtue of the GR property of $v$,
the $C^r$ nature of $e_J$, the $C^{r-1}$ property of the RNCs at $v$ and the $C^r$ semianalyticity of 
the $x_I$ coordinates, it follows that the area of the loop $l_{IJ\e}$ is of $\O(\e^2)$ as measured by the 
RNCs at $v$ and by adjusting  the initial choice of location of ${\tilde v}_J$, we may obtain any desired
area of $\O(\e^2)$. 

To summarise: We do not know if the RNCs are semianalytic. Therefore we choose convenient
semianalytic coordinates  to construct the deformed edges $\{ {\tilde e}_J \}$ so as to ensure that these
edges are themselves semianalytic. Since $r>>1$, the $C^{r-1}$ property of the RNCs ensures  loop
areas of $\O (\e^2)$  both in the RNCs as well as the $x_I$ coordinates. The precise area desired
can then be obtained by adjustment of the kink position ${\tilde v}_J$. 

\subsection{\label{seca2.1} Kink shifting}
In this section we show how to move the kink  ${\tilde v}_J$ along $e_J$ to a location as close as desired to $v$.
For small enough $\e$ choose semianalytic coordinates $\{\vec{\bx}\}\equiv(\bx,\by,\bz) $ around $v$ such that the part of
$e_J$ relevant to our considerations below is along the $\bx$ axis and its outward orientation from $v$ coinciding with increasing $\bx$.

%For the purposes of this section {\em only}, it is useful to employ the following notation.
Denote the 
(unshifted) kink position ${\tilde v}_J$ by $p_0$ so that ${\tilde e}_J$ connects $v_{I,\e}$ with $p_0$.  Let the desired (shifted) position of the kink be $p_1$.
%Denote the $\{x_J\}$ coordinates  simply by $\{{\vec \bx}\}= \{ \bx,\by, \bz \}$. 
Denote the part of $e_J$ between
$p_0$ and $v$ by $\gamma_1$. Let ${\bar p}_2$ be a point on ${\tilde e}_J$ which is a coordinate distance 
 $\delta_1 <<<\e$ away from   $p_0$ in the $\{{\vec \bx}\}$ coordinates. Let $B_{\delta_1}(p_0)$ be a ball 
 of coordinate radius $\delta_1$ centered at $p_0$. Denote ${\tilde e}_J\cap B_{\delta_1}(p_0)$ by $\gamma_2$.
 
 By employing an appropriate $GL(3,R)$ transformation we additionally restrict  $\{\bx\}$   to be such that the outward edge tangent to $\gamma_2$ at $p_0$ is $(\frac{d}{d\by})^a$. Since $\gamma_2$ is $C^r$, it follows
 straightforwardly that 
 $\gamma_2$ is confined within a cone of solid angle $\O(\delta_1)$ around  a line parallel to the $\by$ axis emanating from $p_0$.
 Let ${\tilde \gamma}_1$ be the straight line in the barred coordinates joining $p_1$ to ${\bar p}_2$.
 Choosing $\bx$ to parameterize ${\tilde \gamma}_1$, it is straightforward to see that for small enough $\d_1$,
  the tangent to ${\tilde \gamma}_1$ has coordinate components $(1, \d_1 +O(\d_1^2), O(\d_1^2))$
 in the barred coordinates.  Semianalyticity of ${\tilde \gamma}_1, \gamma_2$  and the fact that 
 these edges are, respectively, almost along the $\bx, \by$ directions then implies that 
 these two edges intersect in at most a  finite number of isolated points all within a distance of $\O(\d_1^2)$ of ${\bar p}_2$.
 Denote the intersection point first encountered  going along ${\tilde \gamma}_1$ from $p_1$ to ${\bar p}_2$ by 
 $p_2$.  
 
 Next, note that, by virtue of its semianalyticity,  for small enough $\delta_1$, $\gamma_2- p_0$ is {\em analytic} in the barred coordinates.  Hence there exists a small enough coordinate sphere  $B_{p_2}(\d )$ around $p_2$ of
 size $\d<<\d_1$ such that ${\tilde \gamma}_1\cap (\gamma_2  \cap B_{p_2}(\d )) = p_2$ and such that 
 $(\gamma_2  \cap B_{p_2}(\d ))$ is analytic in the barred coordinates. For sufficiently small $\d$, from Lemma 3.1 of \cite{qsd}, there exist coordinates $\{{\vec x^{\prime\prime}}\}$ centered at $p_2$ and 
 analytically related to $\{{\vec \bx}\}$ such that within  $B_{p_2}(\d )$,
 $\gamma_1= (x^{\prime\prime}, 0, 0)$ and $\gamma_2 = (0, y^{\prime\prime}, 0)$.
% \footnote{Note that if we paramaterize  $\gamma_1$  by $x^{\prime\prime}$ it acquires an orientation opposite to that of 
 Through an appropriate  $GL(3,R)$ transformation on these `double primed' coordinates we define linearly related coordinates $\{{\vec x}^{\prime}\}$ such that $\gamma_1$ remains along the $x^{\prime}$ axis and $\gamma_2$ is
 a line in the $x^{\prime}$-$y^{\prime}$ plane 
 at an angle of 135 degrees with the $x^{\prime}$ axis. 
 %Choosing $-x^{\prime} \geq 0$ as a parameter for $\gamma_1,\gamma_2$, we have that
 %$\gamma_1= (x^{\prime}, 0, 0)$ and $\gamma_2 = ( x^{\prime}, -x^{\prime}, 0)$. 
 Next, we remove a small ball around
 the origin in  the primed system  of radius  $\sqrt{2} \alpha$, $\alpha < < \d$.  Our task is then to join 
the point $q_2$ with primed coordinates $(-\alpha,\alpha, 0)$ on $\gamma_2$ to the point $q_1$ with primed coordinates
$(\sqrt{2}\alpha, 0, 0)$  by a semianalytic curve which joins in a $C^r$ manner with $\gamma_2$ at $q_2$
and $\gamma_1$ at $q_1$.  This can be done through an application of a lemma which we prove below.\\

%By appropriate translation of  the primed coordinate system,
% these $C^r$ joins can be implemented through an application of the Lemma below. We explain how this 
% is done after we prove the Lemma below. \\
%By replacing $p_1$ in the argumentation above by the vertex $v$, it then  immediately follows that the kink
%$p_0$ can be smoothened.\\

\noindent{\em Lemma}:  Consider a coordinate system $(x,y)$ in ${\bf R}^2$. Let $\gamma_1$  be the line along  and oriented opposite to the $x$-axis with end point at $x=\sqrt{2} \alpha$. Let $\gamma_2$ be the line  at angle 135 degrees with respect to
to the $x$-axis with beginning 
point  $(-\alpha, \alpha)$, $\alpha>0$,  and orientation in the direction of increasing $\alpha$. Then there exists a semianalytic
curve ${\gamma}$ which joins the end point of $\gamma_1$ to the beginning point of $\gamma_2$ such that 
 the curve $\gamma_2 \circ {\gamma} \circ \gamma_1$ is $C^r$.
\\

\noindent{\em Proof}: 
%Denote the straight line connecting the origin to $(-\alpha, \alpha)$ by $l$.
%Let $0<\beta <<\alpha$ and let $l_{\beta}$ be the part of $l$ between the origin and $(-\beta, \beta)$.
%We show that it is possible to replace $l$ by the desired  semianalytic  curve $\tilde{\gamma$ such that 
%$(l-l_{\beta}) \circ \gamma \circ \gamma_1$ is $C^r$ i.e the joins of $\gamma$ with $\gamma_1$ at the origin
%and with $l-l_{\beta}$ at $(-\beta, \beta)$ are $C^r$.
We parameterize $\gamma$ by $-x$, with $ -\sqrt{2}\alpha \leq  -x \leq  \alpha$ and set
\ba
\gamma^x(-x) &= &x  \label{x1}
\\
\gamma^{y}(-x) &=& -f(x)x \label{x2}
\ea
where $f$ is a $C^r$ semianalytic function defined as follows. 
%Let $B_{\tau}(x,y)$ denote the ball of radius $\tau$ centered at the point $(x,y)$.
Let $\tau<<\alpha$ and define $f$ to be such that:
\ba
f(x) = 1, &   -\alpha -\tau \leq  x \leq  -\alpha +\tau, & \\
f(x) = 0, &   -\tau < x  &\\
0\leq f(x) \leq 1, &   -\alpha -\tau \leq x  &
\label{x.04}
\ea
It is straightforward to infer the existence of such $f$ \cite{lost,ttbook} from the fact 
that there always exists a seminanalytic partition of unity subordinate to any finite open cover of a compact manifold $\Sigma$.
The $C^r$ semianalytic nature of $f$ ensures that $\gamma$ is semianalytic and the behavior of $f$ near the end points of $\gamma$ ensures that the joins to $\gamma_1,\gamma_2$ at these end points are $C^r$.
\\

Identifying $x^{\prime}\equiv x$, $y^{\prime}\equiv y$, an application of the Lemma immediately leads to the desired semianalytic curve with $C^r$ join at $q_1,q_2$.

\subsection{\label{seca2.2} Kink Smoothening} 
In this section we show how to smoothen the kink ${\tilde v}_J$. 
Exactly as in section \ref{seca2.1} fix $\{\vec{\bx}\}$ coordinates around $p_0$. Using the same notation as in that section, for small enough $\delta_1$, $B_{2\delta_1}(p_0)$ lies
within the barred coordinate patch so that within this ball ${\tilde e}_J$ is confined within a cone of angle $O(\delta_1)$ about the $\by$ - axis and $e_J$ is along the $\bx$ axis.
Let ${\bar p}_2 \in \partial B_{\delta_1}(p_0) \cap {\tilde e}_J$, let $p_1= \partial B_{\delta_1}(p_0) \cap e_J$, let the straight (coordinate) line joing $p_1$ to ${\bar p}_2$ be ${\bar l}$ and 
let $p_2$ be the first point of intersection between ${\bar l}$ and ${\tilde e}_J$ when moving along ${\bar l}$ from $p_1$. Similar to section \ref{seca2.1} the confinement of 
${\tilde e}_J$  within a cone of angle $O(\delta_1)$ about the $\by$ - axis ensures that this line intersects ${\tilde e}_J$ at $p_2$ transversely. 
Call part of ${\bar l}$ joining $p_1$ to $p_2$ as $l$.

Next, note that semianalyticity of ${\tilde e}_J$ ensures that 
for small enough $\delta_1$,  ${\tilde e}_J \cap B_{2\delta_1}(p_0) -p_0$ is analytic in the barred coordinates. Using the construction in \cite{qsd} augmented with a suitable $GL(3, R)$ transformation,
there exists a small enough neighborhood of $p_2$ with chart $\{ {\vec x}^{\prime}\}$ analytically related to $\{\vec{\bx}\}$  and a (`primed') coordinate ball $B_{\tau}(p_2)$ such that for small enough $\tau <<\delta_1$ :\\
(i) the part of ${\tilde e}_J$ emanating from $p_2$ towards the displaced vertex $v_{I,\e}$ within $B_{\tau}(p_2)$
is a coordinate straight line along the $x^{\prime}$ axis\\
(ii) $l \cap B_{\tau}(p_2)$ is at an angle of 135 degrees with respect to $x^{\prime}$ axis in the $x^{\prime}$-$y^{\prime}$ plane.

A straightforward  application of the Lemma in section \ref{seca2.1} then allows the replacement of $l \cap B_{\tau}(p_2)$ by a semianalytic curve which $C^r$ joins at $p_2$ and at $l \cap \partial B_{\tau}(p_2)$.
Noting that at $p_1$, both $l$ and $e_J$ are straight lines in the barred coordinates, a similar application of this  Lemma allows the replacement of $l$ in a small neighborhood of $p_1$ by a semianalytic curve
with a $C^r$ join at  $p_1$  with $e_J$ and a similar join with the part of $l$ outside this neighborhood.

The final result of these 2 sets of joins in arbitrarily small neighborhoods of $p_2,p_1$ ensures the removal of the kink ${\tilde v}_J$.

\subsection{\label{seca2.3} Upward Conicality}
%For the purposes of this section {\em only}, it is useful to employ the following notation.
Let $B_{\d}(v_{I,\e})$ be coordinate ball of size $\d<<\e$ around $v_{I,\e}$ in the $\{x_I\}$ coordinates 
of section \ref{seca2.0}. 
%Denote the coordinate system $\{x_I\}$ of section \ref{a2.0}  by $\{x\}$. 
It is straightforward  to see that through a suitable translation
and rotation of the coordinate system $\{x_I\}$  we may define, for small enough $\e, \d <<\e$,
a coordinate system in $B_{\d}(v_{I,\e})$ with  origin at $v_{I,\e}$ in which  $e_I \cap B_{\d}(v_{I,\e})$  is along the $y$-axis and ${\tilde e}_{J} \cap B_{\d}(v_{I,\e})$ is a straight line in the $x$-$y$ plane.
 Denoting ${\tilde e}_{J} \cap B_{\d}(v_{I,\e})$ by $l$, this implies that the coordinates $(l^x,l^y,l^z)$ of points along $l$ can be written as:
 \be 
 l^x= \alpha  t\;\;\; l^y = \beta t \;\; l^z =0 \;\;  t\in [0, t_1)  \;\; \alpha \neq 0
 \label{uc1}
 \ee
where $t_1 = \O(\d)$.  
We remove the part of $l_0$ of $l$ between parameter values $0, t_0$, with $t_0< <t_1$. Note that $t=0$
labels the point $v_{I,\e}$.
We seek to replace $l_0$  by a semianalytic curve $\gamma$  in the $x$-$y$ plane which joins in a $C^r$  manner to $l-l_0$
at the point  on $l$ with parameter value $t_0$ and which has the desired tangent at the point  $v_{I,\e}$. 
We define $\gamma: [0,t_0] \rightarrow \Sigma$ through its $x,y$ coordinates $\gamma^x, \gamma^y$ as:
\ba
\gamma^x &=& \alpha t  \label{uc2}\\
\gamma^y &=& \beta t +  b_{r+1} (t-t_0)^{r+1} + b_{r+2}(t- t_0)^{r+2} \label{uc3}.
\ea
Equations (\ref{uc2}), (\ref{uc3}) ensure that the join is $C^r$ at the point $l(t_0)=\gamma (t_0)$.  For upward conicality
at $v_{I,\e}$ we require that:
\ba
\gamma^y(t=0) & = &0 \label{uc4}\\
\frac{d\gamma^y}{dt}|_{t=0} &=&  \eta^{-1} \label{uc5}
\ea
where $0 <\eta << 1$ is a small positive parameter.
It is straightforward to solve for the coefficients $b_{r+1}, b_{r+2}$ so as to impose (\ref{uc4}), (\ref{uc5}) and substitute
back into (\ref{uc3}) to obtain:
\be
\gamma^y (t) = t \{\beta + [(\eta^{-1}- \beta)( 1- \frac{t}{t_0})^{r+1}]\}.
\label{uc6}
\ee
Equation (\ref{uc6}) implies that for sufficiently small $\eta$, 
\be
|\gamma^y(t)| < t_0 (|\beta|+ |\eta^{-1} - \beta |)  <2\eta^{-1} t_0.
\label{uc7}
\ee
Choosing $t_0 <<\eta$ ensures that $\gamma$ is confined within $B_{\d}(v_{I,\e})$
and (\ref{uc2}) ensures that $\gamma$ does not intersect $l- l_0$ except at  $l(t_0)$
and does not intersect $e_I\cap B_{\d}(v_{I,\e})$ except at $v_{I,\e}$. Moreover, the GR property of
$v$ and the setup of section \ref{seca2.0} ensures that no intersections ensue (other than at $v_{I,\e}$) 
with any ${\tilde e}_{K\neq J}$
nor with their upward conical modifications (these modifications being constructed along the lines above).

From (\ref{uc4})-(\ref{uc5}) we have that at $v_{I,e}$
\be
\frac{ {\dot{\gamma}}^y}{{\dot{\gamma}}^x} = (\alpha\eta)^{-1}
\ee
which implies  the same  ratio of the $y$ and $x$ components of the {\em unit}  (with respect to the metric $h_{ab}$) 
tangent ${\hat {\gamma}}^a$ to $\gamma$
at $v_{I,\e}$.%of the 
%components of the tangents along $e_I$ and transverse to it. 
This, together with the fact (see above) that  $e_I \cap B_{\d}(v_{I,\e})$ is along the $y$ axis,
 implies that   the unit tangent  to $\gamma$ 
%${\hat \gamma}^a$ to $\gamma$
at $v_{I,\e}$ takes the form:
\be
{\hat \gamma}^a = {\hat e}^a_I|_{v_{I,\e}} +  \O(\eta) \omega^a
\label{ucfinal}
\ee
where ${\hat e}^a_I|_{v_{I,\e}} $ is the unit (with respect to $h_{ab}$ ) tangent to the edge $e_I$ at the 
point $v_{I,\e}$ and $\omega^a$ is some ($J$ dependent) vector of norm of $\O(1)$.
Clearly we may choose $\eta << \e^m, m>2$ to obtain (\ref{upward}).

\section{\label{seca3}Relative kink placement for $\frac{d_{min}}{d_{max}}= \frac{1}{2} +O(\e^q)$}

We prove a series of subsidiary results and lemmas from which  the desired result follows. 
All points under consideration
are assumed to be in a single Convex Normal Neighborhood (CNN) with respect to the metric $h_{ab}$. RN Coordinates centered at the point $v$ are denoted by $\{x\}$ and any reference to 
coordinates is to these coordinates which we fix once and for all.

\noindent{\bf 1.} Let $v,v_1,v_2$ be vertices of a triangle with sides $vv_1, vv_2, v_1v_2$ of lengths $ar, br, cr$ for some $r>0$. Let the angle between $vv_1, vv_2$ be $\theta$ so that:
\be
c^2= a^2 + b^2 -2ab\cos\theta.
\label{tr1}
\ee
Let $\theta_\d$ be such that 
%an angle which is equal to a $\d$ independent  angle $\theta_0$ to leading order in $\d$ 
\be
\cos \theta_\d = \cos \theta_0 + O(\d) 
\label{tr1.0}
\ee
for some $\d$ independent angle $\theta_0$.\\
\noindent{\em Case (i)}: Let  the range of $\theta_0$ be 
\be
\pi > \theta_0 \geq \cos^{-1}\frac{1}{4} .
\label{tr1.1}
\ee
Let $a=1$ and $b$ be such that:
\be
b^2- 2b\cos\theta_0 =3 .
\label{tr2}
\ee
It follows straightforwardly from (\ref{tr1}) with $\theta := \theta_\d$ that $c=2 +O(\d )$, and from (\ref{tr2}) that $1 <b\leq 2$, from which it follows that 
the ratio $R$ of the smallest to the largest side lengths of the triangle $v_1vv_2$ is:
\be
R= \frac{1}{2} + O(\d ).
\label{tr3}
\ee
\noindent{\em Case (ii)}: 
Let  the range of $\theta_0$ be 
\be
0< \theta_0 <\cos^{-1}\frac{1}{4}.
\label{tr4}
\ee
Let $a=1$, $b=2$.  
It follows straightforwardly from (\ref{tr1}) with $\theta := \theta_\d$ that 
\be 
c^2 = 5-4\cos\theta_0 + O(\d)
\label{rt5}
\ee
It is straightforward to check that in the range (\ref{tr4}), we have that
\be
1 <5-4\cos\theta_0 <4.
\label{rt6}
\ee
From (\ref{rt5}) and (\ref{rt6}) it follows that, identical to (\ref{tr3}),
\be
R= \frac{1}{2} + O(\d ).
\label{tr7}
\ee

\noindent{\bf 2.} {\em Lemma}: Let $e_1, e_2$ be a pair of semianalytic edges emanating from the point $v\in \Sigma$. Let each of $v_1, v_2$ be located at a coordinate distance of $\O(\d)$ from $v$
along $e_1,e_2$ respectively. Let the angle as measured in the coordinates $\{x\}$ between the unit edge tangents ${\vec {\hat e}_1}, {\vec {\hat e}_2}$ at $v$ be $\theta_0$. Then the angle $\theta$ at $v$
between the coordinate straight lines from $v$ to $v_1$ and from $v$ to $v_2$ is such that
\be
\cos \theta_\d = \cos \theta_0 + O(\d) .
\label{rt2.1}
\ee

\noindent{\em Proof}: We are given that: 
\be
\cos \theta_0= \delta_{\mu \nu}{\hat e}_1^{\mu} {\hat e}_2^{\mu}
\label{rt2.2}
\ee
Let $e_i,\;i=1,2$ have semianalytic parameterization by the parameters ${\bar t}_i$
\footnote{Since we do not know if the  RNCs are semianalytic charts, we do not know if they are semianalytic functions of ${\bar t}_i$. Since the proof only uses the $C^{r-1}$ property of RNCs together with the 
$C^r$ property of the semianalytic edges $e_1,e_2$, it is unaffected by this lack of knowledge.}
 with the beginning point   $v$ of $e_i$  located at $t_i=0$.
By appropriate rescaling
by a constant we choose $t_i$ to be such that $\frac{dx^{\mu}}{d{\bar t}_i}|_{{\bar t}_i=0}= {\hat e}^{\mu}_i$.
Let $v_i$ be located at parameter value ${\bar t}_i = t_i$ along $e_i$. Taylor expansion along $e_i$ yields:
\be
x^{\mu}(v_i)  = t_i {\hat e}^{\mu}_i + O(t_i^2)
\label{rt2.10}
\ee
where we have set $x^{\mu}(v)=0$ in the RNCs centered at $v$.
\be
\Rightarrow |{\vec x}(v_i)| := \sqrt{ \delta_{\mu\nu} x^{\mu}(v_i) x^{\nu}(v_i)}
= t_i+O(t_i^2)
\label{rt2.20}
\ee
which implies that:
\be
\O (t_i) = \O (\d ).
\label{rt2.30}
\ee
It then follows straightforwardly from (\ref{rt2.10})- (\ref{rt2.30}) that:
\be
\frac{x^{\mu}(v_i)}{|{\vec x}(v_i)|}= {\hat e}^{\mu}_i + O(\d)
\ee
from which we have, using (\ref{rt2.2})  that:
\be
\cos \theta = \delta_{\mu \nu}\frac{x^{\mu}(v_1)}{|{\vec x}(v_1)|}\frac{x^{\mu}(v_2)}{|{\vec x}(v_2)|}
= \cos \theta_0 + O(\d ).
\ee
\\
\noindent{\bf 3.}  {\em Lemma}: Let $v,v_1,v_2$ be located as in {\bf 2.} above. Let $vv_i, i=1,2$ be coordinate straight lines from $v$ to $v_i$. Let $v_1v_2$ be
the coordinate straight line from $v_1$ to $v_2$ and let $\Delta {\vec x}:= {\vec x}(v_2) - {\vec x} (v_1)$.
Denoting the coordinate length of $\Delta {\vec x}$ by $|\Delta {\vec x}|$, we have  from  {\bf 2.} that 
$\Delta {\vec x} = \O(\d)$.  For small enough $\d$ let the geodesic distance between $v_1$ and $v_2$ be $d$.
Then:
\be
d= |\Delta {\vec x}|( 1+ O(\d )).
\ee

\noindent{\em Proof}: Let the geodesic $e_{12}$ between $v_1,v_2$ be parameterized by its metric length
and let the tangent at $v_1$ to this geodesic so parameterized be ${\hat e}^a_{12}$. Taylor expansion
at $v_1$ implies that:
\be
\Delta { x}^{\mu}= d {\hat e}^{\mu}_{12} +O(d^2).
\label{aa1}
\ee
Using the fact that ${\hat e}^{a}_{12}$ is unit with respect to the metric $h_{ab}$ at $v_1$, 
equation (\ref{aa1}) implies that:
\be
h_{\mu \nu}(v_1) \Delta  x^{\mu} \Delta x^{\nu} = d^2(1 +O(d)).
\label{aa2}
\ee
Taylor expansion of metric components $h_{\mu \nu}$ around $v$ in conjunction with the property of RNCs implies
that
\be
h_{\mu \nu}(v_1)= \delta_{\mu \nu} + O (\d).
\label{aa4}
\ee
Using   $|\Delta {\vec x}|= \O(\d)$ in (\ref{aa1}) implies that 
\be
d= \O (\d).
\label{aa3}
\ee

It is then straightforward to see that (\ref{aa1})-(\ref{aa4}) imply the desired result.
\\

\noindent{\bf 4.}  {\em Lemma}: Denote the geodesic distance between points $p,q$ by $d(p,q)$.
Let $v_3$ be located at geodesic distance of $O(\e^p)$ from $v$ and let $v_1,v_2$ be at a geodesic 
distance of $O(\e^q)$ from $v$ with $p>>q$. Then for $i=1,2$:
\be
d(v_3,v_i)=  d(v,v_i) + O(\e^p).
\ee
\noindent {\em Proof}: This immediately follows from the fact that  for two points in a CNN, the  geodesic connecting them is the shortest
distance path between the  two points.
\\

\noindent {\bf 5.} Recall that in the RNCs centered at a point $p$, the coordinate line from $p$ to any 
point $q$ in the CNN around $p$ is a geodesic and its metric length coincides with the coordinate 
length.

The desired result follows from {\bf 1.}- {\bf 5.}.  To see this, let the three kinks be $v_i, i=1,2,3$  with $v_3$ 
placed at a coordinate distance of $O(\e^p)$ from  the vertex $v$ and $v_1,v_2$ placed in accordance with {\bf 1.} with $r= \e^q$, $p>>q>>1$.  Then {\bf 2.}  holds with $\d= \e^q$.  {\bf 1.} then implies that 
the ratio of the smallest to largest coordinate distances between $v, v_1,v_2$ is $\frac{1}{2} + O (\e^q)$.
{\bf 5.} and {\bf 3.} imply that the same result holds for geodesic distances between these three points.
Finally this together with {\bf 4.}  implies that the same result holds for the geodesic distances between
the points $v_1,v_2,v_3$ which proves   the desired result mentioned in the section header.

\section{\label{seca4}Semianalytic diffeomorphisms of compact support}
Section \ref{sec4.0a} constructs a useful class of semianalytic diffeomorphisms of compact support.  Sections \ref{sec4.0b}-\ref{seca4.2}  apply the construction method to obtain 
the results described by their title headers.
\subsection{\label{sec4.0a} General Construction}

\noindent{\em Lemma}: Equip $R^n$ with a $C^r$ semianalytic structure. Fix, once and for all, a global  semianalytic  chart $\{x\}$ on $R^n$.  Let $U, V$ be open neighborhoods of the origin with ${\bar V} \subset U$. Let $f^{\mu}, \mu=1,.., n$
be an $n$- tuple of semianalytic functions  compactly supported in ${\bar V}$. Define $\phi : R^n\rightarrow R^n$ by
\be
x^{\prime \mu}= \phi_{\lambda} (x)= x^{\mu}   + \lambda f^{\mu}.
\label{defphil}
\ee
Then there exists $\lambda_0 >0 $ such that $\forall \lambda$ such that  $0 \leq \lambda \leq \lambda_0$,  the map $\phi_{\lambda}$ is a $C^r$ semianalytic diffeomorphism of $R^n$.
\\
\noindent{\em Proof}: Since $f^{\mu}$ is $C^r$, we have that $\frac{\partial f^{\mu}}{\partial x^{\nu}}$  is bounded in $U$ and vanishes outside $U$  for $\mu,\nu=1,..n$. 
The Jacobian matrix $J$ between $\{x^{\prime}\}$ defined through (\ref{defphil}) and $\{x\}$ is:
\be
J^{\mu}{}_{\nu} := \frac{\partial x^{\prime \mu}}{\partial x^{\nu}} = \delta^{\mu}_{\nu}  + \lambda \frac{\partial f^{\mu}}{\partial x^{\nu}}.
\ee 
 Since $f^{\mu}$ is $C^r$, we have that $\frac{\partial f^{\mu}}{\partial x^{\nu}}$  is bounded in $U$ and vanishes outside $U$  for $\mu,\nu=1,..,n$, it is straightforward to see that 
 there exists small enough $\lambda_0$ 
such that for all $\lambda \leq \lambda_0$,  $\lambda \geq 0$, and all $x^{\mu}$,   
\be
\det J \neq 0.
\label{a4.01}
\ee
The inverse function theorem then implies that $\phi_{\lambda}$ is `locally' a diffeomorphism. More precisely, given any $p\in R^n$, there exists an open neighborhood $O_p$ of $p$ in which 
$\phi_{\lambda}: O_p \rightarrow \phi_{\lambda} (O_p)$ is a diffeomorphism. 

Next, we note that for any compact set $K\subset R^n$ we have that $\phi_{\lambda}^{-1}(K)$ is compact. To show this we proceed as follows.
Since $K$ is compact, it is closed and bounded. We have that:
%Since $f^{\mu}$ is $C^r$ and of compact support it is bounded. It follows from (\ref{defphil}) that there exists small enough positive $\lambda_0 \leq \lambda_1$ such that 
%for all positive $\lambda \leq \lambda_0$ we have that $\phi_{\lambda} ({\bar V})$ is bounded, and that $\phi_{\lambda} (R^n- {\bar V})$ is the identity. Accordingly we have:
\be
\phi^{-1}_{\lambda} (K) =(\phi^{-1}_{\lambda} (K) \cap (R^n- {\bar V})) \cup  (\phi^{-1}_{\lambda} (K) \cap {\bar V}).
\label{a4.011}
\ee 
From (\ref{defphil}) we have that $\phi_{\lambda}$ is the identity map on  $(R^n- {\bar V})$. It follows that 
$ (\phi^{-1}_{\lambda} (K) \cap (R^n- {\bar V}))  \subseteq K$. Boundedness of $K$ then implies that $ (\phi^{-1}_{\lambda} (K) \cap (R^n- {\bar V}))$ is bounded.
Next, note that 
$(\phi^{-1}_{\lambda} (K) \cap {\bar V})$ is bounded by virtue of the boundedness of ${\bar V}$. Hence equation (\ref{a4.011})implies that 
$\phi^{-1}_{\lambda} (K)$ is bounded.  Since $K$ is closed and $\phi_{\lambda}$ is continuous, $\phi^{-1}_{\lambda} (K)$ is closed.  Hence $\phi^{-1}_{\lambda} (K)$ is closed and bounded
and hence compact as asserted.

The local diffeomorphism nature of $\phi_{\lambda}$ and  the compactness of $\phi_{\lambda}^{-1}(K)$ together with Hadamard's theorem (see Theorem 6.2.3 of \cite{hadamard}) imply 
that $\phi_{\lambda}$  is globally  one to one and onto. Since $\phi_{\lambda}$  is explicitly  a $C^r$ semianalytic map, it is a $C^r$ semianalytic diffeomorphism. 
This completes the proof of the Lemma.\\
%Corollary A.7 of the Appendix of \cite{lost}
%implies that $

Next, note that (\ref{defphil}) implies that the semianalytic diffeomorphism $\phi_{\lambda}$ is the identity outside ${\bar V}$ so that $\phi({\bar V}) ={\bar V}$  from which our final result below 
clearly holds:

\noindent Let $V,U \subset \Sigma$ be open sets with ${\bar V}\subset U$. Let $U$ be such that it is covered by a single $C^r$ semianalytic chart $\{x\}$. Let $f^{\mu}, \mu=1,2,3$ be $C^r$ semianalytic functions
supported exclusively in ${\bar V}$. Then there exists $\lambda_0 >0 $ such that for all postive $\lambda \leq \lambda_0$ we have that $\psi_{\lambda}: \Sigma \rightarrow \Sigma$ is a $C^r$ semianalytic 
diffeomorphism where $\psi_{\lambda}$ is defined through:
\ba
\psi_{\lambda}(p)  &=&  p, \;\; p \notin {\bar V}  \label{a4.02}\\
 x^{\mu}(\psi_{\lambda}(p)) &=& x^{\mu} (p) + \lambda f^{\mu}(p), \;\; p \in {\bar V} .
\label{a4.03}
\ea

\subsection{\label{sec4.0b}Construction of ${\bar \phi}_{I,\e}$ in section \ref{sec5.2}}
From the definition of a semianalytic edge \cite{lost}, there exists a small enough neighborhood $O$ of $v$, and semianalytic patch  $\{x\}$ thereon such that $v$ is at the origin and $e_I\cap O$ is along the 
3rd coordinate (i.e. $z$ axis). Rescale  the $z$ coordinate if so required so  that $(\frac{d}{dz})^a$ coincides with the (metrically) unit edge tangent ${\hat e}^a_{I}$ at $v$:
\be
(\frac{d}{dz})^a|_v= {\hat e}^a_{I}|_v
\label{a4.031}
\ee
%Fix a semianalytic patch $\{x\}$ in an open set $O$ containing $v$ such that
For small enough $\e$ set $V$ in (\ref{a4.02}) to be $V:=B_{2\e}(v) \subset O$, where  $B_{\delta }(v)$ denotes the  coordinate ball of radius $\d$ centered at $v$ in the coordinates $\{x\}$.
Let $f$ be a $C^r$ semianalytic, compactly supported function such that:
\ba
f(p) = 1, & p \in B_{2\e -\tau}(v), \tau<< \e& \\
f(p) = 0, &  p \notin B_{2\e}(v), \tau<<\e &\\
0\leq f(p) \leq 1, & p \in \Sigma . & 
\label{a4.04}
\ea
It is straightforward to infer the existence of such $f$ \cite{lost,ttbook} from the fact 
that there always exists a seminanalytic partition of unity subordinate to any finite open cover of $\Sigma$.
%(see \cite{lost,ttbook} for a proof of this fact). 
Set 
\be
f^{\mu} = f \delta^{\mu}_3.
\label{a4.041}
\ee
The vector field $e^a_{I,\e}$ of section \ref{sec5.2} is then identified with the vector field $f^a$ with components $f^{\mu}$ given by (\ref{a4.041}).
The results of section \ref{sec4.0a} imply that there exists $\lambda <<\tau$ such that $\psi_{\lambda}$ as defined by (\ref{a4.02}) is a semianalytic diffeomorphism which corresponds to 
a point dependent translation in the $z$ direction. In particular $v$ is translated along $e_I$.  Clearly there exist $\lambda$, $N_{\lambda}$ such that 
the repeated action $(\psi_{\lambda})^{N_{\lambda}}$ of $\psi_{\lambda}$ a total of $N_{\lambda}$ times  translates $v$ to the desired position $v_{I,\e}$. 
It follows from (\ref{a4.031}) and  the fact that $z$ provides a semianalytic parameterization of $e_I$ in the neighborhood $O$ of interest, together with (\ref{t4}),
that 
\be
N_{\lambda}{\lambda} = \e + O(\e^2). 
\label{a4.0411}
\ee
The analysis of Reference \cite{p4}, repeated here, indicates that we translate the point $v$ by 
an affine parameter $\e$ along the integral curves of the vector field $e^a_{I,\e}$. 
In this regard (\ref{a4.0411}) implies that the diffeomorphism $(\psi_{\lambda})^{N_{\lambda}}$ translates the point $v$ by an affine amount $\e + O(\e^2)$. Since we are interested in
leading order in $\e$ behavior this choice  of $N_{\lambda}$ is acceptable. 

Accordingly we set
\be
{\bar \phi}_{I,\e} := (\psi_{\lambda})^{N_{\lambda}} .
\label{adefphie}
\ee
It is straightforward to see that, for a GR vertex $v$, the action of ${\bar \phi}_{I,\e}$ produces  loops ${\bar l}_{IJ,\e}, J\neq I$ of coordinate area of $\O(\e^2)$ in the coordinates $\{x\}$. The coordinate area of these loops in the RNCs at $v$ is then also 
$\O(\e^2)$ by virtue of their $C^{r-1}$ nature.  From (\ref{a4.04})- (\ref{adefphie}), the deformed edges lie within $B_{2\e}(v)$.

For our purposes in the main text, we need to be able to change the areas of the loops ${\bar l}_{IJ,\e}, J\neq I$  by $\O(\e^2)$  without altering the position of $v_{I,\e}$. 
%We describe two ways in which this can be done. The first is quite simple and transparent. It consists of 
In what follows we 
construct such diffeomorphisms. These diffeomorphisms  are arrange to be the  identity at the locations of the edges $e_J, e_I$ and 
will move segments  of the deformed edges transverse to themselves either towards or away from the $z$-axis thereby increasing or decreasing the loop areas.
%Next, we construct diffeomorphisms which move the deformed edges transverse to themselves so as to change the areas of the loops $l_{IJ,\e}$ to their desired values.
%These diffeomorphisms are arranged to be the identity in small neighborhoods of $e_J,e_I$.  
We proceed as follows.
Note that for small enough $\e$, $B_{\sqrt{\e}} (v)$ lies within the domain of the coordinate patch $O$. Also note
that by virtue of the GR nature of $v$, each of the loops $l_{IJ,\e}$ can be enclosed by an angular wedge $W_{\alpha_{J,\e}}$ of azimuthal angular width $\alpha_{J,\e}= \O(\sqrt{\e})$ around the $z$- axis.
More precisely, $W_{\alpha_{J,\e}}$ refers to the part of the infinite coordinate wedge which lies in $B_{\sqrt{\e}}(v)$.
\footnote{Since we desire the freedom to change the loop area by $\O(\e^2)$, diffeomorphisms of interest could displace the deformed edge outside $B_{2\e}(v)$. This is the reason we 
introduce the $\sqrt{\e}$ even though $l_{IJ,\e}$ constructed through (\ref{a4.04})- (\ref{adefphie}) are contained in a wedge of smaller angle $\e$.} 
For small enough $\e$, the GR nature of $v$ implies that the angular wedges $\{W_{\alpha_{J,\e}}, J\neq I\}$ intersect each other only along  the $z$-axis.
%Let $C_{\delta}$ be that part of the infinite coordinate cylinder of radius $\delta$ around the $z$-axis which is contained in the domain $O$ of the coordinate chart.
%For $\tau$ small enough that the coordinate ball $B_{\tau}(v)$ is contained in $O$, define:
%\be
%C_{\delta,\tau} := C_{\delta}\cap B_{\tau}(v)
%\label{defcyldt}
%\ee
%Define:
%\ba
%U_J:= W_{2\alpha_{J,\e}} \cap (\Sigma - C_{\delta=\e^m, \tau=3\e})

Let $C_{\e^m}, m>>1$ be the cylinder of radius $\e^m$ around the $z$-axis (more precisely, the part of the cylinder in the domain $O$ of the coordinate chart). Set $U_J= W_J \cap (\Sigma - {\bar C}_{\e^m})$ i.e. $U_J$ is obtained by removing a small (closed) cylindrical neighborhood of the $z$ axis from $W_J$.
The GR nature of $v$ then ensures that for sufficiently small $\e$, 
\be
e_I\cap U_J = \emptyset = U_J\cap U_K, J\neq K.
\label{wedgeinter}
\ee
Let $V_J \subset U_J$ be an open set containing a segment of  ${\bar \phi}_{I,\e} (e_J)$ which is away from its end points and which does not intersect $e_J$.
Let $f_J$ be a semianalytic function compactly supported in ${\bar V}_J$. Similar to (\ref{a4.04}) let $f_J$ take values in $[0,1]$ with $f_J= 1$ in the vicinity of a neighborhood $V^{\prime}_J$ of  this segment of 
${\bar \phi}_{I,\e} (e_J)$,  $V^{\prime}_J \subset V_J$.
Let $v_J^{\mu}$ be a coordinate vector field transverse to ${\bar \phi}_{I,\e} (e_J)$. 
Set $V\equiv V_J, f^{\mu}\equiv f_J^{\mu}= f_Jv_J^{\mu}$ in (\ref{a4.03}) to obtain $\psi_{\lambda}:= \psi_{J,\lambda}$.  With appropriate choices of the segment to be moved, of the open sets $V_J,V^{\prime}_J$, of the parameter
$\lambda = \lambda_J$,
of the iteration number $N_{\lambda_J}$ of $\psi_{J,\lambda}$, and of  $v^{\mu}$,
%and choosing $V_J,V^{\prime}_J,\lambda = \lambda_J$ appropriately
%and iterating $\psi_{J,\lambda_J}$ an appropriate number $N_{\lambda_J}$ times, 
the area of the loop ${\bar l}_{IJ,\e}$ can be decreased or increased in a controlled manner. Combining the action of several 
such diffeomorphisms,  the RNC measured area of the loop ${\bar l}_{IJ,\e}$ can be adjusted to its desired value. Denote the resulting diffeomorphism by $\psi_J$.

%By further acting with enough iterations, $N_\lambda := N_{J,\lambda}$ of suitably defined 
%diffeomorphisms  $\psi_{\lambda}:= \psi_{J,\lambda}$ with appropriate choices of $V\equiv V_J, f^{\mu}\equiv f_J^{\mu}$ in (\ref{a4.02}) 
%which translate segments of $\phi_\e (e_J)$ either toward or away from $e_I$,  we may adjust the RNC measured area of each $l_{IJ,\e}$  to be of the desired magnitude. More in detail, by virtue of the GR nature of $v$, $V_J$ may be chosen as 
%an appropriate part of an angular 
%`wedge' around the $z$-axis containing the segment to be moved and $f_J^{\mu}$ may be chosen as $f_J v^{\mu}$ where $v^{mu}$ is a suitably chosen coordinate vector field transverse to $\phi_\e (e_J)$  and $f_J$ is a suitably chosen semianlytic function
%compactly supported in $V_J$ away from the $z$- axis so that  $e_I\cap V_J = \empty = V_J\cap V_K, J\neq K$.

It follows that ${\bar \phi}_{I,c_I,\e}$ may be constructed as ${\bar \phi}_{I,c_I, \e} =   \left(\prod_{J\neq I} \psi_{J}\right)         {\bar \phi}_{I,\e}    $
where the ordering in the product is irrelevant due to the choice of non-intersecting $\{U_J, J\neq I\}$.

\subsection{\label{seca4.1} Construction of $l_{IJ,\e}$ as  diffeomorphic images of each other}

Given a construction of $l_{IJ,\e_0}$ through a definition of $\phi_{I,\e_0}$ we seek, in this section, to construct $l_{IJ,\e}$ for $\e<\e_0$ as diffeomorphic images of $l_{IJ,\e_0}$ which map the set of undeformed 
edges $\{e_J\}, J=1,..,N$ at $v$ to themselves.  We assume that for all but 3 of these edges, the kinks at the intersection of these edges with their deformed counterparts have been smoothened using the 
the constructions of section \ref{seca2.2}.
Clearly by choosing $f^{\mu}:= -f_I \delta^3_{\mu}$ with $f_I$  supported in the narrow cylindrical neighborhood $C_{\e^m}$ (see section \ref{sec4.0b}), we may 
move $v_{I,\e_0}$  to its desired position $v_{I,\e}$ through a suitable number of iterations of the resulting $\psi_{\lambda}$ for appropriately chosen $\lambda$. The area of the loop $l_{IJ,\e_0}$ is affected only
to $O(\e^{m+1})$. 

Next, recall that 3 of the edges have kinks. Let $e_{J_i}, i=1,2,3$ be these edges. Let ${\tilde v}_{J_i}$ be the kink on $e_{J_i}$ located at the intersection of $e_{J_i}$ and its deformed image ${\tilde e}_{J_i}$.
Using a similar construction as in the paragraph before this, by choosing a small cylindrical neighborhood of each $e_J$ we may move the kink ${\tilde v}_{J_i}$ to its desired position close to $v$ with an arbitrarily small
change in the area of $l_{IJ_i,\e_0}$. Finally, we may decrease the area of the resulting loops  as well as the loops $\{l_{IK,\e_0}, K \neq J_1,J_2, J_3\}$ to their desired values of $O(\e^2)$  
through wedge supported constructions of diffeomorphisms similar  to that of the previous section.

\subsection{\label{seca4.2}Cone Stiffening through diffeomorphisms}
Given a construction of upward conical edge tangents through the $C^r$ join techniques of section \ref{seca2.3} at parameter $\e=\e_0$ we show, in this section, how to obtain
the required `stiffening' of this configuration at any $\e <\e_0$ by the action of diffeormorphisms.
As noted in section \ref{seca4.1}, the vertex $v_{I,\e_0}$ can be moved to its desired position $v_{I,\e}$ by the action of appropriate diffeomorphisms. Since these diffeomorphisms act as rigid translations
in the vicinity of $v_{I,\e_0}$, the upward conicality of the edge tangents is left unaffected.
Using the same coordinate patch as in that section, the $I$th edge emanating from  $v_{I,\e}$ is along the $z$-axis.
For small enough $\tau <<\e$, set $V$ in (\ref{a4.02}) to be $V:=B_{\tau}(v_{I,\e}) \subset O$, where  $B_{\tau }(v_{I,\e})$ denotes the  coordinate ball of radius $\tau$ centered at $v_{I,\e}$ in the coordinates $\{x\}$.
Let $f$ be supported in ${\bar V}$ with $0\leq f \leq 1$ similar to section \ref{sec4.0b} with $f=1$ in a small neighborhood 
$B_{\frac{\tau}{2} }(v_{I,\e})$ of 
$v_{I,\e}$. Set $f^{\mu}= f G^{\mu}{}_{\nu}x^{\nu}$ where the matrix $G$  is defined as
\ba
G^{\mu}{}_{\nu} &=& 0 \;{\rm for}\;\;\nu \notin \{1,2\}\\
&=& -\delta^{\mu}_{j}\;{\rm for} \;\;j \in \{1,2\}. \\
\ea
It is easily  checked that for $p\in B_{\frac{\tau}{2} }(v_{I,\e})$  we have
\be
x(\psi_{\lambda}(p)) = x(p) (1- \lambda ), \;\; 
y(\psi_{\lambda}(p)) = y(p) (1- \lambda), \;\; 
z(\psi_{\lambda}(p)) = z(p). \;\; 
\ee
Note that $\psi_{\lambda} (v_{I,\e}) = v_{I,\e}$ and that $\psi_{\lambda}$ maps $e_I$ to itself.
By choosing $\lambda$ appropriately and iterating the transformation an appropriate number of times, we obtain the desired stiffening of the edge tangents at $v_{I,\e}$ such that the metrically unit edge
tangents at $v_{I,\e}$ satisfy the desired conditions described in section \ref{sec2.3.3}.
%Note the resulting diffeomorphism is identity on the edge $e_I$.
 
\end{document}